%% file: thesis.tex
\documentclass[11pt,a4paper]{book}

\usepackage[T1]{fontenc}
\usepackage{color}
\usepackage{pstricks}
\usepackage{amssymb}
\usepackage{graphicx}        % including PostScript
\usepackage{bbm}
\usepackage{longtable}
\usepackage[small,loose]{subfigure}
\usepackage[small]{caption} % font of captions is different from text
\usepackage{cite}
\usepackage{pifont}
\usepackage{paralist}
\usepackage{latexsym}
\usepackage[centertags]{amsmath}
\usepackage{amsfonts}
\usepackage{amscd}
\usepackage{epsfig}
\usepackage{a4wide}
\usepackage{makeidx}
\usepackage{scrtime}
\usepackage{epic}
\usepackage{eepic}
\usepackage{pst-all}
\usepackage{longtable}
\usepackage{cancel}
\usepackage{lscape}
\usepackage{colortbl}  % adding color to the tables
\usepackage{booktabs}
\usepackage{accents}
\usepackage{cancel}    % to cross (x) text and equations, e.g. \xcancel{bla}
\usepackage{psfrag}    % to substitute text thru eqs. in eps-files
\usepackage{fancyhdr}
\setcaptionwidth{0.9\textwidth}

\definecolor{red}{rgb}{0.9,0,0}
\definecolor{green}{rgb}{0,0.5,0.1}
\definecolor{blue}{rgb}{0,0,0.6}
\definecolor{black}{rgb}{0,0,0}

\DeclareMathOperator{\re}{Re} 
\DeclareMathOperator{\tr}{tr} \DeclareMathOperator{\diag}{diag}

\newcommand{\CenterEps}[2][1]{\ensuremath{\vcenter{\centerline{\hbox{\includegraphics[scale=#1]{#2.eps}}}}}}
\newcommand{\I}{\mathrm{i}}
\newcommand{\BmL}{\ensuremath{B\!-\!L} }

\newcommand{\gut}{{\sc gut}}
\newcommand{\susy}{{\sc susy}}
\newcommand{\sm}{{\sc sm}}
\newcommand{\mssm}{{\sc mssm}}
\newcommand{\vev}{{\sc vev}}
\newcommand{\Gsm}{\ensuremath{G_\mathrm{SM}}}
\newcommand{\E}[1]{\ensuremath{\mathrm{E}_{#1}}} % e.g. \E{8}
\newcommand{\G}[1]{\ensuremath{\mathrm{G}_{#1}}}
\newcommand{\SO}[1]{\ensuremath{\mathrm{SO}(#1)}}
\newcommand{\SU}[1]{\ensuremath{\mathrm{SU}(#1)}}
\newcommand{\U}[1]{\ensuremath{\mathrm{U}(#1)}}
\newcommand{\T}[1]{\ensuremath{\text{#1}}}
\newcommand{\Z}[1]{\ensuremath{\mathbbm{Z}_{#1}}} % Z_N ->\Z{N}
\newcommand{\ZZ}[2]{\ensuremath{\mathbbm{Z}_{#1}\times\mathbbm{Z}_{#2}}} % Z_NxZ_M ->\ZZ{N}{M}

\newcommand{\bs}[1]{\ensuremath{\boldsymbol{#1}}}
\newcommand{\bsb}[1]{\ensuremath{\boldsymbol{\overline{#1}}}}
\newcommand{\cellgrayh}{\cellcolor[gray]{.75}}
\newcommand{\rowgrayh}{\rowcolor[gray]{.75}}

\if@thdraft
\fi

\renewcommand{\d}{\ensuremath{\partial}}

\newcommand{\ket}[1]{|#1\rangle}

\newcommand{\vevof}[1]{\langle #1\rangle}
\newcommand{\lvevof}[1]{\left\langle #1\right\rangle}
\newcommand{\ol}[1]{\overline{#1}}
\newcommand{\ti}[1]{\ensuremath{\widetilde{#1}}}

\newcommand{\h}[2]{\ensuremath{\frac{#1}{#2}}}
\newcommand{\hh}[2]{\ensuremath{\tfrac{#1}{#2}}}
\newcommand{\maR}{\ensuremath{\mathbbm{R}} }
\newcommand{\maZ}{\ensuremath{\mathbbm{Z}} }

\newcommand{\maN}{\ensuremath{\mathcal{N}} }
\newcommand{\maG}{\ensuremath{\mathcal{G}} }
\newcommand{\maO}{\ensuremath{\mathcal{O}} }
\newcommand{\maH}{\ensuremath{\mathcal{H}} }
\newcommand{\calZ}{\ensuremath{\mathcal{Z}} }
\newcommand{\calW}{\ensuremath{\mathcal{W}} }
\newcommand{\calM}{\ensuremath{\mathcal{M}} }
\newcommand{\calA}{\ensuremath{\mathcal{A}} }

\newcommand{\Ra}{\ensuremath{\Rightarrow} }

\newcommand{\al}{\ensuremath{\alpha}}

\newcommand{\depa}[2]{\ensuremath{\frac{\d #1}{\d #2}}}

\newcommand{\id}{\mathbbm{1}}
\newcommand{\tfr}{\ensuremath{\tfrac{1}{2}}}
\newcommand{\2}{\ensuremath{\frac{1}{2}}}
\newcommand{\3}{\ensuremath{\frac{1}{3}}}
\newcommand{\4}{\ensuremath{\frac{1}{4}}}
\newcommand{\6}{\ensuremath{\frac{1}{6}}}
\newcommand{\8}{\ensuremath{\frac{1}{8}}}
\newcommand{\x}{\ensuremath{\times}}

\newcommand{\be}{\begin{equation}}
\newcommand{\ee}{\end{equation}}
\newcommand{\bi}{\begin{itemize}}
\newcommand{\ei}{\end{itemize}}
\newcommand{\ba}{\begin{array}}
\newcommand{\ea}{\end{array}}
\newcommand{\bea}{\begin{eqnarray}}
\newcommand{\eea}{\end{eqnarray}}
\newcommand{\bse}{\begin{subequations}}
\newcommand{\ese}{\end{subequations}}
\newcommand{\bex}{\begin{example}}
\newcommand{\eex}{\end{example}}
\newcommand{\bnr}{\begin{numremark}}
\newcommand{\enr}{\end{numremark}}

\newcommand{\rojo}[1]{\textcolor{red}{#1}}
\newcommand{\azul}[1]{\textcolor{blue}{#1}}
\newcommand{\verde}[1]{\textcolor{green}{#1}}
\newcommand{\negro}[1]{\textcolor{black}{#1}}

\makeindex

\begin{document}
\pagestyle{empty}

\phantom{.}
\vskip 1.5cm
\begin{center}
{\Large \bf Towards Low Energy Physics from the Heterotic String}
\footnote{Based on the Ph.D.~thesis of the author.}\\
\vskip 1.5cm
Sa\'ul Ramos-S\'anchez\\[1.5cm]

{\it Bethe Center for Theoretical Physics and}\\[-0.05cm]
{\it Physikalisches Institut der Universit\"at Bonn,}\\[-0.05cm]
{\it Nussallee 12, 53115 Bonn, Germany}\\[0.15cm]
\vskip 1.5cm
{\tt  ramos@th.physik.uni-bonn.de} 
\end{center}

\vskip 2cm

\begin{center} {\bf Abstract } \end{center}
\vskip 1cm

\noindent
We investigate orbifold compactifications of the heterotic string, addressing
in detail their construction, classification and phenomenological potential.
We present a strategy to search for models resembling the minimal supersymmetric
extension of the standard model (MSSM) in \Z6--II orbifold compactifications.
We find several MSSM candidates with the gauge group and the exact spectrum of
the MSSM, and supersymmetric vacua below the compactification scale.
They also exhibit the following realistic features: $R$-parity,
seesaw suppressed neutrino masses, and intermediate scale of supersymmetry breakdown.
In addition, we find that similar models also exist in other \Z{N} orbifolds
and in the \SO{32} heterotic theory.

\clearpage

\newpage
\phantom{23}
\vspace{16cm}
Angefertigt mit Genehmigung der Mathematisch--Naturwissenschaftlichen Fakult\"at der Universit\"at Bonn

\vspace{2cm}
\begin{tabular}{ll}
Referent:     & Prof.~Dr.~Hans-Peter Nilles\\
Koreferent:   & Prof.~Dr.~Albrecht Klemm\\
Tag der Promotion: & 24. Juni 2008
\end{tabular}

\newpage
\hspace{8cm}\begin{minipage}[h]{8cm}{\it F\"ur die Frauen meines Lebens:\\\\\phantom{.}\quad\qquad\qquad Virginia und Adriana}\end{minipage}

%%%%%%%%%%%%%%%%%%%%%%%%%%%%%%%%%%%%%%%%%%%%%%%%%%%%%%%%%%%%%%%%%%%%%%%%%%
%  Danksagung
%%%%%%%%%%%%%%%%%%%%%%%%%%%%%%%%%%%%%%%%%%%%%%%%%%%%%%%%%%%%%%%%%%%%%%%%%%
\newpage
\phantom{23}
\vspace{12cm}
Mein Dank geb\"uhrt vor allem Prof.~Hans-Peter Nilles f\"ur die Aufnahme in seine Arbeitsgruppe, seine
gute Bretreuung, und die M\"oglichkeit, auf einem hochinteressanten, herausfordernden
Forschungsgebiet meine Doktorarbeit schreiben zu k\"onnen. Ich bedanke mich bei den Mitgliedern
meines Promotionsausschusses, Prof. Frank Bertoldi, Prof. Klaus Desch und Prof. Albrecht Klemm.
Ich danke auch Dr.~Doris Thrun und Frau Petra
Weiss der Bonner Internationalen Graduiertenschule (BIGS) f\"ur die Unterst\"utzung, die ich st\"andig
von ihnen bekam. Ich danke den Mitgliedern der Forschungsgruppe von Prof.~Nilles f\"ur die
hilfreichen Diskussionen, von denen ich viel gelernt habe, und die angenehme Arbeitsatmosph\"are. Au\ss erdem 
bedanke ich mich bei Takeshi Araki, Prof.~Wilfried Buchm\"uller, Dr.~Kang-Sin Choi, Prof.~Stefan
F\"orste, Dr.~David Grellscheid, Prof.~Koichi Hamaguchi, Dr.~Mark Hillenbach, Prof.~Tatsuo Kobayashi,
Prof.~Jisuke Kubo, Dr.~Oleg Lebedev, Prof. Oscar Loaiza Brito, Dr. Andrei Micu,
Prof.~Michael Ratz, Prof.~Stuart Raby, Dr.~Gianmassimo Tasinato, Patrick Vaudrevange und
Dr.~Ak{\i}n Wingerter, f\"ur das gute Zusammenarbeiten und die Beantwortung meiner zahlreichen
Fragen. Ein herzlicher Dank gilt Michael Ratz und Patrick Vaudrevange f\"ur aufschlussreiche Diskussionen
und das Korrekturlesen dieser Arbeit. 

Schlie\ss lich danke ich herzlichst meiner Mutter Virginia S\'anchez und meiner Frau Adria\-na Vergara Gonz\'alez
f\"ur ihre unerm\"udliche Liebe und Unterst\"utzung, ohne die ich nie der Mensch geworden w\"are, der ich bin.

\frontmatter
\pagestyle{fancyplain}

\tableofcontents
\clearpage

\mainmatter

%%%%%%%%%%%%%%%%%%%%%%%%%%%%%%%%%%%%%%%%%%%%%%%%%%%%%%%%%%%%%%%%%%%%%%%%%%
%  Introduction
%%%%%%%%%%%%%%%%%%%%%%%%%%%%%%%%%%%%%%%%%%%%%%%%%%%%%%%%%%%%%%%%%%%%%%%%%%
\chapter{Introduction}
\label{ch:intro}

Almost four centuries ago, Newton's theory of gravity transformed our understanding of Nature. Newton's
idea seems today very simple: the force that makes an apple fall from a tree on the Earth is exactly the
same that describes the movement of the planets around the Sun. All at once, Newton unified the
natural laws on the Earth with those of the cosmos. It was the first time that someone found out that two
phenomena, apparently so different, have indeed a single origin. However, this would not be the last
time. By the end of the 19th century, Maxwell found out that electricity and magnetism are affections
of the same fundamental force. Furthermore, the success of the electroweak theory, a
model that unifies electromagnetic and weak interactions, appears to indicate as well that most of the
phenomena in Nature could have a universal explanation. 

One of the current goals of theoretical physics is to formulate a theory which explains all
observed forces simultaneously. In this sense, the Standard Model ({\sc sm}) of particle
physics~\cite{Glashow:1961tr,Salam:1964ry,Weinberg:1967tq} is one of the major breakthroughs in
physics of the last century. Including three of the four known fundamental forces 
through local $\SU3_c\times\SU2_L\times\U1_Y$ gauge invariance, the {\sc sm} describes with
great precision the interactions between particles at currently probed energies ($\sim100$ GeV). It 
also predicts the existence of an \SU2 doublet, called Higgs boson, which gives masses to all quarks and
leptons once it acquires a vacuum expectation value (\vev). Although the Higgs boson is
still to be discovered, there are good reasons to believe that this will occur at the Large Hadron
Collider (LHC).  

Despite its predictive power, from a theoretical point of view, the \sm\ leaves still some questions unanswered,
such as the stability of the electroweak scale ({\it{hierarchy problem}}), the large number of free 
parameters, the source of the accelerated expansion of the universe ({\it dark energy}), the origin of
the observed repetition of families and, most importantly from the standpoint of unification, it does not
offer a quantum description of gravity.
These issues indicate that the \sm\ is not a fundamental theory, but rather an effective limit of
more general physics at higher energies. Thus, it results imperative to investigate physics beyond the
\sm.  

An appealing extension of the \sm\ is obtained by including a symmetry between bosons and
fermions, known as supersymmetry (\susy)~\cite{Golfand:1971iw,Volkov:1972jx,Wess:1974tw,Nilles:1983ge}. 
\susy\ explains elegantly how a reasonable Higgs mass can be protected from (quadratically divergent)
quantum corrections without fine-tuning the parameters of the theory. Therefore, the Higgs mass remains of
the order of the electroweak scale, ensuring the stability of this scale even if the supersymmetric
theory is valid up to very high energies. Unfortunately, just including \susy\ in the \sm\ is not enough
to obtain a consistent theory as this would yield unwanted baryon ($B$) and lepton ($L$) number violating
interactions such as $q_i\ell_j\bar{d}_k$ and $\bar{u}_i\bar{u}_j\bar{d}_k$, which combined lead to rapid
proton decay. A solution to this problem is demanding the existence of symmetries that do not commute
with supersymmetry, the so-called {\it $R$-symmetries}. The minimal supersymmetric extension of the \sm\
is the \mssm, in which unwanted interactions are suppressed thanks to the additional $R$-parity
\begin{equation}
R_P~=~(-1)^{3(B-L)+2S}\,,
\end{equation}
where $S$ denotes spin. Despite its qualities, \susy\ introduces new particles
associated to those already known. The so-called superpartners of the \sm\ particles differ only by their
spin, so that the superpartner of a fermion is a boson and vice versa. Since no superpartner has been
detected so far, \susy\ must be broken. Yet one can argue that its breakdown occurs in a fashion such
that some of the properties of \susy\ influence low-energy physics.

There exist good reasons to think that all fundamental forces accept a unified description. To mention one,
the running of the couplings and the symmetries of the particle content of the \sm\ suggest a unified
picture of strong and electroweak interactions through grand unified theories
({\gut}s)~\cite{Georgi:1974sy,Pati:1974yy}. The fundamental feature of these theories is that, at a
higher scale $M_{GUT}$, all gauge interactions of the \sm\ are gathered together in a single
and bigger gauge group, such as \SU5, \SO{10} or \E6. Even though this idea is very appealing, the 
renormalization group equations of the couplings in the \sm\ lead to a picture where at most two of 
them can be unified consistently. This situation is greatly improved if physics 
between the electroweak scale and $M_{GUT}$ include \susy. In the \mssm, all gauge couplings do 
meet at $M_{GUT}\sim2\x10^{16}$ GeV, stressing the key role that \susy\ may play in physics beyond the \sm.

One particularly interesting {\gut} is the \SO{10} unified model~\cite{Georgi:1975qb,Fritzsch:1974nn}, in
which one generation of matter is accommodated in a single spinor representation, according to 
\begin{equation}
\ba{rl}
\bs{16}~=& (\bs{3}, \bs{2})_{1/6} + (\bsb{3}, \bs{1})_{-2/3}+(\bsb{3}, \bs{1})_{1/3}+  
(\bs{1}, \bs{2})_{-1/2} + (\bs{1}, \bs{1})_{1}+ (\bs{1}, \bs{1})_{0} \;,\\
&\mbox{\scriptsize${\qquad q\qquad\qquad\qquad \overline{u}\qquad\qquad\qquad \overline{d}
  \qquad\qquad\quad\ \  \ell\qquad\qquad\qquad \overline{e}\qquad\qquad\ \overline{\nu}}$}
\ea
\end{equation}
where quantum numbers with respect to $\SU3_c\times\SU2_L$ are shown in parentheses and the subscript
denotes hypercharge. A remarkable prediction of this theory is the existence of right--handed neutrinos,
which were not expected in the \sm\footnote{Right--handed neutrinos, however, can be naturally embedded
in the \sm.} and can be used to explain left--handed neutrinos with mass. Moreover, the embedding of
the hypercharge in  $\SU5\subset\SO{10}$ predicts the weak mixing angle $\theta_w$ and provides thereby
an explanation of the electric charge quantization. 

Beside their attractive properties, {\gut}s introduce some problems of their own and leave some questions
unsolved. A puzzling feature is that, while matter generations are described by complete \gut\
representations, Higgs and gauge bosons appear only as incomplete or {\it split} \gut\
multiplets. This is known as the {\it doublet-triplet splitting problem} and is present in all
interesting {\gut}s. Other issues include questions like why there are three families in the \sm, why their mass
eigenstates mix as they do, what the explanation for dark energy is, are still riddles that await their
resolution in these scenarios. Some proposals such as incorporating additional discrete (family) symmetries
might answer some of these questions. However, we have still to deal with the fact
that gravity does not admit a description by {\gut}s.

Unification of gravity with the rest of the fundamental forces into a single theory led Kaluza and
Klein to introduce a fifth spatial dimension compactified on a minute
circle~\cite{Kaluza:1921xx,Klein:1926xx}. Their proposal consisted in extending general relativity to a
five-dimensional spacetime. The resulting theory contains surprisingly a set of equations equivalent to
those of general relativity, and another set equivalent to Maxwell's equations. That the fifth dimension
escapes to our observations was justified by arguing that it can be compact and very small. 
If the fifth dimension is compactified, then there must be in addition to the observed particles
an infinite set of heavy particles (modes) which build the so-called {\it Kaluza-Klein
tower}. Despite its beauty, this early attempt revealed soon not to be appropriate for 
the unified description of gravity and electromagnetism, since the resulting theory cannot be quantized. 
Therefore, with the growing success of quantum mechanics, the interest in Kaluza-Klein compactifications
receded considerably. However, this idea came again to life several years later in a theory that, with
its evolution, turned into a good candidate to unify consistently all known forces: string theory. 

\subsection*{String Theory}

String theory arose by attempting to describe the strong interactions, but, once it was noted that string
theory includes a spin 2 particle corresponding to the graviton, it became clear that its purpose was very 
different.

\begin{figure}[h!]
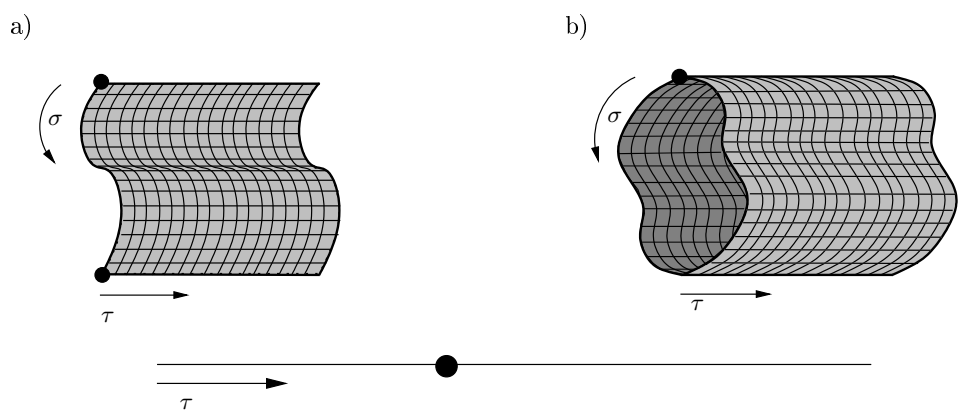

\centerline{\input openclose.pstex_t}
\caption{The worldline (with time-like coordinate $\tau$) of a point particle is replaced by the 2D
  worldsheet (with coordinates $\sigma$ and $\tau$ swept by a) open and b) closed strings.} 
\label{fig:openclose}
\end{figure}

String theory is a perturbative quantum theory in which ordinary point particles are replaced
by one-dimensional objects, whose various vibrational modes at the string scale $M_\mathrm{str}$ can be
identified at low energies with different particles. These extended objects, named strings, cover a two
dimensional space called {\it worldsheet}, in which many of its properties acquire a description through
conformal field theory. Depending on their boundary conditions, they can be closed or open (see
fig.~\ref{fig:openclose}). 

Not only does string theory
contain the graviton as one of the vibration modes of the strings, but  it indeed reduces to Einstein's
theory of gravity at low energies and, due to the extended nature of the string theoretical graviton,
this theory also avoids the ultraviolet divergences of graviton scattering amplitudes. Therefore, a
quantized description of gravity is possible in string theory~\cite{Yoneya:1973ca,Yoneya:1974jg,Scherk:1974ca}.

Several constraints are inherent to the quantum nature of strings. For instance, requiring that quantum
anomalies do not spoil Lorentz invariance of the theory constrains the dimension of the spacetime in which
the strings can consistently propagate. This is a striking theoretical achievement because no theory
before offered a prediction about the dimensionality that our spacetime must have. At the same time, this
poses a major challenge since no consistent string theory describes a
spacetime with four dimensions like the one that is so familiar to us.

Historically, the first string theory discovered was the bosonic string, that is consistent in 26
dimensions. This theory was immediately discarded for it contains unphysical particles with imaginary
rest mass (negative square mass) called tachyons. Furthermore, this theory clearly cannot yield a
description of our universe because the particles composing the observed matter are fermions.

\begin{figure}[t]
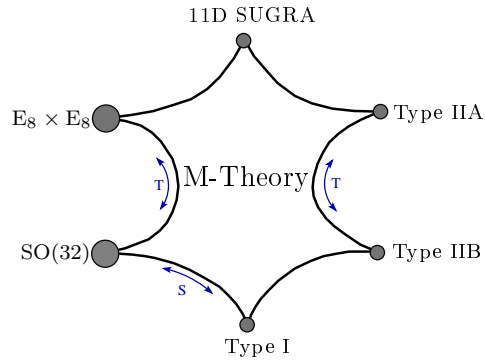

\centerline{\input Duality.pstex_t}
\caption{(Super)string theories are connected by a web of dualities.}
\label{fig:StringDualities}
\end{figure}

Consistent tachyonic-free string theories require (local worldsheet) supersymmetry (\susy)  at very high
energies~\cite{Gliozzi:1976qd,Brink:1976bc} and predict a ten-dimensional spacetime. There exist only
five consistent (super)string theories: type IIA, type IIB, type I, and the \E8\x\E8 and \SO{32}
heterotic theories. Both type II theories present  $\maN=2$ \susy\ whereas the other string theories have
$\maN=1$ in ten dimensions. These theories are connected by a web of (conformal) dualities and thought of as
different limits of an underlying 11-dimensional theory (M-theory)~\cite{Schwarz:1995jq,Townsend:1996xj},
as we depict in fig.~\ref{fig:StringDualities}.

The type I and the heterotic string theories are attached to a remarkable discovery,
namely that the \SO{32} gauge symmetry of the type I and \SO{32} heterotic string as well as the \E8\x\E8
gauge group of the other heterotic theory~\cite{Gross:1984dd,Gross:1985fr} follow from anomaly
cancellation~\cite{Green:1984sg,Green:1984ed}. In contrast to these theories, both type II string
theories are not automatically endowed with gauge groups. This is why most of the early works on string theory
were focused on the \E8\x\E8 heterotic string which seemed from the very beginning the most promising
candidate to be a theory capable to describe physics at low energies. It was only after the discovery of
extended higher dimensional objects called
D--branes~\cite{Polchinski:1995mt,Polchinski:1996na,Polchinski:1998rr} that type II theories regained
interest.

Ever since the discovery of the five consistent string theories, one of the most important tasks of
string theorists has been to make contact with reality. In fact, this is the primary motivation of this
thesis. There are, unfortunately, many aspects of string theory that make difficult to believe that it
has something to do with the observed universe. Apart from the fact that the \sm\ gauge group does not
appear automatically in these theories, the most disturbing feature is that all superstring theories
predict a ten-dimensional spacetime. Happily, the old idea of Kaluza and Klein can be adapted effectively
in string theories to obtain a consistent reduction from ten to four dimensions. 

\subsection*{String Compactifications}

Consistent string theories are formulated in ten dimensions. If string theory has anything to do with the
observable universe, it must provide us with a mechanism to `hide' the six additional dimensions. Such a
mechanism must guarantee that these dimensions become compact and small enough to escape experimental
detection.
Schemes proposed that render four-dimensional theories include Kaluza-Klein (KK)
compactifications and D--brane worlds~\cite{Uranga:2002pg}. Let us spend some words on their general
properties.

\index{Kaluza-Klein}
$\bullet$ {\it KK compactifications} are a natural extension of the five-dimensional KK approach. One
considers the ten dimensional spacetime $\calM^{10}$ of the string to be a direct product of a four-dimensional
flat (Minkowski) spacetime $\mathbbm{M}^{3,1}$ and some unknown compact manifold $\calM^6$, i.e.
$\calM^{10}=\mathbbm{M}^{3,1}\x\calM^6$.
Further, one assumes that the metric of the space is block-diagonal, such that
\be\label{eq:metricAnsatz}
ds^2=g_{\mu\nu}^{(4)}(x)\, dx^\mu dx^\nu+ g_{mn}^{(6)}(y)\, dy^m dy^n\,,
\ee
where $g_{\mu\nu}^{(4)}$ is the Minkowski metric and  $g_{mn}^{(6)}$ is the metric of the
compact internal space.

The expansions of the ten-dimensional fields in the modes of the internal manifold $\calM^6$ yield the
theory in four dimensions. As a result of these expansions, an infinite tower of massive states appears in
the lower-dimensional theory. The masses of these four-dimensional states depend inversely on the size of
the extra dimensions. If one chooses the size of the internal manifold to be sufficiently small, the
massive KK states become heavy and thus decouple from the spectrum at low energies.

\index{Calabi-Yau}
Not every six-dimensional manifold is admissible as internal manifold $\calM^6$. In particular, if one
insists on preserving $\maN=1$ \susy\ in the four-dimensional theory, the internal space must have \SU3
holonomy. 
Furthermore, it is necessary to choose the six-dimensional manifold to be Ricci flat, i.e. such that the
Ricci tensor vanishes everywhere. Manifolds with these properties are called {\it Calabi-Yau
manifolds}~\cite{Candelas:1985en}. Compactifications of this type can lead to models that reproduce
the matter spectrum of the \mssm~\cite{Braun:2005nv}.
Unfortunately, in compactifications on Calabi-Yau manifolds, the computation of relevant physical
quantities of the resulting four-dimensional models can be very difficult (if not impossible). 

A good alternative to circumvent this problem is provided by compactifying on orbifolds, which will be
the main focus of this thesis.
{\it Orbifold compactifications}~\cite{Dixon:1985jw,Dixon:1986jc,Ibanez:1986tp,Ibanez:1987xa,Ibanez:1987sn} are very 
similar to Calabi-Yau manifolds in the sense that both of them can lead to supersymmetric
four-dimensional theories. Orbifolds are defined to be the quotient of a six-dimensional torus divided by
a discrete set of its isometries. In comparison to Calabi-Yau manifolds, the advantage of orbifolds is
that these are Riemann flat, with the exception of a finite set of points, where the curvature of the
space concentrates. Therefore, the metric, which  for (almost) all Calabi-Yau
manifolds is still unknown, can be easily computed in orbifolds. Moreover, in these constructions it
is comparatively straightforward to investigate phenomenological properties, such as the low--energy
gauge symmetry, the particle spectrum, the Yukawa couplings and the K\"ahler potential, among others. For
these reasons, orbifold compactifications are a rich and natural source of inspiration for
phenomenological investigations.

\index{brane worlds}
$\bullet$ In {\it D--brane world} scenarios, D--branes play a crucial role.  They are subspaces of the
ten-dimensional spacetime on which open strings can end. This property equip them in general with a
nonabelian gauge symmetry. Filling the space with several stacks of D--branes intersecting at angles in
the type II string theories can reproduce not only the gauge group of the \mssm, but also its matter content.
Matter then lives on a four-dimensional hypersurface while the mediators of gravity propagate in the full
ten dimensions. In that sense, brane world constructions do not really compactify the spacetime. 
A possible disadvantage of these constructions with respect to KK compactifications is that
{\gut}s like \SO{10} or \E6 cannot be realized in these setups. Nevertheless, these constructions have
recently revealed that, even though the global probability of getting something close to the \mssm\ is rater
low~\cite{Gmeiner:2005vz,Douglas:2006xy,Gmeiner:2007zz}, promising models can also be
found~\cite{Gmeiner:2008xq}.  

\subsection*{Realistic Phenomenology}
\index{local guts@local {\gut}s}

In this thesis, we will focus on orbifold compactifications of the heterotic string theory. As it was
very early noticed, it is preferable to consider the heterotic string with \E8\x\E8 gauge group since, on
the one hand, it includes naturally the so-called chain of {\gut}s
\be
\ba{rl}
& G_{SM} \;\subset\; \SU5 \;\subset\; \SO{10} \;\subset\; \E6 \;\subset\; \E8\\
\T{\footnotesize sometimes also labeled as}& 
  {\scriptstyle{\ ``\E3\T{''}\qquad\ \ ``\E4\T{''}\qquad\quad\  ``\E5\T{''}\qquad\ \,\E6\qquad\; \E8}}
\ea
\ee
and, on the other, the presence of spinors like the \bs{16}--plet of \SO{10} is more frequent than in
the \SO{32} theory. This particular fact facilitates enormously the task of getting models with \sm\
generations. 

In spite of its relative simplicity, orbifolds have not been systematically
studied yet. Furthermore, although it is known that models resembling the \mssm\
exist~\cite{Ibanez:1987sn,Casas:1988se,Kobayashi:2004ya,Buchmuller:2005jr,Kim:2006hv}, 
as yet there is no model that accommodates simultaneously all properties of the \mssm\ and
everything suggests that it is complicated to find accidentally such a model. One is thus encouraged to turn
to a strategy that sets a guiding principle through the search for realistic vacua.

In order to get phenomenological viable models from orbifold compactifications,
one can draw on the insight gained from grand unified theories by introducing the concept of
{\it local {\gut}s}~\cite{Nilles:2004ej,Buchmuller:2004hv,Buchmuller:2005jr,Buchmuller:2005sh}.
In scenarios with local {\gut}s, there are special (fixed) points in the
internal space where the gauge symmetry is locally that of certain {\gut}s  while the four-dimensional gauge
symmetry is that of the \sm\ (up to additional gauge factors that compose the hidden sector). If matter
fields are localized at such special points, they form complete \gut\ 
representations. This applies, in particular, to a $\bs{16}$--plet of a local \SO{10}. On the
other hand, bulk fields form incomplete \gut\ multiplets. In the particular case of \SO{10}, from the
four-dimensional viewpoint, the localized states are complete matter generations whereas the bulk fields
can adopt the form of, say, Higgs doublets (an incomplete representation of \SO{10}). This
might offer an intuitive explanation for the observed family structure of the \sm\ and, at the same time, a
solution to the doublet-triplet splitting problem. 

If there are orbifold models that, at least, have the matter spectrum  of the \mssm, addressing their
phenomenological viability is of utmost importance. A number of questions can be posed in this
direction. One challenge is, for instance, to verify whether these models admit supersymmetric vacua,
that is, whether a combination of fields can attain vacuum expectation values ({\vev}s), such that neither
supersymmetry is borken at very high energies nor there appear inadmissible phenomenological features,
such as unknown particles at observable energies. Models with realistic traits would also provide insights
about the unavoidable breaking of supersymmetry, the origin of neutrino masses, the delicate suppression
of the supersymmetric coupling $h_u\,h_d$ and the absence of proton decay, just to mention some issues.

The ultimate goal of physics beyond the \sm\ is still to identify a theory that can reproduce our current
knowledge and improve our understanding of physics. Unification provides doubtless a framework where
physics beyond the \sm\ takes an appealing form. Adopting this idea into more elaborated theories, such
as orbifold compactifications of the heterotic string, can certainly shed light on some of the puzzles of
contemporary science. In the present work, we utilize the beauty and mathematical consistency of string
theory as a tool in order to build a bridge between confirmed or foreseeable physics and
a theory possibly capable to describe all fundamental forces in a unified way.

\subsection*{Overview}

To guide the reader through the present work, let us outline the discussion of the chapters to follow.
\vskip 5mm

{\it Chapter 2.} After a brief introduction to the heterotic string, we proceed to explain the details of
abelian orbifold compactifications of the heterotic string. This chapter intends to be as general as
possible. Hence, we do not focus particularly on any of its two variants, \E8\x\E8 and \SO{32}. We also
consider both \Z{N} and \ZZ{N}{M} orbifolds on the same footing. The effect of the choice of the six-dimensional
compactification lattice as well as the constraints on the orbifold parameters are given.
We present then in all detail how to compute the massless matter spectrum of orbifold models with and
without Wilson lines and illustrate the method with a simple example. Discrete torsion is introduced as
an additional degree of freedom which, contrary to previous claims, can also appear in \Z{N} orbifolds
even in the absence of Wilson lines. We propose an interpretation of models with discrete torsion and a
special type of gauge embeddings. To close the chapter, we provide a succinct discussion about Yukawa
couplings on heterotic orbifolds.

{\it Chapter 3.} Due to the enormous number of redundancies in string constructions, it is necessary to
implement  a useful method to classify these constructions, that is, to determine 
all (or at least a large number of) different parameters leading to inequivalent orbifold models. We discuss in
this chapter two methods. The first one goes by the name of {\it Dynkin diagram strategy} for it makes
extensive use of the properties of the Dynkin diagram of a Lie algebra to determine all admissible gauge
embeddings. The second method of classification resorts to a suitable ansatz describing
all shifts and/or Wilson lines of a given order, minimizing duplicities.
We discuss their advantages and
drawbacks and give examples of their application. Finally, rather than constructing all gauge embeddings
of certain classes of models, one would prefer to know the total number of models. With that purpose, we
introduce a statistical procedure that, in addition, provides us samples of characteristic models. 
Some of the topics discussed in this chapter were
presented in refs.~\cite{Nilles:2006np,Ploger:2007iq,Lebedev:2008un}. 

{\it Chapter 4.} We describe a general strategy we proposed in ref.~\cite{Lebedev:2006kn} to obtain
orbifold models that resemble the \mssm, using as guiding principle the concept of local {\gut}s. Our
search is performed by compactifying the \E8\x\E8 heterotic string on the \Z6-II orbifold, since it has
shown to house some models with realistic properties. After providing the criteria comprising our search
strategy, we analyze the results obtained. We find that our approach, as opposed to a random scan, is
successful and that a considerable fraction of the models with \SO{10} and \E6 local \gut\ structures
posseses promising features. We consider this to be one of the central results of this thesis.

{\it Chapter 5.} 
The study of some aspects of the phenomenology of our \mssm\ candidates is presented in this
chapter, following our previous discussions from refs.~\cite{Lebedev:2006tr,Buchmuller:2007zd,Lebedev:2007hv}.
To illustrate our results, we describe one characteristic model with the exact spectrum
of the \mssm. Particular attention is given to the search of supersymmetric vacua, supersymmetry breaking
proton decay and neutrino masses.

{\it Chapter 6.}
We extend our search for realistic models. In addition to models with three Wilson lines in the \Z6-II 
orbifold, we also analyze the appearance of models with realistic features in other \Z{N} orbifold 
compactifications of both the \E8\x\E8 and the \SO{32} heterotic string theories.

\newpage

%%%%%%%%%%%%%%%%%%%%%%%%%%%%%%%%%%%%%%%%%%%%%%%%%%%%%%%%%%%%%%%%%%%%%%%%%%
%  Main Chapters
%%%%%%%%%%%%%%%%%%%%%%%%%%%%%%%%%%%%%%%%%%%%%%%%%%%%%%%%%%%%%%%%%%%%%%%%%%

\renewcommand{\arraystretch}{1.1}
\clearpage
\input{./Basics}    % Chapter 2
\clearpage
\input{./OrbiClass} % Chapter 3
\clearpage
\input{./OrbiPheno} % Chapter 4 and 5
\clearpage
\input{./Statistics} % Chapter 6
\clearpage
%%%%%%%%%%%%%%%%%%%%%%%%%%%%%%%%%%%%%%%%%%%%%%%%%%%%%%%%%%%%%%%%%%%%%%%%%%
% Conclusions
%%%%%%%%%%%%%%%%%%%%%%%%%%%%%%%%%%%%%%%%%%%%%%%%%%%%%%%%%%%%%%%%%%%%%%%%%%
\chapter{Conclusions and Outlook}

In this thesis we have discussed orbifold compactifications of the heterotic string. We have emphasized their
relation to low--energy phenomenology and, therefore, its viability as a unified theory of all observed
forces. In particular, we have focused on the phenomenology of abelian orbifold compactifications of the
\E8\x\E8 heterotic string theory.

We have briefly described the heterotic string and provided some recipes for the computation of the
direct properties of symmetric abelian orbifold models, such as their fixed points and their matter
spectra. We have also discussed the role that some additional degrees of freedom denominated discrete
torsion play in theses constructions. In concrete, we have seen that discrete torsion alters the
chiral-matter content of the orbifold. We have shown that, contrary to previous statements, 
\Z{N} orbifolds with Wilson lines as well as \ZZ{N}{M} orbifold models admit discrete torsion. Further,
we have found out that the assignments of discrete torsion can be traded for translations of the gauge
embedding parameters (shifts and Wilson lines) in the gauge lattice $\Lambda$.

We have explained and extended some conventional methods employed to classify orbifold models, paying
special attention to the construction of ans\"atze that allow us to obtain systematically all
inequivalent input parameters of orbifold compactifications. The proposed
classification methods are very flexible and can be adapted to both (\E8\x\E8 and \SO{32}) heterotic string theories. 
Our methods have already been tested in all \Z{N} and \ZZ{N}{M} orbifolds. We have made an exhaustive
classification of \Z{N} orbifolds without background fields and also a complete classification of \Z3\x\Z3
orbifold models, including discrete torsion. Whenever possible, we have compared our results to those of
the literature.

To address the question of whether there are realistic orbifold compactifications,
we have analyzed the heterotic \E8\x\E8 string compactified on a \Z6-II orbifold, allowing
for up to two discrete Wilson lines. Using a search strategy based on the concept of local {\gut}s,
we have obtained about $3\x10^4$ inequivalent models. Out of them, 223 models exhibit the
\mssm\ gauge group structure, three light families and vectorlike exotics. We show that all the
vectorlike exotics can decouple without breaking supersymmetry for 190 of these models. This means that
almost 1\,\% out of $3\x10^4$ models have the gauge group and the chiral matter content of the
\mssm. This result shows that orbifold compactifications of   the  heterotic string  correspond to a
particularly fertile region in the landscape  and that the probability of getting something close to the
\mssm\ is  significantly higher than that in other constructions. It would be interesting to extend
these results to other regions of the landscape where  some promising models also exist.

Furthermore, we have found that requiring realistic features in the set of \mssm\ candidates is
correlated with the supersymmetry breaking scale  such that, in the context of gaugino condensation,
intermediate scale ($\sim 1$ TeV) of \susy\ breaking is favored. This occurs because most of the models with
realistic features have a hidden sector with \SU4 or \SO8 nonabelian symmetry. 

In order to get closer to the \mssm, we defined a successful strategy for obtaining models with an exact
$R$-parity based on spontaneous breaking of a gauge $\U1_{\BmL}$ symmetry. We find 87 models which have a
renormalizable top Yukawa coupling. Out of these, we identify 15 models with an exact $R$-parity, no
light exotics or \U1 gauge bosons and an order one top quark Yukawa coupling.
We would like to remark that the number 15 is a lower bound, mainly since our search is based on a specific
strategy related to \BmL symmetry. Additionally, many of our constraints, such as demanding a
renormalizable coupling for the top quark are just artifacts invoked in order to simplify 
our analysis. Therefore, an interesting question would be to obtain similar results in cases where: a) no
trilinear coupling for the top quark is demanded, b) no vectorlikeness of the \sm\ matter generations
with respect to \BmL is imposed, c) also models with anomalous \BmL are admitted, and d) $R$-parity
results directly from the so-called R-charge conservation (string) selection rule.

On the other hand, we notice that dimension five baryon and lepton number violating operators, like
$q\,q\,q\,\ell$, appear in orbifold models quite regularly even if \BmL is imposed. They are either
generated in the superpotential to some order in \sm\ singlets, or they may also be generated when
integrating out heavy exotics. In some cases, only fine-tuning the {\vev}s of some \sm\ singlets can
alleviate this problem. One is thus encouraged to investigate other possible solutions, such as the
identification of some discrete symmetries.

Another aspect of phenomenology we have studied is the seesaw mechanism. Since there are no
upper bounds on the amount of right--handed neutrinos that might appear, we find that in a scenario with
$\maO(100)$ right--handed neutrinos, the seesaw mechanism is realized from orbifold
constructions. Moreover, (left--handed) neutrino masses are then enhanced (in comparison to the na\"ive estimate
of the neutrino masses in {\gut}s) to more realistic values. 
We consider this setup to be plausible and perhaps even desirable, since some consequences of the
existence of many right--handed neutrinos could shed some light in some open issues, such as
leptogenesis~\cite{Eisele:2007ws,Ellis:2007wz}.
We might even consider the abundance of right--handed neutrinos to be a prediction of string theory. 

We have also studied in detail one model that we have called {\it orbifold-\mssm}. The propertiesof this
model are generic in the sense that other \mssm\ candidates possess very similar qualities: only the
exact matter spectrum of the \mssm\ (up to many \sm\ singlet fields); admissible $\BmL$ symmetry which
is spontaneously broken to matter parity in a supersymmetric vacuum configuration; gaugino condensation
scale leading to \susy\ breaking at an intermediate scale; vanishing $\mu$--term; and proton decay
through dimension five operators can be avoided by some fine-tuning.
The top Yukawa coupling is order $\widetilde{s}^2$ 
due to Higgs doublet mixing in this model.  Further, both the up and down quarks are massless at
order six in \sm\ singlets. However, the up quark becomes massive at order seven and the down quark gets a
mass at order eight. These properties are very interesting, but we believe that the best model awaits
still in some corner of the landscape.

Finally, we have extended our search of models with realistic features to \Z6-II orbifold models with
three Wilson lines. Out of a total of $10^7$, we have found almost 300 inequivalent models with the
\mssm\ spectrum and gauge coupling unification. Most of these models originate from local {\gut}s.
We also found that low energy supersymmetry breaking is also favored in these models.
Further, we have explored \Z3, \Z4, \Z6-I and \Z7 orbifolds searching for promising models. We have found
small sets of realistic models in \Z6-I and \Z7 orbifolds. This suggests that a broader scan could also
help to understand how the \mssm\ may be embedded in string theory.

There are some phenomenological issues that we have not addressed in this thesis. In particular, one
issue concerns proton stability. The examples we studied are challenged by the presence of 
dimension five proton decay operators. Their suppression may require additional (discrete)
symmetries or even a change in the geometry of the orbifold, as suggested in appendix~\ref{ch:Z6IINonFactorizable}.
There are also dimension six operators, generated by \gut\ gauge boson exchange, which we
have not discussed.

Further, we have
not studied precision gauge coupling unification. Although hypercharge is normalized as in
four-dimensional {\gut}s thus allowing gauge coupling unification in the first approximation, there are
various corrections that can be important. First, a detailed analysis would require the calculation of
string threshold corrections in the presence of discrete Wilson lines. However in specific cases these
corrections are known to be small~\cite{Mayr:1993kn}. Second, there are corrections from the vectorlike
exotic states. It is possible that precision gauge coupling unification may require anisotropic
compactifications, leading to an effective orbifold \gut.
\cite{Witten:1996mz,Ibanez:1992hc,Kobayashi:2004ya,Hebecker:2004ce}. These questions have been explored
in ref.~\cite{Nilles:2008gq} and refs.~\cite{Dundee:2008ts,Dundee:2008tr}.

Another question that was not clarified in this thesis concerns moduli stabilization. We have seen that,
e.g. in order to determine the scale of gaugino condensation, stabilizing the dilaton $S$ is crucial. 
The minimum of its potential affects strongly the scale of supersymmetry breakdown in the 
observable sector. However, the dilaton is not the sole modulus in the theory. Beside some geometrical moduli,
many singlet fields (with neither {\vev}s nor masses) appear in the matter spectrum of orbifolds. Their stabilization is also 
important for addressing cosmological issues, such as inflation and baryogenesis.

%%%%%%%%%%%%%%%%%%%%%%%%%%%%%%%%%%%%%%%%%%%%%%%%%%%%%%%%%%%%%%%%%%%%%%%%%%
%  Appendices
%%%%%%%%%%%%%%%%%%%%%%%%%%%%%%%%%%%%%%%%%%%%%%%%%%%%%%%%%%%%%%%%%%%%%%%%%%
\newpage
\begin{appendix}
\clearpage
\input{./Appendices}

\clearpage

\end{appendix}

%%%%%%%%%%%%%%%%%%%%%%%%%%%%%%%%%%%%%%%%%%%%%%%%%%%%%%%%%%%%%%%%%%%%%%%%%%
%  Bibliography and Index
%%%%%%%%%%%%%%%%%%%%%%%%%%%%%%%%%%%%%%%%%%%%%%%%%%%%%%%%%%%%%%%%%%%%%%%%%%
\cleardoublepage
{\small
\addcontentsline{toc}{chapter}{Bibliography}
\bibliography{Orbifold}
\bibliographystyle{ArXiv}
}
\clearpage
\addcontentsline{toc}{chapter}{Index}
\printindex

\end{document}

%% file: openclose.pstex_t
\begin{picture}(0,0)%
\includegraphics{openclose.pstex}%
\end{picture}%
\setlength{\unitlength}{3947sp}%
\begingroup\makeatletter\ifx\SetFigFont\undefined%
\gdef\SetFigFont#1#2#3#4#5{%
  \reset@font\fontsize{#1}{#2pt}%
  \fontfamily{#3}\fontseries{#4}\fontshape{#5}%
  \selectfont}%
\fi\endgroup%
\begin{picture}(5968,2506)(221,-2267)
\put(1288,-2236){\makebox(0,0)[lb]{\smash{{\SetFigFont{9}{10.8}{\rmdefault}{\mddefault}{\updefault}$\tau$}}}}
\put(3706,119){\makebox(0,0)[lb]{\smash{{\SetFigFont{10}{12.0}{\rmdefault}{\mddefault}{\updefault}$\text{b)}$}}}}
\put(4497,-1617){\makebox(0,0)[lb]{\smash{{\SetFigFont{9}{10.8}{\rmdefault}{\mddefault}{\updefault}$\tau$}}}}
\put(3974,-431){\makebox(0,0)[lb]{\smash{{\SetFigFont{9}{10.8}{\rmdefault}{\mddefault}{\updefault}$\sigma$}}}}
\put(221,119){\makebox(0,0)[lb]{\smash{{\SetFigFont{10}{12.0}{\rmdefault}{\mddefault}{\updefault}$\text{a)}$}}}}
\put(795,-1690){\makebox(0,0)[lb]{\smash{{\SetFigFont{9}{10.8}{\rmdefault}{\mddefault}{\updefault}$\tau$}}}}
\put(463,-454){\makebox(0,0)[lb]{\smash{{\SetFigFont{9}{10.8}{\rmdefault}{\mddefault}{\updefault}$\sigma$}}}}
\end{picture}%

%% file: Duality.pstex_t
\begin{picture}(0,0)%
\includegraphics{Duality.pstex}%
\end{picture}%
\setlength{\unitlength}{4144sp}%
\begingroup\makeatletter\ifx\SetFigFont\undefined%
\gdef\SetFigFont#1#2#3#4#5{%
  \reset@font\fontsize{#1}{#2pt}%
  \fontfamily{#3}\fontseries{#4}\fontshape{#5}%
  \selectfont}%
\fi\endgroup%
\begin{picture}(2755,2086)(108,-1387)
\put(1132,-358){\makebox(0,0)[lb]{\smash{{\SetFigFont{11}{13.2}{\rmdefault}{\mddefault}{\updefault}$\text{M-Theory}$}}}}
\put(2387, 40){\makebox(0,0)[lb]{\smash{{\SetFigFont{8}{9.6}{\rmdefault}{\mddefault}{\updefault}Type IIA}}}}
\put(2379,-793){\makebox(0,0)[lb]{\smash{{\SetFigFont{8}{9.6}{\rmdefault}{\mddefault}{\updefault}Type IIB}}}}
\put(1378,-1356){\makebox(0,0)[lb]{\smash{{\SetFigFont{8}{9.6}{\rmdefault}{\mddefault}{\updefault}Type I}}}}
\put(163,-799){\makebox(0,0)[lb]{\smash{{\SetFigFont{8}{9.6}{\rmdefault}{\mddefault}{\updefault}\SO{32}}}}}
\put(1165,624){\makebox(0,0)[lb]{\smash{{\SetFigFont{8}{9.6}{\rmdefault}{\mddefault}{\updefault}11D SUGRA}}}}
\put(108,  8){\makebox(0,0)[lb]{\smash{{\SetFigFont{8}{9.6}{\rmdefault}{\mddefault}{\updefault}$\E8\times\E8$}}}}
\end{picture}%

%% file: Basics.tex
\chapter{Orbifold Compactifications}
\label{ch:OrbifoldCompactifications}

\begin{center}
\begin{minipage}[t]{14cm}
In this chapter, we study the heterotic string theory compactified on orbifolds. We start by reviewing
briefly some aspects of the heterotic string. Then we explain a method to get four-dimensional supersymmetric
models based on abelian orbifold compactifications. Our discussion is abstract at some level, but it is
addressed to people willing to get acquainted with orbifold constructions. We also introduce here the
notation to be used along the entire work. 
\end{minipage}
\end{center}

\section{Heterotic String}
\label{sec:HeteroticString}

It is well known that in closed-string theories left- and
right-moving modes are decoupled~\cite{Gross:1984dd,Gross:1985fr}. This offers the possibility of a new
consistent string theory in which left- and right-movers are of different types. The heterotic
string~\index{heterotic string} arises as the result of combining a ten-dimensional right-moving
superstring~\cite{Green:1987sp,Green:1987mn} (ensuring thereby space-time supersymmetry) with a
26-dimensional left-moving bosonic string.

\subsection*{Right-movers}
\begin{figure}[t]
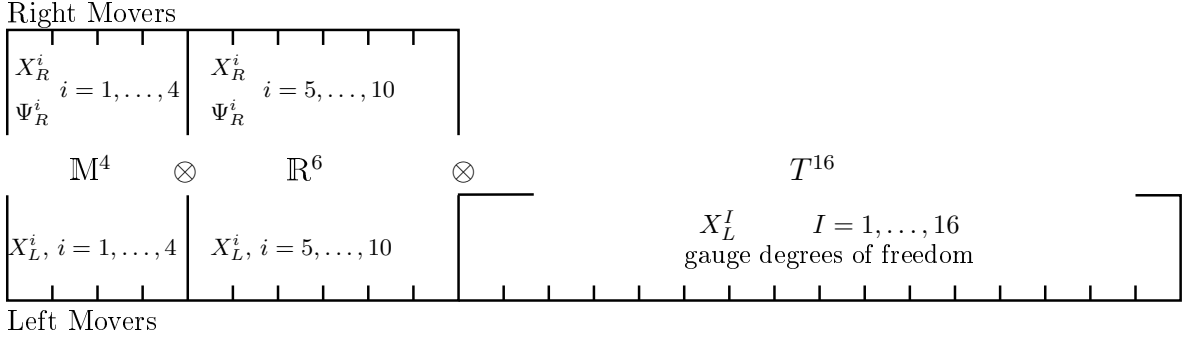

\centerline{\input Heterotic.pstex_t}
\caption{Heterotic string degrees of freedom. The supersymmetric right-movers live in 10 dimensions
  whereas the bosonic left-movers, in 26 dimensions. The 16 additional bosonic coordinates compactified
  on a torus $T^{16}$ give rise to the gauge degrees of freedom. The first four dimensions of the ten dimensional
  heterotic string correspond to the Minkowski space $\mathbbm{M}^4$.} 
\label{fig:Dimensions_in_HeteroticString}
\end{figure}

The right-moving\index{heterotic string!right-movers} bosonic and fermionic degrees of freedom 
of the superstring are denoted by $X^i_R$ and $\Psi^i_R$, respectively, where $i=1,\,\ldots,\,10$. 
We can assume that the first four coordinates correspond to the observed minkowskian spacetime. 
This situation is depicted in fig.~\ref{fig:Dimensions_in_HeteroticString}. Since not all these degrees
of freedom are independent, we choose the light-cone gauge, in which the coordinates corresponding to
$i=1,2$ are fixed.  

The solutions to the motion equations of the heterotic string action are given by the
mode expansions of the bosonic and fermionic degrees of freedom
\begin{eqnarray}
 \label{eq:superstringBoson}
  X_R^i(\tau-\sigma) &=& \frac{1}{2}x^i+\frac{1}{2}p^i(\tau-\sigma)
          +\frac{\I}{2}\sum_{n\neq0}\frac{\alpha_n^i}{n}e^{-2\I n(\tau-\sigma)}\,,\\
  \label{eq:superstringFermionR}
 \Psi_R^i(\tau-\sigma)&=& \sum_{n\in\Z{}}d_n^i e^{-2\I n(\tau-\sigma)} \quad \qquad\mbox{\sc ramond}\,,\\
  \label{eq:superstringFermionNS}
 \Psi_R^i(\tau-\sigma)&=& \sum_{r\in\Z{}+\frac{1}{2}}b_r^i e^{-2\I n(\tau-\sigma)}  \qquad\mbox{\sc neveu-schwarz}\,,
\end{eqnarray}
where the constants $x^i$ and $p^i$ denote the center-of-mass coordinates and momenta, and the
coefficients $\alpha^i,d^i, b^i$ are called {\it oscillators}.
{\sc ramond} and {\sc neveu-schwarz} denote fermionic states with respectively periodic
and antiperiodic boundary conditions, i.e
\begin{eqnarray}
 \Psi^i_R(\tau-(\sigma+\pi))&=&+\Psi^i_R(\tau-\sigma)   \qquad\mbox{\sc ramond}\,,\\
 \Psi^i_R(\tau-(\sigma+\pi))&=&-\Psi^i_R(\tau-\sigma)  \qquad\mbox{\sc neveu-schwarz}\,.
\end{eqnarray}

Right-moving states are then (oscillator) perturbations to the vacuum state $\ket{0}_R$, which is defined
by $b_r\ket{0}_R=d_n\ket{0}_R=0$ for $r,\,n>0$. The masses of these states are given by 
\begin{equation}
 \label{eq:mass_R}
 \h{m_R^2}{4}~=~  N_R - a_R~=~ \left\{
   \begin{array}{lccl}
     \sum_{n=1}^{\infty}\left(\alpha^i_{-n}\alpha^i_n + nd^i_{-n}d^i_n\right) &&&\text{\sc ramond}\,,\\[1mm]
     \sum_{n=1}^{\infty}\alpha^i_{-n}\alpha^i_n + \sum_{r=\frac12}^\infty rb^i_{-r}b^i_r -\2&&&\text{\sc neveu-schwarz}\,,
   \end{array}\right.
\end{equation}
where the constant $a_R$ is called {\it zero point energy} and arises during quantization from the normal-ordering of the
oscillators. In the Ramond sector $a_R=0$ whereas in the Neveu-Schwarz sector $a_R=-\2$.
Note that $N_R$ counts the number of oscillators $\alpha,\,d,\,b$. We are interested in physical states that
are massless at the string scale. From eq.~\eqref{eq:mass_R}, we observe that $m_R=0$ for states with one
oscillator $b_{-1/2}$ acting on the vacuum, $b^i_{-1/2}\ket{0}_R,$ in the Neveu-Schwarz sector, and for
$d^i_0\ket{0}_R$ in the Ramond sector. The 
eight transverse excitations $b^i_{-1/2}\ket{0}_R$ ($i=3,\dots,10$) behave as bosons from the spacetime
perspective and form the vectorial representation \bs{8_v} of \SO8, which is the (transversal) Lorentz
group of the uncompactified space. 

On the other hand, after quantization the oscillators $d^i_0$ obey the (Clifford) algebra 
\be
\left\{\I\sqrt{2}d^i_0,\,\I\sqrt{2}d^j_0\right\} = 2 \eta^{ij}\;,
\ee
the ground state $d^i_0\ket{0}_R$ forms a spinorial representation with 16 real components. In order to
match the number of on-shell fermionic and bosonic degrees of freedom, one has to introduce a {\sc gso}
projection~\cite{Gliozzi:1976qd}, which does not only reduce the number of massless degrees of freedom by
a factor \2, but also ensures an equal number of bosons and fermions at each mass level, as
required by supersymmetry. After the {\sc gso} projection, we are then left with the \bs{8_s}
representation of \SO8 in the Ramond sector.

For convenience, $\ket{q}_R$ will denote the right-moving ground state for both sectors, where $q$ stands
for the weights of the corresponding \SO8 representation in Cartan-Weyl labels
\be
\ket{q}_R ~=~\left\{
\ba{ccl}
\big|\underline{\pm\,1,\quad\,0,\quad\,0,\quad\,0}\big\rangle_R &\sim&\bs{8_v}\;,\qquad\qquad\text{\sc neveu-schwarz}\phantom{a_{A_{A_A}}}\\
\big|\pm\tfr,\,\pm\tfr,\,\pm\tfr,\,\pm\tfr\big\rangle_R&\sim&\bs{8_s}\;,\qquad\qquad\text{\sc ramond}\phantom{a_A^{A^A}}
\ea\right.
\label{eq:masslessright}
\ee
where the spinor representation has an even number of plus signs. The underline denotes permutation of
the entries. In this notation, eq.~\eqref{eq:mass_R} becomes
\be
\h{m_R^2}{4}= \2 q^2 - \2\;.
\label{eq:mass_R2}
\ee

\subsection*{Left-movers}
The coordinates of the bosonic string\index{heterotic string!left-movers} are denoted by
$X_L^i,\,i=1,\,\ldots,\,10$, and $X_L^I,\,I=1,\,\ldots,\,16$. As for the right-movers, the coordinates
corresponding to $i=1,2$ are fixed by the light-cone gauge. The evident mismatch in the number of spatial
dimensions of left- and right-movers is amended by compactifying the coordinates $X_L^I$ on a
16-dimensional torus $T^{16}$ with radii as small as the string scale, as illustrated in
fig.~\ref{fig:Dimensions_in_HeteroticString}. As a result of this compactification, the 16-dimensional
internal momenta $p$ are nonvanishing and proportional to the winding of states in the compactified space. 

The left movers are characterized by the mode expansions
\begin{eqnarray}
 \label{eq:bosonicBosonM}
  X_L^{i}(\tau+\sigma) &=& \frac{1}{2}x^{i}+\frac{1}{2}p^{i}(\tau + \sigma)
          +\frac{\I}{2}\sum_{n\neq0}\frac{\tilde\alpha_n^{i}}{n}e^{-2\I n(\tau+\sigma)}
          \qquad i=3,\dots,10\,,\\
 \label{eq:bosonicBosonI}
  X_L^{I}(\tau+\sigma) &=& \frac{1}{2}x^{I}+\frac{1}{2}p^{I}(\tau + \sigma)
          +\frac{\I}{2}\sum_{n\neq0}\frac{\tilde\alpha_n^{I}}{n}e^{-2\I n(\tau+\sigma)}
          \qquad I=1,\dots,16\,,
\end{eqnarray}
where $\tilde\alpha_n$ are {\it left-moving oscillators}.
As for the right movers, the left-moving states are (oscillator) perturbations to the vacuum $\ket{0}_L$.
The masses of these states after compactifying the 16 internal degrees of freedom $X^I$ are given by
\be
\h{m^2_L}{4} = \2 p^2 + \ti{N} - 1\;,
\label{eq:mass_L}
\ee
where $\ti{N}$ counts left-moving oscillator excitations and $-1$ is the zero point energy of the bosonic
string. 

Gauge and gravitational anomaly cancellation is guaranteed by one-loop modular invariance. It
imposes severe constraints on the theory. In particular, the underlying lattice $\Lambda$
of the 16-dimensional torus $T^{16}$ must be euclidean, even and self-dual. There are only two such
lattices in 16 dimensions: the root lattice of $\E8\times\E8$ and the weight lattice of
Spin(32)/\Z2. Consequently, the (nonabelian) gauge group \maG of rank 16 provided by the compactification can
be either $\E8\times\E8$ or \SO{32}, depending on the choice of $\Lambda$. Each lattice yields an
independent consistent heterotic string theory.

According to eq.~\eqref{eq:mass_L}, at the massless level we have the following left-moving states:
\bse\bea
 \ti\al^i_{-1}\ket{0}_L & &\qquad i=3,\ldots,10\;,\\
 \ti\al^I_{-1}\ket{0}_L & &\qquad I=1,\ldots,16\;,\\
 \ket{p}_L & &\qquad p^2=2.
\eea\label{eq:masslessleft}\ese 
\vskip -5mm
There are 480 internal momenta $p$ fulfilling $p^2=2$. As we will see below, it is not a coincidence that
the adjoint representation of \maG contains 480 charged bosons, too. In fact, the states
$\ket{p}_L$ correspond to the left-moving part of the gauge bosons (and gauginos) of this theory. For that reason, 
we will represent $p$ by the vectors in Cartan-Weyl labels for the corresponding charged bosons
\bse\label{eq:ps}
\bea
\maG=\E8\times\E8:\;\; & &p~\in~\left\{
\ba{l}
\big(\underline{(\pm1)^2,0^6}\big) \big(0^8\big)\;,\;\; \big(0^8\big)\big(\underline{(\pm1)^2,0^6}\big)\;, \\[3mm]
\big((\pm\2)^8\big)\big(0^8\big)\;,\;\;\big(0^8\big)\big((\pm\2)^8\big)\;,\quad\text{even \# of +}\;,
\ea \right.\label{eq:psE8}\\
\maG=\SO{32}:\;\;\;\, & & p~\in~\left(\underline{(\pm1)^2,0^{14}}\right)\;,\label{eq:psO32}
\eea
\ese
where the exponent of an entry counts the number of times that such an entry appears in the 16-dimensional
vector, and the underline stands, as before, for all permutations.

\subsection*{Massless Heterotic Spectrum}
Let us analyze the spectrum of the heterotic string.\index{heterotic string!massless spectrum} Physical
states must fulfill the level matching condition
\be
m_R^2 = m_L^2\;,
\label{eq:levelcond}
\ee
which follows from the variation of the worldsheet metric. This constraint implies that, in contrast to
other string theories, in the heterotic string the {\sc gso} projection is not implemented in order to avoid the
presence of tachyons.\index{heterotic string!tachyons} Since the mass of the lowest energy left-moving
state ($\ti N=0,\;p=0$) is $m_L^2 = -1$ whereas the mass of the right-moving (Neveu-Schwarz) tachyon
($N_R=0$) is $m_R^2=-\2$, then eq.~\eqref{eq:levelcond} enforces the absence of states with
negative mass square in the spectrum of the heterotic string. 

At the massless level, combining the right and left-moving states of eqs.~\eqref{eq:masslessright}
and~\eqref{eq:masslessleft} gives rise to the following states:
\bi
\item a ten-dimensional $\maN=1$ {\bf supergravity multiplet}
\be \ket{q}_R\otimes\ti\alpha_{-1}^j\ket{0}_L\;, \qquad i=3,\ldots,10\,,
\label{eq:SugraMultiplet}
\ee including the graviton $g^{ij}$, the dilaton $\phi$, \index{dilaton}
the antisymmetric tensor $B^{ij}$ and their {\sc susy} partners;\footnote{Notice that $\ket{q}_R$
represents both bosonic and fermionic degrees of freedom, according to
eq.~\eqref{eq:masslessright}. Recall also that the (bosonic) $q$ carries a spacetime index $i$.}
\item 16 {\bf uncharged gauge bosons} (and gauginos)
\be \ket{q}_R\otimes\ti\alpha_{-1}^I\ket{0}_L\;,\qquad I=1,\ldots,16\ee which comprise the set of Cartan
generators $H_I$ of the gauge group $\mathcal{G}$; 
\item 480 {\bf charged gauge bosons} (and gauginos) 
\be \ket{q}_R\otimes\ket{p}_L\;, \qquad p^2=2\;,\ee with $p$ given by eq.~\eqref{eq:ps}. 
\ei

Let us make a couple of remarks on the supergravity multiplet. As $p=0$, it is a
gauge singlet. Further, since the oscillator $\ti\alpha^j_{-1}$ transforms as an \bs{8_v} of \SO8, the
states $\ket{q}_R\otimes\ti\alpha_{-1}^j\ket{0}_L$ can be expressed in group-theory language as
\begin{eqnarray}
  \label{eq:SugraMultipletWeights1}
  &&\begin{array}{rcccccc}
   \bs{8_v}\x \bs{8_v} &=& \bs1 &+&\bs{28}&+&\bs{35_v}\,,\\
                       && \varphi && B^{ij} && g_{ij}
  \end{array}\qquad\quad\text{\bf NS}\\[1mm]
  \label{eq:SugraMultipletWeights2}
  &&\begin{array}{rcccc}
   \bs{8_s}\x \bs{8_v} &=& \bs{8_c}&+&\bs{56_c}\,,\\
                       && \ti\varphi && \ti{g}^{ij}
  \end{array}\qquad\qquad\qquad\qquad\text{\bf R}
\end{eqnarray}
where $\varphi$ ($\ti\varphi$) denotes the dilaton (dilatino), $g^{ij}$ ($\ti{g}^{ij}$) is the graviton
(gravitino), and $B^{ij}$ stands for the antisymmetric two-form.

\index{anomaly cancellation!heterotic string}
Uncharged and charged gauge bosons together form the 496-dimensional adjoint representation\footnote{The
 adjoint representation of $\E8\times\E8$ reads $(\bs{248},\bs{1})\oplus(\bs1,\bs{248})$.} 
of the gauge group $\mathcal{G}$. We notice that the effective theory with the massless content provided
before is $\maN=1$ supergravity in ten dimensions coupled to Yang-Mills. Such a theory has a
gravitational anomaly of 496 units, which can be cancelled by including the 496 gauginos that we have at
hand. This cancellation is not surprising because the heterotic string is, by construction, modular invariant and
thus anomaly free. 

\section{Compactification on Orbifolds}
\label{sec:Orbifolds}\index{orbifold constructions|(}
It is clear that the heterotic string by itself is not a theory which describes the observable
universe. A striking difference of this theory with respect to the four-dimensional spacetime of everyday
experience is that the heterotic string is ten-dimensional. Further, the gauge symmetry group \maG is too
big compared to the one of the {\sc sm} or its minimal supersymmetric version, the {\sc mssm}. Therefore, in order
for the heterotic string theory to make contact with low--energy physics, one has to introduce a mechanism
in the theory ensuring that the additional dimensions are as small as to escape detection.

Compactification of extra dimensions on circles and tori have been discussed
since the pioneering works by Kaluza and Klein~\cite{Kaluza:1921xx,Klein:1926xx}. However, if the
heterotic string is compactified to four dimensions on a six-torus, the resulting theory is far from
being phenomenologically acceptable. In torus compactifications no supersymmetry is broken, so one gets
$\maN=4$ {\sc susy} in four dimensions. In supersymmetric theories with $\maN\ge 2$, both vector and
matter supermultiplets transform according to the same gauge group representation. Hence, these theories
have the undesirable feature of being nonchiral.

Orbifold compactifications~\cite{Dixon:1985jw,Dixon:1986jc} are much more attractive. Orbifolds are very
similar to Calabi-Yau manifolds in the sense that both of them can lead to four-dimensional theories with
$\maN=1$. The advantage of orbifolds is that they are Riemann flat, with the exception of a
finite set of points. Therefore, the metric, which for almost all Calabi-Yau manifolds is still unknown,
can be easily computed in orbifolds. Consequently, in these constructions it is comparatively
straightforward to investigate phenomenological properties, such as the low--energy gauge symmetry, the
particle spectrum, the Yukawa couplings and the K\"ahler potential, among others.

\begin{figure}[t!]
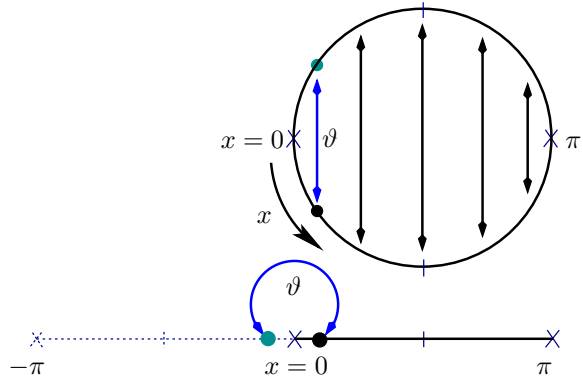

\centerline{\input S1byZ2.pstex_t}
\caption{Compactification of one dimension on an \Z2 orbifold. Points in the compactified dimension $S^1$
  are identified by the \Z2 point group element $\vartheta$. Clearly, the points at $x=0$ and $x=\pi$ are
  left fixed by the \Z2 action. The fundamental region of $S^1/\Z2$ is $x\in[0,\pi]$.}
\label{fig:S1byZ2}
\end{figure}

\index{point group}
In general, an orbifold is defined to be the quotient of a manifold by a discrete set of its isometries,
called the {\it point group} $P$.\index{orbifold constructions!point group} The simplest example is a one dimensional circle $S^1$ divided by
the point group $P=\Z2$. As illustrated in fig.~\ref{fig:S1byZ2}, the points $x$ and $-x$ are identified
by \Z2. This identification originates the space of points describing the complete orbifold to lie on
the interval $x\in[0,\pi]$ (i.e. one half of $S^1$), which we shall call the {\it fundamental region} of
the orbifold. A special feature of orbifolds is the appearance of singularities. In the case of
$S^1/\Z2$, the points $x=0$ and $x=\pi$ are left invariant (or fixed) by the action of \Z2. Although not
obvious in our one dimensional example, points left invariant under any nontrivial element of $P$ map to
singular points of the orbifold. 

\subsection{Toroidal Orbifolds}
\label{subsec:ToroidalOrbifolds}
To construct a heterotic orbifold, one first compactifies six dimensions on a torus $T^6$. The six-torus
is understood as the quotient $\maR^6/\Gamma$, where $\Gamma$ is the lattice of a semisimple Lie algebra
of rank 6. Points in $\maR^6$ differing by lattice vectors are identified on the
torus, i.e.
\be
X^i~\sim~X^i + n_\alpha e^i_\al\,,\;\qquad n_\al\in\maZ\;,\;i=5,\ldots,10\;,
\label{eq:TorusTranslations}
\ee
where $e_\alpha$ denote the basis vectors of the lattice $\Gamma$.

In a second step, one has to conceive a suitable point group $P$.\footnote{In principle, one could freely
choose two different tori, $T^6_L$ and $T^6_R$, for left- and right-movers and, correspondingly, two
different point groups, $P_L$ and $P_R$. Orbifolds containing these elements are called
asymmetric~\cite{Narain:1986qm,Ibanez:1987pj}. In the present work, however, we will focus on the case
of symmetric orbifolds, where $T_L^6=T_R^6$ and $P_L=P_R$.} Considering only abelian point groups, we
are left with the cyclic groups 
\bse\label{eq:PointGroups}
\bea
\Z{N}            &=&\left\{\theta=\vartheta^k\quad\;|\;0\le k <N\right\}\qquad\text{ and } \\
\Z{N}\times\Z{M} &=&\left\{\theta=\vartheta^k\omega^\ell\;|\;0\le k < N;\;\, 0\le \ell< M\right\}\;,
\eea
\ese
where $\vartheta^N=\omega^M=\id$, $M$ is an integer multiple of $N$ and $k,\,\ell\in\maZ$. The point group
generators $\vartheta$ and $\omega$, also called {\it twists}, are discrete rotation generators acting
crystallographically on the torus lattice $\Gamma$. All point groups of these kinds have been already
classified~\cite{Dixon:1985jw,Font:1988mk}. \index{space group} It is common to
combine the action of the point group on $\Gamma$ with the identification of points due to the torus compactification,
eq.~\eqref{eq:TorusTranslations}. The result is the so-called {\it space group}, defined by
\be
S~=~P\ltimes\Gamma ~=~ \big\{g=(\theta,\,n_\al e_\al)\;|\;\theta\in P,\;n_\al\in\maZ \big\}\;,
\label{eq:SpaceGroup} 
\ee
where the sum over $\alpha$ is understood. Due to the properties of the semidirect product `$\ltimes$',
the multiplication of two space group elements is given by
\be
g_1 g_2 ~=~ (\theta_1,\,n_\al e_\al)(\theta_2,\,m_\al e_\al)~=~ (\theta_1\theta_2, n_\al e_\al + m_\al
\theta_1 e_\al)\in S\;.
\label{eq:ProductSpaceGroup}
\ee
One can also verify that the inverse of a space group element is given by
\be
g^{-1}~=~ (\theta,\,n_\al e_\al)^{-1} ~=~ (\theta^{-1},\,-n_\al \theta^{-1}e_\al)\,.
\label{eq:InverseSpaceGroup}
\ee
Moreover, the action of a space group element $g=(\theta,\,n_\al e_\al)$ on the six compact dimensions is
provided by
\be
X^i ~\stackrel{g}{\longrightarrow} (\theta X)^i + n_\al e_\al^i\;,\qquad i=5,\ldots,10\;.
\label{eq:SpaceGroupAction}
\ee

Modular invariance requires the action of the space group $S$ to be embedded into the 16 gauge degrees of
freedom
\be
S\, \hookrightarrow\, G\;,
\ee
where the {\it gauge twisting group} $G$ is, in general, a subgroup of the automorphisms of the
$\E8\times\E8$ or $\SO{32}$ Lie algebras. Space group elements are mapped to elements of the gauge
twisting group, according to\index{shift vector}\index{Wilson lines}
\be\label{eq:SpaceGroupEmbedding}
\ba{rll}
\Z{N}~:~            &(\vartheta^k,\,n_\al e_\al) & \mapsto (k V,\,n_\al A_\al)\\
\Z{N}\times\Z{M}~:~ &(\vartheta^k\omega^\ell,\,n_\al e_\al)&\mapsto (k V_1 + \ell V_2,\,n_\al A_\al)
\ea
\quad k,\,\ell,\,n_\al\in\maZ\,,
\ee
where the 16-dimensional {\it shift vectors} $V,\,V_i$ parametrize the embedding automorphisms of the
respective twists.\footnote{It is possible to embed the action of a space group element $(\theta,e_\al)$
as $(\Theta,\,A_\al)$, where $\Theta$ denotes a rotation in the gauge degrees of
freedom~\cite{Ibanez:1987xa,Forste:2005rs}.} 
The shifts $A_\al$ represent {\it Wilson lines}~\cite{Ibanez:1986tp,Ibanez:1987sn,Bailin:1986pd},
i.e. they are gauge transformations associated to the noncontractible loops generated by $e_\al$. An
element of the gauge twisting group acts on the 16 gauge degrees of freedom of the heterotic string as
\be
\ba{rll}
\Z{N}~:~            &X^I ~\longrightarrow & X^I + k V^I + n_\al A^I_\al\\
\Z{N}\times\Z{M}~:~ &X^I ~\longrightarrow & X^I + k V_1^I + \ell V_2^I + n_\al A^I_\al
\ea
\;\qquad I=1,\ldots,16\;.
\label{eq:GaugeTwistingGroupAction}
\ee

Finally, we are in position to define a heterotic orbifold. A heterotic orbifold is made up of the
product of the quotient spaces of $T^6/P$ and $T^{16}/G$:
\be
\maO~=~T^6 \big/P \,\otimes\, T^{16} \big{/} G~=~\maR^6 \big/S \,\otimes\, T^{16} \big{/} G\;,
\label{eq:DefOrbifold}
\ee
where we have made use of the definition of the space group. 

\subsubsection*{Space Group Conjugacy Classes}
\index{space group!conjugacy class}

Not all elements of the space group describe a distinct action on the orbifold. A useful concept to gather
those elements producing the same effect on the orbifold is that of {\it conjugacy class}. Two space
group elements $g_1,\,g_2\in S$ are {\it conjugate} if there exists another space group element 
$h\in S$, such that
\be
h g_1 h^{-1}~=~g_2 \quad \Leftrightarrow \quad g_1\backsimeq g_2\,.
\label{eq:ConjugacySpaceGroup}
\ee
One says then that both $g_1$ and $g_2$ belong to the same conjugacy class $[g]$, defined by
\be
[g] = \left\{h\,g\,h^{-1}\; |\;h\in S\right\}\,,
\label{eq:DefConjugacyClass1}
\ee
and are therefore equivalent.

With help of eqs.~\eqref{eq:ProductSpaceGroup} and~\eqref{eq:InverseSpaceGroup}, one can easily verify
that conjugation of $g=(\theta,\,n_\al e_\al)$ under an arbitrary group element $h=(\theta_h,\,m_\al
e_\al)$ yields 
\be
\ba{rl}
h\,g\,h^{-1} &=~
(\theta_h,\,m_\al e_\al)\,(\theta,\,n_\al e_\al)\,(\theta_h^{-1},\,-m_\al\theta_h^{-1} e_\al)\\[2mm] 
&=~ (\theta,\,n_\al \theta_h e_\al + (\id - \theta)m_\al e_\al)\,.
\ea
\label{eq:ConjugacyClassGral}
\ee
Thus, the conjugacy class~\eqref{eq:DefConjugacyClass1} of a general space group element $g$ becomes
\be
[g] ~=~ \left\{(\theta,\,n_\al \theta_h e_\al + (\id - \theta)m_\al e_\al)\; |\;\theta,\,\theta_h\,\in P\right\}\,.
\label{eq:DefConjugacyClass2}
\ee

\index{space group!twisted sectors}\index{space group!untwisted sector}
It is evident that there are several different conjugacy classes. It is convenient to organize
them into two categories: a) $\theta~=~\id$ and b) $\theta~\neq~\id$. The conjugacy classes of the former
case compose the so-called {\it untwisted sector}, denoted $U$. Those of the second case constitute one or more {\it
twisted sectors}, denoted $T_{k,\ell}$, depending on the number of nontrivial $\theta$s available. The origin and meaning of the
tags {\it untwisted} and {\it twisted} will be clarified in section~\ref{sec:HeteroticSpectrum}.

Notice that the elements of the untwisted sector are just lattice translations, $(\id,\,n_\al
e_\al)$. The corresponding conjugacy classes, according to eq.~\eqref{eq:DefConjugacyClass2}, acquire
then the form
\be
[(\id,\,n_\al e_\al)] ~=~ \left\{(\id,\,n_\al \theta_h e_\al)\; |\;\theta_h\,\in P\right\}\,.
\label{eq:ConjugacyClassTranslations}
\ee

\subsection{Consistency Conditions}
\label{subsec:ConsistencyConditions}

We have seen in section~\ref{sec:HeteroticString} that the heterotic string has intrinsic theoretical
constraints on its geometry and spectrum. It is then natural to expect some requirements for the
needed parameters in orbifold compactifications. These constraints fall into three classes:
\bi
\item $\maN=1$ {\sc susy};
\item $S\hookrightarrow G$ embedding conditions; and
\item modular invariance.
\ei

\subsubsection*{$\boldsymbol{\maN=1}$ \textbf{\textsc{susy}}}

By compactifying on a six-dimensional space we clearly distinguish between our four-dimensional Minkowski
space-time and the six internal coordinates. Consequently, the transversal \SO8 of the ten-dimensional
Lorentz group will break. The specific form of this breaking depends on the geometry of the internal
space and is directly related to the amount of supersymmetry in four dimensions. Generically, the
breaking is of the form 
\begin{equation}
  \label{eq:SO8breaking}
  \SO8\longrightarrow \SO2\x\SO6\sim\U1\x\SU4\,.
\end{equation}
The \U1 is associated to the uncompactified directions of the Minkowski space-time and
can therefore be interpreted as the four-dimensional helicity.

\index{holonomy group@holonomy group!$\maN\neq0$ {\susy}}
On the other hand, the ten-dimensional gravitino  contains the two helicity states
of the spin-3/2 four-dimensional gravitini, transforming as \bs{4}--plets under the internal \SU4
symmetry: 
\begin{equation}
  \label{eq:4dgravitini}
   \bs{56_c}\supset \bs{4}_{3/2} + \bsb{4}_{-3/2}\,.
\end{equation}
We denote the spinor fields associated to the four-dimensional gravitini by $\eta_i$, $i=1,\dots,4$.
The amount of unbroken \susy\ charges $Q_i$ is given by the number of covariantly constant spinors
$\eta_i$, i.e. by the number of gravitini invariant under the holonomy group ($\nabla_m\eta_i=0$).
We observe that, depending on the holonomy group of the compact space, we can have
\begin{subequations}
 \label{eq:gravitiniTrafo}
 \begin{eqnarray}
  \label{eq:gravitiniTrafoTrivial}
  \mbox{trivial holonomy:} &\phantom{....}&\bs{4}\longrightarrow \bs{1}+\bs{1}+\bs{1}+\bs{1}\qquad\Ra\maN=4\,,\\
  \label{eq:gravitiniTrafoSU2}
  \mbox{\SU2 holonomy:} &&  \bs{4}\longrightarrow \bs{2}+\bs{1}+\bs{1}\qquad\qquad\!\Ra\maN=2\,,\\
  \label{eq:gravitiniTrafoSU3}
  \mbox{\SU3 holonomy:} &&  \bs{4}\longrightarrow \bs{3}+\bs{1}\qquad\qquad\qquad\!\!\Ra\maN=1\,.
 \end{eqnarray}
\end{subequations}

We know that theories with $\maN>1$ are nonchiral and
thus unrealistic and models with no supersymmetry are phenomenologically disfavored as
they cannot alleviate some of the fundamental puzzles of the standard model. Thus, a natural
phenomenological requirement for orbifold models is to have $\maN=1$ {\sc susy}. This can be guaranteed
by an appropriate choice of the point group $P$.

\index{orbifold constructions!compactification lattices}
\begin{table}[!t]
\begin{center}
{\footnotesize
\begin{tabular}{|c|c|c|}
\hline
Point group $P$ & 6D Lattice $\Gamma$      & Twist vector $v$\\
 \hline\hline
\rowgrayh
 $\Z{3}$     & $\SU3^3$                    & $\frac{1}{3}(0,\,1,\,1,\,-2)$  \\[1mm]
 $\Z{4}$     & $\SU4^2$                    & $\phantom{A_{A_{A_A}}}$        \\[1mm]
             & $\SO5\times\SU4\times\SU2$  & $\frac{1}{4}(0,\,1,\,1,\,-2)$  \\[1mm]
             & $\SO5^2\times\SU2^2$        & $\phantom{A^2_{A_{A_A}}}$      \\
\rowgrayh
 $\Z{6}$-I   & $\G2^2\times\SU3$           & $\frac{1}{6}(0,\,1,\,1,\,-2)$  \\[1mm]
 $\Z{6}$-II  & $\G2\times\SU3\times\SO4$   & $\frac{1}{6}(0,\,1,\,2,\,-3)$  \\[1mm]
             & $\SU6\times\SU2$            & $\frac{1}{6}(0,\,2,\,1,\,-3)$  \\[1mm]
             & $\SU3\times\SO8$            &                                \\[1mm]
             & $\SU3\times\SO7\times\SU2$  & $\phantom{A^2_{A_{A_A}}}$  \\
\rowgrayh
 $\Z{7}$     & $\SU7$                      & $\frac{1}{7}(0,\,1,\,2,\,-3)$  \\[1mm]
 $\Z{8}$-I   & $\SO9\times\SO5$            & $\frac{1}{8}(0,\,1,\,-3,\,2)$  \\[1mm]
\rowgrayh
 $\Z{8}$-II  & $\SO{10}\times\SU2$         & $\frac{1}{8}(0,\,1,\,3,\,-4)$  \\[1mm]
\rowgrayh
             & $\SO9\times\SU2^2$          & $\phantom{A^2_{A_{A_A}}}$  \\
 $\Z{12}$-I  & $\E6$                       & $\frac{1}{12}(0,\,1,\,-5,\,4)$ \\[1mm]
             & $F_4\times\SU3$             & $\phantom{A^2_{A_{A_A}}}$  \\
\rowgrayh
 $\Z{12}$-II & $F_4\times\SU2^2$           & $\frac{1}{12}(0,\,1,\,5,\,-6)$ \\[1mm]
\hline
\end{tabular}
}
\end{center}
\caption{Admissible six-dimensional crystallographic lattices~\cite{Kobayashi:1991rp} and twist vectors
  for \Z{N} orbifolds.} 
\label{tab:6DLattices}
\end{table}

\index{point group@point group!\susy\ preserving}
In abelian orbifolds, the point group is a subset of the full holonomy group. 
Insisting on $\maN=1$ amounts to demanding
\be
P\subset \SU3 \subset \SO6\,.
\ee
Allowed point groups are the cyclic groups \Z{N}, with $N=3,4,6,7,8,12$, and $\Z{N}\times\Z{M}$, with
$N,\,M=2,3,4,6$ and $M$ an integer multiple of $N$. As we are about to see, it turns out that the
specific form of the twist $\vartheta$ is also constrained by $\maN=1$.

\index{complex planes}
To simplify the notation, the six compact coordinates of the torus $X^i$, $i=5,\ldots,10$, are
conveniently combined into three complex coordinates
\be
Z^1=\h{1}{\sqrt{2}}\left(X^5+\I X^6\right)\,,\quad Z^2=\h{1}{\sqrt{2}}\left(X^7+\I X^8\right)\,,\quad Z^3=\h{1}{\sqrt{2}}\left(X^9+\I X^{10}\right)\,.
\label{eq:ComplexCoordinates}
\ee 
In the light-cone gauge, the observable spacetime can be represented by $Z^0=\h{1}{\sqrt{2}}\left(X^3+\I X^4\right)$.

\index{nonfactorizable lattice} \index{factorizable lattice}
Further, we will assume the torus $T^6$ to be {\it factorizable},\footnote{Orbifolds in nonfactorizable
tori have been also extensively studied. For details, see e.g.~\cite{Forste:2006wq,Takahashi:2007qc}. A
possible relation between orbifolds on factorizable and nonfactorizable lattices was first conjectured in~\cite{Ploger:2007iq}.}
i.e. it can be written as $T^6=T^2\otimes T^2\otimes T^2$. \index{orbifold constructions!twist vector}In
the basis~\eqref{eq:ComplexCoordinates}, the twist of a \Z{N} orbifold is then a diagonal $3\times
3$-matrix of the form 
\be
\vartheta~=~\diag\left(e^{2\pi\I v^1},\, e^{2\pi\I v^2},\, e^{2\pi\I v^3}\right)\;.
\label{eq:ThetaMatrix}
\ee
where $v=(0,\,v^1,\,v^2,\,v^3)$ is called the {\it twist vector} and carries the full information of the point
group action.\footnote{In $\Z{N}\times\Z{M}$ orbifolds,
there is one twist vector for each of the two point group generators, $\vartheta$ and $\omega$.} 
The action of the point group on the complex compact coordinates is then given by
\be
Z^a \;\stackrel{\vartheta}{\longrightarrow}\; \exp\{2\pi\I v^a\}\, Z^a,\qquad\;\; a=1,2,3\,,
\label{eq:PointGroupActionInZ}
\ee
where we can include the trivial action of the twist on the observable spacetime plane $Z^0$.

\begin{table}[!t]
\begin{center}
{\small
\begin{tabular}{|c|c|c|}
\hline
Point group $P$    &  Twist vector $v_1$       & Twist vector $v_2$ \\
 \hline\hline
\rowgrayh
 $\Z{2}\times\Z2$  & $\2(0,\,1,\,0,\,-1)$      & $\2(0,\,0,\,1,\,-1)$\\[1mm]
 $\Z2\times\Z4$    & $\2(0,\,1,\,0,\,-1)$      & $\h14(0,\,0,\,1,\,-1)$\\[1mm]
\rowgrayh
 $\Z3\times\Z3$    & $\h13(0,\,1,\,0,\,-1)$    & $\h13(0,\,0,\,1,\,-1)$\\[1mm]
 $\Z2\times\Z6$-I  & $\2(0,\,1,\,0,\,-1)$      & $\h16(0,\,0,\,1,\,-1)$\\[1mm]
\rowgrayh
 $\Z2\times\Z6$-II & $\2(0,\,1,\,0,\,-1)$      & $\h16(0,\,1,\,1,\,-2)$\\[1mm]
 $\Z4\times\Z4$    & $\h14(0,\,1,\,0,\,-1)$    & $\h14(0,\,0,\,1,\,-1)$\\[1mm]
\rowgrayh
 $\Z3\times\Z6$    & $\h13(0,\,1,\,0,\,-1)$    & $\h16(0,\,0,\,1,\,-1)$\\[1mm]
 $\Z6\times\Z6$    & $\h16(0,\,1,\,0,\,-1)$    & $\h16(0,\,0,\,1,\,-1)$\\[1mm]
\hline
\end{tabular}
}
\end{center}
\caption{Twist vectors for $\Z{N}\times\Z{M}$ orbifolds leading to $\maN=1$. Six-dimensional
 compactification lattices can be found in e.g.~\cite{Forste:2006wq,Takahashi:2007qc} (see also appendix~\ref{ch:Tables}).}  
\label{tab:ZNxZMTwists}
\end{table}

In $\maN=1$ {\sc susy} there is only one gravitino. So, we need to determine what the form of the twist
vector must be so as to ensure that only one gravitino survives after compactification. The
ten-dimensional gravitino contained in the supergravity multiplet~\eqref{eq:SugraMultiplet} of the
heterotic string splits into four gravitini in four dimensions:
\be
\Big|\;\pm\2\;;\;\;\;\pm\2,\,\pm\2,\,\pm\2\; \Big\rangle_R\otimes \ti\al^\nu_{-1}\Big|\;
0\;\Big\rangle_L\,,\qquad \text{even \# of $+$ signs}\,.
\label{eq:4gravitini}
\ee
The first component of the right-mover provides both chiralities for the gravitini. Due to the
compactification, the last three components are internal indices that account for a multiplicity factor
of four for each chirality. Since the left-mover carries lorentzian index, only the right-mover
transforms under the orbifold action
\be
\big|\,\pm\tfr\;;\;\;\pm\tfr,\,\pm\tfr,\,\pm\tfr\,\big\rangle_R \;\stackrel{\vartheta}{\longrightarrow}\;
e^{-2\pi\I \left(\pm\2 v^1\pm\2 v^2\pm\2 v^3\right)}\big|\,\pm\tfr\;;\;\;\pm\tfr,\,\pm\tfr,\,\pm\tfr\,\big\rangle_R\,.
\ee
The spectrum of an orbifold contains only states that are invariant under the orbifold action. Thus, in
order to get $\maN=1$, the phase $\exp\{-\pi\I \left(\pm v^1\pm v^2\pm v^3\right)\}$ must be 
trivial for one gravitino. One notices that the condition $\pm v_1\pm v_2 \pm v_3=0$ for one combination
of signs assures the presence of solely one gravitino in the orbifold spectrum. Therefore, one can choose
the components of the twist vector to satisfy 
\be
v^1+v^2+v^3=0\,,
\label{eq:TwistSusyCondition}
\ee
so that $|\pm(\2;\,\,\2,\2,\2)\rangle_R\otimes \ti\al^\nu_{-1}|0\rangle_L$ be the surviving gravitino.
Moreover, as the twist vector $v$ of \Z{N} orbifolds corresponds to a twist $\vartheta\in P$ of order $N$
($\vartheta^N=\id$), then its components fulfill in general $N v^a \in \maZ,\;\,a=1,2,3$. Note that if
one $v^a$ is zero (or integer), we obtain $\maN=2$ in four dimensions, implying that all components $v^a$ of a
\Z{N} orbifold twist vector must be nontrivial. In the third column of table~\ref{tab:6DLattices}, we present
our choice of twist vectors for all admissible \Z{N} orbifolds.

\begin{figure}[t!]
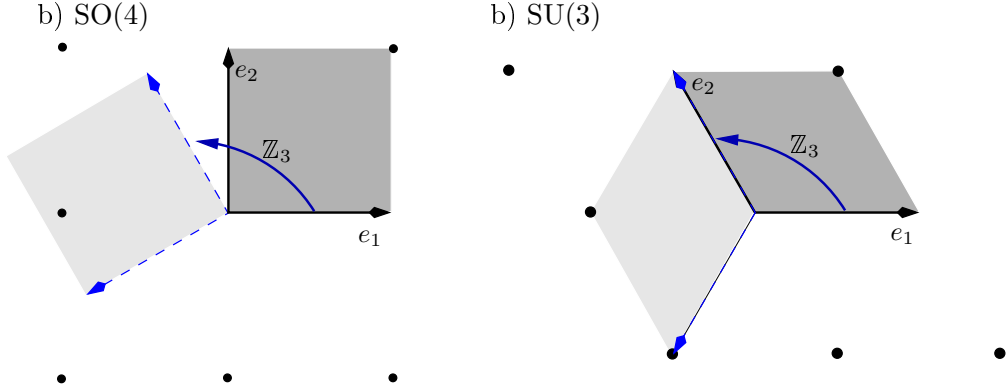

\centerline{\input Z3Cristal.pstex_t}
\caption{A two-dimensional \Z{3} transformation a) does not act crystallographically on the root lattice of
  \SO4. In contrast, b) the root lattice of \SU3 is mapped to itself under \Z3.}
\label{fig:Z3Crystal}
\end{figure}

In $\Z{N}\times\Z{M}$ orbifolds, there are two twist vectors, $v_1$ and $v_2$, that also fulfill
eq.~\eqref{eq:TwistSusyCondition}. Only their combined action must lead to $\maN=1$. In this case, one
can choose the twist vectors as shown in table~\ref{tab:ZNxZMTwists}. Notice that each of the twist
vectors leads to $\maN=2$, but their conjoint action provides a theory with $\maN=1$, as required.

\index{orbifold constructions!compactification lattices}
A secondary effect of the restriction $\maN=1$ occurs in the six-dimensional compactified space.
As already mentioned, the point group $P$ has to act crystallographically on the root lattice $\Gamma$ of a
six-dimensional Lie algebra. In other words, the lattice of the torus $T^6$ must be mapped to itself
under the action of $P$. Provided a point group $P$, not any root lattice is admissible. As an example, consider
fig.~\ref{fig:Z3Crystal}: in two dimensions, the action of \Z3 can be regarded as a rotation by
$2\pi/3$. Clearly, whereas the root lattice of \SU3 is left invariant under \Z3, the root lattice of \SO4
is not. Suitable root lattices for \Z{N} orbifolds are given in table~\ref{tab:6DLattices}. We have also
compiled a list of some allowed lattices for $\Z{N}\times\Z{M}$ orbifolds in table~\ref{tab:ZNxZMWLConstraints}.

\subsubsection*{$\boldsymbol{S\hookrightarrow G}$ Embedding Conditions}
\index{shift vector}\index{Wilson lines}
The embedding of the space group into the gauge degrees of freedom imposes some conditions on the shift
vector(s) $V$ and the Wilson lines $A_\al$. First of all, embedding the twist satisfying
$\vartheta^N~=~\id$ implies that the associated shift vector has to be also of order $N$, i.e.
\be
\label{eq:ShiftInLattice}
N V \in \Lambda\,,
\ee
where $\Lambda$ is the 16-dimensional weight lattice of the corresponding heterotic string.

On the other hand, Wilson lines are subject to certain conditions depending on the compactification
lattice $\Gamma$. Wilson lines are the embeddings of the six lattice generators $e_\al$, therefore one might
be enticed to think that there are six distinct Wilson lines $A_\al$. However, not all directions $e_\al$ are
independent in the orbifold. Consider for example the two-dimensional \SU3 lattice of
fig.~\ref{fig:Z3Crystal}. One sees that a \Z3 transformation maps $e_1$ to $e_2$, implying that the
corresponding Wilson lines $A_1$ and $A_2$ have to be equivalent in the orbifold. Consequently, in \Z3
orbifolds with compactification lattice $\Gamma=\SU3^3$ (cf. table~\ref{tab:6DLattices}), one finds 
\be
\label{eq:Z3WLRelations}
e_\al \stackrel{\vartheta}{\longrightarrow} e_{\al+1}\;\Longrightarrow\; A_\al \approx A_{\al+1}\,,\quad
\al=1,3,5;\;\,\vartheta\in\Z3\,,
\ee
where `$\approx$' indicates that the Wilson lines are identical up to lattice translations in $\Lambda$.

\index{Wilson lines!order of}
In general, relations between the Wilson lines are effect of equivalences between space group
elements. As we have seen before, elements of a given conjugacy class are equivalent.
According to eq.~\eqref{eq:ConjugacyClassTranslations}, the conjugacy class of a lattice translation
$g=(\id,\,n_\al e_\al)$ is given by
\be
[(\id,\,n_\al e_\al)] = \left\{(\id,\,n_\al \theta e_\al)\;|\; \theta\in P\right\}
\ee
and contains elements describing the same orbifold action. For example, in the case of \Z3 orbifolds, the
relation
\be
(\id,\,e_1)\backsimeq (\vartheta,0)(\id,e_1)(\vartheta,0)^{-1}=(\id,\,e_2)\,,
\ee
implies that $(\id,\,e_1)$ and $(\id,\,e_2)$ are indistinguishable from the orbifold perspective.
Then, their embedding into the gauge degrees of freedom should also be identified. This restricts $A_1$
and $A_2$ to be equal up to lattice vectors, as we had already shown. 

It is not hard to realize that also the order of the Wilson lines gets restricted by the choice of
$\Gamma$. Consider in the \Z3 orbifold the following elements of the same conjugacy class:
\be
(\id,\,e_2)\backsimeq (\vartheta,\,0)(\id,\,e_2)(\vartheta,\,0)^{-1}=(\id,\,-e_1-e_2)\,.
\ee
Embedding this relation into the gauge degrees of freedom and using $A_1\approx A_2$ yields
\be
A_2\approx -A_1-A_2 \approx -2 A_2\quad \Leftrightarrow \quad 3A_2 \approx 0\,.
\ee
In other words, $A_2$ (as well as $A_1$) has to be a Wilson line of order 3, $3 A_2 \in \Lambda$.
Similar relations apply also for the other two \SU3 factors of the lattice $\Gamma$. Hence, all in all, we obtain
\be
3A_1\approx 3A_3\approx 3A_5 \approx 0\,;\;\;A_1\approx A_2,\,A_3\approx A_4,\,A_5\approx A_6\,.
\ee
The number of independent Wilson lines $A_\al$ and their order $N_\al$ for admissible choices of
$\Gamma$ in \Z{N} orbifolds are provided in table~\ref{tab:ZNWLConstraints}.\footnote{Notice that several
typos of the literature have been corrected there.} See
table~\ref{tab:ZNxZMWLConstraints} for constraints on Wilson lines of $\Z{N}\times\Z{M}$ orbifolds.

\subsubsection*{Modular Invariance}
\index{orbifold constructions!modular invariance conditions}
\index{modular invariance conditions}
Terms of the one-loop partition function of abelian orbifolds acquire in general a nontrivial phase under modular
transformations~\cite{Minahan:1987ha,Senda:1987pf}. Demanding the partition function to be modular invariant
safeguards the resulting theory from anomalies. Therefore, we are committed to requiring the phase that arises
from modular transformations to vanish. This imposes constraints on the orbifold parameters which, for
\Z{N} orbifolds without Wilson lines, are usually expressed as~\cite{Dixon:1986jc,Vafa:1986wx,Senda:1987pf}
\be
N(V^2-v^2)=0\mod 2\,.
\label{eq:WeakModInvZNwoWilson}
\ee
In including Wilson lines (and a second twist of the point group in the case of $\Z{N}\times\Z{M}$ orbifolds),
eq.~\eqref{eq:WeakModInvZNwoWilson} has to be replaced by~\cite{Ploger:2007iq}  
\begin{subequations}\label{eq:NewModInv}
\begin{eqnarray}
  N\,\left(V_{1}^{2} - v_{1}^{2} \right)  & = & 0 \mod 2\;, \\
  \label{eq:fsmi1}
  M\,\left(V_{2}^{2} - v_{2}^{2} \right)  & = & 0 \mod 2\;, \\
  \label{eq:fsmi2}
  M\,\left(V_{1}\cdot V_{2} - v_{1}\cdot v_{2} \right)  & = & 0 \mod 2\;, \\
  \label{eq:fsmi3}
  N_\al\,\left(A_{\al}\cdot V_{i}\right)  & = & 0 \mod 2\;, \\
  \label{eq:fsmi4}
  N_\al\,\left(A_{\al}^2\right)  & = & 0 \mod 2\;, \\
  \label{eq:fsmi5}
  Q_{\al\beta}\,\left(A_{\al}\cdot A_{\beta}\right)  & = & 0 \mod 2 \quad (\al \neq \beta)\;,
  \label{eq:fsmi6}
\end{eqnarray}
\end{subequations}
where $N_\al$ corresponds to the order of the Wilson line $A_\al$ ($N_\al A_\al \in\Lambda$), and
$Q_{\alpha\beta}\equiv\text{gcd}(N_{\al},N_{\beta})$ denotes the greatest common divisor of $N_{\alpha}$
and $N_{\beta}$.\footnote{In the case of two different \Z2 Wilson lines we find that \eqref{eq:fsmi5} can
be relaxed, i.e.\ $\text{gcd}(N_\al,N_\beta)$ can be replaced by $N_\al\,N_{\beta}=4$, provided there
exists no  $g \in P$ with the property $g\,e_\alpha~\neq~e_\alpha$ but $g\,e_\beta~=~e_\beta$. 
Imposing the weaker condition leads, as we find, to anomaly-free spectra.}
For \Z{N} orbifolds, one has $V_1=V,\;v_1=v$ and eqs.~\eqref{eq:fsmi2}
and~\eqref{eq:fsmi3} are clearly unnecessary. 

\subsection{Orbifold Geometry}
\label{subsec:OrbifoldGeometry}
\index{orbifold constructions!fixed points}
\begin{figure}[t!]
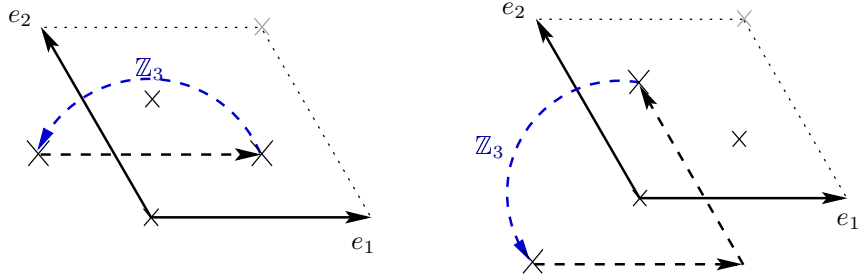

\centerline{\input Z3_fixedpoints.pstex_t}
\caption{Fixed points arise naturally by the action of a discrete symmetry on the torus. In an \SU3
  lattice, the action of \Z3 is counteracted by lattice translations, leaving three points fixed.}
\label{fig:Z3FixedPoints}
\end{figure}

Fixed points appear naturally in orbifold compactifications due to the action of the twist on the compact
space. As an example, consider a one-complex-dimensional \Z3 orbifold on an \SU3 torus lattice. \Z3 acts as a rotation
by $2\pi/3$ on the complex plane. Evidently, the point at the origin is not affected by the \Z3 action
and is therefore fixed. Furthermore, considering torus translations, one finds that there
are two additional fixed points inside the fundamental cell of the torus. Observe the situation depicted in
figure~\ref{fig:Z3FixedPoints}. The discrete rotation of those points is counteracted by translations
in the torus lattice, so that the points remain unaffected in the orbifold. 

Let us make two remarks. First, note that the three fixed points described above can be related neither
by further lattice translations nor by the repeated action of \Z3. This independence characterizes all
fixed points on the orbifold.\footnote{As we will see, there are situations in which some points are
fixed under the action of one point group element, but are connected to other fixed points by the action of
another one. In that case, not all fixed points in the torus are independent fixed points
on the orbifold.} Secondly, if one considers not only the fundamental cell of the torus illustrated in
figure~\ref{fig:Z3FixedPoints}, but the entire (infinite) root lattice of \SU3, clearly, one finds an
infinite set of fixed points. Yet all of them are identified to the three fixed points on the orbifold by
the conjoint action of \Z3 rotations and lattice translations, that is, by the action of the space
group. Hence, the fixed points belong to only three different classes and, for describing all of them, it
suffices to take one representative out of each of these classes. 

\begin{figure}[t!]
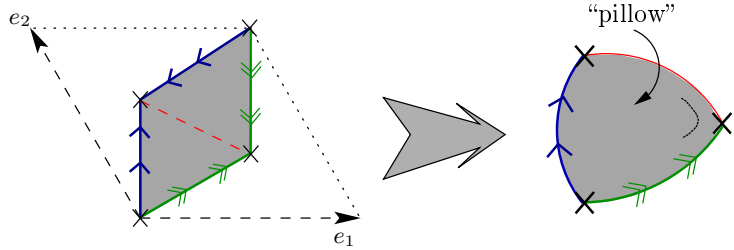

\centerline{\input Z3_pillow.pstex_t}
\caption{The fundamental region of a one-complex-dimensional \Z3 orbifold is reduced to one third of the
 fundamental cell of the \SU3 root lattice. Folding appropriately the fundamental region, one can see
 that the \Z3 orbifold is a flat manifold everywhere except at the three singularities corresponding to the
 fixed points.}
\label{fig:Z3Pillow}
\end{figure}

Orbifolds are generalizations of manifolds in the sense that they are smooth (almost) everywhere,
with exception of a constrained set of points, where the curvature concentrates; i.e. orbifolds admit
singularities. That can be realized in the last example, by observing the space resulting after moding 
out the discrete group \Z3. In that case, the entire space can be described by one third of the area of
the torus, the so-called fundamental region, as sketched in figure~\ref{fig:Z3Pillow}. Then, one notices
that points along the boundaries of the fundamental region of the orbifold are identified under the
action of the space group. To visualize the orbifold space, one has to fold the fundamental region and
paste the edges together. The outcome is a triangular pillow-like object with sharp corners located at
the fixed points. One can proof that such corners are conical singularities that concentrate the
curvature of the orbifold. This means that only at the fixed points the holonomy group is nontrivial
(generically, it is a subgroup of \Z{N}). 

In any six-dimensional orbifold, fixed points are determined by considering the underlying lattice
and the space group action. To be more precise, consider an arbitrary space group element $g=(\theta,\,n_\al
e_\al)$. Following from eq.~\eqref{eq:SpaceGroupAction}, its action on the complex coordinates $Z^a$ is
given by  
\be
Z\;\stackrel{g}{\longrightarrow}\;\theta Z + n_\al e_\al\,,
\label{eq:SpaceGroupActionInZ}
\ee
where the basis vectors $e_\al$ of the six-torus are now expressed in the complexified basis
eq.~\eqref{eq:ComplexCoordinates}, and $\theta$ denotes an arbitrary point group element taking the form
$\theta=\vartheta^k$ for \Z{N} orbifolds or $\theta=\vartheta^k\,\omega^\ell$ for $\Z{N}\times\Z{M}$
orbifolds. A point $Z_f$ in the compact space is said to be fixed in the orbifold if it is invariant under the
action of a particular space group element $g_f$. This means that fixed points satisfy
\be
Z_f = g_f Z_f = \theta Z_f + n_\al e_\al \quad\Leftrightarrow\quad (\id-\theta)\,Z_f\,\in\, \Gamma
\label{eq:FixedPoints}
\ee
for a given $g_f\in S$. 

\index{space group!constructing elements}
It is convenient to label a fixed point by the corresponding space group element $g_f$ instead of by its
spatial coordinates $Z_f^a$. The space group elements $g_f$ will be called {\it constructing elements}.
This notation is rather convenient for several reasons. We know that, even
though the number of solutions of eq.~\eqref{eq:FixedPoints} is infinite, only a reduced finite number of
points are inequivalent in the orbifold. In fact, inequivalent points in the compact space are related to
space group elements from different conjugation classes. Secondly, we can say that points expressed by
$g_f=(\id,\,n_\al e_\al)$ belong to the {\it untwisted} sector. Further, fixed points represented by
$g_f=(\theta,\,n_\al e_\al)$ with  $\theta\neq\id$ `live' in one of the {\it twisted} sectors. For example,
fixed points left invariant under the action of $\theta=\vartheta^k\neq\id$ of a \Z{N}
orbifold are said to belong to the $k$-th twisted sector ($k=1,\ldots,N-1$). Analogously, invariant
points under $\theta = \vartheta^k\omega^\ell \neq \id$ in $\Z{N}\times\Z{M}$ are called fixed points of
the $(k,\,\ell)$-th twisted sector.  

The number of distinct (conjugacy classes of) fixed points varies for different sectors of an
orbifold. One first notices that in the untwisted sector, every point of the space is evidently
invariant and $g_f=(\id,\,0)$. Thus, we end up with a six-dimensional fixed torus without singularities in the untwisted
sector. A less boring situation appears in the twisted sectors. There, the solutions of
eq.~\eqref{eq:FixedPoints} are either isolated fixed points or one-complex-dimensional invariant surfaces, commonly
called {\it fixed tori}. The former case applies to points fixed under $\theta$ such that $\id-\theta$ is
nonsingular (i.e. $\det(\id-\theta) \neq 0$). The latter appears when $\id-\theta$ is singular. This is
easy to understand because $\id-\theta$ is singular only if one of the eigenvalues of $\theta$ is one or,
stated differently, only if one complex plane is left invariant under $\theta$. 

In case that $\det(\id-\theta)\neq 0$ and the lattice of the compact space is
factorizable,\footnote{i.e. the six-dimensional lattice can be written as the product of three
two-dimensional sublattices, each of which corresponds to a complex plane with coordinate $Z^a$ and $a$
fixed.} the number of isolated fixed 
points in the twisted sector corresponding to $\theta$ is given by an (over)simplified version of the
Lefschetz fixed point theorem~\cite{Giffiths:1978xx}  
\be
\#Z_f ~=~\det(\id-\theta)~=~ \prod_{a=1}^3 4\,\sin^2(\pi v^a)\,,
\label{eq:SimpleLefschetz}
\ee
where $v^a$ are the entries of the twist vector. Formula~\eqref{eq:SimpleLefschetz} is, at first sight,
very appealing, since it does not depend on the particular geometry of the underlying lattice
$\Gamma$. Nonetheless, there are too few cases for which eq.~\eqref{eq:SimpleLefschetz} applies. For
example, in most of the \Z{N} orbifolds, only the number of fixed points in the first twisted sector
($\theta=\vartheta^1$) are determined by that formula. Few other twisted sectors of both \Z{N} and
$\Z{N}\times\Z{M}$ orbifolds can also be addressed in this way. 

In presence of fixed tori, that is, when $\det(\id-\theta) = 0$, one might conjecture that it suffices to
extract the nontrivial two-complex-dimensional part of $~\theta$ and then to apply
formula~\eqref{eq:SimpleLefschetz}. Unfortunately, the result obtained in that way is, in general, wrong.
The reason can be traced back to the origin of formula~\eqref{eq:SimpleLefschetz}. In a more complex version, the
Lefschetz fixed point theorem\footnote{See e.g. appendix A of ref.~\cite{Wingerter:2005xx}.} states that
the number of fixed points (or fixed tori) is given by the index of the space of elements $g_f$
(associated to the fixed points) divided by the largest set of (sub)symmetries of the space group 
relating fixed points among each other. Only if the set of modded out symmetries comprises {\it exclusively} lattice
translations on the lattice $\Gamma$ (and $\Gamma$ is factorizable), then
formula~\eqref{eq:SimpleLefschetz} (or a lower dimensional version of it) yields a correct result.

\begin{figure}[t!]
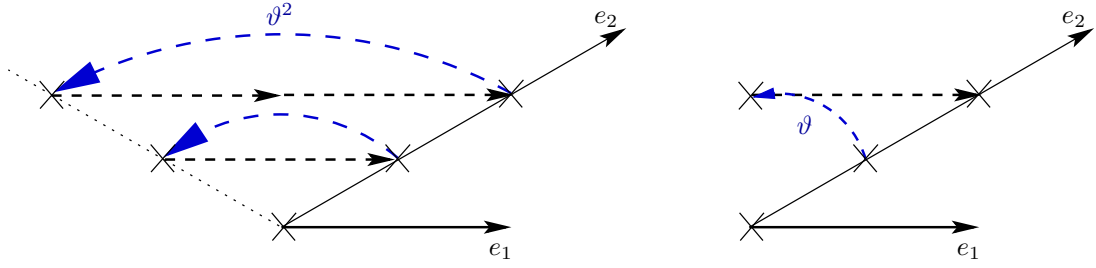

\centerline{\input Z6II_fixedpoints_G2.pstex_t}
\caption{Fixed points of the $\vartheta^2$ twisted sector on the torus spanned by the root lattice of \G2. The
 point group generator $\vartheta$ acts as a rotation by $2\pi/6$. Some points invariant under
 $\vartheta^2$ on the torus are identified on the orbifold by the action of $\vartheta$; such points are
 equivalent in the orbifold.}
\label{fig:Z6IIFixedPointsG2}
\end{figure}

Let us examine more closely this situation in an example. In \Z{N} orbifolds, $\id-\theta$ is singular
only for higher twisted sectors corresponding to $\theta=\vartheta^k$ with $1<k<N$. Suppose 
$\vartheta$ to be a \Z6 generator and the compact space to be spanned by the root lattice of
$\G2\times\SU3\times\SO4$. As it will be detailed shortly, the action of
$\vartheta^2$ is trivial in the sublattice spanned by \SO4, so that we get fixed tori in the second
twisted sector. Now, focus on the sublattice spanned by \G2. The point group generator $\vartheta$ acts
as a rotation by $2\pi/6$ on the \G2 plane; therefore, fixed points (tori) of the $\vartheta^2$ twisted
sector are those points left invariant under a rotation by 120 degrees. In
figure~\ref{fig:Z6IIFixedPointsG2}, we present the three points of the \G2 sublattice left invariant by
this rotation. Notice that a one-complex-dimensional version of formula~\eqref{eq:SimpleLefschetz} also
leads to three invariant points:
\be
\#Z_f^{\G2} ~=~ \left|\,1-e^{2\times 2\pi\I/6}\,\right| ~=~ \left|\,\h32 -
  \I\h{\sqrt{3}}{2}\,\right|~=~\h94 +\h34 ~=~3\,. 
\label{eq:PseudoFixedPointsZ6IIG2}
\ee
The points invariant under $\vartheta^2$ in the \G2 sublattice are represented by the space group
elements $g_1=(\vartheta^2,\,0)$, $g_2=(\vartheta^2,\,e_1)$ and $g_3=(\vartheta^2,\,2
e_1)$. Nevertheless, to conclude that there are three fixed points in the \G2 plane of the second twisted
sector is wrong. It is not hard to verify that 
\be
g_3 ~=~ (\vartheta,\,e_1)\, g_2 \,(\vartheta,\,e_1)^{-1}\,,
\ee
situation that is also depicted in figure~\ref{fig:Z6IIFixedPointsG2}. Then $g_2$ and $g_3$ belong to the
same conjugacy class, implying that only the conjugacy classes of $g_1$ and $g_2$ are
independent. Therefore, only two points (tori) are truly fixed under the action of the orbifold in the
\G2 sublattice. 

Before proceeding to the details of two important examples, let us add a remark. The compactification
lattice plays a very important role in the number of fixed points. For example, the usual $\Z2\times\Z2$
orbifold compactification on the factorizable lattice of $\SU2^6$ admits twice the amount of fixed tori of the same
orbifold on the nonfactorizable root lattice of \SO{12}~\cite{Faraggi:2006bs}. In fact, it is possible to state
that the number of fixed points (or tori) of orbifolds with factorizable compact space is, in general, bigger than
that of the same orbifold with a nonfactorizable lattice. Our discussion here will restrict to the
factorizable case.

\subsubsection{Standard Example: The \bs{\Z3} Orbifold}
\index{Z3 geometry@\Z3 geometry}

\begin{figure}[t!]
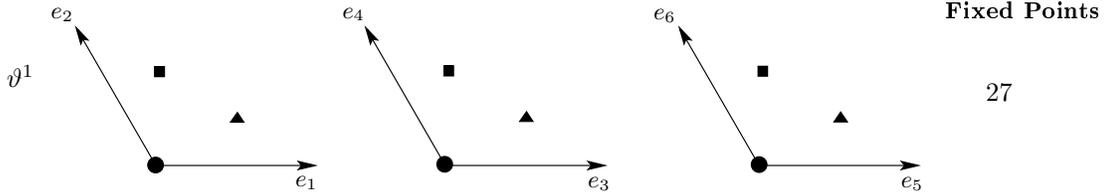

\centerline{\input Z3_geometry.pstex_t}
\caption{The geometry of \Z3 orbifolds compactified on an $\SU3^3$ lattice. The 27 fixed points of the
  $\vartheta^1$ and $\vartheta^2$ twisted sectors are identical.}
\label{fig:Z3Geometry}
\end{figure}

The \Z3 orbifold has been long studied since the
mid-eighties~\cite{Dixon:1985jw,Ibanez:1986tp,Casas:1989wu,Giedt:2000bi} mainly because it is the
simplest orbifold and because, even in that scope, there are chances to get semirealistic
models~\cite{Ibanez:1987sn,Casas:1988se}.

The compact space of the \Z3 orbifold is spanned by the root lattice of $\SU3^3$. The
twist vector preserving $\maN=1$ {\sc susy} is given by 
\be
v~=~(0,\,1/3,\,1/3,\,-2/3)\,,
\label{eq:Z3twist}
\ee 
implying that the \Z3 point group generator $\vartheta$ acts as a simultaneous rotation by $2\pi/3$ on
all three \SU3 sublattices. The \Z3 orbifold has three sectors: the untwisted sector ($\vartheta^0=\id$)
and two twisted sectors ($\vartheta^1$ and $\vartheta^2$). In the untwisted sector the action of the
point group is trivial and, therefore, all points are invariant.

To find the points fixed under $\vartheta$, we use the fact that the underlying lattice is factorizable.
We have already studied the \Z3 action on a single \SU3 lattice. We have seen that three independent
points are left invariant by \Z3. As an extension of that case, we find that each of the three
sublattices has three independent fixed points, so that the six-dimensional space of the first twisted
sector contains a total of 27 fixed points, displayed in figure~\ref{fig:Z3Geometry}. This result can be
verified by using formula~\eqref{eq:SimpleLefschetz}.

The fixed points of the $\vartheta^2$ twisted sector are identical to those of the first twisted
sector. Therefore, there is no need to consider these two sectors separately. We will see in
section~\ref{sec:HeteroticSpectrum} that this structure will be reflected in the matter spectrum of the
orbifold. In any orbifold, one can show that the fixed point structure of the sector $\theta$ is
equal to that of the sector $\theta^{-1}$. Note that in the \Z3 case $\vartheta^2~=~\vartheta^{-1}$.

\subsubsection{The \bs{\Z6}-II Orbifold}
\index{Z6-II geometry@\Z6-II geometry}

\begin{figure}[t!]
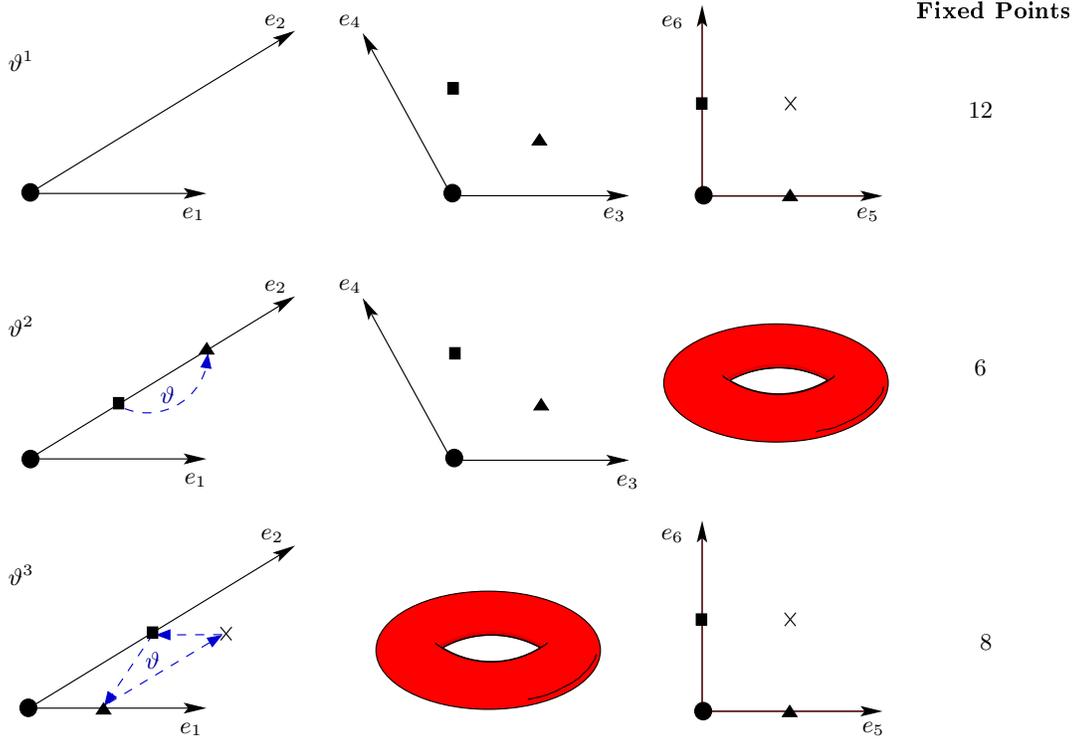

\centerline{\input Z6II_geometry.pstex_t}
\caption{The geometry of \Z6-II orbifolds compactified on a $\G2\times\SU3\times\SO4$ lattice. Fixed
points of the $\vartheta^1$, $\vartheta^2$ and $\vartheta^3$ twisted sectors are presented. The fixed
points of the $\vartheta^4$ and $\vartheta^5$ twisted sectors are identical to those of the $\vartheta^1$
and $\vartheta^2$ twisted sectors, respectively.}
\label{fig:Z6IIGeometry}
\end{figure}

The \Z6-II orbifold was first studied in detail in ref.~\cite{Kobayashi:2004ud,Kobayashi:2004ya}. In those works,
the structure of the fixed points was suggested as a tool to get models with phenomenologically
acceptable features. Since the present thesis is based on the \Z6-II orbifold, we discuss in detail the
structure of its fixed points. We illustrate our results in figure~\ref{fig:Z6IIGeometry}.

We will consider the compact space of the \Z6-II orbifold to be spanned by the factorizable
lattice\footnote{\Z6-II orbifolds on nonfactorizable lattices can also lead to interesting results. Some
 of their properties are briefly discussed in appendix~\ref{ch:Z6IINonFactorizable}.} 
\be
\Gamma_{\Z6-II} ~=~ \G2\times\SU3\times\SO4\,.
\label{eq:Z6IILattices}
\ee
An advantage of the lattice $\Gamma_{\Z6-II}$ being factorizable, is that we can find the fixed points
independently for each sublattice without loss of information, and then put them all together
in order to obtain the entire fixed point structure.

The point group \Z6-II is generated by $\vartheta$ which acts simultaneously as a rotation by $2\pi/6$ on
the \G2 plane, a rotation by $2\pi/3$ on the \SU3 plane, and a reflection on the origin of the \SO4
sublattice. The action of $\vartheta$ is then described by the twist vector 
\be
v~=~(0,\,1/6,\,1/3,\,-1/2)\,.
\label{eq:Z6IITwist}
\ee
There are five twisted sectors corresponding to the different powers of $\vartheta$.

Let us consider the first twisted sector. Formula~\eqref{eq:SimpleLefschetz} tells us that there are 12
fixed points. Their precise location in the compact space can however only be found by means of
eq.~\eqref{eq:FixedPoints}. The constructing elements corresponding to the fixed points are given by
\be
g_f^{\vartheta}~\in~\big\{(\vartheta,\, n_5 e_5 + n_6 e_6),\,(\vartheta,\,e_3+ n_5 e_5 + n_6 e_6),\,(\vartheta,\,e_3+e_4+ n_5 e_5 + n_6 e_6)\big\}
\label{eq:Z6IIFixedPointsT1}
\ee 
with $n_5,n_6=0,1$. 

The action of $\vartheta^2$ is encoded in $2 v = (0,\,1/3,\,2/3,\,-1)$. This means particularly that all
points of the \SO4 sublattice are left invariant. Therefore, the action of $\vartheta^2$ on the compact space
introduces fixed tori. In this case, formula~\eqref{eq:SimpleLefschetz} does not count correctly the
number of fixed points. However, from previous discussions we know that, under a rotation of $2\times
2\pi/3$ on the \SU3 sublattice, three points are left fixed. Further, we have also seen that there are
only two inequivalent fixed points in the \G2 sublattice. All in all, we find $2\times 3$ fixed tori in
the second twisted sector. The constructing elements are
\be
g_f^{\vartheta^2}~\in~\big\{(\vartheta^2,\,0),\,(\vartheta^2,\,e_1),\,(\vartheta^2,\,e_4),\,(\vartheta^2,\,e_1+e_4),\,
              (\vartheta^2,\,e_3+e_4),\,(\vartheta^2,\,e_1+e_3+e_4)\big\}\,.
\label{eq:Z6IIFixedPointsT2}
\ee

In the third twisted sector, the point group action, described by $3 v = (0,\,1/2,\,1,\,-3/2)$, acts
trivially in the \SU3 sublattice, hence, we obtain fixed tori in this case too. It is easy to see that
eq.~\eqref{eq:FixedPoints} leads to four invariant points under $\vartheta^3$ in each of the other two
sublattices. However, the points out of the origin of the \G2 sublattice are equivalent. Let us denote
the associated space group elements by $g_1=(\vartheta^3,\,e_1)$, $g_2=(\vartheta^3,\,e_1+e_2)$ and
$g_3=(\vartheta^3,\,e_2)$. It is easy to verify that
\be
g_1\backsimeq (\vartheta,\,e_1)\ g_1\ (\vartheta,\,e_1)^{-1}=g_2\quad\text{ and }\quad g_2\backsimeq
(\vartheta,\,2e_1-e_2)\ g_2\ (\vartheta,\,2e_1+e_2)^{-1}=g_3\,, 
\ee
whence it follows that all three elements belong to the same conjugacy class. We are then
left with a total of $2\times 4$ fixed tori, described by the following constructing elements:
\be
g_f^{\vartheta^3}~\in~\big\{(\vartheta^3,\, n_5 e_5 + n_6 e_6),\,(\vartheta^3,\,e_2+ n_5 e_5 + n_6 e_6)\big\}
\label{eq:Z6IIFixedPointsT3}
\ee 
with $n_5,n_6=0,1$. 

The fourth and fifth twisted sectors posses the structure of the second and first twisted sectors,
respectively. Therefore, it is enough to study the three twisted sectors depicted in figure~\ref{fig:Z6IIGeometry}.

\section{Strings on Heterotic Orbifolds}
\label{sec:HeteroticSpectrum}
\index{orbifold spectrum}

\begin{figure}[t!]
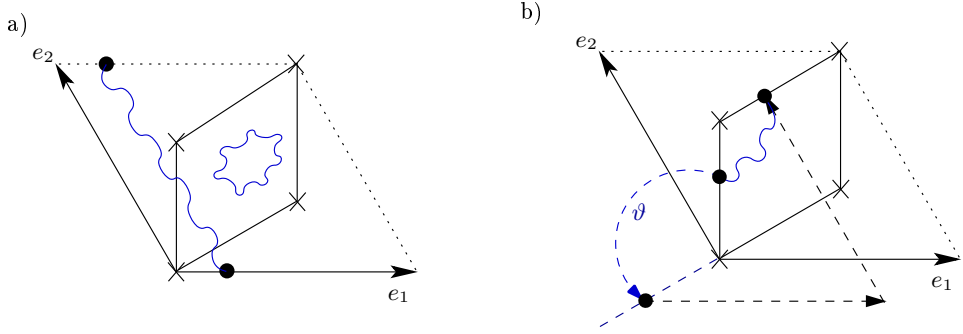

\centerline{\input StringTypes.pstex_t}
\caption{Closed strings in orbifold compactifications. a) Untwisted strings are closed on the torus,
  whereas b) twisted strings only close through the action of the twist $\vartheta$.}
\label{fig:StringTypes}
\end{figure}

\index{untwisted strings}\index{twisted strings}
The final ingredient of orbifold compactifications is their spectrum of matter. Matter in heterotic orbifolds
is described by closed strings. A special feature of orbifold compactifications is that they admit two
types of closed strings: {\it untwisted} and {\it twisted} strings (see
figure~\ref{fig:StringTypes}). Untwisted strings fulfill the following boundary conditions:
\be
Z(\tau,\,\sigma + \pi) = Z(\tau,\,\sigma) +n_\al e_\al\,. 
\label{eq:UntwistedBC}
\ee
This indicates that they are closed already on the torus and free to propagate in the compact space. 
In contrast, twisted strings are closed only after identifying points by the twist $\theta$, according to
their boundary conditions 
\be
Z(\tau,\,\sigma + \pi) = \theta Z(\tau,\,\sigma) +n_\al e_\al\,,
\label{eq:TwistedBC}
\ee
whence it follows that they are bound to the fixed points. The relation between the types of strings in
an orbifold and the action of the point group is the reason why we have called untwisted to the space
group elements with a trivial point group action, and twisted to those other which include a nontrivial
element of the point group.

Since massive strings have masses of the order of $M_{str}$, they are too heavy to contribute directly to
low--energy physics. Therefore, we will consider the spectrum of heterotic orbifolds to be composed only of
massless string states, satisfying additionally the level matching condition for right- and left-movers,
eq.~\eqref{eq:levelcond}. 

Massless states differ for the several constructing elements $g$. On the one hand, untwisted states
(with constructing element $g=(\id,\,0)$) are written in general as 
\be
\ket{q}_R \otimes \ti\al_{-1}^x \ket{p}_L\,,
\label{eq:UntwistedStates}
\ee
where $q$ is a weight of \SO8 and $p\in\Lambda$ as given in eq.~\eqref{eq:ps}. $\ti\al_{-1}^x$ represents an
oscillator excitation in one direction of either the Minkowski space ($\ti\al^\mu_{-1}$) or the compact
space ($\ti\al^a_{-1},\ti\al^{\bar{a}}_{-1}$) or the gauge degrees of freedom ($\ti\al^I_{-1}$).

On the other hand, massless states associated to a twisted constructing element $g=(\vartheta^k,\,n_\al e_\al)$
(with $k=1,\ldots,N-1$) are in general expressed as  
\be
\ket{q_{\text{sh}}}\otimes\ti\al\ \ket{p_{\text{sh}}}\equiv\ket{q + v_g}_R \otimes \ti\al\ \ket{p + V_g}_L\,,
\label{eq:TwistedStates}
\ee
where $\ti\al$ denotes in this case a product of oscillators of the form
$\ti\al^a_{-\eta^a}$ (or $\ti\al^{\bar{a}}_{-1+\eta^a}$) in the complex directions $a=1,2,3$ (or their
conjugates $\bar{a}=\bar{1},\bar{2},\bar{3}$), with $\eta^a = k\,v^a\mod 1$, such that $0<\eta^a\leq 1$. 
Here we also defined the {\it local twist} and the {\it local shift} vectors associated to the
constructing element $g$ as \index{local twist vector}\index{local twist vector}
\be\ba{rl}
 v_g =& k\, v \,\quad\quad\text{ and }\\[1mm]
 V_g =& k\, V +  n_\al\, A_\al\,,
\ea
\label{eq:LocalTwistAndShift}
\ee
respectively. One can trivially extend these
results to $\Z{N}\times\Z{M}$ orbifolds  by including a second twist vector $v_2$ and a second shift
vector $V_2$ in eq.~\eqref{eq:LocalTwistAndShift}.

\subsubsection{Orbifold Projections}
As we have already mentioned, the matter spectrum of orbifolds will be composed only of those massless states
which are invariant under the action of the space group $S$ and its counterpart, the gauge twisting group
$G$, that is, under the orbifold group $O\subset S\otimes G$. The problem here is to find a prescription to figure
out whether a massless state is projected out by the orbifold action. Let us make some general observations.

\index{orbifold spectrum!Hilbert spaces}
The boundary conditions for closed strings on orbifolds, eqs.~\eqref{eq:UntwistedBC}
and~\eqref{eq:TwistedBC}, are summarized by
\be
Z(\tau,\,\sigma + \pi) = g Z(\tau,\,\sigma)
\label{eq:AllBC_Z}
\ee
for a constructing element $g$. The set of all states with compact coordinates fulfilling eq.~\eqref{eq:AllBC_Z}
define a {\it Hilbert space} $\maH_g$. 

To ensure compatibility of the states from $\maH_g$ with the orbifold, we let an arbitrary space group
element $h\in S$ act on the coordinates describing the strings. The boundary condition becomes
\be\ba{rl}
(hZ)(\tau,\,\sigma + \pi) &= hgZ(\tau,\,\sigma)\\
                          &= hgh^{-1}(hZ)(\tau,\,\sigma)\,,
\ea
\label{eq:AllBC_hZ}
\ee
where we have made use of eq.~\eqref{eq:AllBC_Z}. To interpret this, we have to distinguish between two
cases: a) $h$ commutes with $g$, and b) $h$ does not commute with $g$.

{\bf a)} Let us consider first that $h$ commutes with $g$, i.e. 
\be
[g,\,h] =0\,.
\ee
In this case, eq.~\eqref{eq:AllBC_hZ} translates to
\be
(hZ)(\tau,\,\sigma + \pi) = g (hZ)(\tau,\,\sigma)
\label{eq:AllBC_CommutinghZ}
\ee
This boundary condition indicates that states described by the coordinates $h\ Z$ belong to the same Hilbert
space $\maH_g$. Consequently, $h$ must act trivially on the states from $\maH_g$:\index{orbifold projection phase}
\be
\label{eq:ProjectionCondition}
\ket{q_\text{sh}}_R \otimes \ti\al\ \ket{p_\text{sh}}_L
~\stackrel{h}{\longmapsto}~
\Phi\, \ket{q_\text{sh}}_R \otimes \ti\al\ \ket{p_\text{sh}}_L
~\stackrel{!}{=}~ \ket{q_\text{sh}}_R \otimes \ti\al\ \ket{p_\text{sh}}_L\,.
\ee
States from $\maH_g$ that do not fulfill eq.~\eqref{eq:ProjectionCondition} have to be projected out.

\index{centralizer}
Let us define the {\it centralizer} $\calZ_g$ of a constructing element $g$ as the set of all space group elements $h$
commuting with $g$:
\be
\calZ_g ~=~ \left\{ h\in S\ |\ [g,\, h]=0 \right\}\,.
\ee
The massless matter spectrum of orbifold compactifications is then formed by the massless states which
are invariant under all elements of the centralizer.

{\bf b)} Consider now a noncommuting space group element $h$,
\be
[g,\,h] \neq 0\,.
\ee
In this case, eq.~\eqref{eq:AllBC_hZ} indicates that $h$ maps states from a given Hilbert space $\maH_g$
onto a different Hilbert space $\maH_{h\,g\,h^{-1}}$. Subsequent application of $h$ then leads to the
sequence 
\begin{equation}
\mathcal{H}_g ~ \xrightarrow{h}~ \mathcal{H}_{h\,g\,h^{-1}}~ \xrightarrow{h}~ \mathcal{H}_{h^2\,g\,h^{-2}}~ \xrightarrow{h}~
\mathcal{H}_{h^3\,g\,h^{-3}} ~ \xrightarrow{h}~ \ldots\;.
\end{equation}
The crucial point is now that, since $g$ and $hgh^{-1}$ belong to the same conjugacy class, $h\,Z$ and
$Z$ are identified on the orbifold. This means that, on the orbifold, the different Hilbert spaces
$\mathcal{H}_{h^n\,g\,h^{-n}}$ are to be combined into a single orbifold Hilbert space. Invariant states
are then linear combinations of states from all $\mathcal{H}_{h^n\,g\,h^{-n}}$. Such linear combinations
involve, in general, relative phase factors (often called \emph{$\gamma$--phase})~\cite{Lebedev:2007hv}.

Let us emphasize here that the action of noncommuting space group elements on physical states of $\maH_g$
{\it does not project out} any state from the spectrum.

\subsection{Untwisted Sector}
\label{subsec:UntwistedSector}
In the untwisted sector $U$,
the level matching condition for massless states of orbifolds coincides with that of the uncompactified heterotic
string, that is,\index{masslessness condition, orbifold!untwisted sector}
\be
\label{eq:MasslessnessUntwisted}
 \h{m_L^2}{4} =\2p^2 + \ti N  - 1 = 0 = \2q^2-\2=\h{m_R^2}{4}\,,
\ee
where $p$ is a root of either $\E8\times\E8$ or \SO{32} (see eq.~\eqref{eq:ps}), $q$ denotes the \SO8
weight vector of the right-mover and $\ti N$ counts the number of oscillator excitations. As in the
uncompactified heterotic string, eq.~\eqref{eq:MasslessnessUntwisted} has solution only if $q^2 = 1$ for
the right-movers and either $p^2=0,\,\ti N=1$ or $p^2=2,\,\ti N=0$ for the left-movers.
The spectrum of the untwisted sector, nevertheless, is different from that
of the heterotic string discussed in section~\ref{sec:HeteroticString}. The reason being that
some states are projected out by the action of elements of the centralizer.

\subsubsection{Transformation Phase}
\index{transformation phase!untwisted sector}
\index{orbifold projection phase!untwisted sector}

In the projection condition, eq.~\eqref{eq:ProjectionCondition}, we have considered that massless states
of $\maH_g$ acquire a phase $\Phi$ under the action of an arbitrary element $h$ of the centralizer
$\calZ_g$. In fact, every element of the massless states transform differently under the action of an
element $h$ of the centralizer. Embedding $h$ into the gauge degrees of freedom shifts the bosonic
coordinates $X^I$ by $V_h^I$, where $V_h$ is the local shift vector of $h$ (see
eq.~\eqref{eq:LocalTwistAndShift}). In the momentum space, this accounts for a phase that depends on the
momentum $p$ of the state and the local shift vector $V_h$:
\be
\ket{p}_L ~\stackrel{h}{\longmapsto}~ e^{2\pi\I p\cdot V_h} \ket{p}_L\;.
\label{eq:pPhase}
\ee
The action of $h$ provides the right-moving states with a similar phase
\be
\ket{q}_R ~\stackrel{h}{\longmapsto}~ e^{2\pi\I q\cdot v_h} \ket{q}_R\,,
\label{eq:qPhase}
\ee
where $v_h$ is the local twist vector associated to $h$. Finally, oscillators are transformed as
\be\ba{rl}
\ti\al^I_{-1}        &\stackrel{h}{\longmapsto}~  \ti\al^I_{-1}\,,\qquad\qquad I=1,\ldots,16\,,\\
\ti\al^\mu_{-1}      &\stackrel{h}{\longmapsto}~  \ti\al^\mu_{-1}\,,\qquad\qquad \mu=2,\,3\,,\\
\ti\al^a_{-1}        &\stackrel{h}{\longmapsto}~ e^{2\pi\I  v_h^a}\;\ti\al^a_{-1}\,,\\
\ti\al^{\bar{a}}_{-1}&\stackrel{h}{\longmapsto}~ e^{-2\pi\I  v_h^a}\;\ti\al^{\bar{a}}_{-1}\,.
\ea
\label{eq:OscillatorPhase}
\ee

Left- and right-moving momenta lie on an even, self-dual lattice of lorentzian signature
(22,6)~\cite{Narain:1985jj}, implying a relative sign between the phases of the left- and right
movers. Therefore, the phase acquired by untwisted massless states under the action of $h$ is given by 
\be\ba{rl}
\ket{q}_R \otimes \ket{p}_L :&\quad \Phi~=~ e^{2\pi\I \left[p\cdot V_h\,-\,q\cdot v_h \right]}\\
\ket{q}_R \otimes \ti\al_{-1}^x\ket{0}_L :&\quad \Phi~=~ e^{2\pi\I \left[-\,q\cdot v_h +(\delta_{x,a}-\delta_{x,\bar{a}})v_h^a\right]}\,,
\ea
\label{eq:UntwistedTrafo}
\ee
where $x$ stands for the direction in which the oscillator excitations acts.

\subsubsection{Untwisted Spectrum}
\index{orbifold spectrum!untwisted sector}
The constructing element of the untwisted sector is $g=(\id,\,0)$ and, therefore, the associated
centralizer contains all elements of the space group. The spectrum of the untwisted sector
is composed by those massless states of the uncompactified heterotic string which are invariant
under all elements of the space group. In the following, we evaluate the effect of the orbifold
projection on the massless spectrum of the heterotic string, discussed in section~\ref{sec:HeteroticString}.

The {\bf ten-dimensional supergravity multiplet} of the heterotic string splits into:\index{model-independent axion}\index{aMI@$a_{MI}$}
\begin{itemize}\index{dilaton}
\item a {\bf four-dimensional graviton $\bs{g^{\mu\nu}}$, dilaton $\bs{\varphi}$, antisymmetric tensor
 $\bf{B^{\mu\nu}}$} (whose dual is the model-independent axion $a_{MI}$), and their superpartners.
They are given by the space-group-invariant components of
\be
\ket{q}_R\otimes \al_{-1}^\nu\ket{0}_L\,.
\label{eq:4DSUGRAmultiplet}
\ee
Since the left-movers $\al_{-1}^\nu\ket{0}_L$ do not transform under any $h$, then the right-movers
$\ket{q}_R$ must transform trivially too. From eq.~\eqref{eq:qPhase}, we see that this occurs only if the
six-dimensional momenta $q$ in eq.~\eqref{eq:4DSUGRAmultiplet} correspond to\footnote{The SO(8) weights $q$
in eq.~\eqref{eq:q4DSUGRAmultiplet} carry implicitly minkowskian index $\mu$.} 
\be
q=\left\{
\ba{l}
\pm\left(\2,\,\2,\,\2,\,\2\right)\\
\pm\left(1,\,0,\,0,\,0\right)\,.
\ea
\right.
\label{eq:q4DSUGRAmultiplet}
\ee
This is a trivial consequence of requiring to preserve $\maN=1$ {\sc susy} in orbifold compactifications
(cf. section~\ref{subsec:ConsistencyConditions});

\item some {\bf geometrical moduli} given by states of the type
\be
\ket{q}_R\otimes\ti\al_{-1}^{a/\bar{a}}\;\ket{0}_L
\label{eq:4DModuli}
\ee
satisfying the invariance condition
\be
q\cdot v_h \pm v_h^a = 0 \mod 1\,,
\label{eq:InvarianceOf4DModuli}
\ee
where the relative sign $-$ ($+$) is associated with an oscillator carrying holomorphic index $a$
(antiholomorphic index $\bar{a}$). These states are gauge singlet fields, arising from the compact
components of the ten-dimensional graviton and antisymmetric tensor (and their superpartners). In
particular, the symmetric combinations are the moduli for the flat metric of the compact space, which can
be written as\index{model-dependent axion}\index{aMD@$a_{MD}$}
\be
G_{\al\beta}~=~e_\al\cdot e_\beta\,,
\label{eq:6DCompactMetric}
\ee
where $e_\al$ correspond to the basis vectors of the compact space. The surviving components of the
antisymmetric tensor give rise to the so-called model-dependent axions $a_{MD}$. %Note that, additionally
%to those states that do not satisfy the invariance condition eq.~\eqref{eq:InvarianceOf4DModuli}, also
%states such as $\ket{q}_R\otimes\ti\al_{-1}^{\nu/I}$ are projected out by the orbifold action.
\end{itemize}

The action of $h$ on the 16 {\bf ten-dimensional uncharged gauge bosons} leaves invariant only their
four-dimensional components specified by 
\be
\ket{q}_R\otimes\ti\al^I_{-1}\ket{0}_L
\label{eq:4DCartans}
\ee
with the right-mover momenta $q$ also given by eq.~\eqref{eq:q4DSUGRAmultiplet} due to the
invariance of $\ti\al^I_{-1}$. These states are the {\bf 16 Cartan generators of} the four-dimensional
gauge group $\bs{\maG_{4D}}$. Therefore, the rank of the gauge group cannot be reduced by compactifying
on this kind of orbifolds.\footnote{It is possible to reduce the rank of the algebra by embedding the
point group generator $\theta$ into the gauge degrees of freedom as a rotation $\Theta$ instead of a
shift vector $V$.}

The {\bf 480 charged gauge bosons} are of the form $\ket{q}_R\otimes\ket{p}_L$. Those states left
invariant under the action of the space group acquire different natures depending on their transformation
properties: 
\begin{itemize}
\item the {\bf charged gauge bosons} (and gauginos) of the four-dimensional gauge group $\maG_{4D}$ are
those states where both left- and right-movers transform trivially under the action of any element $h$ of
the space group, that is, where $q\cdot v_h=0\mod 1$ and $p\cdot V_h=0\mod 1$, independently. The only
right-moving momenta satisfying the former constraint are those provided in
eq.~\eqref{eq:q4DSUGRAmultiplet}, which are also the right-movers of the Cartan generators of the
unbroken gauge group. The condition for the left-moving momenta must be fulfilled for any $h$, then it
can be restated neatly as 
\bse\label{eq:4DChargedBosonsCond}
\bea
p\cdot V &=&0\mod 1\,,\qquad\text{(for $V_1$ and $V_2$ in $\Z{N}\times\Z{M}$ orbifolds)} \label{eq:4DChargedBosonsCondA}\\
p\cdot A_\al &=&0\mod 1\,,\qquad\al=1,\ldots,6\,, \label{eq:4DChargedBosonsCondB}
\eea
\ese
where $A_\al$ are Wilson lines. These states transform in the adjoint representation of $\maG_{4D}$.
Provided that not all 480 left-moving momenta $p$ of the
ten-dimensional charged bosons satisfy eqs.~\eqref{eq:4DChargedBosonsCond}, even though the rank is not
reduced, the gauge symmetry can be broken; 

\item those states whose left- and right-moving components transform nontrivially and satisfy the
invariance condition
\be
p\cdot V_h - q\cdot v_h = 0\mod 1
\label{eq:4DUntwistedMatterCond}
\ee
constitute the so-called {\bf untwisted charged matter} of orbifold models. The various matter states
form several {\sc susy} chiral-multiplets. Their gauge transformation properties with respect to
$\maG_{4D}$ depend on their momenta $p$.
\end{itemize}

\subsection{Twisted Sectors}
\label{subsec:TwistedSectors}
\index{orbifold spectrum!twisted sectors}
Zero modes of twisted sectors $T_{k(,\ell)}$ are associated to the constructing elements $g$ of the fixed
points. Requiring the states to be massless accounts for the following conditions on the left- and
right-moving momenta:\index{masslessness condition, orbifold!twisted sectors}
\be\ba{rl}
\h14 m_L^2 =& \2p^2_\text{sh} + \ti N  - 1 + \delta c \stackrel{!}{=} 0\,,\\[2mm]
\h14 m_R^2 =& \2q^2_\text{sh}-\2 + \delta c \stackrel{!}{=} 0\,,
\ea
\label{eq:MasslessnessTwisted}
\ee
where $\delta c$ corresponds to a change in the zero point energy related to the appearance of twisted
oscillators $\ti\al^{a}_{-\eta^a}$, $\ti\al^{\bar a}_{-1+\eta^a}$. It is expressed by
\be
\delta c = \2\sum_{a} \eta^a(1-\eta^a)\;
\label{eq:DefDeltaC}
\ee
with $\eta^a=v_g^a\mod 1$, such that $0<\eta^a \leq 1$. For massless states, one can write the twisted
(fractional) oscillator number $\ti N$ as
\be
\label{eq:OscillatorNumber}
\ti N= \sum_{a=1}^3 \eta^a \ti N^a_g + \bar\eta^a\ti N^{*a}_g\,,
\ee
Here, $\bar\eta^a=-v_g^a\mod 1$ such that $0<\bar\eta^a \leq 1$, and
$\ti N^a_g$ and $\ti N^{*a}_g$ are integer oscillator numbers, counting respectively the number of
excitations in the holomorphic $a$ and antiholomorphic $\bar{a}$ directions.

\subsubsection{Transformation Phase}
\index{transformation phase!twisted sectors}
\index{orbifold projection phase!twisted sectors}

The transformation of left- and right-moving states $\ket{p_\text{sh}}_L,\,\ket{q_\text{sh}}_R$ under the
action of an arbitrary centralizer element $h$ can also be read off from eqs.~\eqref{eq:pPhase}
and~\eqref{eq:qPhase}, where we have only to substitute $p$ for $p_\text{sh}$ and $q$ for $q_\text{sh}$.

Further, just as in the untwisted sector, only the oscillator excitations on the compact directions $a$
and $\bar{a}$ transform nontrivially. Their transformations are given by
\be\ba{rl}
\ti\al^a_{-\eta^a}        &\stackrel{h}{\longmapsto}~ e^{2\pi\I  v_h^a}\;\ti\al^a_{-\eta^a}\\
\ti\al^{\bar{a}}_{-1+\eta^a}&\stackrel{h}{\longmapsto}~ e^{-2\pi\I  v_h^a}\;\ti\al^{\bar{a}}_{-1+\eta^a}
\ea
\label{eq:TwistedOscillatorPhase}
\ee

Putting everything together, the complete transformation phase of a massless twisted states reads
\begin{equation}\label{eq:TwistedTrafo}
 \Phi ~=~  e^{2\pi\I\,[p_\text{sh}\cdot V_h- q_\text{sh} \cdot v_h
 + (\ti{N}_g - \ti{N}_g^*)\cdot v_h]}\, \Phi_\mathrm{vac}\;,
\end{equation}
where the vacuum phase
\begin{equation}
\Phi_\mathrm{vac} ~=~ e^{2\pi\I\,[\,-\2(V_g\cdot V_h - v_g\cdot v_h)\,]}\;
\label{eq:VacuumPhase}
\end{equation}
arises as consequence of the geometrical properties of twisted strings (cf. appendix of
ref.~\cite{Ploger:2007iq}).

\subsection[A \Z3 Example]{A $\bs{\Z3}$ Example}
\label{subsec:Z3Example}
\index{Z3 spectrum@\Z3 spectrum, example|(}

The model studied here was presented in refs.~\cite{Ibanez:1987sn,Casas:1988se}. In those works, the
$\E8\times\E8$ heterotic string was compactified on a \Z3 orbifold with abelian embedding and two Wilson
lines. Here, let us first study the spectrum of the model in the absence of Wilson lines and then
consider the effect of these background fields. 

The embedding of the point group into the gauge degrees of freedom is chosen to be given by the shift
vector 
\be
V~=~\left(\,\h13^4,\,\h23,\,0^3\,\right)\left(\,\h23,\,0^7\,\right)\,,
\label{eq:Z3ExampleShift}
\ee
which, together with the twist vector of a \Z3 orbifold, $v=(0,\,1/3,\,1/3,\,-2/3)$,
satisfies the modular invariance condition, eq.~\eqref{eq:WeakModInvZNwoWilson}. 

\subsubsection{Untwisted Sector}
\index{Z3 spectrum2@ !without Wilson lines}

Aside from the fields $G^{\mu\nu}$, $B^{\mu\nu}$ and $\varphi$ (and their superpartners), we have nine
geometrical moduli of the type
\be
\ket{q}_R\otimes\ti\al^{\bar{a}}\ket{0}_L\,.
\label{eq:Z34DModuli}
\ee
They arise as follows. In the absence of Wilson lines, the transformation properties of the states in the
orbifold spectrum are determined by the action of the point group generated by $\theta$. Therefore,
denoting $\varrho=e^{2\pi\I/3}$, we can classify the right-movers according to their
eigenvalues with respect to the space group element $h=(\vartheta,\,0)$:
\be\ba{rll}
\varrho^0: &\ket{q}_R \ \text{ with }&
 q = \pm(\phantom{-}\2,\,\phantom{-}\2,\,\phantom{-}\2,\,\phantom{-}\2),\;\pm(1,\,\phantom{-}0,\,0,\,0)\,,\\[2mm]
\varrho^1: &\ket{q}_R \ \text{ with }&
 q = \phantom{-}(\phantom{-}\2,\,\phantom{-}\underline{\2,\,-\2,\,-\2}),\;\phantom{-}(0,\,\underline{\phantom{-}1,\,0,\,0})\,,\\[2mm]
\varrho^2: &\ket{q}_R \ \text{ with }&
 q = \phantom{-}(-\2,\,\underline{-\2,\,\phantom{-}\2,\,\phantom{-}\2}),\;\phantom{-}(0,\,\underline{-1,\,0,\,0})\,,
\ea 
\label{eq:qTrafos}
\ee
where, as usual, the underscore denotes all permutations. On the other hand, left-moving oscillators with
internal indices also transform:
\be\ba{rl}
\varrho^1: & \ti\al_{-1}^a\,,\\
\varrho^2: & \ti\al_{-1}^{\bar{a}}\,.
\ea
\label{eq:oscTrafos}
\ee
Therefore, for each complex direction $a$ we obtain six $(\varrho^1)_R(\varrho^2)_L$ invariant states
and other six with $(\varrho^2)_R(\varrho^1)_L$. However, note that the former states are conjugate to the latter
ones. Together, they enter into a single physical state, so that it is enough to count {\it left-chiral}
states, i.e. states where the Ramond (half-integer) \SO8 weight has $q^0=-\2$.\footnote{Note that the
designation of chirality is arbitrary.} In the present example, we count only the six
$(\varrho^2)_R(\varrho^1)_L$ invariant states. Furthermore, since {\sc susy} is preserved, a physical
state of the spectrum must contain a Ramond weight along with an Neveu-Schwarz (integer) weight. In this
way, we come up to three invariant states for each of the three complex directions, as given in
eq.~\eqref{eq:Z34DModuli}. Combinations of these nine moduli correspond to the nine independent
deformation parameters of the compact space of a \Z3 orbifold~\cite{Casas:1991ac}.

The unbroken gauge group has rank sixteen because all Cartan generators given in eq.~\eqref{eq:4DCartans}
are invariant. The explicit breaking of $\E8\times\E8$ is obtained through the left-moving momenta $p$
(i.e. the roots of $\E8\times\E8$ in eq.~\eqref{eq:psE8})
satisfying $p\cdot V=0\mod 1$. There are 72 invariant momenta $p$ from the first \E8 and 84 from the second
\E8 factor. Adding to this number the eight Cartan generators of each \E8 provides the dimensionality of
the adjoint of the unbroken gauge group. In this case $72+8\rightarrow\bs{80}$ corresponds to the adjoint of \SU9
and $84+8\rightarrow\bs{91}\oplus\bs{1}$ correspond to the adjoint of $\SO{14}\times\U1$. The
corresponding momenta $p$ constitute the roots of the unbroken gauge group. The left-moving states
with these momenta $\ket{p}_L$ (transforming as $\varrho^0$) tensor together with the invariant
($\varrho^0$) right-movers of eq.~\eqref{eq:qTrafos}. Since the number of left-chiral states gives the
multiplicity of states, we find that these gauge bosons appear, as expected, only once. Therefore, the orbifold
action with the shift vector eq.~\eqref{eq:Z3ExampleShift} breaks the gauge group as
\be
\maG=\E8\times\E8\quad\longrightarrow\quad \maG_{4D}=\SU9\times\SO{14}\times\U1\,,
\label{eq:Z3ExampleGG}
\ee
where the \U1 generator is given by the vector
\be
\mathsf{t}~=~\left(\,0^8\,\right)\,\left(\,-18,\,0^7\,\right)\,.
\label{eq:Z3ExampleNoWLU1}
\ee

More precisely, in order to determine the unbroken gauge group, one has to take the $72+84$ invariant
roots $p$ and chose a basis that fixes a semiordering in the weight space~\cite{Cahn:1985wk}. From the
semiordering, one can find the simple roots $\al_i$, defined as those roots which are
positive and cannot be written as the sum of two positive roots. Provided that the roots have
squared length 2, one can then compute the Cartan matrix $A_{ij} = \al_i\cdot\al_j$ for each gauge group
factor $G_a$ (e.g. for \SU9), which describes uniquely the algebra. A more detailed discussion on how to
obtain the unbroken gauge group in orbifold compactifications can be found in section 3.2 of
ref.~\cite{Vaudrevange:2004xx}.  

The untwisted charged matter is formed by tensoring together left-moving $\ket{p}_L$ and
right-moving $\ket{q}_L$ states which, separately, acquire nontrivial transformation phases under the
action of $h=(\vartheta,\,0)$. There are $2\,[84 +(14 + 64)]$ momenta $p$ from
eq.~\eqref{eq:psE8} which lead to nontrivial phases for the left-moving states: $84 +(14 + 64)$
of them transform with a phase $\varrho^1$ and the remaining $84 +(14 + 64)$ with $\varrho^2$. Moreover, 
$2\times 84$ momenta come from the first \E8 whereas $2\times(14+64)$ come from the second one. 
The gauge properties of the states are encoded in the momenta $p$. The nonabelian gauge quantum
numbers with respect to $\maG_{4D}$, eq.~\eqref{eq:Z3ExampleGG}, are obtained by rewriting the
momenta in Dynkin labels:
\be
p\;\rightarrow\;p_{DL(G_a)}=(\al_1\cdot p,\ldots, \al_n\cdot p )\quad\text{ for each }G_a\,,
\label{eq:DynkinLabels}
\ee
where $\al_i$ are the roots of the gauge factor $G_a\subset \maG_{4D}$ of rank $n$. Comparing the momenta
in Dynkin labels $p_{DL}$ with the results in the tables of ref.~\cite{Slansky:1981yr}, one identifies
the representations. In our example, we summarize the properties of the left-moving states $\ket{p}_L$ by
\be\ba{rl}
\ket{p}_L\;\T{ with }\;\varrho^1 \quad &\sim (\bsb{84},\,\bs{1})_0\oplus(\bs{1},\,\bs{14})_{+18}\oplus(\bs{1},\,\bsb{64})_{-9}\,,\\[2mm]
\ket{p}_L\;\T{ with }\;\varrho^2 \quad &\sim (\bs{84},\,\bs{1})_0\oplus(\bs{1},\,\bs{14})_{-18}\oplus(\bs{1},\,\bs{64})_{+9}\,.
\ea
\label{eq:Z3ExampleNoWLNontrivialp}
\ee
The \U1 charge, denoted by the subscript, is given by $t\cdot p$, with $t$ representing the \U1 generator
given in eq.~\eqref{eq:Z3ExampleNoWLU1}. Finally, the physical states are the result of combining 
the left-movers described in eq.~\eqref{eq:Z3ExampleNoWLNontrivialp} with the right movers given in
eq.~\eqref{eq:qTrafos}, which provide the states with a multiplicity factor of three. Again, one finds
that $(\varrho^1)_R(\varrho^2)_L$ left-chiral whereas $(\varrho^2)_R(\varrho^1)_L$ are right-chiral
invariant states, thus they enter together into physical states. The untwisted charged matter spectrum is
\begin{center}
\begin{tabular}{|c|}
\hline
\qquad$3(\bs{84},\,\bs{1})_0\oplus 3(\bs{1},\,\bs{14})_{-18}\oplus 3(\bs{1},\,\bs{64})_{+9}\phantom{^{A^{A^A}}}$\\[2mm]
\hline
\end{tabular}
\end{center}

\subsubsection{Twisted Sectors}
\begin{figure}[t!]
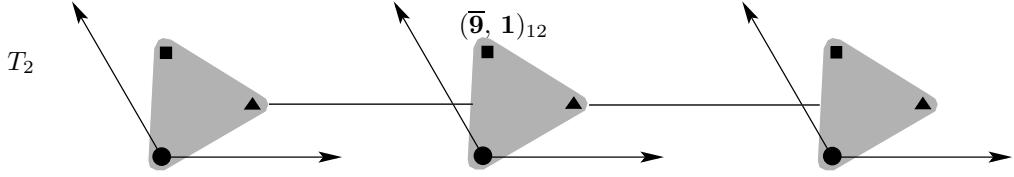

\centerline{\input Z3_example_noWL.pstex_t}
\caption{In the absence of Wilson lines, all fixed points are degenerate. The twisted matter content
  correspond then to 27 copies of $(\bsb{9},\,\bs{1})_{12}$.}
\label{fig:Z3ExampleNoWL}
\end{figure}

As discussed in section~\ref{subsec:OrbifoldGeometry}, the \Z3 orbifold has 27 fixed points on each of
its two twisted sectors, labeled $T_1$ and $T_2$ for the action of $\vartheta^1$ and $\vartheta^2$,
respectively. Twisted states are attached to the fixed points. They are described by the momenta
satisfying the masslessness conditions eq.~\eqref{eq:MasslessnessTwisted}, $q_\text{sh}= q + v_g$ and
$p_\text{sh}=p+V_g$, where $p\in\Lambda$, and $v_g$ and $V_g$ depend on the constructing element $g$ (see
eq.~\eqref{eq:LocalTwistAndShift}). 

Let us compute the spectrum of the first twisted sector.
Consider the fixed point at the origin. Its constructing element is $g=(\vartheta,\,0)$;
therefore, the local twist vector is $v_g=v$ and the local shift vector is $V_g=V$. Since the Wilson
lines are trivial in this example, notice that the local twist and shift vectors are the same for all
constructing elements of this sector. Then it follows that the matter spectrum at each of the fixed
points is the same, so that the spectrum of the sector is given by 27 copies of the matter at the origin.

The conditions on the left- and right-moving momenta for massless states read
\be\ba{l}
 p^2_\text{sh} = (p + V)^2 \stackrel{!}{=} 2 - 2\ti N  - 2\,\delta c = \h43-2\ti N\,,\\[2mm]
 q^2_\text{sh} = (q + v)^2 \stackrel{!}{=} 1 - 2 \delta c =\h13\,,
\ea
\label{eq:Z3ExampleMasslessnessTwisted}
\ee
where we have used eq.~\eqref{eq:DefDeltaC} to compute $\delta c=1/3$. Further, since $V^2=4/3$ and 
$\ti N$ is nonnegative, eq.~\eqref{eq:Z3ExampleMasslessnessTwisted} has solution only for $\ti
N=0$. There are only nine $p_\text{sh}$ and two $q_\text{sh}$ that solve
eq.~\eqref{eq:Z3ExampleMasslessnessTwisted}:
\begin{center}
\begin{tabular}{rlrl}
$p_\text{sh}$ : & $\left(\underline{-\h23,\,\h13^3},\,-\h13,\,0^3\right)\Big(\,\h23,\,0^7\,\Big)$\,,
                & \quad $q_\text{sh}$ : &  $\left(0,\,\h13,\,\h13,\,\h13\right)$\,,\\
 & $\left(\;\;\h13^4,\,\phantom{\h23+\ \ \;}\h23,\,0^3\right)\Big(\,\h23,\,0^7\,\Big)$\,,  
                &  &  $\left(\h12,\,-\h16,\,-\h16,\,-\h16\right)$\,.\\
 & $\left(-\h16^4,\,\h16,\,\underline{\h12^2,\,-\h12}\right)\Big(\,\h23,\,0^7\,\Big)$\,,  & & \\
 & $\left(-\h16^4,\,\h16,\,-\h12^3\phantom{\ \;\h12}\right)\Big(\,\h23,\,0^7\,\Big)$\,.  & &  
\end{tabular}
\end{center}
Notice that the states $\ket{q_\text{sh}}_R\otimes\ket{p_\text{sh}}_L$ of the first twisted sector are
right-chiral ($q^0_\text{sh}=+1/2$ for the Ramond momentum).

It turns out that in the $T_1$ (and $T_{N-1}$) sector of \Z{N} orbifolds all solutions to the mass equation
enter into an invariant state. In other words, {\bf all states from the first twisted sector are invariant
under the orbifold action}. The reason is as follows. The
elements $h$ of the centralizer $\calZ_g$ can also be expressed as $g^i$ with $i=0,\ldots,N-1$. In this
case, one can verify that momenta $p_\text{sh}$ and $q_\text{sh}$ solving the masslessness conditions
always satisfy 
\be
p_\text{sh}\cdot V_h - q_\text{sh}\cdot v_h +(N_g-N^*_g)\cdot v_h -\2 (V_g\cdot V_h-v_g\cdot v_h) = 0\mod 1
\label{eq:InvCondT1}
\ee
for any centralizer element $h$. 

Therefore, all the momenta $p_\text{sh}$ and $q_\text{sh}$ given above survive the orbifold projection,
providing, in considering all the fixed points of the $T_1$ sector, 27 right-chiral supermultiplets which
transform as 
\be
27 (\bs{9},\,\bs{1})_{-12}
\label{eq:Z3ExampleRightChiralT1}
\ee
under gauge transformations.

In the $T_2$ sector, the situation is similar. As in the $T_1$ sector, it suffices to consider the
constructing element of the fixed point at the origin $g=(\theta^2,\,0)$ and count 27 copies of the
associated matter content, as illustrated in fig.~\ref{fig:Z3ExampleNoWL}. The associated massless
momenta are given by 
\begin{center}
\begin{tabular}{rlrl}
$p_\text{sh}$ : & $\left(\underline{\h23,\,-\h13^3},\ \;\h13,\ 0^3\ \right)\Big(-\h23,\,0^7\,\Big)$\,,
                & \quad $q_\text{sh}$ : &  $\left(0,\,-\h13,\,-\h13,\,-\h13\right)$\,,\\
 & $\left(-\h13^4,\,\phantom{\h23\;}-\h23,\,0^3\right)\Big(-\h23,\,0^7\,\Big)$\,,  
                &  &  $\left(-\h12,\,\h16,\,\h16,\,\h16\right)$\,.\\
 & $\left(\h16^4,\,-\h16,\,\underline{-\h12^2,\,\h12}\right)\Big(-\h23,\,0^7\,\Big)$\,,  & & \\
 & $\left(\h16^4,\,-\h16,\ \ \h12^3\phantom{\ \ \;\h12}\right)\Big(-\h23,\,0^7\,\Big)$\,.  & &  
\end{tabular}
\end{center}
Since these momenta differ from those of the $T_1$ by a sign, the states
$\ket{q_\text{sh}}_R\otimes\ket{p_\text{sh}}_L$ of the $T_2$ sector correspond to the conjugate of the
states in the $T_1$ sector. One can verify that these states have the following gauge transformations
\be
27 (\bsb{9},\,\bs{1})_{+12}\,.
\label{eq:Z3ExampleLeftChiralT2}
\ee
Notice that these states are left-chiral and, therefore, combine with those states from the $T_1$ sector
to form complete {\sc susy} multiplets. 

In summary, the matter content of the present \Z3 orbifold model is given in terms of the gauge
representations of the states by
\begin{center}
\begin{tabular}{|cl|}
\hline
Sector & Matter content\\
\hline\hline
$U$    & $\ 3(\bs{84},\,\ \bs{1})_0$\\
       & $\ 3(\ \bs{1},\,\bs{14})_{-18}$\\
       & $\ 3(\ \bs{1},\,\bs{64})_9$\\
\hline
$T_2$  & $27 (\;\bsb{9},\,\ \bs{1})_{12}$ \\
\hline
\end{tabular}
\end{center}
We omit here the gravity multiplet, the moduli and gauge bosons. 

\subsubsection{Including Wilson Lines}
\index{Z3 spectrum2@ !with Wilson lines}
We introduce now the two Wilson lines given by the vectors~\cite{Ibanez:1987sn,Casas:1988se}
\be\ba{rl}
A_1 &=\Big(0^7,\,\h23\Big)\,\left(0,\,\h13^2,\,0^5\right)\,,\\
A_3 &=\left(\h13^3,\,\h23,\,\h13,\,0,\,\h13^2\right)\,\left(\h13^2,\,0^6\right)\,.
\ea
\label{eq:Z3ExampleWLs}
\ee
Wilson lines $A_\al$ are gauge transformations associated to the noncontractible cycles in
the directions $e_\al$ of the compact space. We have seen before that, due to the structure of the compact
space, the Wilson lines of \Z3 orbifolds satisfy $A_\al=A_{\al+1}$ with $\al=1,3,5$
(cf. eq.~\eqref{eq:Z3WLRelations}).  

Geometrical moduli, the graviton, the dilaton and the antisymmetric tensor fields do not feel the
presence of the Wilson lines because their transformation properties only depend on the geometry of the
compact space (in other words, on the twist vector $v$ and the right-moving momenta $q$).
Gauge bosons with left-moving momenta $p$, on the other hand, have to satisfy additionally 
\be
p\cdot A_\al=0\mod 1
\label{eq:Z3ExampleWLUntwistedCond}
\ee
in order to be invariant (cf. eq.~\eqref{eq:4DChargedBosonsCondB}). In this way, the gauge
symmetry is further broken. 

In the present example, the action of Wilson lines leave invariant only eight momenta $p$ from the first
\E8 and 40 from the second \E8. We find that, together with the 16 invariant Cartan generators, they
constitute the adjoint representations of the unbroken gauge group in four dimensions
\be
 \maG_{4D}\quad\stackrel{A_\al}{\longrightarrow}\quad\maG_{4D}'=\SU3\times\SU2\times[\U1^5\times\SO{10}\times\U1^3]\,.
\label{eq:Z3ExampleWLGG}
\ee
Here, we have separated symbolically what we will call {\it the observable sector} from the {\it hidden
sector}, choice justified on the appearance of the $\SU3\times\SU2$ gauge factors. Optimistically, one could at
this point say that a model with such a gauge group is a good candidate for describing the standard model
of particle physics. The eight \U1 generators are labeled $\mathsf{t}_i$ with $i=1,\ldots,8$.

Another effect of the presence of nontrivial Wilson lines is the change of the matter spectrum. Many of
the formerly invariant momenta $p$ and $p_\text{sh}$ are not invariant any more with respect to all
elements of the centralizer(s). Let us focus first on the untwisted sector. The centralizer of the
constructing element $g=(\id,\,0)$ is, as mentioned before, the complete space group. A valid basis of
$\calZ_g$ is given by
\be
\calZ_g=\left\{(\vartheta,\,0),\,(\id,\,e_1),\,(\id,\,e_2),\,(\id,\,e_3),\,
             (\id,\,e_4),\,(\id,\,e_5),\,(\id,\,e_6)\right\}\,.
\label{eq:Z3ExampleBasisU}
\ee
In particular, the action of the space group elements $h=(\id,\,e_\al)$ imposes new constraints on the
momenta $p$ of the untwisted charged matter. From eq.~\eqref{eq:4DUntwistedMatterCond} (with
$V_h=A_\al$), it follows that these constraints are also given by
eq.~\eqref{eq:Z3ExampleWLUntwistedCond}. We are then left with $2[(3+3\times2+2)+16]$ momenta $p$. After
tensoring left- and right-moving states together and computing their gauge quantum numbers, we find the
following matter representations of $\SU3\times\SU2\times\SO{10}$ (we omit the \U1 charges):
\begin{center}
\begin{tabular}{|c|}
\hline
\qquad$3(\bs{3},\,\bs{1},\,\bs{1})\oplus 3(\bsb{3},\,\bs{1},\,\bs{1})\oplus 3(\bs{1},\,\bs{2},\,\bs{1})
       \oplus 3(\bs{1},\,\bs{1},\,\bsb{16})\phantom{^{A^{A^A}}}$\\[2mm]
\hline
\end{tabular}
\end{center}

\begin{figure}[t!]
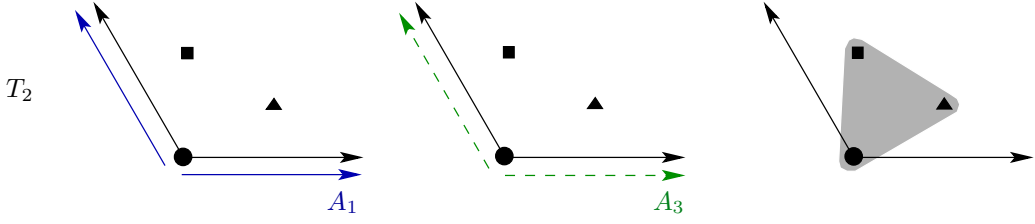

\centerline{\input Z3_example_2WL.pstex_t}
\caption{The action of the Wilson lines $A_1$ and $A_3$ lifts the degeneracy of the fixed points of the
  first and second complex planes. Note that the Wilson line $A_\al$ is related to the directions $e_\al$ and
$e_{\al+1}$ of the compact space.}
\label{fig:Z3Example2WL}
\end{figure}

The computation of the matter states in the twisted sectors gets more involved than before. In the
previous discussion, we have used the fact that, in absence of Wilson lines, the solutions to mass
equations and the projection condition are identical at any fixed point of a given twisted
sector. Therefore, in that case, all fixed points are degenerate. This situation changes in presence of
Wilson lines. Consider, for example, two different constructing elements of the $T_2$ sector:
$g_1=(\vartheta^2,\,0)$ and $g_2=(\vartheta^2,\,e_2)$. The massless momenta $p_\text{sh}=p+V_{g_1}$ for
$g_1$ correspond to those found in the case without Wilson lines because $V_{g_1}=2V$. The right-moving
momenta is not altered by the introduction of Wilson lines. The states
$\ket{q_\text{sh}}_R\otimes\ket{p_\text{sh}}_L$ at $g_1$ form the gauge representations
\be
1 (\bs{1},\,\bs{2},\,\bs{1})\oplus 1(\bsb{3},\,\bs{1},\,\bs{1})\oplus 4(\bsb{1},\,\bs{4},\,\bs{1})
\label{eq:Z3ExampleSpectrumg1}
\ee
under the gauge group $\maG_{4D}'$ given in eq.~\eqref{eq:Z3ExampleWLGG} (excluding the \U1 charges).

For $g_2$ the local shift vector changes to $V_{g_2}=2V+A_2=2V+A_1$. Hence, the solutions to the mass
equation for the left-movers take a different form:
\begin{center}
\begin{tabular}{rl}
$p_\text{sh}$ : & $\left(-\h13^4,\,-\h23,\,\quad 0^3\quad\right)\Big(\underline{\h13^2,\,-\h23},\,0^5\,\Big)$\,,\\
 & $\left(\h16^4,\,-\h16,\,\underline{-\h12,\,\h12},\,\h16\right)\Big(\underline{-\h23,\,\h13^2},\,0^5\,\Big)$\,.   
\end{tabular}
\end{center}
These nine momenta along with the right-moving momenta $q_\text{sh}$ of the orbifold without Wilson lines
comprise the matter spectrum of the fixed point with constructing element $g_2$, which is written down
in terms of the corresponding gauge quantum numbers (excluding \U1 charges):
\be
9 (\bs{1},\,\bs{1},\,\bs{1})\,.
\label{eq:Z3ExampleSpectrumg2}
\ee

\index{Z3 spectrum2@ !local spectra}
\begin{table}[!t!]
\begin{center}
\begin{tabular}{|l|c|}
\hline
Constructing element(s) $g$   &  Matter content \\
\hline\hline
\rowgrayh
$(\vartheta^2,\,n_5e_5+n_6e_6)$       & $1(\bs{1},\,\bs{2},\,\bs{1})\oplus 1(\bsb{3},\,\bs{1},\,\bs{1})\oplus 4(\bs{1},\,\bs{1},\,\bs{1})$\\[1mm]
$(\vartheta^2,\,e_3+n_5e_5+n_6e_6)$   & \\[1mm]
$(\vartheta^2,\,e_1+e_2+e_3+n_5e_5+n_6e_6)$ &  $2(\bs{1},\,\bs{2},\,\bs{1})\oplus 1(\bsb{3},\,\bs{1},\,\bs{1})
                                        \oplus 1(\bs{3},\,\bs{1},\,\bs{1})\oplus 8(\bs{1},\,\bs{1},\,\bs{1})$ \\[1mm]
$(\vartheta^2,\,e_1+e_2+e_3+e_4+n_5e_5+n_6e_6)$  &  \\[1mm]
\rowgrayh
$(\vartheta^2,\,e_3+e_4+n_5e_5+n_6e_6)$ & $1(\bs{1},\,\bs{2},\,\bs{1})\oplus 1(\bs{3},\,\bs{1},\,\bs{1})\oplus 4(\bs{1},\,\bs{1},\,\bs{1})$\\[1mm]
\rowgrayh
$(\vartheta^2,\,e_2+e_3+e_4+n_5e_5+n_6e_6)$  &  \\[1mm]
$(\vartheta^2,\,e_2+n_5e_5+n_6e_6)$ & $9(\bs{1},\,\bs{1},\,\bs{1})$\\[1mm]
$(\vartheta^2,\,e_1+e_2+n_5e_5+n_6e_6)$  &  \\[1mm]
\rowgrayh
$(\vartheta^2,\,e_2+e_3+n_5e_5+n_6e_6)$ & $3(\bs{1},\,\bs{2},\,\bs{1})\oplus 3(\bs{1},\,\bs{1},\,\bs{1})$\\[1mm]
\hline
\end{tabular}
\end{center}
\caption{The matter content attached to the fixed points of the $T_2$ sector of a \Z3 orbifold model with
  two Wilson lines.}  
\label{tab:Z3Example2WLFP}
\end{table}

The difference between the matter content for $g_1$ and $g_2$, eqs.~\eqref{eq:Z3ExampleSpectrumg1}
and~\eqref{eq:Z3ExampleSpectrumg2}, makes manifest that Wilson lines lift the degeneracy of the fixed
points. Only the points with coordinates in the last torus conserve the 
degeneracy as shown in fig.~\ref{fig:Z3Example2WL}. The degeneracy of the last torus gives then a
multiplicity factor of three for the twisted states. In principle, we could expect a
different matter content at each of the nine fixed points with different coordinates in the first two
complex planes.  
We list in table~\ref{tab:Z3Example2WLFP} the matter content associated to the different
fixed points of the $T_2$ sector. Notice that there are only five classes of matter content. As we shall
discuss shortly, the class of matter content of a constructing element $g$ is related to its local shift $V_g$.

The matter states in the $T_1$ sector are, as before, the right-chiral partners of the states in the
$T_2$ sector. Hence, in terms of left-chiral states, the matter spectrum of this models is summarized as follows:
\begin{center}
\begin{tabular}{|cl|cl|}
\hline
Sector & Matter content & Sector & Matter content\\
\hline\hline
$U$    & $\ 3(\bs{3},\,\bs{1},\,\ \bs{1})$  & $T_2$ & $\ 15(\bs{3},\,\bs{1},\,\bs{1})$ \\
       & $\ 3(\bsb{3},\,\bs{2},\,\ \bs{1})$ &       & $\ 12(\bsb{3},\,\bs{1},\,\bs{1})$\\
       & $\ 3(\bs{1},\,\bs{2},\,\ \bs{1})$  &       & $\ 36(\bs{1},\,\bs{2},\,\bs{1})$\\
       & $\ 3(\bs{1},\,\bs{1},\,\bsb{16})$  &       & $171(\bs{1},\,\bs{1},\,\bs{1})$  \\
\hline
\end{tabular}
\end{center}

Let us spend few words on some phenomenological properties of this model. If a string compactification is
to reproduce the standard model of particle physics, it must have the gauge group
$\Gsm=\SU3\x\SU2\x\U1_Y$. It has been shown that in the present model it is possible to find 
a combination of \U1's producing the correct spectrum of the \sm\ plus additional particles, which
turn out to be vectorlike with respect to \Gsm\ (see table 1 of ref.~\cite{Casas:1988se}). The vectorlike 
character of the exotic particles insinuates that such particles can acquire large masses provided that
adequate couplings exist in the theory. As we will see in section~\ref{sec:ourstringselectionrules},
ensuring that such couplings do not vanish is quite nontrivial, for allowed nonvanishing couplings
in string theory have to satisfy strong constraints. 

Other qualities of this model include preservation
of $\maN=1$ \susy\ and the spontaneous breaking of the hidden sector. Yet issues such as correct Yukawa-mass
textures, proton decay, neutrino masses and compatibility with \gut\ theories require further investigation.

\subsection{Local Shift Vectors and Local Spectra}
\label{subsec:LocalShiftsSpectra}
\index{local shifts and local spectra}

We have seen that there are five classes of local spectra at the different fixed points of a \Z3 orbifold
model with Wilson lines. It turns out that this is a model-independent statement for all \Z3 orbifold
models, as we will explain in the following.

Consider first an arbitrary \Z{N} or \ZZ{N}{M} orbifold model.
Locally, at a fixed point with constructing element $g$, the 480 gauge bosons of the ten dimensional
gauge group \maG are affected only by the action of the local shift $V_g$. Therefore, the local gauge
group $\maG_g$ and the local matter spectrum coincide with the four-dimensional gauge group and the
spectrum at one arbitrary fixed point of an orbifold model with global shift vector $V=V_g$ (and no
Wilson lines). The roots of the local gauge group will be those roots $p$ of the ten-dimensional gauge
group $\maG$ satisfying
\be
p\cdot V_g = 0\mod1\;\qquad p^2 = 2\;.
\label{eq:RootsLocalGG}
\ee
Clearly, matter states transform {\it locally} under $\maG_g$ rather than under the four-dimensional
gauge group. This restricts the number of local gauge groups (and local spectra) to be the number of
(inequivalent) admissible shift vectors of the studied orbifold.

The intersection of the local gauge groups of the different fixed points in a particular model
yields the (global) four-dimensional gauge group $\maG_{4D}$:
\be
\maG_{4D}=\maG_{g_1}\cap\maG_{g_2}\cap\maG_{g_3}\cap\ldots\,,
\label{eq:4DGGversuslocalGG}
\ee
whence it follows that the local symmetry group is, in general, bigger than the $\maG_{4D}$, i.e.
\be
\maG_{4D}\subset\maG_{g_i}\,,\quad i=1,2,3,\ldots\,.
\label{eq:4DGGversuslocalGG2}
\ee
From a global perspective in four dimensions, local matter states form representations of $\maG_{4D}$.
These can be derived by the branching rule of the local representations under the breaking
$\maG_g\rightarrow\maG_{4D}$. In higher twisted sectors, some of the local states can be projected out from
a global perspective, so that only incomplete local representation survive in the full orbifold. 

We can now return to our \Z3 example.
Since, in the case of \Z3 orbifolds, there are only five admissible shift 
vectors $V$ (see table~\ref{tab:Shifts_Z3E8}), any local shift vector $V_g$ must be equivalent to
one of them. It follows then that there are also five classes of different local spectra, in agreement
with table~\ref{tab:Z3Example2WLFP}. For concreteness, consider the constructing element
$g_2=(\vartheta^2,\,e_2)$. The corresponding local shift vector is
\be
V_{g_2}=2V+A_1=\left(\h23^4,\,\h43,\,0^2,\,\h23\right)\left(\h43,\,\h13^2,\,0^5\right)\,
\stackrel{\calW+\Lambda}{\longrightarrow}\,\left(\h23,\,\h13^2,\,0^6\right)\left(\h23,\,\h13^2,\,0^6\right)\,,
\label{eq:Z3ExampleShiftg2}
\ee
where $\calW+\Lambda$ denotes Weyl rotations accompanied by $\E8\x\E8$ lattice translations. One can verify that
$V_{g_2}$ corresponds to the shift vector $V^{(2)}$ listed in table~\ref{tab:Z3Example2WLFP}. Therefore,
the local gauge group is $\maG_{g_2}=\E6\x\SU3\x\E6\x\SU3$ and the local matter content is given the
bifundamental representation $(\bs{1},\,\bs{3},\,\bs{1},\,\bs{3})$. It is not difficult to confirm that
the nine singlets of eq.~\eqref{eq:Z3ExampleSpectrumg2} under $\maG_{4D}'$ arise from the breaking
$\maG_{g_2}\rightarrow\maG_{4D}'$.

We will see in chapter~\ref{ch:MLI} that the local picture proves to be a useful tool in the search
after orbifold models with realistic properties, such as grand unification.
\index{Z3 spectrum@\Z3 spectrum, example|)}

\subsection{Anomaly Cancellation}
\label{subsec:AnomalyCancellation}
\index{anomaly cancellation!orbifold compactifications}

Orbifold models present generically one anomalous \U1 symmetry, $\U1_A$~\cite{Witten:1984dg}. However, it
is reasonable to expect that the theory in four dimensions should be anomaly free as it arises from the
heterotic string, where anomaly cancellation is achieved by the Green-Schwarz 
mechanism~\cite{Green:1984sg}. It has been shown~\cite{Schellekens:1986xh} that modular invariance of the
orbifold guarantees that the anomaly polynomial in four dimensions factorizes as $\tr F^2-\tr
R^2$. From there it follows that the anomaly can be canceled by the generalized Green-Schwarz
mechanism~\cite{Dine:1987xk,Sagnotti:1992qw,Berkooz:1996iz,Blumenhagen:2005pm}. Besides, the so-called
$\U1_A$ universality condition holds automatically~\cite{Casas:1987us,Kobayashi:1996pb}
\be
\h{1}{2}\tr\,(\ell\, Q_A) = \h{1}{2|\mathsf{t}_j|^2}\tr\,Q_j^2 Q_A =\h{1}{6 |\mathsf{t}_A|^2}\tr\,Q_A^3
= \h{1}{24}\tr\,Q_A \,\big(\equiv~8\pi^2 \delta_{GS}\big)\quad j\neq A\,,
\label{eq:U1AUniversalityCond}
\ee
where $\ell$ denotes the index of a given representation under a nonabelian gauge group. Furthermore,
$\mathsf{t}_j$ are the generators of the \U1 factors ($\mathsf{t}_A$ corresponds to $\U1_A$) that define the charge $Q_j$ as:
\begin{equation}
Q_j \ket{p_\text{sh}}_L~=~(\mathsf{t}_j \cdot p_\text{sh})\, \ket{p_\text{sh}}_L\,.
\nonumber
\end{equation}
The constant $\delta_{GS}$ enters in the transformation of the dilaton in the Green-Schwarz
mechanism. Orbifold models contain at most one anomalous \U1. Hence the remaining \U1's satisfy
\be
\h{1}{2}\tr\,(\ell\, Q_i) = \h{1}{2|\mathsf{t}_j|^2}\tr\,Q_j^2 Q_i =\h{1}{6 |\mathsf{t}_i|^2}\tr\,Q_i^3
= \h{1}{24}\tr\,Q_i = 0 \quad i,j\neq A\,.
\label{eq:U1otherUniversalityCond}
\ee

The universality condition, eq.~\eqref{eq:U1AUniversalityCond}, states also that pure abelian, mixed
abelian--nonabelian, and mixed abelian--gravitational anomalies are not independent of each
other. Therefore, all of them cancel simultaneously. In ref.~\cite{Araki:2008ek} it is shown that this
holds also for discrete symmetries present in orbifold constructions.
Other anomalies, such as pure nonabelian anomalies or Witten's anomaly~\cite{Witten:1982fp}, vanish
automatically in orbifolds. The conditions~\eqref{eq:U1AUniversalityCond}
and~\eqref{eq:U1otherUniversalityCond} have been confirmed for all models appearing in the present work.

\index{orbifold constructions|)} 

\section{Discrete Torsion in Orbifold Models}
\label{sec:DiscreteTorsion}
\index{discrete torsion|(}

It is well known that discrete torsion~\cite{Vafa:1986wx} introduces an additional degree of freedom in
\ZZ{N}{M} orbifold constructions~\cite{Font:1988mk}. Nonetheless, there are some features of discrete
torsion on orbifolds which have received little attention. Only recently we have
noted~\cite{Ploger:2007iq} that \Z{N} orbifolds also admit discrete torsion and, moreover, that discrete
torsion can add more than one degree of freedom to orbifolds with Wilson lines. In this section, we
study briefly \ZZ{N}{M} and \Z{N} orbifold models with discrete torsion and introduce the concept of {\it
generalized discrete torsion}. Then we analyze an interesting equivalence between models with generalized
discrete torsion and torsionless {\it brother models}. For a more detailed discussion,
ref.~\cite{Vaudrevange:2008sm} is recommended.

The one-loop partition function $Z$ for orbifold compactifications has the overall structure 
\begin{equation}
 Z~=~\sum_{\substack{g,h \\ [g,h]=0}} \varepsilon (g,h)\, Z(g,h)\;,\qquad g,h\in S\,.
\end{equation}
The relative phases $\varepsilon (g,h)$ are called {\it discrete torsion phases}.
Their values can vary between the different terms in the partition function and thus between the
different sectors. Different assignments of phases lead, in general, to different orbifold models. The
arbitrariness of $\varepsilon (g,h)$ corresponds to the freedom of turning on a background antisymmetric
field on the torus~\cite{Vafa:1986wx}.

Although the discrete torsion phases appear to be arbitrary, modular invariance at one loop 
and factorizability of the partition function at two loops impose certain constraints on the torsion
phases. They are given by
\begin{subequations}\label{eq:phaseconstraints}
\begin{eqnarray}
 \varepsilon(g_{1} g_{2},g_{3}) &=& \varepsilon(g_{1},g_{3})\, \varepsilon(g_{2},g_{3})\;,
 \label{eq:phaseconstraints1} \\
 \varepsilon(g_{1},g_{2}) &=& \varepsilon(g_{2},g_{1})^{-1}\;. 
 \label{eq:phaseconstraints2} \\
 \varepsilon(g,g) &=& 1 \;.
\label{eq:phaseconstraints3}  
\end{eqnarray} 
\end{subequations}
The last equation is a convention, rather than a constraint, and can be seen as a sort of normalization of
the phases. 

\subsection{Discrete Torsion without Wilson Lines}
\label{subsec:OldDiscreteTorsion}

In orbifolds without Wilson lines, $g$ and $h$ are chosen to be elements of the point group $P$. It
follows then that for \Z{N} orbifolds the solution of eqs.~\eqref{eq:phaseconstraints} is trivial:
\begin{equation}
\varepsilon(g,h)~=~1 \qquad \forall g,h \in P\;.
\end{equation}
Therefore, in the case of \Z{N} orbifolds without Wilson lines, non-trivial discrete torsion
cannot be introduced. 

In \ZZ{N}{M} orbifolds, still without Wilson lines, the situation is different because there are
independent pairs of elements which commute with each other.  If we take two point group elements
$g=\vartheta^{k_1}\omega^{\ell_1}$ and $h=\vartheta^{k_2}\omega^{\ell_2}$, then the discrete torsion
phase is determined by eqs.~\eqref{eq:phaseconstraints} to be~\cite{Font:1988mk}
\begin{equation}
 \varepsilon(g,h)
 ~=~ \varepsilon(\theta^{k_1}\omega^{\ell_1},\theta^{k_2}\omega^{\ell_2})
 ~=~  \exp\left\{\tfrac{2\pi\I\, a}{N}(k_1 \ell_2-k_2 \ell_1)\right\}\,,\quad\;  a=0,1,\ldots,N\,,
\label{eq:TorsionPhaseZNZMNoWL}
\end{equation}
where $N$ is the order of the twist $\vartheta$.\footnote{Our convention for \ZZ{N}{M} orbifolds is that
$M=n\,N$ with $n\in\maZ$.} Note that there are only $N$ inequivalent assignments of $\varepsilon$.  

\subsection{Generalized Discrete Torsion}
\label{subsec:NewDiscreteTorsion}
\index{discrete torsion!generalized}
More recently, the concept of discrete torsion has been extended by introducing a {\it generalized
discrete torsion} phase in the context of type IIA/B string theory~\cite{Gaberdiel:2004vx}. This generalized
torsion phase depends on the fixed points rather than on the sectors of the orbifold. Therefore, one has
to consider $g$ and $h$ to be elements of the space group $S$.

Considering the space group elements $g=(\vartheta^{k_1}\omega^{\ell_1},\,n_\al e_\al)$ and
$h=(\vartheta^{k_2}\omega^{\ell_2},\,m_\al e_\al)$, the general solution to
eqs.~\eqref{eq:phaseconstraints} for the discrete torsion phase is written down as
\begin{equation}
\varepsilon(g,h)~=~e^{2 \pi \I\,[a\, (k_1\, \ell_2 - k_2\, \ell_1) 
+ b_{\alpha}\, (k_1\, m_{\alpha} - k_2\, n_{\alpha}) 
+ c_{\alpha}\, (\ell_1\, m_{\alpha} - \ell_2\, n_{\alpha}) 
+ d_{\alpha \beta}\, n_{\alpha}\, m_{\beta}]}\;.
\label{eq:generalizedtorsionphase}
\end{equation}
where the sum over $\al,\,\beta$ is understood.
The values of $a,b_{\alpha},c_{\alpha},d_{\alpha\beta}$ are required by modular invariance and the
geometry of the lattice $\Gamma$ to satisfy 
\begin{equation}
\ba{c}
N\,a,\,N_\al\,b_\al,\,N_\al\,c_\al,\,N_{\al\beta}\,d_{\al\beta} ~=~ 0\mod 1\,,\\
d_{\al\beta} ~=~ -d_{\beta\al}\,,
\ea
\label{eq:TorsionParametersConstraints}
\end{equation}
for each $\al,\beta=0,\ldots,6$. Here $N$ is the order of the twist $\vartheta$, $N_\al$ the order of the
Wilson line $A_\al$, and $N_{\alpha\beta}$ is the greatest common divisor of $N_\alpha$ and $N_\beta$
(compare with the conditions for modular invariance, eqs.~\eqref{eq:NewModInv}). Additional constraints
on the parameters $b_\alpha$, $c_\alpha$, $d_{\alpha\beta}$ due to the choice of the lattice $\Gamma$
appear in a similar fashion as those on the order of Wilson lines, explained in
section~\ref{subsec:ConsistencyConditions}. It is not hard to see that if $e_\alpha \simeq e_\beta$ on
the orbifold, then $b_\al=b_\beta,\, c_\alpha=c_\beta$ and $d_{\alpha\beta}=0$ must hold. 

The generalized discrete torsion is not restricted only to \ZZ{N}{M} orbifolds, as it was commonly
believed, but will likewise appear in the \Z{N} case. Clearly, since in \Z{N} orbifolds 
there is only one shift, the parameters $a$ and $c_\al$ vanish.

\subsubsection{Role of Discrete Torsion on Orbifolds}

The most important consequence of nontrivial $\varepsilon$-phases for our discussion is that they modify
the boundary conditions for twisted states and thus change the twisted spectrum. This can be seen from
the transformation phase $\Phi$ of eq.~\eqref{eq:ProjectionCondition}, which is modified in the presence of
discrete torsion according to 
\begin{equation}
 \Phi \longmapsto \Phi'=\varepsilon(g,h) \,\Phi\;.
\label{eq:modifiedProjectionPhase}
\end{equation}
Clearly, the phases $\Phi'$ depend on the constructing element $g$ and an element $h$ of its centralizer
$\calZ_g$, i.e.\ $\Phi' = \Phi'(g,h)$. The projection phases $\Phi'$ and $\Phi$ project out
different twisted matter states. Very frequently all the states located at some fixed point are
projected out by the effect of the modified phase $\Phi'$. One could say that these `empty'
fixed points disappear from the spectrum or, in other words, that they are nonphysical (at massless level).
This feature is interesting because it allows to interpret the effect of discrete torsion in terms of a
change in the metric of the compact space, as is discussed in section~\ref{subsec:DTvsNonfactorizable}.

\subsubsection{Examples}

$\bs{\Z3}${\bf Orbifolds}. Let us consider the \Z3 orbifold compactified on an $\SU3^3$ lattice. 
As we have seen section~\ref{subsec:ConsistencyConditions}, the lattice vectors of
$\Gamma$ are related by the action of the point group generator. In particular, we have that $e_\al
\simeq e_{\al+1}$, for $\al=1,3,5$, on the orbifold. This implies that there are only three
independent $b_\alpha$, namely $b_1,\,b_3,\,b_5$, while the other $b$-parameters satisfy 
$b_2 = b_1,\,b_4= b_3,\,b_6 = b_5$. Further,the antisymmetric matrix $d_{\al\beta}$ takes the form 
\begin{equation}
 d_{\alpha\beta}~=~ \left(
 \begin{array}{cccccc}
 0&0&d_1&d_1&d_2&d_2\\
 0&0&d_1&d_1&d_2&d_2\\
 -d_1&-d_1&0&0&d_3&d_3\\
 -d_1&-d_1&0&0&d_3&d_3\\
 -d_2&-d_2&-d_3&-d_3&0&0\\
 -d_2&-d_2&-d_3&-d_3&0&0
 \end{array}
 \right)\;.
 \label{eq:dmatrixZ3}
\end{equation}
Therefore, there are six independent discrete torsion parameter, which can take the values
$0$, $\tfrac{1}{3}$ or $\tfrac{2}{3}$.

\noindent$\bs{\Z3\x\Z3}${\bf Orbifolds}. The \Z3\x\Z3 orbifold is very similar to \Z3. Since the compactification
lattice is that of the \Z3 orbifold, the six discrete torsion parameters $b_\al$ and
$d_{\al\beta}$ (eq.~\eqref{eq:dmatrixZ3}) are also admissible in this case. Additionally, the parameters
$c_\al$ appear in the theory. In the same way as $b_\al$, they are restricted 
by the geometry, so that only $c_\al$ with $\al=1,3,5$ are independent. Including $a$, we obtain ten
independent discrete torsion parameters with values $0$, $\tfrac{1}{3}$ or $\tfrac{2}{3}$.

\noindent$\bs{\Z6}${\bf-II Orbifolds}. For the \Z6-II orbifold on a \G2\x\SU3\x\SO4 lattice an analogous
consideration shows that there are only few nontrivial discrete torsion parameters. We find that the only
discrete torsion parameters that accept nonzero values are $b_3=b_4=0,\h13,\h23$, $b_5,b_6=0,\2$ and
$d_{56}=-d_{65}=0,\2$; that is, only four independent parameters. However, very frequently the factors
accompanying these parameters in eq.~\eqref{eq:generalizedtorsionphase} (such as $(k_1\, m_{\alpha} -
k_2\, n_{\alpha})$ for the parameters $b_\al$) vanish, implying that the corresponding parameters are
nonphysical. This is explained by taking into account the space group elements $g,\,h$ entering the
$\varepsilon$-phases, that is, all constructing and centralizer elements.
It turns out that this happens for all discrete torsion parameters of \Z6-II
orbifolds. Hence, {\it discrete torsion is irrelevant in \Z6-II orbifolds.} This 
will be important in chapter~\ref{ch:MLI} for the classification of orbifolds of this type.

\subsection{Brother Models}
\label{subsec:BrotherModels}
\index{discrete torsion!brother models}
\index{brother models}

In most of the studies of orbifold models, it is claimed that two models whose parameters $(V_1,V_2,A_\al)$
differ only by lattice translations are equivalent. This is, in general, not true. The reason being that
lattice translations influence the projection condition of twisted states, eq.~\eqref{eq:ProjectionCondition}.

A (torsionless) model M is defined by $(V_1,V_2,A_\al)$. A brother model M$'$ appears by adding lattice
vectors to the shifts and Wilson lines, i.e.\ M$'$ is defined by 
\begin{equation}
 (V'_1, V'_2, A'_\alpha)~=~(V_1+\Delta V_1, V_2+\Delta V_2, A_\alpha+\Delta A_\alpha)\;,
\end{equation}
with $\Delta V_i,\Delta A_\alpha \in \Lambda$. From the conditions~\eqref{eq:NewModInv}, the choice of
lattice vectors $(\Delta V_i,\Delta A_\alpha)$ is constrained by
\begin{subequations}
\begin{eqnarray}
M \left(V_1\cdot\Delta V_2 + V_2\cdot\Delta V_1 + \Delta V_1 \cdot\Delta V_2 \right) & = & 0\text{ mod }
2~\equiv~ 2\,x \;,\label{eq:newModularInvforDeltaVV}\\
N_\alpha \left(V_i\cdot\Delta A_\alpha + A_\alpha\cdot\Delta V_i + \Delta V_i \cdot\Delta A_\alpha
\right) & = &  0\text{ mod }
2~\equiv~ 2\,y_{i\alpha} \;,\label{eq:newModularInvforDeltaVA}\\
Q_{\alpha\beta} \left(A_\alpha\cdot\Delta A_\beta + A_\beta\cdot\Delta A_\alpha + \Delta A_\alpha\cdot\Delta A_\beta
\right) & = &  0\text{ mod }
2~\equiv~ 2\,z_{\alpha\beta}\;,\label{eq:newModularInvforDeltaAA}
\end{eqnarray}
\label{eq:newModularInvforDelta}
\end{subequations}
where $x,\,y_{i\alpha},\,z_{\alpha\beta}\,\in{\mathbb Z}$.

One can verify that the inclusion of the lattice vectors $(\Delta V_1,\Delta V_2,\Delta A_\al)$ alters the projection phase
of brother models as
\be
 \Phi \longmapsto \Phi'=\ti\varepsilon(g,h) \,\Phi\;,
\label{eq:brotherProjectionPhase}
\ee
where the `brother phase' $\ti\varepsilon$ is given by
{\small
\begin{eqnarray}
 \widetilde{\varepsilon}(g,h)&=&\exp\left\{-2\pi \I\,\left[ 
   (k_1\, \ell_2-k_2\, \ell_1)\left(V_{2} \cdot \Delta V_{1} - \frac{x}{M}\right)
 + (k_1\, m_{\alpha} - k_2\, n_{\alpha})\left(A_{\alpha} \cdot \Delta  V_{1}-\frac{y_{1\alpha}}{N_\alpha}\right)\right.\right.
 \nonumber\\*
 & &  \hphantom{\left\{\right.}{}\left.\left.
  + (\ell_1\, m_{\alpha}-\ell_2\, n_{\alpha})\left(A_{\alpha} \cdot \Delta V_{2}-\frac{y_{2\alpha}}{N_\alpha}\right)
  + n_{\alpha}\, m_{\beta}\left( A_{\beta} \cdot \Delta
    A_{\alpha}-\frac{z_{\alpha\beta}}{Q_{\alpha\beta}}\right)\right]
 \right\}\;,
 \label{eq:generalbrotherphase} 
\end{eqnarray}}
corresponding to the constructing element
$g=(\theta^{k_1}\omega^{\ell_1},n_{\alpha}e_{\alpha})$ and the centralizer element
$h=(\theta^{k_2}\omega^{\ell_2},m_{\alpha}e_{\alpha})$. One can see that $D_{\alpha\beta}\equiv A_\beta\cdot\Delta
A_\alpha - z_{\alpha\beta}/Q_{\alpha\beta}$ is (almost) antisymmetric in $\alpha,\,\beta$,
\begin{equation}
D_{\alpha\beta} = -D_{\beta\alpha}\text{ mod }1\,.
\end{equation}
Like the discrete torsion phase, the brother phase is not the same for all fixed points; hence the
local spectrum is changed.

The brother phase $\ti\varepsilon$ and the generalized discrete torsion phase
eq.~\eqref{eq:generalizedtorsionphase} are not only very similar, but can, in fact, be made coincide. This
implies an unexpected connection between lattice translations of the parameters $(\Delta V_1,\Delta
V_2,\Delta A_\al)$ and discrete torsion. In other words, we find that models with discrete torsion can be
mimicked by torsionless models with modified shift vector(s) and background fields. We find however that
the noninteger values of the parameter $d_{\al\beta}$ do not allow an interpretation in terms of lattice
translations in models with trivial Wilson lines. Therefore, discrete torsion is, in this sense, more
general than the concept of brother models.

As an illustration of the relation between brother models and orbifolds with discrete torsion, we have
investigated the distinct \ZZ{N}{M} orbifold models with standard embedding that one can find for
different nonzero values of the discrete torsion parameter $a$. We have found that one can trade the
parameter $a$ for a pair of lattice vectors $(\Delta V_1,\,\Delta V_2)$ which, added the standard
embedding shift vectors, lead to the spectra obtained in the models with discrete torsion. Our findings
are listed in table~\ref{tab:DiscreteTorsionBrothers}.

\subsection{Discrete Torsion and Nonfactorizable Lattices}
\label{subsec:DTvsNonfactorizable}

Let us comment on one last interesting observation. We have found that in many cases orbifold models M with
certain geometry, i.e.\ compactification lattice $\Gamma$, and generalized discrete torsion switched on
are equivalent to torsionless models M$'$ based on a different lattice $\Gamma'$. Model M$'$ has less
fixed points than M, and the mismatch turns out to constitute precisely the `empty' fixed points of model
M due to the discrete torsion phase.

The simplest examples are based on $\mathbbm{Z}_2\times\mathbbm{Z}_2$ orbifolds with standard embedding
and without Wilson lines. By varying the allowed discrete-torsion parameters of this orbifold
(especially, $d_{\al\beta}$), we have found eight different models with nonzero net number of
$\bs{27}$-plets of \E6 (see ref.~\cite{Ploger:2007iq}). Interestingly, these models have already been
discussed in the literature, but in a different context. They appeared first in
ref.~\cite{Donagi:2004ht} in the context of free fermionic string models related to the
$\mathbbm{Z}_2\times\mathbbm{Z}_2$ orbifold with an additional freely acting shift. More recently, new
$\mathbbm{Z}_2\times\mathbbm{Z}_2$ orbifold constructions have been found in studying orbifolds of
non-factorizable six-tori~\cite{Faraggi:2006bs,Forste:2006wq}. 
(More recently, further details of these and similar models have been studied~\cite{Donagi:2008xy,Kiritsis:2008mu}.)
For each of the models found by adding
nonvanishing discrete torsion phases, there is a corresponding `non-factorizable' model M$'$ with the
following properties: 
\begin{enumerate}
\item
 Each `non-empty' fixed point, i.e.\ each fixed point with local zero-modes, in the model M can be mapped
 to a fixed point with the same spectrum in model M$'$. 
\item  
 The number of `non-empty' fixed points in M coincides with the total number of  fixed points in M$'$.
\end{enumerate}

These relations are not limited to $\mathbbm{Z}_2\times\mathbbm{Z}_2$ orbifolds,
rather we find an analogous connection also in other
$\mathbbm{Z}_N\times\mathbbm{Z}_M$ cases ($\mathbbm{Z}_N\times\mathbbm{Z}_M$
orbifolds based on non-factorizable compactification lattices have recently
been discussed in~\cite{Takahashi:2007qc}). This result hints towards an
intriguing impact of generalized discrete torsion on the interpretation of
orbifold geometry. What the (zero-mode) spectra concerns, introducing
generalized discrete torsion (or considering generalized brother models) is
equivalent to changing the geometry of the underlying compact space,
$\Gamma\to\Gamma'$. To establish complete equivalence between these models would
require to prove that the couplings of the corresponding states are the same,
which is beyond the scope of the present study. It is, however, tempting to
speculate that nonresolvable singularities (fixed points with no states attached)
do not `really' exist as one can always choose (for a given spectrum) the 
compactification lattice $\Gamma$ in such a way that there are no `empty' fixed points.
\index{discrete torsion|)}

\section{String Interactions: Yukawa Couplings}
\label{sec:YukawaCouplings}
To close this chapter, let us examine one crucial element necessary in order to study the low--energy
field theory limit of orbifold compactifications: field interactions. In contrast to pure field theory,
where couplings between matter fields are chosen {\it ad hoc}, in orbifold compactifications they are
determined by strict rules derived from string theory.

\subsection{String Selection Rules}
\label{sec:ourstringselectionrules}
\index{Yukawa coupling selection rules}

Consider the $n$--point correlation function of two fermions and $n-2$ bosons.
The corresponding physical states shall be denoted by $\Psi_i$, $i = 1,\ldots,n$. Then, in the field
theory limit, a non--vanishing correlation function induces the following term in the superpotential
\begin{equation}
W~\supset~\Psi_1\, \Psi_2\, \Psi_3 \ldots \Psi_n\;.
\end{equation}
A complete evaluation of the correlation function has only been performed for 3--point couplings and
yields a moduli dependent coupling strength~[\citen{Dixon:1986qv,Hamidi:1986vh},
\citen{Casas:1991ac,Erler:1992gt}]. Recently, the correlation function of $n$--point couplings has been
discussed at some extent~\cite{Choi:2007nb}. 

On the other hand, symmetries of the correlation function give rise to the so-called string selection
rules. These rules determine whether a given coupling vanishes or not. We use the following
notation: the constructing elements of $\Psi_i$ are denoted by $g_i=(\theta_i,n_\al^ie_\al) \in S$ and their left- and
right-moving shifted momenta, by $p_{\mathrm{sh},i}$ and $q_{\mathrm{sh},i}$, respectively. Then, the
string selection rules read:

\vskip 2mm
\noindent{\large \bf $\bullet$ Gauge invariance. }\index{Yukawa coupling selection rules!gauge invariance}\\
Since the 16-dimensional left--moving momenta describe the gauge quantum numbers, the sum over all
left--moving shifted momenta $p_{\mathrm{sh},i}$ must vanish: 
\begin{equation}
\sum_i p_{\mathrm{sh},i} = 0
\label{eq:gaugeinvariance} 
\end{equation}
This translates to the field theoretic requirement of gauge invariance for allowed terms in the
superpotential. However, note that summing all momenta is, in practice, very cumbersome. Instead, one can
verify gauge invariance directly by computing all abelian and nonabelian representations corresponding to
the particles in the spectrum and then using well-known rules to form gauge invariant combinations of
(super)fields. We apply this second approach.

\vskip 2mm
\noindent{\large \bf $\bullet$ Conservation of R--charge. }\index{Yukawa coupling selection rules!conservation of R--charge}\\
In orbifold constructions, R--symmetries are discrete symmetries in the six-dimensional space inherited
from Lorentz invariance of the ten-dimensional theory. Basically, they arise from demanding invariance of
the compact space under the twist. Generically, there are three such symmetries --one for each complex
plane-- whose quantum numbers can be identified with the last three components of the \SO8 weight momenta
$q_\text{sh}$ of the right movers. Since $q_\text{sh}$ is not invariant under the ghost picture
changing~\cite{Kobayashi:2004ya}, the R--charges have to be amended by some oscillator contributions,
resulting in
\begin{equation}\label{eq:RChargeConservation}
R_i~=~q_{\mathrm{sh},i} - \ti N_{g,i} + \ti N_{g,i}^{*} 
\end{equation}
which lie in the \SO8 weight lattice. Here $\ti N_{g,i}$ and $\ti N_{g,i}^{*}$ are (vectors of) integer oscillator
numbers, counting the number of holomorphic and antiholomorphic oscillator excitations,
respectively. From the definition of the charges~\eqref{eq:RChargeConservation}, one notices that
the corresponding R--symmetries do distinguish between bosons and fermions.\footnote{Recall that fermions
and bosons have different right--moving momenta.}

In the untwisted sector, the bosonic R--charges, eq.~\eqref{eq:RChargeConservation}, have only three
different values.  This allows to split the untwisted sector $U$ in further untwisted sectors $U_1$,
$U_2$ and $U_3$ comprised by (super)fields with R--charges $(0,1,0,0)$, $(0,0,1,0)$ and
$(0,0,0,1)$, respectively.

Invariance of the theory under R--symmetries constrains the superpotential of the theory. Since these
symmetries are discrete (as they arise from discrete rotations), the invariance conditions can be stated as
\begin{equation}
\sum_i R_i^a~=~0 \mod  N^a \quad\text{ for }\ a=1,2,3\;,
\label{eq:rchargeconservation} 
\end{equation}
where $N^a$ denotes the order of the twist action on the $a^\mathrm{th}$ complex plane, i.e.\  it is the
smallest integers such that $N^a v^a \in \mathbb{Z}$ (no summation). Here, two of the $R_i$ come from
fermions and the rest from bosons in order to be allowed in the superpotential. For computational
purposes, it is more convenient to use the purely bosonic notation, where
eq.~\eqref{eq:rchargeconservation} becomes  
\be
\sum_i R_i^a = -1 \text{ mod } N^a\ .
\label{eq:BosonicRChargeConservation}
\ee

A caveat is in order here: in all moment we have assumed that the lattice $\Gamma$ of the compact space
is factorizable, that is, that $\Gamma$ can be written as the product of three one-complex-dimensional
sublattices, each of which is embedded in a complex plane $Z^a$ of the internal dimensions. If $\Gamma$
is nonfactorizable, eq.~\eqref{eq:rchargeconservation} has to be modified. A brief discussion on this
issue is provided in appendix~\ref{ch:Z6IINonFactorizable}.

\begin{figure}[t!]
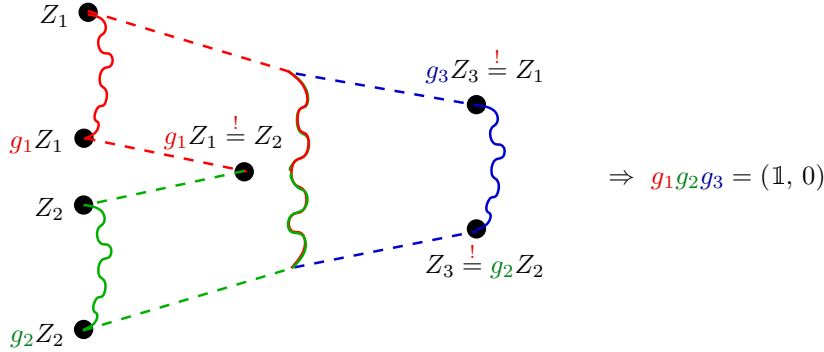

\centerline{\input SG_rule.pstex_t}
\caption{The space group selection rule can be visualized as the ability of twisted strings to
  join. $g_i$ denote three constructing elements.}
\label{fig:SGRule}
\end{figure}

\vskip 2mm
\noindent
{\large \bf $\bullet$ Space group selection rule. }\index{Yukawa coupling selection rules!space group selection rule}\\
Since the states entering the superpotential carry also some information about their space group properties,
the product of constructing elements $g_i$ must lie in the same conjugacy class as the identity, i.e.
\begin{equation}
\prod_i g_i \simeq (\mathbbm{1}, 0) \ .
\end{equation}
In terms of conjugate elements $h_i g_i h_i^{-1}$ of $g_i$, this condition can be reformulated
as~\cite{Kobayashi:1990mc}
\be
\prod_i h_ig_ih_i^{-1}=(\mathbbm{1}, v)\quad\text{ with }\quad v\in\sum_i(\mathbbm{1}-\theta_i)\Lambda\,.   
\ee
Notice that this rule implies in particular that
\be\label{eq:PointGroupRule}
\prod_i \theta_i = \id\,,
\ee
condition which is known in the literature as the {\it point group selection rule}. Stated in this way,
the point group selection rule can be reinterpreted as a discrete symmetry~\cite{Araki:2008ek}. In fact, also
the translational part on the space group rule ($v\in\sum_i(\mathbbm{1}-\theta_i)\Lambda$) can be seen as
a discrete (flavor) symmetry.

This selection rule can be visualized as the geometrical ability of twisted strings to join (see fig.~\ref{fig:SGRule}).

\subsubsection*{On the $\bs{\gamma}$--Rule(s)}
In the literature, there exists an additional selection rule, usually called $\gamma$--rule. In our notation, it
reads~\cite{Casas:1991ac, Kobayashi:2004ya}\index{Yukawa coupling selection rules!gamma rule@$\gamma$--rule}
\begin{equation}
\sum_i \gamma_i = 0\mod1\;,
\label{eq:traditionalgamma}
\end{equation}
where $\gamma_i$ denotes the so-called $\gamma$--phase of $\Psi_i$. If we suppose that the states
$\Psi_i$ are associated to the conjugacy classes of $g_i\in S$, a phase $e^{2\pi\I\gamma_i(h)}$ arises from the action
of a noncommuting element $h$ ($[g_i,h]\neq0$) on the geometrical part of $\Psi_i$. However, since the physical
states $\Psi_i$ are, by definition, space group invariant, this phase comes always along with a phase
$\Phi(p_\text{sh},q_\text{sh}, h)$, such that
\be
\Psi_i\,\stackrel{h}{\longrightarrow}\,e^{2\pi\I\gamma_i(h)}\,\Phi_i(p_\text{sh},q_\text{sh},h)\;\Psi_i ~=~ \Psi_i\,.
\ee
It follows then that the gamma phases $\gamma_i(h)$ can also be written in terms of $p_\text{sh},q_\text{sh}$
and the embedding of $h$ in the gauge degrees of freedom. Considering a coupling
$\Psi_1\Psi_2\Psi_3\ldots$, it turns out that this relation implies that
\be
e^{-2\pi\I\sum_i\gamma_i(h)}=\prod_i\Phi_i(p_\text{sh},q_\text{sh},h)\,.
\ee
Further, we have shown~\cite{Lebedev:2007hv} that $\prod_i\Phi_i(p_\text{sh},q_\text{sh},h)=1$ follows from
the selection rules listed above. Therefore, the expression
\be
\sum_i \gamma_i(h) = 0\mod1
\label{eq:generalizedgamma}
\ee
with an arbitrary space group element $h\in S$, is always true for those couplings allowed by the previous
rules. Notice that if we consider that the elements $h_1=(\vartheta,0)$, $h_2=(\omega,0)$ and $h_{\al+2}=(\id,e_\al)$
form a basis of $S$, we can have at most eight $\gamma$--``rules''. The one traditionally considered
corresponds to that associated with the space group element $h=(\vartheta,0)$ of \Z{N} orbifolds. For a very
detailed discussion on the derivation and triviality of the $\gamma$-rule(s), we refer
to~\cite{Vaudrevange:2008sm}.

%% file: Heterotic.pstex_t
\begin{picture}(0,0)%
\includegraphics{Heterotic.pstex}%
\end{picture}%
\setlength{\unitlength}{4144sp}%
\begingroup\makeatletter\ifx\SetFigFont\undefined%
\gdef\SetFigFont#1#2#3#4#5{%
  \reset@font\fontsize{#1}{#2pt}%
  \fontfamily{#3}\fontseries{#4}\fontshape{#5}%
  \selectfont}%
\fi\endgroup%
\begin{picture}(7064,2010)(-21,-1006)
\put( 46,299){\makebox(0,0)[lb]{\smash{{\SetFigFont{9}{10.8}{\rmdefault}{\mddefault}{\updefault}$\Psi_R^i$}}}}
\put( 46,569){\makebox(0,0)[lb]{\smash{{\SetFigFont{9}{10.8}{\rmdefault}{\mddefault}{\updefault}$X_R^i$}}}}
\put(1216,299){\makebox(0,0)[lb]{\smash{{\SetFigFont{9}{10.8}{\rmdefault}{\mddefault}{\updefault}$\Psi_R^i$}}}}
\put(1216,569){\makebox(0,0)[lb]{\smash{{\SetFigFont{9}{10.8}{\rmdefault}{\mddefault}{\updefault}$X_R^i$}}}}
\put(1531,434){\makebox(0,0)[lb]{\smash{{\SetFigFont{9}{10.8}{\rmdefault}{\mddefault}{\updefault}$i=5,\ldots,10$}}}}
\put(316,434){\makebox(0,0)[lb]{\smash{{\SetFigFont{9}{10.8}{\rmdefault}{\mddefault}{\updefault}$i=1,\ldots,4$}}}}
\put(  1,-511){\makebox(0,0)[lb]{\smash{{\SetFigFont{9}{10.8}{\rmdefault}{\mddefault}{\updefault}$X_L^i,\,i=1,\ldots,4$}}}}
\put(1216,-511){\makebox(0,0)[lb]{\smash{{\SetFigFont{9}{10.8}{\rmdefault}{\mddefault}{\updefault}$X_L^i,\,i=5,\ldots,10$}}}}
\put(4816,-376){\makebox(0,0)[lb]{\smash{{\SetFigFont{10}{12.0}{\rmdefault}{\mddefault}{\updefault}$I=1,\ldots,16$}}}}
\put(4051,-556){\makebox(0,0)[lb]{\smash{{\SetFigFont{10}{12.0}{\rmdefault}{\mddefault}{\updefault}$\text{gauge degrees of freedom}$}}}}
\put(4141,-376){\makebox(0,0)[lb]{\smash{{\SetFigFont{10}{12.0}{\rmdefault}{\mddefault}{\updefault}$X_L^I$}}}}
\put(  1,-961){\makebox(0,0)[lb]{\smash{{\SetFigFont{11}{13.2}{\rmdefault}{\mddefault}{\updefault}$\text{Left Movers}$}}}}
\put(  1,884){\makebox(0,0)[lb]{\smash{{\SetFigFont{11}{13.2}{\rmdefault}{\mddefault}{\updefault}$\text{Right Movers}$}}}}
\put(2656,-61){\makebox(0,0)[lb]{\smash{{\SetFigFont{12}{14.4}{\rmdefault}{\mddefault}{\updefault}$\otimes$}}}}
\put(1666,-61){\makebox(0,0)[lb]{\smash{{\SetFigFont{12}{14.4}{\rmdefault}{\mddefault}{\updefault}$\mathbbm{R}^6$}}}}
\put(991,-61){\makebox(0,0)[lb]{\smash{{\SetFigFont{12}{14.4}{\rmdefault}{\mddefault}{\updefault}$\otimes$}}}}
\put(361,-61){\makebox(0,0)[lb]{\smash{{\SetFigFont{12}{14.4}{\rmdefault}{\mddefault}{\updefault}$\mathbbm{M}^4$}}}}
\put(4681,-61){\makebox(0,0)[lb]{\smash{{\SetFigFont{12}{14.4}{\rmdefault}{\mddefault}{\updefault}$T^{16}$}}}}
\end{picture}%

%% file: S1byZ2.pstex_t
\begin{picture}(0,0)%
\includegraphics{S1byZ2.pstex}%
\end{picture}%
\setlength{\unitlength}{4144sp}%
\begingroup\makeatletter\ifx\SetFigFont\undefined%
\gdef\SetFigFont#1#2#3#4#5{%
  \reset@font\fontsize{#1}{#2pt}%
  \fontfamily{#3}\fontseries{#4}\fontshape{#5}%
  \selectfont}%
\fi\endgroup%
\begin{picture}(3330,2281)(46,-1501)
\put(1531,-556){\makebox(0,0)[lb]{\smash{{\SetFigFont{10}{12.0}{\rmdefault}{\mddefault}{\updefault}$x$}}}}
\put( 46,-1456){\makebox(0,0)[lb]{\smash{{\SetFigFont{10}{12.0}{\rmdefault}{\mddefault}{\updefault}$-\pi$}}}}
\put(1576,-1456){\makebox(0,0)[lb]{\smash{{\SetFigFont{10}{12.0}{\rmdefault}{\mddefault}{\updefault}$x=0$}}}}
\put(3376,-106){\makebox(0,0)[lb]{\smash{{\SetFigFont{10}{12.0}{\rmdefault}{\mddefault}{\updefault}$\pi$}}}}
\put(1921,-99){\makebox(0,0)[lb]{\smash{{\SetFigFont{10}{12.0}{\rmdefault}{\mddefault}{\updefault}$\vartheta$}}}}
\put(1314,-106){\makebox(0,0)[lb]{\smash{{\SetFigFont{10}{12.0}{\rmdefault}{\mddefault}{\updefault}$x=0$}}}}
\put(1704,-983){\makebox(0,0)[lb]{\smash{{\SetFigFont{10}{12.0}{\rmdefault}{\mddefault}{\updefault}$\vartheta$}}}}
\put(3204,-1456){\makebox(0,0)[lb]{\smash{{\SetFigFont{10}{12.0}{\rmdefault}{\mddefault}{\updefault}$\pi$}}}}
\end{picture}%

%% file: Z3Cristal.pstex_t
\begin{picture}(0,0)%
\includegraphics{Z3Cristal.pstex}%
\end{picture}%
\setlength{\unitlength}{4144sp}%
\begingroup\makeatletter\ifx\SetFigFont\undefined%
\gdef\SetFigFont#1#2#3#4#5{%
  \reset@font\fontsize{#1}{#2pt}%
  \fontfamily{#3}\fontseries{#4}\fontshape{#5}%
  \selectfont}%
\fi\endgroup%
\begin{picture}(5996,2272)(104,-1538)
\put(4802,-158){\makebox(0,0)[lb]{\smash{{\SetFigFont{10}{12.0}{\rmdefault}{\mddefault}{\updefault}$\Z3$}}}}
\put(1644,-188){\makebox(0,0)[lb]{\smash{{\SetFigFont{10}{12.0}{\rmdefault}{\mddefault}{\updefault}$\Z3$}}}}
\put(3013,614){\makebox(0,0)[lb]{\smash{{\SetFigFont{11}{13.2}{\rmdefault}{\mddefault}{\updefault}$\text{b) \SU3}$}}}}
\put(2222,-691){\makebox(0,0)[lb]{\smash{{\SetFigFont{10}{12.0}{\rmdefault}{\mddefault}{\updefault}$e_1$}}}}
\put(1486,299){\makebox(0,0)[lb]{\smash{{\SetFigFont{10}{12.0}{\rmdefault}{\mddefault}{\updefault}$e_2$}}}}
\put(4218,220){\makebox(0,0)[lb]{\smash{{\SetFigFont{10}{12.0}{\rmdefault}{\mddefault}{\updefault}$e_2$}}}}
\put(5407,-675){\makebox(0,0)[lb]{\smash{{\SetFigFont{10}{12.0}{\rmdefault}{\mddefault}{\updefault}$e_1$}}}}
\put(303,614){\makebox(0,0)[lb]{\smash{{\SetFigFont{11}{13.2}{\rmdefault}{\mddefault}{\updefault}$\text{b) \SO4}$}}}}
\end{picture}%

%% file: Z3_fixedpoints.pstex_t
\begin{picture}(0,0)%
\includegraphics{Z3_fixedpoints.pstex}%
\end{picture}%
\setlength{\unitlength}{4144sp}%
\begingroup\makeatletter\ifx\SetFigFont\undefined%
\gdef\SetFigFont#1#2#3#4#5{%
  \reset@font\fontsize{#1}{#2pt}%
  \fontfamily{#3}\fontseries{#4}\fontshape{#5}%
  \selectfont}%
\fi\endgroup%
\begin{picture}(5325,1694)(333,-949)
\put(333,567){\makebox(0,0)[lb]{\smash{{\SetFigFont{10}{12.0}{\rmdefault}{\mddefault}{\updefault}$e_2$}}}}
\put(2386,-793){\makebox(0,0)[lb]{\smash{{\SetFigFont{10}{12.0}{\rmdefault}{\mddefault}{\updefault}$e_1$}}}}
\put(1088,246){\makebox(0,0)[lb]{\smash{{\SetFigFont{10}{12.0}{\rmdefault}{\mddefault}{\updefault}$\azul{\Z3}$}}}}
\put(3121,-225){\makebox(0,0)[lb]{\smash{{\SetFigFont{10}{12.0}{\rmdefault}{\mddefault}{\updefault}$\azul{\Z3}$}}}}
\put(5289,-649){\makebox(0,0)[lb]{\smash{{\SetFigFont{10}{12.0}{\rmdefault}{\mddefault}{\updefault}$e_1$}}}}
\put(3288,613){\makebox(0,0)[lb]{\smash{{\SetFigFont{10}{12.0}{\rmdefault}{\mddefault}{\updefault}$e_2$}}}}
\end{picture}%

%% file: Z3_pillow.pstex_t
\begin{picture}(0,0)%
\includegraphics{Z3_pillow.pstex}%
\end{picture}%
\setlength{\unitlength}{4144sp}%
\begingroup\makeatletter\ifx\SetFigFont\undefined%
\gdef\SetFigFont#1#2#3#4#5{%
  \reset@font\fontsize{#1}{#2pt}%
  \fontfamily{#3}\fontseries{#4}\fontshape{#5}%
  \selectfont}%
\fi\endgroup%
\begin{picture}(4345,1474)(402,-785)
\put(3871,569){\makebox(0,0)[lb]{\smash{{\SetFigFont{10}{12.0}{\rmdefault}{\mddefault}{\updefault}$\text{``pillow''}$}}}}
\put(402,561){\makebox(0,0)[lb]{\smash{{\SetFigFont{9}{10.8}{\rmdefault}{\mddefault}{\updefault}$e_2$}}}}
\put(2344,-754){\makebox(0,0)[lb]{\smash{{\SetFigFont{9}{10.8}{\rmdefault}{\mddefault}{\updefault}$e_1$}}}}
\end{picture}%

%% file: Z6II_fixedpoints_G2.pstex_t
\begin{picture}(0,0)%
\includegraphics{Z6II_fixedpoints_G2.pstex}%
\end{picture}%
\setlength{\unitlength}{4144sp}%
\begingroup\makeatletter\ifx\SetFigFont\undefined%
\gdef\SetFigFont#1#2#3#4#5{%
  \reset@font\fontsize{#1}{#2pt}%
  \fontfamily{#3}\fontseries{#4}\fontshape{#5}%
  \selectfont}%
\fi\endgroup%
\begin{picture}(6685,1569)(718,-1059)
\put(4235,378){\makebox(0,0)[lb]{\smash{{\SetFigFont{10}{12.0}{\rmdefault}{\mddefault}{\updefault}$e_2$}}}}
\put(6404,-1010){\makebox(0,0)[lb]{\smash{{\SetFigFont{10}{12.0}{\rmdefault}{\mddefault}{\updefault}$e_1$}}}}
\put(7034,378){\makebox(0,0)[lb]{\smash{{\SetFigFont{10}{12.0}{\rmdefault}{\mddefault}{\updefault}$e_2$}}}}
\put(3605,-1010){\makebox(0,0)[lb]{\smash{{\SetFigFont{10}{12.0}{\rmdefault}{\mddefault}{\updefault}$e_1$}}}}
\put(5446,-286){\makebox(0,0)[lb]{\smash{{\SetFigFont{10}{12.0}{\rmdefault}{\mddefault}{\updefault}$\azul{\vartheta}$}}}}
\put(2272,370){\makebox(0,0)[lb]{\smash{{\SetFigFont{10}{12.0}{\rmdefault}{\mddefault}{\updefault}$\azul{\vartheta^2}$}}}}
\end{picture}%

%% file: Z3_geometry.pstex_t
\begin{picture}(0,0)%
\includegraphics{Z3_geometry.pstex}%
\end{picture}%
\setlength{\unitlength}{3729sp}%
\begingroup\makeatletter\ifx\SetFigFont\undefined%
\gdef\SetFigFont#1#2#3#4#5{%
  \reset@font\fontsize{#1}{#2pt}%
  \fontfamily{#3}\fontseries{#4}\fontshape{#5}%
  \selectfont}%
\fi\endgroup%
\begin{picture}(6878,1307)(160,-542)
\put(6399,629){\makebox(0,0)[lb]{\smash{{\SetFigFont{9}{10.8}{\rmdefault}{\mddefault}{\updefault}$\text{\bf Fixed Points}$}}}}
\put(6669, 84){\makebox(0,0)[lb]{\smash{{\SetFigFont{10}{12.0}{\sfdefault}{\mddefault}{\updefault}$27$}}}}
\put(455,630){\makebox(0,0)[lb]{\smash{{\SetFigFont{9}{10.8}{\rmdefault}{\mddefault}{\updefault}$e_2$}}}}
\put(2392,633){\makebox(0,0)[lb]{\smash{{\SetFigFont{9}{10.8}{\rmdefault}{\mddefault}{\updefault}$e_4$}}}}
\put(4459,624){\makebox(0,0)[lb]{\smash{{\SetFigFont{9}{10.8}{\rmdefault}{\mddefault}{\updefault}$e_6$}}}}
\put(6105,-493){\makebox(0,0)[lb]{\smash{{\SetFigFont{9}{10.8}{\rmdefault}{\mddefault}{\updefault}$e_5$}}}}
\put(4023,-492){\makebox(0,0)[lb]{\smash{{\SetFigFont{9}{10.8}{\rmdefault}{\mddefault}{\updefault}$e_3$}}}}
\put(2079,-489){\makebox(0,0)[lb]{\smash{{\SetFigFont{9}{10.8}{\rmdefault}{\mddefault}{\updefault}$e_1$}}}}
\put(160,177){\makebox(0,0)[lb]{\smash{{\SetFigFont{10}{12.0}{\rmdefault}{\mddefault}{\updefault}$\vartheta^1$}}}}
\end{picture}%

%% file: Z6II_geometry.pstex_t
\begin{picture}(0,0)%
\includegraphics{Z6II_geometry.pstex}%
\end{picture}%
\setlength{\unitlength}{3729sp}%
\begingroup\makeatletter\ifx\SetFigFont\undefined%
\gdef\SetFigFont#1#2#3#4#5{%
  \reset@font\fontsize{#1}{#2pt}%
  \fontfamily{#3}\fontseries{#4}\fontshape{#5}%
  \selectfont}%
\fi\endgroup%
\begin{picture}(6750,4925)(324,-4146)
\put(6747,-1714){\makebox(0,0)[lb]{\smash{{\SetFigFont{9}{10.8}{\sfdefault}{\mddefault}{\updefault}$6$}}}}
\put(6783,-3541){\makebox(0,0)[lb]{\smash{{\SetFigFont{9}{10.8}{\sfdefault}{\mddefault}{\updefault}$8$}}}}
\put(6362,659){\makebox(0,0)[lb]{\smash{{\SetFigFont{9}{10.8}{\rmdefault}{\mddefault}{\updefault}$\text{\bf Fixed Points}$}}}}
\put(6705, -5){\makebox(0,0)[lb]{\smash{{\SetFigFont{9}{10.8}{\sfdefault}{\mddefault}{\updefault}$12$}}}}
\put(2522,614){\makebox(0,0)[lb]{\smash{{\SetFigFont{9}{10.8}{\rmdefault}{\mddefault}{\updefault}$e_4$}}}}
\put(2522,-1150){\makebox(0,0)[lb]{\smash{{\SetFigFont{9}{10.8}{\rmdefault}{\mddefault}{\updefault}$e_4$}}}}
\put(1234,-3665){\makebox(0,0)[lb]{\smash{{\SetFigFont{9}{10.8}{\rmdefault}{\mddefault}{\updefault}$\azul{\vartheta}$}}}}
\put(1333,-1898){\makebox(0,0)[lb]{\smash{{\SetFigFont{9}{10.8}{\rmdefault}{\mddefault}{\updefault}$\azul{\vartheta}$}}}}
\put(2026,612){\makebox(0,0)[lb]{\smash{{\SetFigFont{9}{10.8}{\rmdefault}{\mddefault}{\updefault}$e_2$}}}}
\put(2026,-1152){\makebox(0,0)[lb]{\smash{{\SetFigFont{9}{10.8}{\rmdefault}{\mddefault}{\updefault}$e_2$}}}}
\put(4366,-2450){\makebox(0,0)[lb]{\smash{{\SetFigFont{9}{10.8}{\rmdefault}{\mddefault}{\updefault}$e_3$}}}}
\put(1494,-2439){\makebox(0,0)[lb]{\smash{{\SetFigFont{9}{10.8}{\rmdefault}{\mddefault}{\updefault}$e_1$}}}}
\put(1464,-4097){\makebox(0,0)[lb]{\smash{{\SetFigFont{9}{10.8}{\rmdefault}{\mddefault}{\updefault}$e_1$}}}}
\put(2002,-2808){\makebox(0,0)[lb]{\smash{{\SetFigFont{9}{10.8}{\rmdefault}{\mddefault}{\updefault}$e_2$}}}}
\put(1480,-669){\makebox(0,0)[lb]{\smash{{\SetFigFont{9}{10.8}{\rmdefault}{\mddefault}{\updefault}$e_1$}}}}
\put(4279,-671){\makebox(0,0)[lb]{\smash{{\SetFigFont{9}{10.8}{\rmdefault}{\mddefault}{\updefault}$e_3$}}}}
\put(5962,-671){\makebox(0,0)[lb]{\smash{{\SetFigFont{9}{10.8}{\rmdefault}{\mddefault}{\updefault}$e_5$}}}}
\put(4666,607){\makebox(0,0)[lb]{\smash{{\SetFigFont{9}{10.8}{\rmdefault}{\mddefault}{\updefault}$e_6$}}}}
\put(4663,-2807){\makebox(0,0)[lb]{\smash{{\SetFigFont{9}{10.8}{\rmdefault}{\mddefault}{\updefault}$e_6$}}}}
\put(6000,-4089){\makebox(0,0)[lb]{\smash{{\SetFigFont{9}{10.8}{\rmdefault}{\mddefault}{\updefault}$e_5$}}}}
\put(324,307){\makebox(0,0)[lb]{\smash{{\SetFigFont{9}{10.8}{\rmdefault}{\mddefault}{\updefault}$\vartheta^1$}}}}
\put(324,-3113){\makebox(0,0)[lb]{\smash{{\SetFigFont{9}{10.8}{\rmdefault}{\mddefault}{\updefault}$\vartheta^3$}}}}
\put(324,-1457){\makebox(0,0)[lb]{\smash{{\SetFigFont{9}{10.8}{\rmdefault}{\mddefault}{\updefault}$\vartheta^2$}}}}
\end{picture}%

%% file: StringTypes.pstex_t
\begin{picture}(0,0)%
\includegraphics{StringTypes.pstex}%
\end{picture}%
\setlength{\unitlength}{3729sp}%
\begingroup\makeatletter\ifx\SetFigFont\undefined%
\gdef\SetFigFont#1#2#3#4#5{%
  \reset@font\fontsize{#1}{#2pt}%
  \fontfamily{#3}\fontseries{#4}\fontshape{#5}%
  \selectfont}%
\fi\endgroup%
\begin{picture}(6509,2194)(75,-1400)
\put(2603,-1198){\makebox(0,0)[lb]{\smash{{\SetFigFont{9}{10.8}{\rmdefault}{\mddefault}{\updefault}$e_1$}}}}
\put(240,402){\makebox(0,0)[lb]{\smash{{\SetFigFont{9}{10.8}{\rmdefault}{\mddefault}{\updefault}$e_2$}}}}
\put(6215,-1114){\makebox(0,0)[lb]{\smash{{\SetFigFont{9}{10.8}{\rmdefault}{\mddefault}{\updefault}$e_1$}}}}
\put(3852,486){\makebox(0,0)[lb]{\smash{{\SetFigFont{9}{10.8}{\rmdefault}{\mddefault}{\updefault}$e_2$}}}}
\put(3485,674){\makebox(0,0)[lb]{\smash{{\SetFigFont{9}{10.8}{\rmdefault}{\mddefault}{\updefault}$\text{b)}$}}}}
\put( 75,592){\makebox(0,0)[lb]{\smash{{\SetFigFont{9}{10.8}{\rmdefault}{\mddefault}{\updefault}$\text{a)}$}}}}
\put(4226,-669){\makebox(0,0)[lb]{\smash{{\SetFigFont{9}{10.8}{\rmdefault}{\mddefault}{\updefault}$\azul{\vartheta}$}}}}
\end{picture}%

%% file: Z3_example_noWL.pstex_t
\begin{picture}(0,0)%
\includegraphics{Z3_example_noWL.pstex}%
\end{picture}%
\setlength{\unitlength}{4144sp}%
\begingroup\makeatletter\ifx\SetFigFont\undefined%
\gdef\SetFigFont#1#2#3#4#5{%
  \reset@font\fontsize{#1}{#2pt}%
  \fontfamily{#3}\fontseries{#4}\fontshape{#5}%
  \selectfont}%
\fi\endgroup%
\begin{picture}(6025,1041)(226,-443)
\put(2926,398){\makebox(0,0)[lb]{\smash{{\SetFigFont{10}{12.0}{\rmdefault}{\mddefault}{\updefault}$(\bsb{9},\,\bs{1})_{12}$}}}}
\put(226,177){\makebox(0,0)[lb]{\smash{{\SetFigFont{10}{12.0}{\rmdefault}{\mddefault}{\updefault}$T_2$}}}}
\end{picture}%

%% file: Z3_example_2WL.pstex_t
\begin{picture}(0,0)%
\includegraphics{Z3_example_2WL.pstex}%
\end{picture}%
\setlength{\unitlength}{4144sp}%
\begingroup\makeatletter\ifx\SetFigFont\undefined%
\gdef\SetFigFont#1#2#3#4#5{%
  \reset@font\fontsize{#1}{#2pt}%
  \fontfamily{#3}\fontseries{#4}\fontshape{#5}%
  \selectfont}%
\fi\endgroup%
\begin{picture}(6159,1309)(92,-711)
\put(3962,-662){\makebox(0,0)[lb]{\smash{{\SetFigFont{10}{12.0}{\rmdefault}{\mddefault}{\updefault}$\verde{A_3}$}}}}
\put(2010,-662){\makebox(0,0)[lb]{\smash{{\SetFigFont{10}{12.0}{\rmdefault}{\mddefault}{\updefault}$\azul{A_1}$}}}}
\put( 92, 12){\makebox(0,0)[lb]{\smash{{\SetFigFont{10}{12.0}{\rmdefault}{\mddefault}{\updefault}$T_2$}}}}
\end{picture}%

%% file: SG_rule.pstex_t
\begin{picture}(0,0)%
\includegraphics{SG_rule.pstex}%
\end{picture}%
\setlength{\unitlength}{4144sp}%
\begingroup\makeatletter\ifx\SetFigFont\undefined%
\gdef\SetFigFont#1#2#3#4#5{%
  \reset@font\fontsize{#1}{#2pt}%
  \fontfamily{#3}\fontseries{#4}\fontshape{#5}%
  \selectfont}%
\fi\endgroup%
\begin{picture}(5055,2088)(18,-1418)
\put(2506,210){\makebox(0,0)[lb]{\smash{{\SetFigFont{10}{12.0}{\rmdefault}{\mddefault}{\updefault}$\azul{g_3}Z_3\stackrel{\rojo{!}}{=}Z_1$}}}}
\put( 23,-1369){\makebox(0,0)[lb]{\smash{{\SetFigFont{10}{12.0}{\rmdefault}{\mddefault}{\updefault}$\verde{g_2}Z_2$}}}}
\put( 18,-210){\makebox(0,0)[lb]{\smash{{\SetFigFont{10}{12.0}{\rmdefault}{\mddefault}{\updefault}$\rojo{g_1}Z_1$}}}}
\put(2503,-958){\makebox(0,0)[lb]{\smash{{\SetFigFont{10}{12.0}{\rmdefault}{\mddefault}{\updefault}$Z_3\stackrel{\rojo{!}}{=}\verde{g_2}Z_2$}}}}
\put(167,-617){\makebox(0,0)[lb]{\smash{{\SetFigFont{10}{12.0}{\rmdefault}{\mddefault}{\updefault}$Z_2$}}}}
\put(198,538){\makebox(0,0)[lb]{\smash{{\SetFigFont{10}{12.0}{\rmdefault}{\mddefault}{\updefault}$Z_1$}}}}
\put(3601,-421){\makebox(0,0)[lb]{\smash{{\SetFigFont{10}{12.0}{\rmdefault}{\mddefault}{\updefault}$\Rightarrow\ \rojo{g_1}\verde{g_2}\azul{g_3}=(\id,\,0)$}}}}
\put(941,-183){\makebox(0,0)[lb]{\smash{{\SetFigFont{10}{12.0}{\rmdefault}{\mddefault}{\updefault}$\rojo{g_1}Z_1\stackrel{\rojo{!}}{=}Z_2$}}}}
\end{picture}%

%% file: OrbiClass.tex
\chapter{Classification of Orbifolds}
\label{ch:ClassificationOfOrbifolds}

\begin{center}
\begin{minipage}[t]{14cm}
This chapter is devoted to the techniques used to arrive systematically to inequivalent orbifold
compactifications of the heterotic string. We review first the frequently called {\it Dynkin diagram
method}, based on a theorem by Ka\v{c}. This method is used mainly to classify models without
background fields (Wilson lines). Models with Wilson lines can be classified more effectively
by using a proper ansatz that characterizes Wilson lines of a given order, as discussed in
section~\ref{sec:ClassAnsatz}. Finally, we introduce {\it the C++ Orbifolder}, a computer program
developed to classify orbifold models and compute their properties.
\end{minipage}
\end{center}

\section[\Z{N} Orbifolds without Wilson Lines]{$\bs{\Z{N}}$ Orbifolds without Wilson Lines}
\label{sec:DynkinMethod}
\index{Dynkin diagram!SO(32)@\SO{32} and \E8}

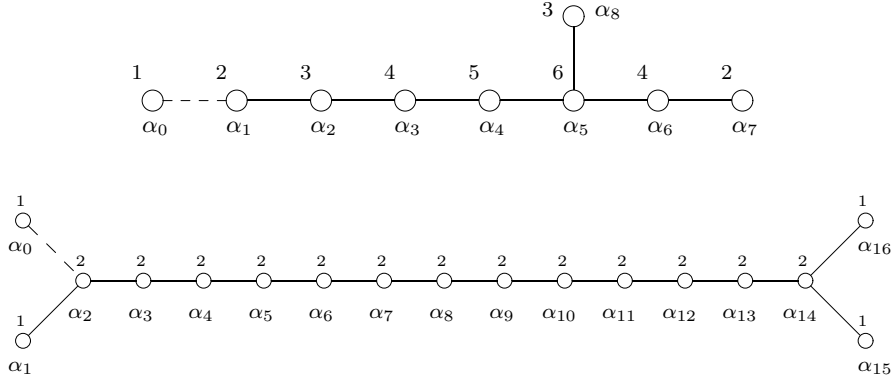
\begin{figure}[t]
\begin{center}
\subfigure{
\setlength{\unitlength}{1.4mm} 
\begin{picture}(70,13) 

\put(8,4){\circle{2}}
\put(16,4){\circle{2}}
\put(24,4){\circle{2}}
\put(32,4){\circle{2}}
\put(40,4){\circle{2}}
\put(48,4){\circle{2}}
\put(56,4){\circle{2}}
\put(64,4){\circle{2}}
\put(48,12){\circle{2}}

\put(9,4){\dashline{1}(0,0)(6,0)}
\put(17,4){\line(1,0){6}} 
\put(25,4){\line(1,0){6}} 
\put(33,4){\line(1,0){6}} 
\put(41,4){\line(1,0){6}} 
\put(49,4){\line(1,0){6}} 
\put(57,4){\line(1,0){6}} 
\put(48,5){\line(0,1){6}} 

{\footnotesize
\put(7,1){$\alpha_0$}
\put(15,1){$\alpha_1$}
\put(23,1){$\alpha_2$}
\put(31,1){$\alpha_3$}
\put(39,1){$\alpha_4$}
\put(47,1){$\alpha_5$}
\put(55,1){$\alpha_6$}
\put(63,1){$\alpha_7$}
\put(50,12){$\alpha_8$}
}

\put(6,6){${\scriptstyle 1}$}
\put(14,6){${\scriptstyle 2}$}
\put(22,6){${\scriptstyle 3}$}
\put(30,6){${\scriptstyle 4}$}
\put(38,6){${\scriptstyle 5}$}
\put(46,6){${\scriptstyle 6}$}
\put(54,6){${\scriptstyle 4}$}
\put(62,6){${\scriptstyle 2}$}
\put(45,12){${\scriptstyle 3}$}

\end{picture} 
}

\subfigure{
\setlength{\unitlength}{1mm} 
\begin{picture}(130,22) 

\put(10,19){\circle{2}}
\put(10,3){\circle{2}}
\put(18,11){\circle{2}}
\put(26,11){\circle{2}}
\put(34,11){\circle{2}}
\put(42,11){\circle{2}}
\put(50,11){\circle{2}}
\put(58,11){\circle{2}}
\put(66,11){\circle{2}}
\put(74,11){\circle{2}}
\put(82,11){\circle{2}}
\put(90,11){\circle{2}}
\put(98,11){\circle{2}}
\put(106,11){\circle{2}}
\put(114,11){\circle{2}}
\put(122,19){\circle{2}}
\put(122,3){\circle{2}}

\put(10.7071,18.2929){\dashline{2}(0,0)(6,-6)} 
\put(10.7071,3.7071){\line(1,1){6.55}}
\put(19,11){\line(1,0){6}} 
\put(27,11){\line(1,0){6}} 
\put(35,11){\line(1,0){6}} 
\put(43,11){\line(1,0){6}} 
\put(51,11){\line(1,0){6}} 
\put(59,11){\line(1,0){6}} 
\put(67,11){\line(1,0){6}} 
\put(75,11){\line(1,0){6}} 
\put(83,11){\line(1,0){6}} 
\put(91,11){\line(1,0){6}} 
\put(99,11){\line(1,0){6}} 
\put(107,11){\line(1,0){6}} 
\put(114.7071,11.7071){\line(1,1){6.55}} 
\put(114.7071,10.2929){\line(1,-1){6.55}} 

\put(8,15){${\scriptstyle\alpha_0}$}
\put(8,-1){${\scriptstyle\alpha_1}$}
\put(16,6){${\scriptstyle\alpha_2}$}
\put(24,6){${\scriptstyle\alpha_3}$}
\put(32,6){${\scriptstyle\alpha_4}$}
\put(40,6){${\scriptstyle\alpha_5}$}
\put(48,6){${\scriptstyle\alpha_6}$}
\put(56,6){${\scriptstyle\alpha_7}$}
\put(64,6){${\scriptstyle\alpha_8}$}
\put(72,6){${\scriptstyle\alpha_9}$}
\put(79,6){${\scriptstyle\alpha_{10}}$}
\put(87,6){${\scriptstyle\alpha_{11}}$}
\put(95,6){${\scriptstyle\alpha_{12}}$}
\put(103,6){${\scriptstyle\alpha_{13}}$}
\put(111,6){${\scriptstyle\alpha_{14}}$}
\put(121,-1){${\scriptstyle\alpha_{15}}$}
\put(121,15){${\scriptstyle\alpha_{16}}$}

\put(9,5){${\scriptscriptstyle 1}$}
\put(9,21){${\scriptscriptstyle 1}$}
\put(17,13){${\scriptscriptstyle 2}$}
\put(25,13){${\scriptscriptstyle 2}$}
\put(33,13){${\scriptscriptstyle 2}$}
\put(41,13){${\scriptscriptstyle 2}$}
\put(49,13){${\scriptscriptstyle 2}$}
\put(57,13){${\scriptscriptstyle 2}$}
\put(65,13){${\scriptscriptstyle 2}$}
\put(73,13){${\scriptscriptstyle 2}$}
\put(81,13){${\scriptscriptstyle 2}$}
\put(89,13){${\scriptscriptstyle 2}$}
\put(97,13){${\scriptscriptstyle 2}$}
\put(105,13){${\scriptscriptstyle 2}$}
\put(113,13){${\scriptscriptstyle 2}$}
\put(121,5){${\scriptscriptstyle 1}$}
\put(121,21){${\scriptscriptstyle 1}$}

\end{picture} 
}
\end{center}
\vskip -7mm
\caption{Extended Dynkin diagram of~\E{8} and~\SO{32} and the associated Coxeter or Ka\v{c} labels.}
\label{fig:DynkinSO32andE8}
\end{figure}

\index{Dynkin diagram!orbifold classification|(}
In the absence of Wilson lines, the {\it Dynkin diagram
method}~\cite{Kac:1986gs,Choi:2003pq,Wingerter:2005xx} is the standard strategy to obtain the 
gauge embeddings of the point group $P$. It consists in identifying the inner automorphisms of the Lie
algebra of the gauge group $\maG$ (either \E8\x\E8 or \SO{32}), as we will now describe.

Consider the extended Dynkin diagram of \E{8} and \SO{32} given in
fig.~\ref{fig:DynkinSO32andE8}. The numbers attached to the nodes are the Coxeter or Ka\v{c} labels
$k_i$, which are by definition the expansion coefficients of the highest root $\alpha_H$ 
in terms of the simple roots, that is
\begin{equation}
\alpha_H = k_1 \alpha_1 + \ldots + k_r \alpha_{r}\,,
\label{eq:def_kac_labels}
\end{equation} 
where $r$ is the rank\footnote{The rank of \E8 is $r=8$ whereas the rank of \SO{32} is $r=16$.} 
of the algebra. For convenience, the Ka\v{c} label of the {\it most negative root} 
$\alpha_0 \equiv -\alpha_H$ is set to $k_0 = 1$. Then, by a theorem due to 
Ka\v{c}~\cite{Kac:1969xx}, all order-$N$ inner automorphisms of an algebra up to conjugation are given by  
\begin{equation}
\sigma_{s}(E_{\alpha_i}) = \exp\left(2\pi \I s_i/N\right) E_{\alpha_i}, \quad i = 0, \ldots, r,
\label{eq:action_kac_on_operators}
\end{equation}
where $E_\al$ are the step operators of the Lie algebra of \maG and
the sequence $s = (s_0, \ldots, s_r)$ may be chosen arbitrarily subject to the conditions that
the coefficients $s_i$ be nonnegative, relatively prime integers and 
\begin{equation}
\sum_{i=0}^{r} k_i s_i = N.
\label{eq:order_N_intermsof_s_i}
\end{equation}

\subsubsection{The Shift Vector}

The embedding of the point group in the gauge degrees of freedom, described by $X^I\mapsto X^I+V^I$
(see eq.~\eqref{eq:SpaceGroupEmbedding}), induces the transformations
\begin{equation}
\sigma_V(H_i) = H_i, \qquad \sigma_V(E_{\al}) = \exp\left(2\pi \I \,\al\cdot V \right) E_{\al}
\label{eq:action_shift_on_operators}
\end{equation}
on the Cartan generators $H_i$ and step operators $E_\al$ of the Lie algebra of $\mathcal{G}$, with $\al$
being a root of $\mathcal{G}$, and $V$ the shift vector. These transformations clearly describe an
automorphism of the algebra. 

To derive the shift vector corresponding to a given automorphism is now particularly easy. Comparing
eq.~\eqref{eq:action_kac_on_operators} to eq.~\eqref{eq:action_shift_on_operators}, it follows that 
\begin{equation}
\alpha_i \cdot V = \frac{s_i}{N}, \quad i=1, \ldots, r,
\label{eq:lineq_for_V}
\end{equation}
for the $r$ linearly independent simple roots $\alpha_i$. Using that the simple roots $\al_i$ and the
fundamental weights (their duals) satisfy $\al_i\cdot\al^*_j=\delta_{ij}$, one can expand the shift
vector $V$ in terms of the fundamental weights as
\begin{equation}
V = \sum_{i=1}^r \h{s_i}{N} \alpha_i^*\,,
\label{eq:V_in_Dynkin_basis}
\end{equation}
i.e. the integers $s_i$ divided by the order $N$ are the Dynkin labels of $V$. A direct calculation
reveals that this shift vector also gives the correct transformation for the step operator
corresponding to the most negative root $\al_0$, 
$\sigma_V(E_{\al_0}) = \exp\left(2\pi \I \,\al_0\cdot V \right) E_{\al_0}$.

Constructing shift vectors of \E8\x\E8 requires to find two sequences $s^a=(s^a_0,\ldots,s^a_8)$ and
$s^b=(s^b_0,\ldots,s^b_8)$, each of which leads to independent automorphism, $V^a$ and $V^b$, acting
differently on each of the two \E8 gauge factors. One can show that the combination
\be
V~=~(V^a)\,(V^b)
\ee
form an inner automorphism of \E8\x\E8.

To compute the explicit form of the shift vectors, we use tables~\ref{tab:RootsSO32}
and~\ref{tab:RootsE8}, where we provide our choice of the basis for the simple
roots and fundamental weights of \SO{32} and \E8, respectively.

\subsubsection{The Unbroken Gauge Group $\bs{\maG_{4D}}$}
\index{Dynkin diagram!unbroken gauge group@unbroken gauge group $\maG_{4D}$}

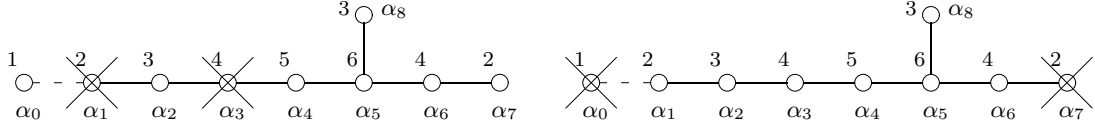
\begin{figure}[t]
\setlength{\unitlength}{1.13mm} 
\begin{minipage}{0.404\textwidth}
\begin{picture}(70,13) 
\put(8,4){\circle{2}}
\put(16,4){\circle{2}} \put(13,1){\line(1,1){6}} \put(13,7){\line(1,-1){6}}
\put(24,4){\circle{2}}
\put(32,4){\circle{2}} \put(29,1){\line(1,1){6}} \put(29,7){\line(1,-1){6}}
\put(40,4){\circle{2}}
\put(48,4){\circle{2}}
\put(56,4){\circle{2}}
\put(64,4){\circle{2}}
\put(48,12){\circle{2}}

\put(9,4){\dashline{1}(0,0)(6,0)}
\put(17,4){\line(1,0){6}} 
\put(25,4){\line(1,0){6}} 
\put(33,4){\line(1,0){6}} 
\put(41,4){\line(1,0){6}} 
\put(49,4){\line(1,0){6}} 
\put(57,4){\line(1,0){6}} 
\put(48,5){\line(0,1){6}} 

{\footnotesize
\put(7,0){${\alpha_0}$}
\put(15,0){${\alpha_1}$}
\put(23,0){${\alpha_2}$}
\put(31,0){${\alpha_3}$}
\put(39,0){${\alpha_4}$}
\put(47,0){${\alpha_5}$}
\put(55,0){${\alpha_6}$}
\put(63,0){${\alpha_7}$}
\put(50,12){${\alpha_8}$}
}

\put(6,6){${\scriptstyle 1}$}
\put(14,6){${\scriptstyle 2}$}
\put(22,6){${\scriptstyle 3}$}
\put(30,6){${\scriptstyle 4}$}
\put(38,6){${\scriptstyle 5}$}
\put(46,6){${\scriptstyle 6}$}
\put(54,6){${\scriptstyle 4}$}
\put(62,6){${\scriptstyle 2}$}
\put(45,12){${\scriptstyle 3}$}
\end{picture}
\end{minipage}
\hspace{1cm}
\begin{minipage}{0.44\textwidth}
\begin{picture}(70,13) 
\put(8,4){\circle{2}} \put(5,1){\line(1,1){6}} \put(5,7){\line(1,-1){6}}
\put(16,4){\circle{2}}
\put(24,4){\circle{2}}
\put(32,4){\circle{2}}
\put(40,4){\circle{2}}
\put(48,4){\circle{2}}
\put(56,4){\circle{2}}
\put(64,4){\circle{2}} \put(61,1){\line(1,1){6}} \put(61,7){\line(1,-1){6}}
\put(48,12){\circle{2}}

\put(9,4){\dashline{1}(0,0)(6,0)}
\put(17,4){\line(1,0){6}} 
\put(25,4){\line(1,0){6}} 
\put(33,4){\line(1,0){6}} 
\put(41,4){\line(1,0){6}} 
\put(49,4){\line(1,0){6}} 
\put(57,4){\line(1,0){6}} 
\put(48,5){\line(0,1){6}} 

{\footnotesize
\put(7,0){${\alpha_0}$}
\put(15,0){${\alpha_1}$}
\put(23,0){${\alpha_2}$}
\put(31,0){${\alpha_3}$}
\put(39,0){${\alpha_4}$}
\put(47,0){${\alpha_5}$}
\put(55,0){${\alpha_6}$}
\put(63,0){${\alpha_7}$}
\put(50,12){${\alpha_8}$}
}

\put(6,6){${\scriptstyle 1}$}
\put(14,6){${\scriptstyle 2}$}
\put(22,6){${\scriptstyle 3}$}
\put(30,6){${\scriptstyle 4}$}
\put(38,6){${\scriptstyle 5}$}
\put(46,6){${\scriptstyle 6}$}
\put(54,6){${\scriptstyle 4}$}
\put(62,6){${\scriptstyle 2}$}
\put(45,12){${\scriptstyle 3}$}
\end{picture}
\end{minipage} 
\caption{Symmetry breaking induced by an \E8\x\E8 inner automorphism of order $N=6$ described by $s^a=(0,1,0,1,0,0,0,0,0)$
  and $s^b=(4,0,0,0,0,0,0,1,0)$.}
\label{fig:ShiftVSO(10)1}
\end{figure}

To determine the unbroken gauge group $\maG_{4D}$, one has to verify the action of an inner
automorphism on the Cartan generators $H_i$ and the step operators $E_\al$ for the simple roots $\al_i$
of the algebra of $\maG$. Their transformations due to the shift vector $V$ are given by
eq.~\eqref{eq:action_shift_on_operators}. We see first that the Cartan generators do not transform,
confirming that the rank is not reduced by the action of the shift vector, as we learnt in 
section~\ref{subsec:UntwistedSector}. Besides, from the transformation properties of the step operators,
eq.~\eqref{eq:action_shift_on_operators}, we note that the only step operators that are invariant under
the automorphism are those that satisfy 
\be
\al_i\cdot V = 0\mod 1\,.
\label{eq:ProjectionSimpleRoots}
\ee
Using that $\al_i\cdot\al_j^*=\delta_{ij}$, we conclude then that in the extended Dynkin diagram(s) depicted
in fig.~\ref{fig:DynkinSO32andE8}, the simple root $\alpha_i$ ($i=0,\ldots,r$) is projected out, if and
only if the coefficient $s_i$ in eq.~\eqref{eq:order_N_intermsof_s_i} is nonzero. 

This allows an interpretation in terms of the Dynkin diagram of the corresponding Lie algebra. Given the
parameters $(s_1,\ldots,s_r)$, the Dynkin diagram of the unbroken gauge group $\maG_{4D}$ is obtained after deleting
the nodes for which $s_i\neq 0$ from the Dynkin diagram of the original gauge group $\maG$. 

To illustrate our discussion, let us consider the \E8\x\E8 Lie algebra. The parameters
\be
s^a=(0,1,0,1,0,0,0,0,0) \quad\text{ and }\quad s^b=(4,0,0,0,0,0,0,1,0) 
\label{eq:sForVSO(10)1}
\ee
acting on the first and second \E8 factors, respectively, describe an automorphism of order $N=6$ because
\be
\sum_{i=0}^8 k_i s^a_i = \sum_{i=0}^8 k_i s^b_i = 6\;,
\label{eq:ExampleAutomorphism}
\ee
where $k_i$ are the mentioned Ka\v{c} labels. By using the basis given in table~\ref{tab:RootsE8}, we
find that the corresponding eight-dimensional shift vectors are given by
\be
V^a~=~\h16\left(\al_1^*+\al_3^*\right)~=~\h16\left(4,2,1^2,0^4\right)\ \quad\text{and}\quad\ V^b~=~\h16\al_7^*~=~\h16\left(2,0^7\right)\,.
\label{eq:ComponentsVSO(10)1}
\ee
These vectors are the two components of a shift vector acting on the 16-dimensional degrees of freedom of
the heterotic string. Therefore, we obtain
\be
V^{\SO{10},1}~=~\h16\left(4,2,1^2,0^4\right)\left(2,0^7\right)\,.
\label{eq:VSO(10)1}
\ee
The unbroken gauge group is found from fig.~\ref{fig:ShiftVSO(10)1} to be
\be
\maG~=~\E8\x\E8\ \stackrel{V^{\SO{10},1}}{\longrightarrow}\ \maG_{4D} ~=~ [\SO{10}\x\SU2\x\SU2\x\U1]\x[\SO{14}\x\U1]\,.
\label{eq:GGVSO(10)1}
\ee
The \U1's appear because the rank is not reduced by $V^{\SO{10},1}$. 

As a side remark, we would like to point out that the shift vector $V^{\SO{10},1}$ turns out to be of
particular interest. We will see in chapter~\ref{ch:MLI} that the phenomenology of models with this
shift vector is very promising. This feature will be associated to the existence of an \SO{10} gauge
group in the resulting $\maG_{4D}$ and certain properties of the spectrum produced by $V^{\SO{10},1}$.

\subsubsection{Restrictions on the Shift Vector}

Not every shift vector $V$ which describes an automorphism of the algebra is an admissible choice for
model construction. We have already seen in section~\ref{subsec:ConsistencyConditions} that for a twist
$\vartheta \in P$ of order $N$, $\vartheta^N = \id$ implies that $N\,V$ should act as the identity on
the gauge degrees of freedom. Hence, a consistency condition on the shift vector $V$ is
\begin{equation}
N\,V \in \Lambda.
\label{eq:gauge_embedding_hom_property}
\end{equation}
Further, modular invariance of the partition function requires that
\begin{equation}
N\left( V^2 - v^2 \right) = 0 \text{ mod } 2
\label{eq:mod_inv_general}
\end{equation}
has to be satisfied, where $v$ is the twist vector, acting in the complexified coordinates of the compact
space.  

From eq.~\eqref{eq:gauge_embedding_hom_property} it is clear that for a given order $N$ of the twist
$\vartheta$, all shifts $V$ of order $N'$ are also admissible, as long as $N'$ divides $N$. In principle, we
could determine the admissible shifts for each $N'$ separately, but a more practical approach is to run
through the outlined procedure for $N$, dropping the condition on the relative-primeness of the sequence
$s = (s_0, \ldots, s_r)$. It is not hard to verify that, by dropping that condition, the order of the
shift can be some $N'$ which is smaller than $N$. 

In the previous example, the shift vector $V^{\SO{10},1}$ of order 6 given in eq.~\eqref{eq:VSO(10)1}
fulfills trivially eq.~\eqref{eq:gauge_embedding_hom_property} because, by construction,
$6\,V\in\Lambda^*$ and the lattice $\Lambda$ is self-dual. In general, all shift vectors generated by the
Dynkin diagram method satisfy $N\,V\in\Lambda^*=\Lambda$. Further, we see that this shift vector combined
with the \Z6-II twist vector $v=(1/6,\,1/3,\,-1/2)$ comply with the modular invariance condition:
\be
6\left((V^{\SO{10},1})^2 - v^2\right)~=~6 \left(\tfrac{13}{18}-\tfrac{7}{18}\right)~=~2 ~=~0\mod 2\ \ \checkmark\,.
\label{eq:ModInvSO(10)1}
\ee
Hence, $V^{\SO{10},1}$ is an allowed \Z6-II shift vector.

\subsubsection{Discriminating Equivalent Models}

Notice that the form of the shift $V$ is not unique because the choice of the basis for the simple roots
$\al_i$ and their duals is not unique. This implies the possibility that two apparently different shift
vectors can lead to identical orbifold models. The reason is that the automorphisms $\sigma_V$ of a Lie algebra
respect certain symmetries. Let us first enumerate them.

\begin{figure}[!t!]
\begin{center}
\subfigure[Breaking due to V$^{(3)}$ of table ~\ref{tab:Shifts_Z4SO32}.]{
\setlength{\unitlength}{0.75mm} 
\begin{picture}(130,22) 
\thicklines
\put(10,19){\circle{2}}
\put(10,3){\circle{2}}
\put(18,11){\circle{2}}
\put(26,11){\circle{2}}
\put(34,11){\circle{2}}
\put(42,11){\circle{2}}
\put(50,11){\circle{2}}
\put(58,11){\circle{2}}
\put(66,11){\circle{2}}
\put(74,11){\circle{2}}
\put(82,11){\circle{2}}
\put(90,11){\circle{2}}
\put(98,11){\circle{2}}
\put(106,11){\circle{2}}
\put(114,11){\circle{2}}
\put(122,19){\circle{2}}
\put(122,3){\circle{2}}

\put(10.7071,18.2929){\dashline{2}(0,0)(6,-6)} 
\put(10.7071,3.7071){\line(1,1){6.55}}
\put(19,11){\line(1,0){6}} 
\put(27,11){\line(1,0){6}} 
\put(35,11){\line(1,0){6}} 
\put(43,11){\line(1,0){6}} 
\put(51,11){\line(1,0){6}} 
\put(59,11){\line(1,0){6}} 
\put(67,11){\line(1,0){6}} 
\put(75,11){\line(1,0){6}} 
\put(83,11){\line(1,0){6}} 
\put(91,11){\line(1,0){6}} 
\put(99,11){\line(1,0){6}} 
\put(107,11){\line(1,0){6}} 
\put(114.7071,11.7071){\line(1,1){6.55}} 
\put(114.7071,10.2929){\line(1,-1){6.55}} 

\put(80,14){\line(2,-3){4.1}}
\put(80,8){\line(2,3){4.1}}
\put(120,22){\line(2,-3){4.1}}
\put(120,16){\line(2,3){4.1}}

\put(4,1){${\scriptstyle\alpha_0}$}
\put(4,20){${\scriptstyle\alpha_1}$}
\put(16,6){${\scriptstyle\alpha_2}$}
\put(24,6){${\scriptstyle\alpha_3}$}
\put(32,6){${\scriptstyle\alpha_4}$}
\put(40,6){${\scriptstyle\alpha_5}$}
\put(48,6){${\scriptstyle\alpha_6}$}
\put(56,6){${\scriptstyle\alpha_7}$}
\put(64,6){${\scriptstyle\alpha_8}$}
\put(72,6){${\scriptstyle\alpha_9}$}
\put(79,6){${\scriptstyle\alpha_{10}}$}
\put(87,6){${\scriptstyle\alpha_{11}}$}
\put(95,6){${\scriptstyle\alpha_{12}}$}
\put(103,6){${\scriptstyle\alpha_{13}}$}
\put(111,6){${\scriptstyle\alpha_{14}}$}
\put(126,1){${\scriptstyle\alpha_{15}}$}
\put(126,20){${\scriptstyle\alpha_{16}}$}

\end{picture} 
}
\vskip -2mm
\subfigure[Breaking due to V$^{(4)}$ from table~\ref{tab:Shifts_Z4SO32}.]{
\setlength{\unitlength}{0.75mm} 
\begin{picture}(130,22) 
\thicklines
\put(10,19){\circle{2}}
\put(10,3){\circle{2}}
\put(18,11){\circle{2}}
\put(26,11){\circle{2}}
\put(34,11){\circle{2}}
\put(42,11){\circle{2}}
\put(50,11){\circle{2}}
\put(58,11){\circle{2}}
\put(66,11){\circle{2}}
\put(74,11){\circle{2}}
\put(82,11){\circle{2}}
\put(90,11){\circle{2}}
\put(98,11){\circle{2}}
\put(106,11){\circle{2}}
\put(114,11){\circle{2}}
\put(122,19){\circle{2}}
\put(122,3){\circle{2}}

\put(10.7071,18.2929){\dashline{2}(0,0)(6,-6)} 
\put(10.7071,3.7071){\line(1,1){6.55}}
\put(19,11){\line(1,0){6}} 
\put(27,11){\line(1,0){6}} 
\put(35,11){\line(1,0){6}} 
\put(43,11){\line(1,0){6}} 
\put(51,11){\line(1,0){6}} 
\put(59,11){\line(1,0){6}} 
\put(67,11){\line(1,0){6}} 
\put(75,11){\line(1,0){6}} 
\put(83,11){\line(1,0){6}} 
\put(91,11){\line(1,0){6}} 
\put(99,11){\line(1,0){6}} 
\put(107,11){\line(1,0){6}} 
\put(114.7071,11.7071){\line(1,1){6.55}} 
\put(114.7071,10.2929){\line(1,-1){6.55}} 

\put(80,14){\line(2,-3){4.1}}
\put(80,8){\line(2,3){4.1}}
\put(120,6){\line(2,-3){4.1}}
\put(120,0){\line(2,3){4.1}}

\put(4,1){${\scriptstyle\alpha_1}$}
\put(4,20){${\scriptstyle\alpha_0}$}
\put(16,6){${\scriptstyle\alpha_2}$}
\put(24,6){${\scriptstyle\alpha_3}$}
\put(32,6){${\scriptstyle\alpha_4}$}
\put(40,6){${\scriptstyle\alpha_5}$}
\put(48,6){${\scriptstyle\alpha_6}$}
\put(56,6){${\scriptstyle\alpha_7}$}
\put(64,6){${\scriptstyle\alpha_8}$}
\put(72,6){${\scriptstyle\alpha_9}$}
\put(79,6){${\scriptstyle\alpha_{10}}$}
\put(87,6){${\scriptstyle\alpha_{11}}$}
\put(95,6){${\scriptstyle\alpha_{12}}$}
\put(103,6){${\scriptstyle\alpha_{13}}$}
\put(111,6){${\scriptstyle\alpha_{14}}$}
\put(126,1){${\scriptstyle\alpha_{15}}$}
\put(126,20){${\scriptstyle\alpha_{16}}$}

\end{picture} 
}
\caption{Extended Dynkin diagram of \SO{32} corresponding to the breaking due to the shifts V$^{(3)}$ and
  V$^{(4)}$ of the  \Z4 orbifold}
\label{fig:V3_V4_breaking}
\end{center}
\end{figure}
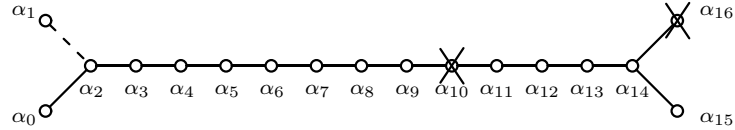
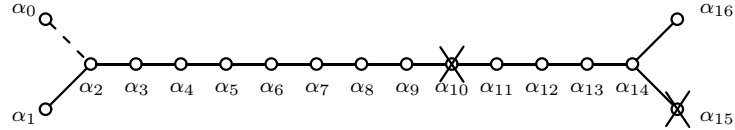

{\bf Weyl reflections.}
The action of an element $w$ of the Weyl group $\calW$ on a shift vector $V$ is given by
\be
V\ \stackrel{w}{\longrightarrow}\ w(V)~=~V - 2\,\h{\al\cdot V}{\al^2}\al\,,
\label{eq:WeylReflectionV}
\ee
where \al\ is an arbitrary root. The Weyl reflections corresponding to $\al_1,\ldots,\al_r$ form a basis
$w_1,\ldots,w_r$, so that any other element of the Weyl group is described by products of these.

In \SO{32}, it is not hard to verify that the group of Weyl reflections alters the shift vector just by
permuting its entries or changing simultaneously the sign of two entries. This follows from the fact that
all simple roots have integer entries $0,\,\pm1$. Weyl reflections in \E8\x\E8 are more complicated. Those 
related to roots with integer entries act on the shift permuting or pairwise sign-flipping its
entries. However, if $\al$ is a half-integer weight of \E8\x\E8 (also called spinorial), the shift 
vector can change nontrivially. Consider, for instance, the Weyl element $w$ associated to the \E8 root
$\al=\left(\2,\,\2,\,\2,\,\2,\,\2,\,\2,\,\2,\,\2\right)$ applied to both eight-dimensional components of
the shift vector $V^{\SO{10},1}$ given in eq.~\eqref{eq:VSO(10)1}. The result reads
\be
w\left(V^{\SO{10},1}\right) ~=~ \h16\ \Big(\,2,\,0,\,-1,\,2,\,1,\,-2^3\,\Big)\,.
\left(\,\h32,\,-\4^7\;\right)
\label{eq:VSO(10),1WeylTransformed}
\ee
Despite the difference between the original shift $V$ and its rotated counterpart
$w(V)$, both have the same effect on the algebra. This can be seen as follows. A Weyl rotation does not
affect only the shift, but the complete space, including the simple roots of the algebra. In fact, the
Weyl group is an isomorphism of the algebra. Hence, scalar products are not modified. This implies
particularly that $V\cdot\al_i=w(V)\cdot w(\al_i)$, i.e.\ the unbroken gauge group is
not modified by Weyl reflections. A less obvious consequence is that the orbifold spectrum is not altered
either. 

{\bf Lattice translations.} We have seen in section~\ref{subsec:BrotherModels} that lattice translations
can have an influence in orbifold models. Namely, models whose gauge embedding is modified by lattice
vectors, i.e.\ brother models, can mimic the effect of discrete torsion. We also pointed out, nevertheless,
that in the case of \Z{N} orbifold models without Wilson lines, this cannot occur. It follows then that
adding lattice vectors to \Z{N} shift vectors is an allowed symmetry of the automorphisms $\sigma_V$.

\begin{table}[t!]
\centering
\begin{tabular}{|c|p{3.6cm}|p{3.5cm}|p{4.2cm}c|}
\hline
$V^{(\#)}$ & \centering{$U$ Sector} & \centering{$T_1$ Sector} & \centering{$T_2$ Sector} &\\
\hline\hline
\rowgrayh
3 & 
\centering{$2(\bs{20},\bsb{6})_{\text{-}\2}$,\, $1(\bs{1},\bs{15})_{1}$,\, $1(\bs{1},\bsb{15})_{\text{-}1}$} &
\centering{$16(\bs{1},\bs{15})_{\4}$,\, $80(\bs{1},\bs{1})_{\text{-}\h{3}{4} }$} & 
\centering{$10(\bs{1},\bs{15})_{\text{-}\2}$,\, $6(\bs{1},\bsb{15})_{\2}$,\, $10(\bs{1},\bs{1})_{\h{3}{2}}$,\, $6(\bs{1},\bs{1})_{\text{-}\h{3}{2}}$}&\\
4 & 
\centering{$2(\bs{20},\bs{6})_{\text{-}\2}$,\, $1(\bs{1},\bs{15})_{\text{-}1}$,\, $1(\bs{1},\bsb{15})_{1}$} & 
\centering{$16(\bs{20},\bs{1})_{\text{-}\h{3}{4}}$,\, $32(\bs{1},\bsb{6})_{\text{-}\4 }$} & 
\centering{$10(\bs{1},\bsb{15})_{\text{-}\2}$,\, $6(\bs{1},\bs{15})_{\2}$,\, $10(\bs{1},\bs{1})_{\h{3}{2}}$,\, $6(\bs{1},\bs{1})_{\text{-}\h{3}{2}}$}&\\
\hline    
\end{tabular}
\caption{Spectra of two \Z4 orbifold models of the \SO{32} heterotic string with similar symmetry
  breakdown patterns. The \U1 charges are written  as subindices.}
\label{tab:V3V4Spectra}
\end{table}

{\bf Symmetries of the Dynkin diagram.} 
It is usually argued that symmetries of the (extended) Dynkin
diagram can help to avoid redundancies. Whereas \E8\x\E8 does not have any symmetry of this type, \SO{32}
has at least the symmetry defined by the operation $\al_i\leftrightarrow\al_{16-i}$, i.e. a reflection on
a vertical axis crossing $\al_8$. The models described by automorphisms obeying this symmetry lead to
identical spectra and gauge group, confirming thereby this symmetry. 

However, {\it not all symmetries of the Dynkin diagram are symmetries of the automorphisms},
as we now explain. Another symmetry of the \SO{32} Dynkin diagram is spinor conjugation,
that is, the redefinition of the simple roots $\al_{15}\leftrightarrow\al_{16}$. Even though this
operation is a symmetry of the Dynkin diagram, it is {\it not} a symmetry of orbifolds. As an example,
consider the \Z4 shifts $V^{(3)}$ and $V^{(4)}$ of table~\ref{tab:Shifts_Z4SO32}. They both induce the
unbroken gauge group $\maG_{4D}=\SO{20}\x\SU6\x\U1$ (see fig.~\ref{fig:V3_V4_breaking}), but their
twisted matter spectra differ, as evident from table~\ref{tab:V3V4Spectra}.

By using effectively these symmetries, one can discriminate the number of inequivalent \Z{N} models
without Wilson lines.

\subsubsection{Advantages and Disadvantages}

The classification strategy described in this section possesses certain features which make it useful 
to obtain all shift embeddings of \Z{N} orbifolds. By construction, the only requirement on the shift
vectors is imposed by modular invariance, eq.~\eqref{eq:mod_inv_general}. One can directly determine the
unbroken gauge group without need of further information about the massless spectrum. It is also
straightforward to eliminate redundancies in the classification. However, this method also has some
drawbacks. In considering \ZZ{N}{M} models or including Wilson lines, a complete classification requires
to invoke additional mechanisms, such as adding linear combinations of \U1 directions to the shift
vector(s) and Wilson lines (see e.g.~\cite{Wingerter:2005xx}). In that case, the elegance 
of this method is lost. 

\subsection{The Classification}
\label{subsec:ineq_models}

\begin{table}[!t!]
\begin{center}
\begin{tabular}{|l|r|r|}
\hline
$\mathbb{Z}_N$      & $\E8\x\E8$ &  \SO{32} \\
\hline
\hline
\rowgrayh
$\mathbb{Z}_3$      &   5        &    6     \\ 
$\mathbb{Z}_4$      &  12        &   16     \\
\rowgrayh
$\mathbb{Z}_6$-I    &  58        &   80     \\
$\mathbb{Z}_6$-II   &  61        &   75     \\
\rowgrayh
$\mathbb{Z}_7$      &  40        &   56     \\
$\mathbb{Z}_8$-I    & 145        &  196     \\
\rowgrayh
$\mathbb{Z}_8$-II   & 146        &  194     \\
$\mathbb{Z}_{12}$-I &1669        & 2295     \\
\rowgrayh
$\mathbb{Z}_{12}$-II&1663        & 2223     \\
\hline
\end{tabular}
\end{center}
\caption{Comparison between the number of inequivalent \Z{N} shifts in the \E8\x\E8
heterotic string and in the \SO{32} heterotic string~\cite{Nilles:2006np}.}
\label{tab:compareSO32andE8xE8}
\end{table}

We have used the method described in section~\ref{sec:DynkinMethod} to classify all shift embeddings of
\E8\x\E8 and \SO{32} heterotic orbifolds. In table~\ref{tab:compareSO32andE8xE8} we list the number of
models without Wilson lines found for each \Z{N} orbifold. With exception of the number of models for
\Z8-II and \Z{12} orbifolds, our results for the \E8\x\E8 agree with those presented in
ref.~\cite{Katsuki:1989bf}. We have verified that the results listed here are correct. Since
not all \SO{32} orbifold models were known, we have thought it would be useful to display them in a web
page~\cite{WebTables:2006xx}, where the following details of each of these models are provided:
\begin{itemize}
\item the twist vector $v$ and the root lattice $\Gamma$, which specifies the geometry,
\item the shift vector $V$ and the corresponding gauge group,
\item the matter content, listed by sectors, including all \U1 charges, where we have denoted
  the anomalous one by $\text{U}_{1A}$.
\end{itemize}

Comparing the numbers of inequivalent \SO{32} models to those of \E8\x\E8, we find that the \SO{32}
heterotic string leads to a larger amount of models. This difference can become important if Wilson lines
are present. We have learnt in section~\ref{subsec:LocalShiftsSpectra} that the local spectra at the
fixed points of orbifold models with Wilson lines correspond always to the twisted spectra of the models
without Wilson lines. In other words, if one defines a local shift vector $V_g$ for each of the
constructing elements denoting the fixed points, then each $V_g$ (which includes Wilson lines) has to be
equivalent to one of the shift vectors $V$ we have before nontrivial Wilson lines are switched on.
In this sense, \SO{32} orbifolds lead to a richer variety of models.

A remark is in order. In classifying orbifold models based on the \SO{32} heterotic string, we were
somehow surprised by the common presence of spinors in the spectra. We give some details about this
feature in appendix~\ref{ch:SpinorsInSO32}. This situation contrasts with the
popular notion that only \E8\x\E8 orbifolds admit, say, \SO{10} spinors and therefore \sm\ families of
quarks and leptons. Yet we have to admit, that spinors are more frequently encountered in \E8\x\E8 than
in \SO{32} orbifolds, what renders \E8\x\E8 more attractive.

Our classification of \Z{N} orbifolds closes one of the unfinished tasks started already in the late
eighties. Nonetheless, a classification of all \Z{N} orbifolds is far from being useful if one does not
include Wilson lines, for they trigger further symmetry breakdown and, hence, the appearance of models
resembling the \sm, which is, by the way, the true aim of any study of this kind. Therefore, one is
forced to find a useful way to study systematically models with Wilson lines.

\index{Dynkin diagram!orbifold classification|)}

\section[Including Wilson Lines and \ZZ{N}{M} Orbifolds]{Including Wilson Lines and $\bs{\ZZ{N}{M}}$ Orbifolds}
\label{sec:ClassAnsatz}

A convincing approach to classify orbifold models with Wilson lines was first proposed in the context of
\Z3 orbifolds in ref.~\cite{Giedt:2000bi}. The proposal consists in a generic ansatz that describes any
(shift vector or) Wilson line of order three. As explained in more detail in appendix~\ref{ch:ansatz}, we
find that this approach can easily be extended to any order. Let us summarize here our findings.

By identifying those models whose shifts and Wilson lines match after the action of lattice translations
and Weyl reflections, one finds the shifts of order $N,M$ and Wilson lines of order $N_\al$ take a block-form:
\bse\label{eq:BlockViAal}\index{shift vector, ansatz}\index{Wilson line, ansatz}
\bea
\label{eq:BlockVi}
V_j &=& \left(V_{j,\text{block 1}},\,V_{j,\text{block 2}},\,V_{j,\text{block 3}},\ldots\right)\,,\\
\label{eq:BlockAal}
A_\al &=& \left(A_{\al,\text{block 1}},\,A_{\al,\text{block 2}},\,A_{\al,\text{block 3}},\ldots\right)\,,
\eea
\ese

\noindent
where the blocks $V_{j,\text{block i}}$ and $A_{\al,\text{block i}}$ are $m_i$-dimensional vectors. The
specific form of these blocks depends on the order of the respective shift vector $V_j$ and Wilson line 
$A_\al$. Let us denote by $X_\text{block i}$ an arbitrary block of a shift or Wilson line of order
$\ti N$. The form of this block is then expressed by (see appendix~\ref{ch:ansatz})
\be \mbox{\footnotesize $
X_\text{block i}=\left\{
\ba{l}
\frac{1}{\ti{N}}\left( (\pm j)^{\alpha}, \T-(\ti{N}\T-j)^{\beta}, 0^{n_0}, 1^{n_1}, \ldots ,
  (\ti{N}\T-j)^{n_{(\ti{N}-j)}-\alpha-\beta},\ldots, \left(\frac{\ti{N}}{2}\right)^{n_{\frac{\ti{N}}{2}}}
\right)\qquad\quad\T{\bf \ \ case a)}\\
\frac{1}{2\ti{N}}\left((\pm j)^{\alpha}, \T-(2\ti{N}\T-j)^{\beta}, 1^{n_1}, 3^{n_3}, \ldots ,
  (2\ti{N}\T-j)^{n_{(2\ti{N}-j)}-\alpha-\beta},\ldots, (\ti{N}\T-1)^{n_{\ti{N}-1}} \right)\quad\T{\bf case b)}\\
\frac{1}{\ti{N}}\left( (\pm j)^\al, \T-(\ti{N}\T-j)^\beta, 0^{n_0}, 1^{n_1},\ldots,(\ti{N}\T-j)^{n_{(\ti{N}-j)}-\al-\beta},\ldots,
 \left(\frac{\ti{N}\T-1}{2}\right)^{n_{\left(\frac{\ti{N}-1}{2}\right)}}\right)\quad\T{\bf \,case c)}
\ea
\right.$}
\label{eq:AllAnsaetze}
\ee
where $\al$ and $\beta$ are either $0$ or $1$ such that $\al+\beta=0,1$ and $\sum_k n_k = m_i$.  
Furthermore, $j$ depends on the value of $\ti N$ and the kind of building block, according to the
following cases:
\begin{itemize}
\item {\bf a) even order $\bs{\ti{N}}$, `vectorial' block.} `Vectorial' means here that the entries have a maximal
  denominator of $\ti{N}$. In this case $j$ takes the values $\{\h{\ti{N}}{2}+1,\ldots,\ti{N}-1\}$;
\item {\bf b) even order $\bs{\ti{N}}$, `spinorial' block.} `Spinorial' means that the entries have a maximal
  denominator of $2\ti{N}$ and odd numerators. We have $j\in\{\ti{N}+1,\ti{N}+3,\ldots,2\ti{N}-1\}$;
\item {\bf c) odd order $\bs{\ti{N}}$.} In this case `vectorial' and `spinorial' blocks are equivalent; so it
  suffices to give one ansatz, where $j\in\{\frac{\ti{N}+1}{2},\frac{\ti{N}+3}{2},\ldots,\ti{N}\}$.
\end{itemize}

Shift vectors and Wilson lines obtained by the repeated use of the ansatz~\eqref{eq:AllAnsaetze} have to
satisfy the modular invariance conditions, eq.~\eqref{eq:NewModInv}, and the embedding constraint
$N V_1$, $M V_2$, $N_\al A_\al\in\Lambda$. In contrast to the Dynkin diagram strategy, the latter requirement
is not automatically fulfilled. Indeed, that requisite implies that all blocks of a shift or Wilson line
must be of the same type (either a, b or c). Once both consistency conditions are met by a set of
parameters, one has encountered an admissible orbifold model. In this way, one arrives at a more general
classification of orbifold models.

There is however a subtlety. As we have explained in section~\ref{subsec:BrotherModels}, {\it brother
models} can appear by considering lattice translations of the shift(s) and/or Wilson lines. The matter
spectrum of a brother to an orbifold model differs in the twisted sectors. Therefore, a model and its
brother are not equivalent. Since in the derivation of the
ans\"atze~\eqref{eq:AllAnsaetze} we have taken lattice translations as a symmetry
of the theory, brother models have been disregarded and thus a classification based on this method is not
comprehensive. On the other hand, it is worth to mention that models with the same matter spectrum of the
brother models can be found by varying the discrete torsion parameters. In a sense, then, a
classification based on the proposed ans\"atze provides us with the full set of inequivalent torsionless
models. 

\subsubsection{Discriminating Equivalent Models}\index{inequivalent models}

In this method, we have tried to avoid redundancies by disregarding those gauge embeddings differing by
lattice translations, and pairwise sign-flips and permutations of their entries. However, in \E8\x\E8
models, as mentioned in section~\ref{sec:DynkinMethod}, there are additional Weyl reflections (those based
on the spinorial simple root), which modify in general shifts and Wilson lines in a very unpredictable
way. In other words, our present strategy has not projected out all duplicities. Inspecting the
symmetries due to those other Weyl symmetries is hopeless due to the huge amount of elements of the Weyl
group $\calW$. Hence, we have to look for an alternative to eliminate equivalent models.

It turns out more practical to use shifts and Wilson lines to compute the massless spectrum of each model
and, then, consider two models as inequivalent if and only if their massless spectra are different.
Of course, this job gets too ambitious if one has at hand a great many models and one lacks computer
support. That might be one of the reasons why a comprehensive classification of orbifolds has been
postponed for several years. 

To perform the comparison of the spectra in a reasonable time, we compare the non-Abelian massless spectra
and the number of singlets, that is, we avoid a comparison of \U1 charges. This underestimates the true
number of models somewhat. Ideally, we should not only compare the full massless spectra (including \U1
charges), but also the localization of the particles and the couplings among them. Furthermore, it might
be also necessary to consider the Kaluza-Klein tower of nonzero modes. This is evidently unfeasible.

We would like to point out that the problem of identifying inequivalent models does not appear only in
this method. Also Dynkin diagram techniques suffer of the same problem when considering Wilson lines. An
interesting question would be to find out whether there is another way to discriminate duplicities in a
classification. 

\subsubsection{Advantages and Disadvantages}

The classification method presented in this section is an alternative to classify orbifold models with
Wilson lines. By means of computer programs, one can generate all models of a given orbifold in a direct
manner. However, this method neglects brother models or, analogously, models with discrete
torsion. Therefore, one is forced to vary the discrete torsion parameters in order to arrive to a
complete set of models. Further, one cannot avoid to overcount some models, which can only be
identified by directly comparing the states of their matter spectra.

\subsection[Sample Classification of \ZZ33]{Sample Classification of $\bs{\ZZ33}$ without Wilson Lines}
\label{sec:ClassifZ3xZ3}\index{classification, Z3xZ3@classification, \Z3\x\Z3}

As a concrete application, let us describe the classification of $\Z3\times\Z3$
orbifolds without Wilson lines. Here we make use of both classification methods introduced in this
chapter. A $\Z3\times\Z3$ orbifold model is described by the twist vectors
\be
v_1 = (0,\,1/3,\,0,\,-1/3)\quad\text{ and }\quad v_2 = (0,\,0,\,1/3,\,-1/3)\;,
\ee
two shift vectors $V_1$ and $V_2$ of order three and three Wilson lines, which are not important for our
current discussion. We aim here basically at a classification of all admissible shift embeddings.

\begin{table}[t!]
\centering
\begin{tabular}{|c|cl|c|}
\hline
$V^{(\#)}$ & \phantom{.}& Shift vector $V_1$ & 6D gauge group $\maG_{6D}$\\
\hline\hline
\rowgrayh
1 && $\3\left( 0^6,\, 1^2\right)\left(\,0^8\,\right)$ & \E7\x\E8 \\[1mm]
2 && $\3\left( 0^6,\, 1^2\right)\left( 0^5,\,1^2,\,2\right)$ &  \E7\x\E6\x\SU3 \\[1mm]
\rowgrayh
3 && $\3\left( 0^3,\, 1^4,\, 2\right)\left(\,0^8\,\right)$ & \SU9\x\E8 \\[1mm]
4 && $\3\left( 0^6,\, 1^4,\, 2\right)\left( 0^5,\,1^2,\,2\right)$ &  \SU9\x\E6\x\SU3 \\[1mm]
\rowgrayh
5 && $\3\left( 0^6,\, 1^6,\,2\right)\left( 0^6,\, 1^6,\,2\right)$ & \SO{14}\x\SO{14} \\[1mm]
\hline
\end{tabular}
\caption{Five inequivalent \E8\x\E8 shift vectors $V_1$ of the \ZZ33 orbifold and their induced gauge
symmetry breakdown.} 
\label{tab:Z3xZ3ShiftsV1}
\end{table}

First, by using the Dynkin diagram strategy depicted in section~\ref{sec:DynkinMethod}, one finds that
there are only five consistent shift vectors $V_1$, which can be written in the generic form
\begin{equation}\label{eq:MinForm1}
V_1~=~\frac{1}{3}(0^{n_0},1^{n_1},2^\alpha)\,(0^{n'_0},1^{n'_1},2^\beta),
\end{equation}
where $\alpha$ and $\beta$ can be either 0 or 1, and $n_i,n'_i\in\mathbbm{Z}$, such that
$n_0+n_1+\alpha=n'_0+n'_1+\beta=8$. The admissible shifts (listed in table~\ref{tab:Z3xZ3ShiftsV1}) can
be obtained by verifying the consistency conditions
\be\ba{rcl}
3(V_1^2 - v_1^2)&=& 0\mod 2\qquad \text{ and } \\
3 V_1 &\in& \Lambda\,.\ea\nonumber
\ee

\noindent
Note that the modular invariance condition forbids the trivial shift $V_1=(0^{8})(0^{8})$. This is a
difference with respect to the \Z3 orbifold.

From the ansatz~\eqref{eq:AllAnsaetze}, it follows that the second shift $V_2$ of order three has the
generic form
{\footnotesize
\begin{eqnarray}\label{eq:MinForm2}
V_2~=~ &\dfrac{1}{3}\left( \left( 
\begin{array}{c}
3\\
\vdots \\
\T-2
\end{array}
 \right),  \left( 
\begin{array}{c}
1\\
0
\end{array}
 \right)^{n_0-1}, \left( 
\begin{array}{c}
1\\
0\\
\T-1
\end{array}
 \right)^{n_1+\alpha}\;  \right) % \nonumber \\
% &\quad
\left( \left( 
\begin{array}{c}
3\\
\vdots\\
\T-2
\end{array}
 \right), \left( 
\begin{array}{c}
1\\
0
\end{array}
 \right)^{n'_0-1}, \left( 
\begin{array}{c}
1\\
0\\
\T-1
\end{array}
 \right)^{n'_1+\beta}\;  \right)\;.
\end{eqnarray}}
Together with one of the shift vectors $V_1$ from table~\ref{tab:Z3xZ3ShiftsV1}, the shift vectors
generated by this ansatz are subject to the usual lattice conditions and modular invariance,
eq.~\eqref{eq:NewModInv}. Disregarding those models that are equivalent at massless level, we get 109
shift embeddings. 

\subsubsection{Generalized Discrete Torsion}

We use now the set of shift embeddings to generate all admissible models. As discussed in the examples of
section~\ref{subsec:NewDiscreteTorsion}, there are ten independent generalized discrete torsion
parameters, whose values are $0,\,\frac{1}{3}$ or $\frac{2}{3}$. One might be tempted to deduce that the
total number of models is $109\x 3^{10}$, but out of them only 1082 models (or more precisely, massless
spectra) are inequivalent. These models comprise the complete set of admissible models without Wilson
lines. The model definitions and the resulting spectra are detailed in our web page~\cite{WebTables:2006xx}.

\section{A Statistical Method}
\label{sec:StatisticalMethod}
\index{orbifold classification!statistical method}

Neither the Dynkin method nor the use of a suitable ansatz is free of redundancies because
shifts and Wilson lines can be related by elements of the Weyl group which has an enormous number of
elements (cf.\ the discussion in \cite{Giedt:2000bi}). This means that, as we have already commented,
several models appear more than once in the classification. If the total number of models of a class is
large (say, of the order of several millions), constructing all models and discriminating the
inequivalent ones is an extremely time--consuming task.

Instead of the complete classification of models, we take a statistical
approach~\cite{Lebedev:2008un}. To understand the basic idea,  consider a simple example. Suppose we have
a set of $M$ models out of which $N$ are inequivalent ($M,N \gg 1$). Assume also that each
inequivalent model is represented $M/N$ times in the set $M$, which corresponds to
a ``flat distribution''. The probability that 2 randomly chosen models are
equivalent is $1/N$. Take now a larger random selection  of models  $n$, $1 \ll
n \ll N$. The probability that there are equivalent models in this set is
\begin{equation}
 p(n,N)~\simeq~\binom{n}{2} \frac{1}{N}~\simeq~\frac{n^2}{2 N} \;.
\end{equation}
For $n=\sqrt{N}$, this probability is 1/2. Thus, in a sample of $\sqrt{N}$ out
of a total of $N$ models, there is order 1 probability that at least 1 model is
redundant. This observation  allows us to estimate the number of inequivalent
models by studying a sample of order $\sqrt{N}$ models.

\begin{figure}[t!]
  \psfrag{PnN}{\footnotesize$P(n,N)$}
  \psfrag{n}{\footnotesize$n$}
  \psfrag{sqrtN}{\footnotesize$\sqrt{N}$}
  \CenterEps[0.8]{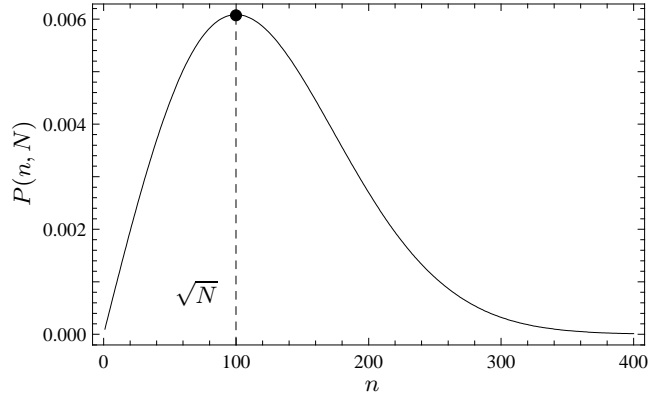}
  \caption{Probability distribution for $N=10000$. The maximum occurs at $\sqrt{N}$.}
  \label{fig:statisticsTheory}
\end{figure}

In order to estimate the total number of models of a certain class and, simultaneously, to obtain
a large sample of them, we propose the following algorithm. Start with a random
model in the sample. Then 

\ding{202} generate another model; and

\ding{203} compare it to the model(s) in the sample. If it is equivalent to any of them, the procedure
is stopped. Otherwise, the model is included in the sample and we repeat step \ding{202}.

In this procedure, the probability that a new model is equivalent to one of the previous models is
$k/N$. Therefore, the probability that we arrive at a sample of size $n$ or larger is
\begin{equation}
  \prod_{k=0}^{n-1} \left( 1-\frac{k}{N} \right)\,.
\end{equation}
The probability  that this algorithm terminates at size $n$ is 
\begin{equation}
 P(n,N)~=~\frac{n}{N}\, \prod_{k=0}^{n-1} \left( 1-\frac{k}{N} \right)
       ~=~ \frac{n(N-1)!}{N^n(N-n)!} \;.
\end{equation}
It is easy to verify that this probability function is well-defined, i.e. $\sum_{n=1}^NP(n,N)=1$.
The maximum of this function is at $n=\sqrt{N}$, as depicted in fig.~\ref{fig:statisticsTheory}.
This implies that, if we produce various sets of models with different sizes $n$ and plot how common a
particular $n$ is, the total number of models can be estimated by $N\simeq n_0^2$, where $n_0$ is the
value at which the plot has a maximum, that is, the most frequent $n$.

An important assumption in this analysis is that all  inequivalent models are equally likely to be
generated. In practice, this is not the case and some models appear more often than the others. This has
to do with the specifics of the  model--generating routine. For example, if the Wilson lines are
generated by selecting random entries constrained to a minimal region of the lattice (as in the
ans\"atze proposed in sec.~\ref{sec:ClassAnsatz}), trivial Wilson lines (with all entries set to zero)
are very unlikely to be generated. 
To take this factor into account, we introduce a fudge parameter $t$ defined by $n_0\, t  \simeq \sqrt{N}$,
where $n_0$ is the predominant size of the sample and $N$ is the true number of inequivalent models. The
parameter $t$ can be determined ``experimentally'' when both $n$ and $N$ are known. 

As we shall see in chapter~\ref{ch:MLII}, this method is successful and let us determine the total amount
of inequivalent models in cases that are untreatable otherwise.

\section{The C++ Orbifolder}
\label{sec:Orbifolder}

The extensive work that an orbifold classification requires, would not be feasible without computer
programs. In collaboration with P.K.S.~Vaudrevange and A.~Wingerter, we have written 
{\it the C++ Orbifolder}. This program, which will be available to the scientific community, 
has been thought to cover the needs of a researcher interested in the phenomenological consequences of 
orbifold compactifications without dealing with the time-consuming and cumbersome computation.

Provided the shift vector(s) and Wilson lines, the C++ Orbifolder can compute the massless spectrum of
the \E8\x\E8 or \SO{32} heterotic string compactified on any symmetric \Z{N} and \ZZ{N}{M} orbifold in
much less than a second on a 2.66 GHz computer. Its results are given in a human readable format, which
can be saved in ``txt'' or ``TEX'' format. The underlying compact lattice $\Gamma$ as well as the active
parameters of generalized discrete torsion can be chosen at will. In computing orbifold spectra, the
program verifies all consistency constraints, such as the conditions for cancellation of anomalies and
modular invariance.  

One can profit of the speed of the C++ Orbifolder in a classification of orbifold models. Some routines
of this program use the classification methods outlined in this chapter. Thus, one can generate a large
amount of models with and without discrete torsion in a reasonable time. In particular, the comparison
of two different spectra takes less than a second, independently of how complicated they are.

A very useful feature of this program is that it is capable to determine all nonvanishing couplings
entering in the superpotential of a given model, according to the string selection rules enumerated in
section~\ref{sec:ourstringselectionrules}. This process has to be performed order by order in the
superpotential and therefore is very demanding. As yet the program can determine the superpotential for
\Z6-II orbifolds up to order eight in the fields within one week, in average. For other orbifolds, like
e.g. \Z3 or \Z{12}, this computation can take significantly less time because there are less particles in
the spectra and/or less allowed couplings. 

A phenomenological viable orbifold model must have at least the \sm\ gauge group and all possible exotic
particles have to be decoupled from the low--energy spectrum. To chose a nonanomalous linear combination
of \U1's that play the role of the hypercharge is nontrivial. We have developed routines to perform this
task in the case that hypercharge is embedded in an \SU5 \gut\ theory. Moreover, the Orbifolder
can identify the \sm\ particles and determine whether additional exotics acquire large masses provided that
some \sm\ singlets attain {\vev}s.

The C++ Orbifolder is still under development. One might be interested also
in considering asymmetric orbifold constructions and a systematic study of discrete
accidental symmetries of the superpotential, which can shed light, for instance, in the solution to the
problem of proton stability.

%% file: OrbiPheno.tex
%%%%%%%%%%%%%%%%%%%%%%%%%%%%%%%%%%%%%%%%%%%%%%%%%%%%%%%%%%%%
%CHAPTER 4
%%%%%%%%%%%%%%%%%%%%%%%%%%%%%%%%%%%%%%%%%%%%%%%%%%%%%%%%%%%%
\chapter{A Mini-Landscape of $\bs{\Z6}$-II Orbifolds}
\label{ch:MLI}

\begin{center}
\begin{minipage}[t]{14cm}
Inspired by the works by Kobayashi et al.~\cite{Kobayashi:2004ud} and Buchm\"uller et
al.~\cite{Buchmuller:2006ik}, we aim at promising models of the \Z6-II orbifold. We perform a
systematic search of models with two Wilson lines guided by the concept of {\it local}
\textit{\textsc{gut}s}, which is introduced in section~\ref{sec:OrbiGUT}. It turns out that out of $3\x 10^4$ \Z6-II
orbifold models with local {\gut}s, about 200 models have the exact matter spectrum of the \mssm\ and some other
realistic properties. This chapter focuses mainly on the methodology of our search. We list the main
results here and the discussion of the phenomenology of the models is left for chapter~\ref{ch:OrbifoldPhenomenology}.
\end{minipage}
\end{center}

\section[Orbifold Local {\gut}s]{Orbifold Local \textbf{\textsc{gut}}s}
\label{sec:OrbiGUT}
\index{local guts@local {\gut}s}

Let us focus on \Z{N} orbifold compactifications. Consider a fixed point or fixed torus with constructing element
$g=(\vartheta^k,\,n_\al e_\al)$. Thus, the associated local gauge shift and local twist vector are given
by $V_g=kV+n_\al A_\al$ and $v_g=k v$, respectively. We have seen in
section~\ref{subsec:LocalShiftsSpectra} that at the 
fixed point $g$, the local gauge group $\maG_g$ is larger than the four-dimensional unbroken
gauge group $\maG_{4D}$. Furthermore, the matter states living at the fixed point $g$ furnish complete
representations under $\maG_g$, which can nonetheless be partly projected out when considering the
complete orbifold by the action of Wilson lines. On the other hand, massless modes of untwisted (bulk)
fields also build representations under $\maG_g$, but they are always partly projected out (even in the
absence of background fields). Therefore, bulk states are, in general, incomplete (also called `split')
multiplets from the four-dimensional perspective.

A useful concept in orbifold compactifications is that of local grand unification, or {\it local
{\gut}s}, which appear when the local symmetry $\maG_g$ at some fixed point(s) is a \gut\ symmetry, 
such as \SO{10} or \E6. In presence of local {\gut}s, the \sm\ gauge symmetry can arise as the
intersection of different local \gut\ groups.

\begin{figure}[t!]
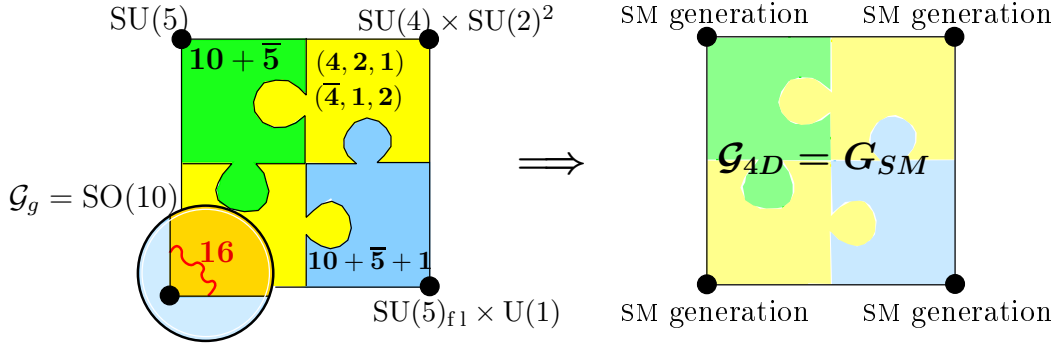

\centerline{\input localGuts.pstex_t}
\caption{Local {\gut}s at different fixed points: local vs. global picture. The intersection of all local
groups is $\maG_{4D}=G_{SM}$.} 
\label{fig:localGuts}
\end{figure}

If we suppose, for example, that the local gauge symmetry is \SO{10} whereas the unbroken gauge group in
four dimensions is 
$\maG_{4D}=G_{SM}=\SU3_{c}\x\SU2_{L}\x\U1_Y$, then the states attached to the fixed points
can transform locally as $\bs{16}$--plets, which under the \sm\ gauge group, form the representations
\begin{equation}\index{spinor, SO(10)}
\boldsymbol{16}~=~ 
(\boldsymbol{3}, \boldsymbol{2})_{1/6} + 
(\boldsymbol{\overline{3}}, \boldsymbol{1})_{-2/3}+
(\boldsymbol{\overline{3}}, \boldsymbol{1})_{1/3}+  
(\boldsymbol{1}, \boldsymbol{2})_{-1/2} + 
(\boldsymbol{1}, \boldsymbol{1})_{1}+ 
(\boldsymbol{1}, \boldsymbol{1})_{0}\;.
\label{eq:16pletSplitting}
\end{equation}
Similarly, $\bs{10}$--plets of \SO{10} can appear in the same (or in a different) local spectrum. They
correspond to the \sm\ representations 
\be
\bs{10}~=~ (\bs{3},\bs{1})_{-1/3}+ (\bs{1},\bs{2})_{1/2}+ (\bsb{3},\bs{1})_{1/3}+ (\bs{1},\bs{2})_{-1/2}\;,
\label{eq:10pletSplitting}
\ee
which are those of the Higgs doublets and Higgs triplets. As we already mentioned, once we abandon the
local picture and consider the orbifold globally, some parts of the $\bs{16}$-- or $\bs{10}$--plets might
not be present in the global spectrum due to the orbifold projections.
In the best of the cases, the parts of a local $\bs{10}$--plet which are projected out will be
those corresponding to Higgs triplets, whereas the complete $\bs{16}$--plets of the matter generation will
survive. This situation would then endow the theory with a natural explanation of the structure of the \sm\
families~\cite{Buchmuller:2004hv,Buchmuller:2005jr,Buchmuller:2005sh,Buchmuller:2006ik}. It would
additionally provide a solution to the doublet-triplet splitting problem of 4D {\gut}s~\cite{Kawamura:2000ir}. 

It is noteworthy that such an ideal case as the example just presented, including other \gut\ symmetries,
can be realized in orbifold constructions. We have to demand that matter generations appear from local
{\gut}s localized at specific points, where the action of the orbifold group on the associated states is
trivial. Furthermore, it is necessary that the Higgs representations come from the bulk or from fixed
points where the orbifold projections can partly alter their structure. Since solely higher
twisted-sector fields are affected by orbifold projections, the first requirement is met when local
{\gut}s appear at the fixed points of the first twisted sector {\it and} the local matter spectrum
includes \gut\ generations (such as \bs{27}--plets of \E6 or \bs{16}--plets of \SO{10}).\footnote{That
the states in the first twisted sector are not affected by the orbifold projection follows from the
fact that the projection phase, eq.~\eqref{eq:TwistedTrafo}, is trivial for $T_1$ (and $T_5$) twisted states
satisfying the masslessness condition, eq.~\eqref{eq:MasslessnessTwisted}.}
    
In section~\ref{subsec:LocalShiftsSpectra}, we have concluded that a local shift vector $V_g$ has to be
equivalent to one of the shift vectors $V$ that define \Z{N} orbifold models without Wilson
lines. Therefore, finding local {\gut}s with \sm\ matter generations at fixed points of the $T_1$ sector 
amounts to first identifying those shifts leading to the desired \gut\ symmetry as the gauge group of
the theory in four dimensions and \gut\ generations in the $T_1$ sector, and then selecting a suitable
set of Wilson lines inducing the symmetry breakdown $\maG_{GUT}\rightarrow G_{SM}$ (see
fig.~\ref{fig:localGuts}). 

One of the qualities of this method is that it ensures the existence of standard \gut\ hypercharge, 
which is consistent with gauge coupling unification, although no \gut\ appears in 4D. Another advantage is
that, even though matter fields form locally complete \gut\ representations, interactions generally break
\gut\ relations since different local {\gut}s are supported at different fixed points. In particular,
mixing of the localized generations with vectorlike bulk states breaks the unwanted \gut\ relations for
the fermion masses~\cite{Asaka:2003iy,Buchmuller:2006ik}. As already mentioned, this approach can explain
why the \sm\ gauge and Higgs bosons do not form complete \gut\ multiplets, while the matter fields do.

In the following sections, the concept of local {\gut}s is used as a guiding principle in the search
for orbifold models with realistic features. 

\section[\mssm\ Search Strategy]{\textbf{\textsc{mssm}} Search Strategy}
\label{sec:SearchStrategy}

It is well known that with a suitable choice of Wilson lines it is not difficult to obtain the \sm\ gauge
group up to \U1 factors. The real challenge is to get the exact matter spectrum of the \sm\ or, more
precisely, of its minimal supersymmetric extension, the \mssm. Demanding additionally gauge coupling
unification constrains possible models.\footnote{In
refs.~\cite{Ibanez:1987sn,Casas:1988se}, only models without standard hypercharge are investigated. We
have verified that, in fact, \Z3 and \Z4 orbifolds do not allow for \gut\ hypercharge normalization
(cf. chapter~\ref{ch:MLII}).} In order to find models which combine both properties, we base our strategy
on the concept of local {\gut}s described in the previous section.

An \SO{10} {\gut} has very compelling features which make it appropriate for model
building~\cite{Nilles:2004ej}. Among other reasons, it incorporates the success of Georgi-Glashow and
Pati-Salam {\gut}s, provides a single gauge coupling, predicts the existence of right-handed neutrinos
and gathers a complete \sm\ generation within a single representation. Hence, we shall
be mostly interested in the gauge shifts $V$ which allow for a local \SO{10}. Also \E6 local {\gut}s can
be reasonable promising as they embed the \SO{10} structure. That is, the shift vectors $V$ we will
consider are such that the  left--moving momenta $p$ satisfying 
\begin{equation}
 p \cdot V ~=~ 0 \mod 1\;,\quad p^2~=~2
\label{eq:LocalGaugeSymmetry}
\end{equation}
are roots of \SO{10} or \E6 (up to extra group factors). Furthermore, the massless states of the $T_1$
sector are required to contain $\boldsymbol{16}$--plets of \SO{10} at the fixed points with \SO{10}
symmetry or $\boldsymbol{27}$--plets of \E6 at the fixed points with \E6 symmetry.   

Since these massless states from $T_1$ are automatically invariant under the orbifold action, they all
survive in 4D and appear as complete \gut\ multiplets. In the case of \SO{10}, that gives one complete
\sm\ generation, while in the case of \E6 we have
$\boldsymbol{27}=\boldsymbol{16}+\boldsymbol{10}+\boldsymbol{1}$ under \SO{10}. It is thus necessary to
decouple all (or part) of the $\boldsymbol{10}$--plets from the low--energy theory.

\begin{table}[t!]
\centering
\begin{tabular}{|c|cl|}
\hline
Shift &\phantom{.}& 4D gauge group $\maG_{4D}$     \\
\hline\hline
\rowgrayh
$V^{\SO{10},1}$ && $\SO{10}\x\SU2^2\x\SO{14}\x\U1^2$ \\
$V^{\SO{10},2}$ && $\SO{10}\x\SU3\x\E7\x\U1^2$       \\
\rowgrayh
$V^{\E6,1}$     && $\E6\x\E8\x\U1^2$               \\
$V^{\E6,2}$     && $\E6\x\SU3\x\E7\x\U1$           \\
\hline
\end{tabular}
\caption{Gauge group associated to shift vectors with local \gut\ structure. Their matter spectrum in the
$T_1$ (or equivalently $T_5$) sector includes 12 \gut\ families.}
\label{tab:GGLocalGUTShifts}
\end{table}

The Wilson lines will be chosen such that the standard model gauge group is embedded into the local
\gut\ as 
\begin{equation}
 G_{SM}~ \subset~ \SU5 \subset \SO{10} ~\text{or}~ \E6 \;.
\end{equation}
It follows then that the hypercharge is that of standard {\gut}s and thus consistent with gauge coupling
unification. Hypercharge will appear as a linear combination of several \U1s and, thus, the generator of
hypercharge (in a standard basis),
\be
\mathsf{t}_Y ~=~ \left(-\2,\,-\2,\,\3,\,\3,\,\3\right)\,,
\label{eq:StdY}
\ee
will be embedded in a 16-dimensional vector acting on the gauge degrees of freedom.
Note that gauge coupling unification in orbifold models is due to the fact that the 10D (not 4D) theory is 
described by a single coupling.

\begin{figure}[t!]
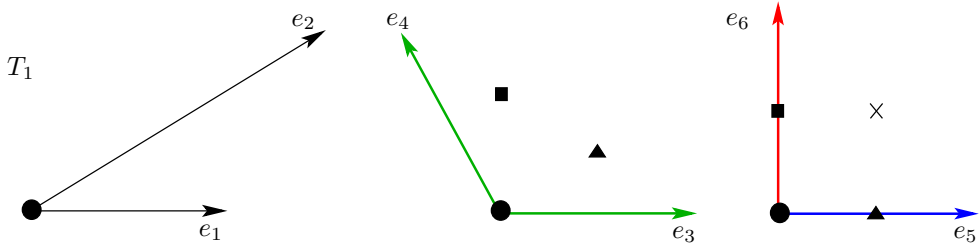

\centerline{\input Z6II_geometryT1.pstex_t}
\caption{$T_1$ sector of the \Z6-II orbifold on a \G2\x\SU3\x\SO4 lattice. The \SU3 torus admits one
  Wilson line $A_3$ of order 3 associated to the directions $e_3$ and $e_4$ while the \SO4
  torus can allocate up to two Wilson lines, $A_5$  and $A_6$, of order 2.}
\label{fig:Z6IIgeometryT1}
\end{figure}

Our model search is carried out in the \Z6-II orbifold compactification of the \E8\x\E8 heterotic
string. Some properties of this orbifold were described in section~\ref{subsec:OrbifoldGeometry} (see
the discussion around eq.~\eqref{eq:Z6IILattices}). 
In this construction, there are two gauge shifts leading to a local
\SO{10} \gut\ \cite{Katsuki:1989cs, Wingerter:2005xx}, 
\begin{eqnarray}
V^{ \SO{10},1}= &
\left(\tfrac{1}{3},\,\tfrac{1}{2},\,\tfrac{1}{2},\,0,\,0,\,0,\,0,\,0\right)&\left(\tfrac{1}{3},\,0,\,0,\,0,\,0,\,0,\,0,\,0\right) \;,
\nonumber \\
V^{ \SO{10},2 }= &
\left(\tfrac{1}{3},\,\tfrac{1}{3},\,\tfrac{1}{3},\,0,\,0,\,0,\,0,\,0\right)&\left(\tfrac{1}{6},\,\tfrac{1}{6},\,0,\,0,\,0,\,0,\,0,\,0\right) \;,
\label{eq:so10shifts}
\end{eqnarray}
and 2 shifts leading to a local \E6 \gut,
\begin{eqnarray}
 V^{\E6 , 1}= &
\left(\tfrac{1}{2},\,\tfrac{1}{3},\,\tfrac{1}{6},\,0,\,0,\,0,\,0,\,0\right)&\left(0,\,0,\,0,\,0,\,0,\,0,\,0,\,0\right)\;,
\nonumber \\
 V^{ \E6 ,2}= &
\left(\tfrac{2}{3},\,\tfrac{1}{3},\,\tfrac{1}{3},\,0,\,0,\,0,\,0,\,0\right)&\left(\tfrac{1}{6},\,\tfrac{1}{6},\,0,\,0,\,0,\,0,\,0,\,0\right).\label{eq:e6shifts}
\end{eqnarray}
The symmetry breakdown of \E8\x\E8 due to these shift vectors is given in table~\ref{tab:GGLocalGUTShifts}.
The shift $V^{\SO{10},1}$ of eq.~\eqref{eq:VSO(10)1} and the one presented here are equivalent. They differ
because we have used Weyl transformations to reduce the number of nonzero entries in order to facilitate
later analysis.
We will focus on these shifts and scan over possible Wilson lines, employing
the classification method of section~\ref{sec:ClassAnsatz}, to get the \sm\ gauge group.
Due to its geometry, 
the \Z6-II orbifold with lattice \G2\x\SU3\x\SO4 allows for up to two Wilson lines of order 2, $A_5$ and
$A_6$, and for one Wilson line of order 3, $A_3$, as illustrated in fig.~\ref{fig:Z6IIgeometryT1}.

The next question is how to get three matter generations. A geometric explanation of the origin of the
three generations is somehow attractive and can be realized in orbifold compactifications. As depicted in
fig.~\ref{fig:Z6IIgeometryT1}, there 
are $3\x 4$ fixed points in the $T_1$ sector, which, in the absence of Wilson lines, are degenerate, that
is, are furnished with identical matter spectrum. Wilson lines lift this degeneracy in the direction they
act. For example, if the Wilson line $A_5$ related to the torus direction $e_5$ is nontrivial, then in
the \SO4 torus the
points $\blacksquare$ and $\bullet$ as well as $\bs{\times}$ and $\blacktriangle$ are
equivalent, but the horizontal symmetry is lost.\footnote{The remaining vertical symmetry is usually
identified with $D_4$ or $S_2$, which in adequate models can be interpreted as a family
symmetry.} The points that keep their local \gut\ structure are in this case the six ones related to
$\blacksquare$ and $\bullet$ in the \SO4 plane. 
Hence, the simplest possibility to get three matter generations is to use three equivalent fixed points with
\bs{16}--plets~\cite{Buchmuller:2004hv} which appear in  models with two Wilson lines of order 2. If the
extra states are vectorlike and can be given large masses, the low--energy spectrum will contain three matter
families. However, if one starts from the shifts given above, this strategy fails since all such models
contain chiral exotic states~\cite{Buchmuller:2006ik}. In the case of \E6, it does not work either since
a simple analysis of the Dynkin diagram of \E6 reveals that one cannot obtain
$G_\mathrm{SM}~\subset~\SU5~\subset~\E6$ with two Wilson lines of order 2.

The next--to--simplest possibility is to use two equivalent fixed points which give rise to two matter
generations. This situation can be obtained in models with one Wilson line of order 3 and another of
order 2. The third generation would then have to come from other twisted or untwisted
sectors. The appearance of the third {\it complete} family can be linked to the \sm\
anomaly cancellation. Indeed, since the shift vectors of eqs.~\eqref{eq:so10shifts}
and~\eqref{eq:e6shifts} lead to \gut\ families in the untwisted sector,
that sector contains part of a $\boldsymbol{16}$--plet after the inclusion of Wilson lines. 
Then the simplest options consistent
with the \sm\ anomaly cancellation are that the remaining matter
either completes the $\boldsymbol{16}$--plet or provides
vectorlike partners of the untwisted sector. In more complicated
cases, additional $\boldsymbol{16}$-- or \bsb{16}--plets can appear. The localized
$\boldsymbol{16}$-- and $\boldsymbol{27}$--plets are true \gut\
multiplets, whereas the third or ``bulk''   generation  only has the
\sm\ quantum numbers of an additional $\boldsymbol{16}$--plet.

\begin{figure}[t!]
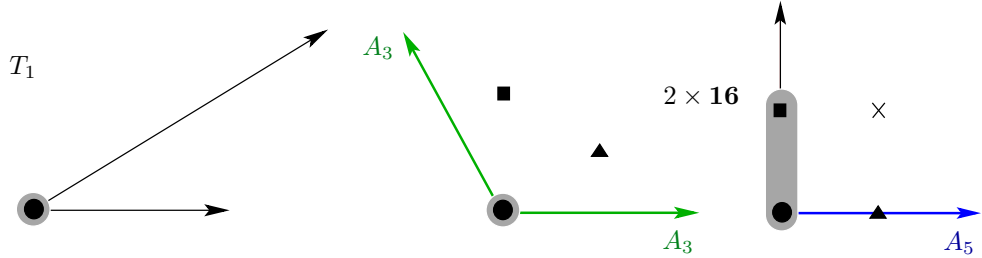

\centerline{\input Z6II_gutsT1.pstex_t}
\caption{A successful approach to get models with three matter generations. Two \gut\ generations are
  located at fixed points of the $T_1$ sector while the third one comes from other untwisted or twisted
  sectors.} 
\label{fig:Z6IIgutsT1}
\end{figure}

We find that the above strategy is successful. First, we notice that the model presented in
refs.~\cite{Buchmuller:2005jr,Buchmuller:2006ik} adjusts precisely to this scheme. Secondly, also in the
case of \E6 {\gut}s we observe that it is not difficult to obtain the \sm\ gauge group. As we will
detail in section~\ref{sec:TowardsRealistic}, in fact, other massless states present in the spectra of
models based on this strategy are often vectorlike with respect to the \sm\ gauge group and can be given
large masses consistent with string selection rules. 

Once established the basis of our study, let us specify now all other details of our search strategy. 
We enumerate the steps of our search as follows:

\ding{"0C0} Generate all models with two Wilson lines. \vskip 1mm

\ding{"0C1} Identify ``inequivalent'' models.\vskip 1mm

\ding{"0C2} Select models with $G_{SM} \subset \SU5 \subset \SO{10}$ or \E6.\vskip 1mm

\ding{"0C3} Select models with three net $(\bs{3},\bs{2})$.\vskip 1mm

\ding{"0C4} Select models with nonanomalous $\U1_{Y} \subset \SU5$.\vskip 1mm

\ding{"0C5} Select models with net three \sm\ families + Higgses + vectorlike.\vskip 1mm

Notice that in step~\ding{"0C0} we have decided not to specialize to the case of one Wilson line of order
2 and one of order 3. Even though we already know that only in that case we can obtain realistic models,
we would like also to know how frequently those models appear in the landscape patch that we have
chosen. In other words, it might be useful to count with the total number of models in order to draw some
reliable statistical results out of our search. To do so, we apply the classification method described in
section~\ref{sec:ClassAnsatz}, that is, we propose an ansatz describing generically Wilson lines of order
2 and 3, based on eqs.~\eqref{eq:BlockAal} and~\eqref{eq:AllAnsaetze}.

Not all models obtained in step~\ding{"0C0} are inequivalent. In step~\ding{"0C1}, we consider two models to be
``equivalent'' if they have identical spectra with respect to nonabelian gauge groups and have the same
number of nonabelian singlets. Thus, models differing only in \U1 charges are treated as
equivalent. In addition, some models differ only by the localization of states on the different fixed
points. We know that these ambiguities occur and it is possible that in some cases Yukawa couplings are
affected. Hence our criterion may underestimate somewhat the number of truly inequivalent models.

The criterion~\ding{"0C2} is crucial for obtaining viable models. We require particularly that \SU3\x\SU2
appears embedded in an \SU5 subgroup of the local {\gut}s. This constraint is imposed in order for the
hypercharge to be of Georgi-Glashow type and, hence, consistent with gauge coupling unification.

Of course, we could try immediately to identify an appropriate hypercharge. It proves however to be more
practical to first reduce further the number of models available. Criterion~\ding{"0C3} amounts to finding
those models where three copies of quark-doublet-like particles exist. Let us just remark that, in
general, both representations $(\bs{3},\bs{2})$ and $(\bsb{3},\bs{2})$ under \SU3\x\SU2 appear in the
spectra. They form some vectorlike pairs that will eventually decouple from the low--energy
spectrum. Thus, only models satisfying $|\#(\bs{3},\bs{2})-\#(\bsb{3},\bs{2})|=3$ will lead to realistic
phenomenology. 

After the above procedure, it is then straightforward to identify the hypercharge. In an ideal case, the
hypercharge generator would be given by~\eqref{eq:StdY} embedded in a 16-dimensional vector. However,
this form is basis-dependent. In general, what one does is to compare\\[0.5mm]
\begin{minipage}{0.55\textwidth}
the simple roots of the \SU5 embedded in the \gut\ group to those of \SU3\x\SU2. Since the roots of \SU5 span a
four-dimensional space whereas those of \SU3\x\SU2 only span a three-dimensional space, the hypercharge
generator $\mathsf{t}_Y$ is uniquely
\end{minipage}
\begin{minipage}{0.35\textwidth}
\setlength{\unitlength}{0.95mm} 
\begin{center}
\begin{picture}(60,22) 
\thicklines
\put(5,8){\circle{2}}
\put(13,8){\circle{2}}
\put(29,8){\circle{2}}

\put(5,18){\circle{2}}
\put(13,18){\circle{2}}
\put(21,18){\circle{2}}
\put(29,18){\circle{2}}

\put(6,8){\line(1,0){6}} 

\put(6,18){\line(1,0){6}} 
\put(14,18){\line(1,0){6}} 
\put(22,18){\line(1,0){6}} 

\put(3,3){${\scriptstyle\alpha_1}$}
\put(11,3){${\scriptstyle\alpha_2}$}
\put(19,6){$\mathsf{t}_Y$}
\put(27,3){${\scriptstyle\alpha_4}$}

\put(3,13){${\scriptstyle\alpha_1}$}
\put(11,13){${\scriptstyle\alpha_2}$}
\put(19,13){${\scriptstyle\alpha_3}$}
\put(27,13){${\scriptstyle\alpha_4}$}

\put(32,6){: \SU3\x\SU2\x$\U1_Y$}
\put(32,16){: \SU5}
\end{picture}
\end{center}
\end{minipage}\\[0.8mm]
determined by the sole direction orthogonal to the simple roots of
\SU3\x\SU2 which lies in the space of \SU5. In other words, the hypercharge results directly from the
symmetry breakdown $\SU5\longrightarrow G_{SM}$.
Ambiguities arise in certain seldom cases when $\U1_Y$ can
be defined in different ways. In this case, we have to count the model twice.
On the other hand, the hypercharge generator can unfortunately mix with the anomalous \U1 of the
orbifold, what leads to the undesirable consequence that it has to be broken at very high
energies. Therefore, at step~\ding{"0C4}, one is forced to disregard those models where the hypercharge appears anomalous. 

We are in position to identify the \mssm\ candidates. In step~\ding{"0C5}, we have to verify that the
resulting spectrum contains exactly the spectrum of the \mssm, i.e. three net matter generations and, at
least, one pair of Higss-doublets with the correct quantum numbers under
$G_{SM}=\SU3_c\x\SU2_L\x\U1_Y$. Other states in the spectra of admissible models, generically called
exotics must form vectorlike pairs.

\section{A Fertile Patch in the Landscape}
\label{sec:Results}

The results of our search strategy are presented in table~\ref{tab:Summary}. Interestingly enough, this
strategy has allowed us to find 223 explicit examples with the \mssm\ matter content plus additional
vectorlike particles. These models represent a major result in the context of string compactifications and
are considered one of the central results of this work. 
We see that most of the \mssm\ candidates arise from a local \SO{10} \gut. Na\"ively, we can
consider this result to be another argument why \SO{10} {\gut}s might be preferred to other {\gut}s. On
the other hand, it might just reflect the fact that two Wilson lines cannot break \E6 enough to get a
comparable amount of models with $G_{SM}$ gauge symmetry. One might then expect that adding a third
Wilson line, many more models with promising properties arise from the \E6 shifts.

We find that the properties of the models with the chiral \mssm\ matter content, such as the number and type of
vectorlike exotics and \sm\ singlets, are so similar that there is no model which can be preferred {\it a
priori}. This implies that, even though one is also interested in studying the phenomenological
features of each model, some interesting conclusions can be drawn in a generic form. 
The details of the models are listed in~\cite{WebTables:2006xx}.

\renewcommand{\arraystretch}{1.3}
\begin{table}[t!]
\centerline{
\begin{tabular}{|l||r|r||r|r|}
\hline
Criterion & $V^{\SO{10},1}$ & $V^{\SO{10},2}$ & $V^{\E6,1}$ & $V^{\E6,2}$\\
\hline\hline
\rowgrayh
 \ding{"0C1}  inequivalent models with 2 Wilson lines
  &$22,000$ & $7,800$  &$680$  &$1,700$ \\
  \ding{"0C2} \sm\ gauge group $\subset$ \SU{5} $\subset$ \SO{10} ({\rm or}~\E6)
  &$3,563$ & $1,163$ &$27$ &$63$\\
\rowgrayh
  \ding{"0C3} 3 net $(\boldsymbol{3},\boldsymbol{2})$ 
  &$1,170$ &$492$ &$3$ &$32$\\
  \ding{"0C4} nonanomalous $\U1_{Y}\subset \SU5 $
  &$528$ &$234$ &$3$ &$22$\\
\rowgrayh
  \ding{"0C5}  spectrum $=$ 3 generations $+$ vectorlike
  &$128$ &$90$ &$3$ &$2$  \\
\hline
\end{tabular}
}
\caption{Statistics of \Z6-II orbifold models based on the shifts
$V^{\SO{10},1},V^{\SO{10},2},V^{\E6,1},V^{\E6,2}$ with two Wilson lines.
\label{tab:Summary} }
\end{table}
\renewcommand{\arraystretch}{1.1}

It is instructive to compare our model scan to others. In certain types of intersecting D--brane models,
it was found that the probability of obtaining  the \sm\ gauge group and three generations of quarks and
leptons, while allowing for chiral exotics, is $10^{-9}$~\cite{Gmeiner:2005vz,Douglas:2006xy}.
More recently, in ref.~\cite{Gmeiner:2007zz} intersecting D--branes
on \Z6-II orbifolds have lead to the conclusion that the probability of finding such matter configurations
can be enhanced to $10^{-8}$. The criterion which comes closest to the requirements imposed in
refs.~\cite{Gmeiner:2005vz,Douglas:2006xy,Gmeiner:2007zz} is \ding{"0C3}. We find that within our sample
the corresponding probability is 5\,\%. 

In refs.~\cite{Dijkstra:2004cc,Anastasopoulos:2006da}, orientifolds of Gepner models were scanned for
chiral \mssm\ matter spectra, and it was found that the fraction of such models is $10^{-14}$. In our set
of models, the corresponding probability, i.e.\ the fraction of models passing criterion \ding{"0C5}, is
of order 1\,\%. Note also that, in all of our models, hypercharge is normalized as in standard {\gut}s
and thus consistent with  gauge coupling unification. 

This comparison shows that our sample of heterotic orbifolds is unusually ``fertile'' compared to other
constructions.  The probability of finding something close to the \mssm\ is much higher than that in
other patches of the landscape analyzed so far.
It would be interesting to extend these results to other regions of the
landscape where promising models exist~\cite{Braun:2005nv,Bouchard:2005ag,Kim:2006hv,Cleaver:1998sa}.

We would like to make an additional remark. Our strategy to determine the hypercharge is, of course, not
the only choice. One could instead express $\U1_Y$ as an arbitrary linear combination of all \U1's (not
only of those embedded in the local \gut\ symmetry) which gives the correct values of hypercharge to the
\mssm\ particles. This approach was followed in ref.~\cite{Raby:2007yc}. The authors of
ref.~\cite{Raby:2007yc} find that the majority of the models at step~\ding{"0C3} allow for a definition
of a nonanomalous $\U1_Y$. However, 
only in 12\,\% of those models, hypercharge is in harmony with coupling unification. That means, in
particular, that even in a more general scheme, relaxing the demand $\U1_{Y}\subset \SU5$, (almost)
only those 223 models at step~\ding{"0C5} of our search meet all the phenomenological properties we
require. 

\section{Towards Realistic String Models}
\label{sec:TowardsRealistic}

Taking as a base the \mssm\ candidates passing criteria \ding{"0C2}-\ding{"0C5}, there are now many ways
to address the question of their phenomenological viability. One could, for example, opt for a
model--by--model approach, in an attempt to identify one model with many features that match the known
low--energy physics and, simultaneously, to find the solution to as many as possible of the currently
open questions in theoretical high energy physics. By that approach, certainly one might find one model 
with some beautiful properties that also gives answer to some puzzles. However, a different model will
have some other nice features and will solve some other problems.

We would rather follow an alternative approach. Provided that the \mssm\ candidates are very similar, one
could ask general questions, such as whether the vectorlike exotics decouple, or whether \susy\
preserving vacuum configurations are realizable from these models, among other matters. The answers to
those questions are, of course, model dependent, but, on a statistical footing, they will probably yield
some predictions or, at least, exclude some regions of the landscape.

\subsection[Coupling Selection Rules in \Z6-II]{Coupling Selection Rules in \bs{\Z6}-II}
\label{subsec:SelectionRulesZ6II}\index{selection rules, Z6@selection rules, \Z6-II}

Let us start by stating the string selection rules which will be repeatedly used in our study. 
Assume a coupling between a number of states $\Psi_i$. Based on our discussion of
section~\ref{sec:ourstringselectionrules}, couplings entering the superpotential must satisfy the
following rules.  

{\bf Gauge invariance.} The usual rule on the left-moving momenta $p_{\mathrm{sh},i}$ of the states
$\Psi_i$, $\sum_i p_{\mathrm{sh},i} = 0$, applies. 

{\bf R--charge conservation.}\footnote{R--symmetries do distinguish between fermions and bosons,
therefore, in some approaches~\cite{Araki:2007ss}, the remaining discrete subgroup after breaking these
symmetries can well be identified with an $R$--parity. We do not follow this approach.} 
Since the twist vector reads $v=(0,1/6,1/3,-1/2)$, the order of the corresponding (discrete)
R--symmetries is respectively $6$, $3$ and $2$. Therefore, a coupling in the superpotential is invariant
if the R--charges of the states involved fulfill the conditions
\be
\sum_i R^1_i = -1\mod 6\,,\;\;\;\sum_i R^2_i = -1\mod 3\,,\;\;\;\sum_i R^3_i = -1\mod 2\;.
\label{eq:Z6IIRChargeConservation}
\ee

{\bf Space group selection rule.} In general, this rule reads $\prod_ig_i\simeq(\id,\,0)$ with
$g_i=(\vartheta^i,n_\al^ie_\al)\in S$ denoting the constructing elements of the states entering the
coupling. It can be restated in \Z6-II orbifolds as the following set of constraints:
\bse\label{eq:SpaceGroupZ6II}\index{point group rule, Z6@point group rule, \Z6-II}
\index{flavor/discrete symmetries, Z6@flavor/discrete symmetries, \Z6-II}
\bea
\6\sum_i k_i &=& 0\mod1\,,\label{eq:SpaceGroupZ6IIk}\\
\3\sum_i(n_3^i+n_4^i)&=&0\mod1\,,\\
\2\sum_i n_5^i&=&0\mod1\,,\\
\2\sum_i n_6^i&=&0\mod1\,.
\eea
\ese
Notice that, in this notation, these rules can be identified with conservation of certain discrete
\Z{n} symmetries with (integer) charges $k$, $n_3+n_4$, $n_5$ and $n_6$. Some consequences of the
appearance of these symmetries have been studied in ref.~\cite{Araki:2008ek}.

\subsection{Decoupling Exotic Particles}

\index{decoupling exotics}
Now we are in position to investigate the decoupling of vectorlike extra matter $\{ x_i \}$. One could argue that
the existence of these exotics in the spectra of our \mssm\ candidates constitutes a 
problem by itself since vectorlikeness does not guarantee immediately that such exotics get large masses. 
Therefore, we have to corroborate whether the extra matter can be given a large mass by explicitly
computing the couplings that endow exotics with masses. The mass terms for such states are provided by
the superpotential 
\begin{equation}\label{eq:decoupling}
 W~=~  x_i\, \bar x_j \,\calM^{ij}_{x\bar x}~\equiv~ x_i\, \bar x_j\, \langle s_1\, s_2\dots \rangle \;,
\end{equation}
where $s_1,s_2,\dots$ are \sm\ singlets. Some singlets are required to get large (close to
$M_\mathrm{str}$) {\vev}s in order to cancel the Fayet--Iliopoulos (FI) term of the anomalous \U1
intrinsic to the majority of orbifold compactifications. Then, if the relevant Yukawa couplings are
allowed by the string selection rules discussed above, mass terms
as that of eq.~\eqref{eq:decoupling} appear, making the vectorlike matter heavy. Thus the exotic
particles decouple from the low--energy theory. 

Clearly, one cannot switch on the singlet {\vev}s at will. Instead, one has to ensure that they are
consistent with \susy. Supersymmetry requires vanishing $F$-- and $D$--terms. The number of the
$F$--term equations equals the  number of complex singlet fields $s_i$, therefore there are in general
nontrivial singlet configurations with vanishing $F$--potential. The $D$--terms can be made zero by
complexified gauge transformations~\cite{Ovrut:1981wa} if each field enters a gauge invariant
monomial~\cite{Buccella:1982nx}. Thus, to ensure that the decoupling of exotics is consistent with
\susy, one has to show that all \sm\ singlets appearing in the mass matrices for the exotics enter gauge
invariant monomials involving only \sm\ singlets and carrying anomalous charge. However, for simplicity,
we will assume momentarily that all \sm\ singlets entering the mass matrices develop large supersymmetric
{\vev}s.\footnote{This assumption will be confirmed in the next chapter by examining carefully the
decoupling of exotics in \susy\ preserving vacua.} 

In many cases, vectorlike exotics happen to have the same quantum numbers as the \sm\ particles. Hence,
in the process of decoupling, the vectorlike states can mix with the localized $\boldsymbol{16}$-- and
$\boldsymbol{27}$--plets (as long as it is allowed by the \sm\ quantum numbers) such that the physical
states at low energies are \emph{neither} localized \emph{nor} ``true'' \gut\ multiplets. 
Even though part of the beauty of a geometrical explanation of the family structure dissolves thereby,
it is clear that whatever the mixing, in the end exactly three \sm\ families will be left if the mass
matrices $\calM_{x\bar x}$ have maximal rank. 

Decoupling of exotics in \susy\ preserving vacua has additional effects on the low--energy theory.
Generally, there are supersymmetric vacua in which all or most of the \sm\ singlets get large {\vev}s. As
a consequence, many of the (abelian and nonabelian) gauge group factors get spontaneously broken, such
that the low--energy gauge group can be $G_{SM}$ up to a hidden sector:
\begin{equation}
G_{SM} \times G_\mathrm{hidden}\;,
\end{equation}
where the \sm\ matter is neutral under $G_\mathrm{hidden}$. This (true) hidden sector is important to
deal with the problem of low--energy \susy\ breaking, as we shall outline in section~\ref{sec:SUSYBreaking}.

In practice, to show that the decoupling of exotics is consistent with string selection rules is a technically
involved and time consuming issue. In order to simplify the task and to reduce the number of models, we
first impose an additional condition. We require that the models possess a renormalizable top--Yukawa
coupling as motivated by phenomenology (namely, by the large mass of the top--quark). We point out that this
requirement, nevertheless, is not imperative since a large top--Yukawa coupling can also be obtained as a
result of a conspiracy of the {\vev}s of the fields or by other means.

As a second technical simplification, we consider only superpotential couplings up to order
eight. Thus, the next two steps in our selection procedure are:\vskip 1.5mm

\ding{"0C6} Select models with a ``heavy top''. \vskip 1.5mm

\ding{"0C7} Select models in which the exotics decouple at order 8.\vskip 1.5mm

In step~\ding{"0C6}, we require a renormalizable $\maO(1)$ Yukawa coupling of the type\footnote{Notice
that at this level it is impossible to distinguish between lepton doublets $\bar\ell$ and up--type
Higgses $h_u$.}
\be
q\,\bar u\,h_u \sim(\bs{3}, \bs{2})_{1/6}\,(\bsb{3},\bs{1})_{-2/3} \,(\bs{1}, \bs{2})_{1/2}\,.
\label{eq:TopCoupling}
\ee
To accomplish this condition, we have first to verify whether such a coupling is allowed by string
theory. In the \Z6-II orbifold, combining the point group selection rule, eq.~\eqref{eq:SpaceGroupZ6IIk},
with R--charge conservation, one finds that only renormalizable couplings of fields from the sectors
\begin{equation}\index{point group rule, Z6II@point group rule, \Z6-II, trilinear}
 U_1\,U_2\,U_3\;,\quad U_3\,T_2\,T_4\;,\quad U_2\,T_3\,T_3\;,\quad T_1\,T_2\,T_3 \;,\quad T_1\,T_1\,T_4 \;
\label{eq:Z6IIRenormCoupl}
\end{equation}
are nonvanishing. On the other hand,
the coupling strength of string interactions is given by their correlation function, which for
renormalizable couplings has been computed analytically~\cite{Dixon:1986qv,Hamidi:1986vh}. It turns out
that the coupling strength has a dependence on the localization of the fields involved. In particular, the coupling for
twisted fields is proportional to $e^{-\calA}$, where $\calA$ denotes the area of the
triangle formed by the fixed points where the strings are located; hence, it can be highly suppressed if
the strings are located far apart from each other. Therefore, couplings including an untwisted field are
unsuppressed whereas $T\,T\,T$-couplings are significant only when the twisted fields are localized at the
same fixed point. We discard models in which the above couplings vanish or are suppressed.

In the next step \ding{"0C7}, we select models in which the  mass matrices $\calM_{x\bar x}$ for the
exotics  (cf.\ eq.~\eqref{eq:decoupling}) have a maximal rank such that no exotic states appear at low
energies. Here, we consider only superpotential couplings up to order eight and for this analysis we
assume that all \sm\ singlets can obtain supersymmetric {\vev}s.

In table~\ref{tab:Summary2} we summarize our results (see ref.~\cite{WebTables:2006xx} for further
details). We make a distinction between those having a ``heavy top'', as in step~\ding{"0C6}, and those
without it, \rojo{\xcancel{\negro{\ding{"0C6}}}}. We identify 93 models that pass requirements
\ding{"0C6} and \ding{"0C7}. This means that a significant fraction of our models can serve as an
ultraviolet completion of the \mssm\ in string theory. 

There are also many models which do not fulfill criterion~\ding{"0C6}. As already mentioned, although
our na\"ive approach indicates that no heavy top can be found directly in these models, it can be that
further analysis reveals that such conclusion is incorrect. For that reason, we have analyzed also the
decoupling of the exotics in these \mssm\ candidates. In this set, exotic particles can get large masses
in 103 models. All in all, we see that if the criterion~\ding{"0C6} is skipped, out of 223 \mssm\
candidates, 196 do not present exotics at low energies.

\renewcommand{\arraystretch}{1.3}\index{mssm candidates@\mssm\ candidates}
\begin{table}[t!]
\centerline{
\begin{tabular}{|l||rr|rr||r|r|}
\hline
           & \multicolumn{2}{c|}{$V^{\SO{10},1}$} & \multicolumn{2}{c||}{$V^{\SO{10},2}$} & $V^{\E6,1}$ & $V^{\E6,2}$\\
\cline{2-7}
 Criterion & \ding{"0C6}  & \rojo{\xcancel{\negro{\ding{"0C6}}}} 
           & \ding{"0C6}  & \rojo{\xcancel{\negro{\ding{"0C6}}}} & \ding{"0C6} & \ding{"0C6}\\
\hline\hline
\rowgrayh
 \ding{"0C5} spectrum $=$ 3 generations $+$ vectorlike & 72 & 56 & 37 & 53 & 3 & 2 \\
 \ding{"0C7} exotics decouple at order 8               & 56 & 50 & 32 & 53 & 3 & 2 \\
\hline
\end{tabular}
}
\caption{A subset of the \mssm\ candidates. The number of models with ``heavy top'' are listed
under~\ding{"0C6}. \rojo{\xcancel{\negro{\ding{"0C6}}}} denotes no ``heavy top''.
\label{tab:Summary2}}
\end{table}
\renewcommand{\arraystretch}{1.1}

\index{exotica}
A word on possible light exotics is in order. In ref.~\cite{Raby:2007hm} it has been considered the
possible presence of a special kind of vectorlike exotics (baptized {\it exotica}) in the low--energy
spectrum. These exotic particles, which appear in some of our models, would not affect gauge coupling
unification and could have a fractional electric charge. Although unlikely, one wonders whether such
particles might appear in future colliders. If it happens, we would have to reconsider not to decouple
all exotics.

We would like to point out an observation. We find that there are in our models several pairs of
Higgs doublets with a matrix of $\mu$--like mass terms. Generally, all Higgs doublets can acquire large
masses just as the (other) vectorlike exotics do. Hence, finding a massless pair of Higgses that
eventually triggers the Higgs mechanism of the \mssm\ is quite nontrivial. To get only one pair of
massless Higgs doublets usually requires fine-tuning in the {\vev}s of the \sm\ singlets such that the
mass matrix for the $({\bf 1}, {\bf 2})_{-1/2}, ({\bf 1}, {\bf 2})_{1/2} $ states gets a zero
eigenvalue. This is the notorious supersymmetric $\mu$--problem. The fine-tuning can be ameliorated if
the vacuum respects certain (approximate) symmetries~\cite{Kim:1983dt,Giudice:1988yz}. We will find in
section~\ref{sec:RParity} that such symmetries might appear regularly in promising orbifold constructions.

In conclusion, we have found a very interesting set of semi-realistic models. In the most optimistic case,
we find that any of 196 models could very well house the physics of the \sm.
However, we are still far of making any phenomenological conjecture from string theory about physics at
low energies.
To verify whether our \mssm\ candidates are consistent with phenomenology requires
addressing several questions. Some important issues we must still clarify include \vskip 0.8mm

$\bullet$ \susy\ preserving vacua at high energies, \vskip 0.8mm  %D-flatness

$\bullet$ hierarchically small \susy\ breaking, \vskip 0.8mm      %Gaugino condensation

$\bullet$ realistic flavor structures, and\vskip 0.8mm            %Seesaw

$\bullet$ absence of fast proton decay.\\[0.8mm]                  %R-parity
These and some other phenomenological questions will be discussed in the next chapter.

%%%%%%%%%%%%%%%%%%%%%%%%%%%%%%%%%%%%%%%%%%%%%%%%%%%%%%%%%%%%
%CHAPTER 5
%%%%%%%%%%%%%%%%%%%%%%%%%%%%%%%%%%%%%%%%%%%%%%%%%%%%%%%%%%%%
\chapter[Low Energy Physics from Orbifolds]{Low Energy Physics from Heterotic Orbifolds}
\label{ch:OrbifoldPhenomenology}

\begin{center}
\begin{minipage}[t]{15cm}
In this chapter we examine the phenomenological properties of the models found in the previous chapter.
First, we discuss the matter spectra of a characteristic model with the exact spectrum of
the \mssm, which serves as example all through this chapter. Then we proceed to evaluate the implications
of the hidden sectors of the \mssm-like models. 
We investigate a correlation between the realistic properties of these models and \susy\ breaking via
gaugino condensation. A surprising result is that the breakdown of \susy\ occurs generically at an
intermediate energy scale. Another interesting problem is whether it is possible to impose $R$-parity in
order to guarantee proton stability. We find a chance, defining a matter parity from the spontaneous
breaking of $\U1_{\BmL}$ which however can only be defined in some models. Other plausible possibilities are not
discussed here. Finally, we address the question of the viability of the seesaw mechanism in
orbifold models. Contrary to previous statements, we find that the presence of $\maO(100)$ right-handed
neutrinos in \mssm-like models supports the success of the seesaw mechanism in these models. 
\end{minipage}
\end{center}

\section[An Orbifold-\mssm]{An Orbifold-\textbf{\textsc{mssm}}}
\label{sec:SeesawModel}

In the last chapter we found more than 200 models with the \mssm\ matter spectrum and vectorlike
exotics. By demanding that the vectorlike exotics decouple from the low--energy spectrum, we have
seen that the number of models is barely reduced. We then opted to take a more aggressive approach: we
disregarded those models where a trilinear Yukawa coupling $q\,\bar u\,h_u$ is vanishing or suppressed
according to string selection rules. Note that this requirement is not imperative and was implemented
only as a technical simplification.
It turns out that this criterion is rather strict and reduces the amount of realistic models by
a factor $1/2$. The remaining $\maO(100)$ models have many common properties and will lead us to
interesting phenomenological conclusions along the lines of this chapter.

\index{orbifold-mssm@orbifold-\mssm}
We study now one generic model out of the set of promising \mssm\ candidates with ``heavy top'', which
will be referred to as {\it orbifold-\mssm}. The model is based on the gauge shift
\begin{eqnarray}
V^{ {\rm SO(10)},1} & =
&\left(\tfrac{1}{3},\,-\tfrac{1}{2},\,-\tfrac{1}{2},\,0,\,0,\,0,\,0,\,0\right)\left(\tfrac{1}{2},\,-\tfrac{1}{6},\,-\tfrac{1}{2},\,-\tfrac{1}{2},\,-\tfrac{1}{2},\,-\tfrac{1}{2},\,-\tfrac{1}{2},\,\tfrac{1}{2}\right)
\;.
\end{eqnarray}
where we have added an \E8\x\E8 lattice vector to simplify computations. The Wilson lines are chosen as
\begin{eqnarray}
A_{3} & = &\left(-\tfrac{1}{2},\,-\tfrac{1}{2},\,\tfrac{1}{6},\tfrac{1}{6},\,\tfrac{1}{6},\,\tfrac{1}{6},\tfrac{1}{6},\,\tfrac{1}{6}\right)\left(\tfrac{10}{3},\,0,\,-6,\,-\tfrac{7}{3},\,-\tfrac{4}{3},\,-5,\,-3,\,3\right)\,,\nonumber\\
A_{5} & = &\left(\tfrac{1}{4},\,-\tfrac{1}{4},\,-\tfrac{1}{4},-\tfrac{1}{4},\,-\tfrac{1}{4},\,\tfrac{1}{4},\tfrac{1}{4},\,\tfrac{1}{4}\right)\left(1,\,-1,\,-\tfrac{5}{2},\,-\tfrac{3}{2},\,-\tfrac{1}{2},\,-\tfrac{5}{2},\,-\tfrac{3}{2},\,\tfrac{3}{2}\right)\,.
\end{eqnarray}
The hypercharge generator is identified with the \SU5 standard form
\begin{eqnarray}
\mathsf{t}_{Y} & = &\left(0,\,0,\,0,\,-\tfrac{1}{2},\,-\tfrac{1}{2},\,\tfrac{1}{3},\,\tfrac{1}{3},\,\tfrac{1}{3}\right)
\big(0,\,0,\,0,\,0,\,0,\,0,\,0,\,0\big)\,.
\end{eqnarray}
The gauge group after compactification is
\begin{equation}\label{eq:Seesaw4DGG}
 G_{SM} \times \SO8 \times \SU2 \times \U1^7 \;,
\end{equation}
while the massless spectrum is given in table~\ref{tab:spectrum}. From there, we see that the three \sm\
particles $\ell_i$ and $\bar d_i$ come accompanied with additional vectorlike pairs $\ell-\bar\ell$ and
$\bar d-d$, respectively. Exotics with nonzero hypercharge, such as $x^+_i$ and $x^-_i$ , are also
vectorlike. Since the exotics with no hypercharge $m_i$ and $y_i$ are doublets under $\SU2_L$, they can
allow couplings such as $m_i m_j$ and decouple from the low--energy spectrum. The remaining particles are
either \sm\ singlets or the three \sm\ matter generations which need not be decoupled. 

Note that Higgs doublets and lepton doublets cannot be distinguished in a supersymmetric theory,
therefore both classes of particles 
are denoted by $\ell$. That implies particularly that the down--type Higgs $h_d$ is given by a linear
combination of some $\ell_i$, whereas the up--type Higgs $h_u$ (required in supersymmetric theories) is a
superposition of $\bar\ell_i$.

\begin{table}[!t!]
\begin{minipage}{0.7\textwidth}
\begin{tabular}{|rlc||rlc|}
\hline
  \#  &  {\small Representation}  & {\footnotesize Label}     &  \#  & {\small (Anti-)Repr.} & {\footnotesize Label} \\
\hline\hline
\rowgrayh
3 & $( {\bf 3},  {\bf 2};  {\bf 1},  {\bf 1})_{1/6}$  & $q_i$       &    &   & \\
8 & $(  {\bf 1},  {\bf 2};  {\bf 1},  {\bf 1})_{-1/2}$ & $\ell_i$   &  5 & $(  {\bf 1},  {\bf 2};  {\bf 1},  {\bf 1})_{1/2}$  &  $\bar{\ell}_i$\\
\rowgrayh
3 & $( {\bf 1},  {\bf 1};  {\bf 1},  {\bf 1})_{1}$    & $\bar{e}_i$ &    &   & \\
3 & $( \bsb{3},   {\bf 1};  {\bf 1},  {\bf 1})_{-2/3}$ & $\bar{u}_i$ &    &   & \\
\rowgrayh
7 & $( \bsb{3},   {\bf 1};  {\bf 1},  {\bf 1})_{1/3}$  & $\bar{d}_i$ &  4 & $(  {\bf 3},  {\bf 1};  {\bf 1},  {\bf 1})_{-1/3}$ & $d_i$  \\
\hline
4 & $( {\bf 3},  {\bf 1};  {\bf 1},  {\bf 1})_{1/6}$  & $v_i$       &  4 & $(  {\bsb{3}}, {\bf 1};  {\bf 1},  {\bf 1})_{-1/6}$ & $\bar{v}_i$ \\
\rowgrayh
20& $( {\bf 1},  {\bf 1};  {\bf 1},  {\bf 1})_{1/2}$  & $s^+_i$     & 20 & $(  {\bf 1},  {\bf 1};  {\bf 1},  {\bf 1})_{-1/2}$ & $s^-_i$   \\
2 & $( {\bf 1},  {\bf 1};  {\bf 1},  {\bf 2})_{1/2}$  & $x^+_i$     &  2 & $(  {\bf 1},  {\bf 1};  {\bf 1},  {\bf 2})_{-1/2}$  & $x^-_i$ \\
\hline
\end{tabular}
\end{minipage}
\hskip -5mm
\begin{minipage}{0.25\textwidth}
\begin{tabular}{|rlc|}
\hline
  \#  &  {\small Representation}           & {\footnotesize Label}   \\
\hline\hline
\rowgrayh
 4 & $(  {\bf 1},  {\bf 2};  {\bf 1},  {\bf 1})_{0}$ & $m_i$\\
 2 & $(  {\bf 1},  {\bf 2};  {\bf 1},  {\bf 2})_{0}$ & $y_i$\\
\rowgrayh
47 & $(  {\bf 1},  {\bf 1};  {\bf 1},  {\bf 1})_{0}$ & $s_i^0$\\
26 & $(  {\bf 1},  {\bf 1};  {\bf 1},  {\bf 2})_{0}$ & $h_i$\\
\rowgrayh
 9 & $(  {\bf 1},  {\bf 1};  {\bf 8},  {\bf 1})_{0}$ & $w_i$\\
\hline
\end{tabular}
\vskip 1.5cm\phantom{.}
\end{minipage}
\caption{Massless spectrum. The quantum numbers are shown with respect to
$\SU3_c\x \SU2_L \times \SO8 \times \SU2$, the hypercharge is
given by the subscript.}
\label{tab:spectrum}
\end{table}

\subsection{Renormalizable Couplings, ``Heavy Top'' and Proton Decay}
At trilinear level, string selection rules (see section~\ref{subsec:SelectionRulesZ6II}) allow only
few couplings of \sm\ particles. There is one coupling of the type $q_i\,\bar{\ell}_j\,\bar{u}_k$ with
only untwisted fields, which we consider to be the Yukawa coupling that gives mass to the
top--quark. Notice that this coupling allows us to identify the right--handed top, the up--type Higgs
doublet $h_u$ and the quark doublet of the third generation. 

There are also four couplings of the type $q_i\, \ell_j\, \bar{d}_k$ and four of the type $\bar{e}_i\,
\ell_j\, \ell_k$. They can produce the down--type quark and lepton masses as well as lepton number
violating interactions. Clearly, depending on the specifics of the vacuum configuration and,
particularly, on the choice of the down--type Higgs $h_d$, it can be that lepton violating operators do not
appear in this model.

\begin{figure}[t]
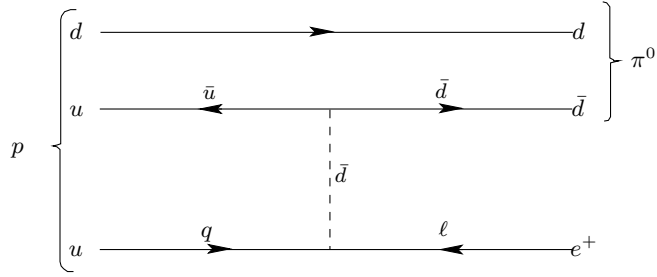

\centerline{\input R-parity_violating_decay.pstex_t}
\caption{Dangerous contribution to proton decay through exchange of the scalar component of $\bar d$ due
  to dimension four operators of the type $q_i\, \ell_j\, \bar{d}_k$ and $\bar{u}_i \bar{d}_j\bar{d}_k$.
  The proton decays rapidly as $p\rightarrow\pi^0 e^+$.}
\label{fig:4DProtonDecay}
\end{figure}

Further, we note that lepton number violating interactions $q_i\, \ell_j\, \bar{d}_k$ are harmless to the
stability of the proton as long as they do not come accompanied by quark interactions, $\bar{u}_i
\bar{d}_j\bar{d}_k$ (see fig.~\ref{fig:4DProtonDecay}).
Thus, in this model, due to the absence of $\bar{u}_i \bar{d}_j\bar{d}_k$ operators, the proton is stable at
trilinear level, so that dangerous effective dimension four operators can appear only suppressed by
different powers of $M_\mathrm{str}$.

\subsection{Spontaneous Symmetry Breaking and Decoupling of Exotics}

Let us suppose that all \sm\ singlets $s_i^0$ develop nonzero {\vev}s. Since they are only charged under
some hidden \U1's, the gauge group gets spontaneously broken to
\begin{equation}
G_{SM} \times G_\mathrm{hidden} \;,
\end{equation}
where $G_\mathrm{hidden}=\SO8\x\SU2$. Note that if more than one fields $h_i$ were allowed to develop {\vev}s, the
unbroken hidden gauge group would certainly be $G_\mathrm{hidden}=\SO8$.

Additionally, the vectorlike states get large masses. In order to verify this, one has to obtain the mass
matrices $\calM_{x\bar x}$ of the exotics and assume that the {\vev}s of the singlets $s_i^0$ are of
the same order $\langle s_i^0\rangle\approx s\approx M_\mathrm{str}$. Replacing the singlets by their {\vev}s,
one can then compute the rank of the effective mass matrices. A mass matrix with maximal rank ensures that
all exotics of the corresponding class obtain masses proportional to the {\vev}s $s$.
We make at this point a strong assumption, namely that all singlets acquire {\vev}s consistent with
supersymmetry, that is, along $D$-- and $F$--flat directions. This assumption will be confirmed later.

We have checked that the rank of all the mass matrices is maximal, such that the exotics do
decouple from the effective low-energy theory. Below we present most of the matrices. There, $s^n$ indicates the dominant mass term for each entry,
i.e. $s^n$ denotes that the corresponding coupling appears first when $n$ singlets are involved. 
Each entry usually contains many terms and involves different singlets as well as coupling strengths,
which are presumed to be of order 1 in string units.\\
\begin{minipage}{0.45\textwidth}
\bea
\mathcal{M}_{m  m}&=& \left(
\begin{array}{cccc}
0   & s^5 & s^6 & s^6\\
s^5 & 0   & s^6 & s^6\\
s^6 & s^6 & 0 & s^5\\
s^6 & s^6 & s^5 & 0\\
\end{array}
\right)\,,\nonumber\\
\mathcal{M}_{v \bar v}&=&
\left(
\begin{array}{cccc}
s^5 & s^5 & s^5 & s^5\\
s^5 & s^5 & s^5 & s^5\\
s^6 & s^6 & s^1 & s^5\\
s^6 & s^6 & s^5 & s^1\\
\end{array}
\right)\,,\nonumber\\
\mathcal{M}_{x^+x^-} &=&
\left(
\begin{array}{cc}
s^5  & s^5 \\
s^5  & s^5 \\
\end{array}
\right)\,,\nonumber\\
\mathcal{M}_{yy} &=&
\left(
\begin{array}{cc}
s^1  & s^5 \\
s^5  & s^1 \\
\end{array}
\right)\,,\nonumber
\eea
\end{minipage}
\begin{minipage}{0.48\textwidth}
\bea
 \mathcal{M}_{d \bar d} &=&\left(
\begin{array}{ccccccc}
s^6 & s^6 & s^3 & s^6 & s^6 & s^1 & s^1\\
s^6 & s^6 & s^3 & s^6 & s^6 & s^1 & s^1\\
s^3 & 0 & 0 & s^3 & 0 & s^6 & s^6\\
s^6 & s^3 & 0 & s^6 & s^3 & s^6 & s^6\\
\end{array}
\right)\,,\nonumber\\
\mathcal{M}_{\ell \bar\ell} &=&\left(
\begin{array}{ccccc}
s^3 & s^1 & s^1 & s^1 & s^1\\
s^1 & s^3 & s^3 & s^3 & s^3\\
s^1 & s^3 & s^3 & s^3 & s^3\\
s^1 & s^3 & s^3 & s^3 & s^3\\
s^1 & s^3 & s^3 & s^3 & s^3\\
s^1 & s^3 & s^6 & s^6 & s^3\\
s^4 & s^2 & s^6 & s^2 & s^2\\
s^4 & s^2 & s^6 & s^2 & s^2\\
\end{array}
\right)\,.\nonumber
\eea
\end{minipage}\vskip 2mm

Similarly, the $20\x20$ mass matrix $\calM_{s^+s^-}$ has also maximal rank. The $d\bar d$ mass matrix
is $4\x7$ such that there are three massless $\bar d$ states. The $\ell \bar\ell$ mass matrix is $8\x 5$,
so there are effectively three massless lepton doublets. 

As we already mentioned, without additional information, it is not possible to distinguish between lepton
doublets and Higgs doublets. Therefore, since the mass matrix $\calM_{\ell \bar\ell}$ has maximal rank,
all possible Higgs pairs are massive. If one wishes to recover the Higgs mechanism, it is necessary to choose
a special vacuum configuration where the rank of the matrix $\calM_{\ell \bar \ell}$ is reduced. This
corresponds precisely to the supersymmetric $\mu$--problem, that in a first approach is not automatically
solved. Note that, since the matrix $\calM_{\ell \bar \ell}$ is not diagonal {\it per se}, the resulting
Higgs doublets will be indeed linear combinations of the fields $\ell_i$ and $\bar\ell_i$. 

Summing up all the ingredients just described, we end up with the exact \mssm\ spectrum.

Let us stress that these properties, although particular to this orbifold-\mssm, are quite similar to
those of other models. That is, all 93 \mssm\ candidates with ``heavy top'' accept this
description. Perhaps one of the most relevant differences will be the number of lepton doublets $\ell$
and $\bar\ell$ that can be interpreted as Higgs doublets. In some cases, there are only four \susy\ fields
$\ell_i$ and one $\bar\ell_i$, which may be seen as three lepton doublets and one pair of Higgses.
This can lead to considering a particular model more predictive than another. 

Let us stop our discussion on this model here and proceed to explain how to guarantee \susy\ at high
energies in these models. This model will appear regularly in the following sections in order to
exemplify some of the new elements introduced.

\section{Supersymmetric Vacua}
\label{sec:SUSYVacua}

In the previous section, we have assumed that all \sm\ singlets can acquire nonvanishing {\vev}s without
destroying some of the properties of our orbifold-\mssm. However, this is in general not true. In this
section we discuss the constraints to get supersymmetric vacuum configurations, their consequences and a
method to obtain such vacua. 

Momentarily, let us assume only global supersymmetry.
We would like to verify whether the \mssm\ candidates admit supersymmetric vacuum configurations.
This amounts to inspecting whether a combination of fields can attain {\vev}s, such that the  $D$-- and
$F$--terms vanish. This is explained as follows.
The scalar potential in a supersymmetric gauge theory is given, in terms of the auxiliary fields $D_a$
and $F_i$ of the gauge and chiral multiplets, by
\be\label{eq:ScalarPotential}\index{scalar potential}
V(\phi_i,\phi_i^\dagger)=\2\sum_a D_a(\phi_i,\phi_i^\dagger)^2 + \sum_i|F_i|^2\,,
\ee
where the auxiliary fields are expressed in terms of the fields $\phi_i,\phi_i^*$ and the gauge group
generators $\mathsf{t}_a$ in the representation of $\phi_i$. The auxiliary fields are given
by\footnote{Formally $-F_i^\dagger=\depa{W}{\phi_i}$, but, as we shall demand
 $\vev{F_i}=0$, it is equivalent to use this definition.} 
\bea\index{D-term}\index{F-term}\index{auxiliary fields, D- and F-}
D_a&=& \sum_i \phi_i^\dagger\mathsf{t}_a \phi_i\,,\label{eq:AuxFieldsD}\\
F_i&=& \depa{W}{\phi_i}\label{eq:AuxFieldsF}
\eea
with $W=W(\phi_i)$ being the superpotential. From a field theoretical perspective,
solutions of eqs.~\eqref{eq:AuxFieldsD} and~\eqref{eq:AuxFieldsF} represent supersymmetric vacua.
However, string theory introduces a new element, namely an anomalous \U1 symmetry, $\U1_A$ (such that
$\tr\mathsf{t}_A \neq 0$). In presence of $\U1_A$, the corresponding $D$--term analogous to
eq.~\eqref{eq:AuxFieldsD} gets an additional contribution\footnote{To maximize confusion,
normally $\xi$ is called FI term. Note that $D_A$ is called here FI $D$--term.}
$\xi$~\cite{Atick:1987gy}, giving rise to the so-called FI $D$--term 
\be\label{eq:AuxFieldsFID}\index{D-term!Fayet-Iliopoulos term}
D_A = \sum_i \phi_i^\dagger\mathsf{t}_A \phi_i + \xi \equiv \sum_i q^A_i|\phi_i|^2 + \h{g\
  M_\mathrm{Pl}^2}{192\pi^2}\tr\mathsf{t}_A\,, 
\ee
where $\mathsf{t}_A$ is the generator of $\U1_A$ and $g$ is the four-dimensional coupling constant,
given by the \vev\ of the four-dimensional dilaton $\varphi$ as $g=e^{\vevof{\varphi}}$. In our
conventions, $\tr\mathsf{t}_A>0$ and thus $\xi>0$. Note that the orbifold point vacuum, i.e. $\vevof{\phi_i}=0$,
is not supersymmetric as (for $g\neq0$) the extra term $\xi$
induces \susy\ breakdown at a scale too large to be realistic.\index{coupling constant! string coupling}

Supersymmetry remains unbroken if and only if the scalar potential admits a minimum $\phi_i=\vevof{\phi_i}$ such that
$V(\vevof{\phi_i},\vevof{\phi_i^\dagger})=0$. Therefore, any supersymmetric ground state has to satisfy
the $D$-- and $F$--flatness conditions
\bse\label{eq:susyVacuumCond}
\bea\index{susy constraints@\susy\ constraints}
\vevof{D_a} &=& 0\qquad\forall\, a\,,\label{eq:susyVacuumCondD}\\
\vevof{F_i} &=& 0\qquad\forall\, i\,.\label{eq:susyVacuumCondF}
\eea\ese
Note that there are as many $F_i$--constraints as fields $\phi_i$.

Provided a set of fields $\phi_i$, supersymmetric field configurations are given by the sets of {\vev}s
$\vevof{\phi_i}$ satisfying eqs.~\eqref{eq:susyVacuumCond}. Na\"ively, it appears  that the number of
constraints is larger than the number of variables, so that the system seems to be overconstrained. 
However, this is  not the case. As well known, complexified gauge transformations along ($F$--)flat
directions allow us to eliminate the $D$--term constraints\footnote{For a field-theoretic discussion, see
e.g. ref.~\cite[p.57-58]{Wess:1992cp}. Details on heterotic orbifolds can be found in
ref.~\cite{Buchmuller:2006ik}}, such that the number of variables equals the number of equations. 

\subsection[$D$--Flatness]{$\bs{D}$--Flatness}
\label{subsec:DFlat}\index{D-flatness}

Let us review first the issue of $D$--flatness and cancellation of the FI
term~\cite{Buccella:1982nx,Gatto:1986bt,Font:1988tp,Dine:1987xk,Cleaver:1997jb}.
Since, in particular, $D_A$ must vanish in a supersymmetric vacuum, at least some of the scalars are
forced to attain large {\vev}s, typically not far below the string scale. This sets the scale of the
breaking of $\U1_A$ and other gauge symmetries, under which the scalars might be charged. In other words,
vanishing of the FI $D$--term triggers spontaneous symmetry breaking at a very high scale. On the other
hand, the condition $D_A=0$ also implies that there must be at least one field whose anomalous charge
$q^A$ is (in our convention) negative (i.e. opposite in sign to $\tr\mathsf{t}_A$).

\index{D-flatness!invariant monomials (HIM)}
$D$--flatness can be achieved by the noticeable observation that, in general, every holomorphic gauge invariant
monomial (HIM) represents a $D$--flat direction~\cite{Buccella:1982nx}.
Particularly, in theories without an anomalous \U1, the condition~\eqref{eq:susyVacuumCondD} is satisfied if there
exists a monomial $I(\phi_i)$. Then, the {\vev}s
of the scalar fields in a $D$--flat vacuum configuration are read off from
\be\label{eq:BuccellaCondI}
\lvevof{\depa{I}{\phi_i}}=c\vevof{\phi_i^\dagger},
\ee
where $c$ is a complex dimensional constant, $c\neq0$. This remarkable result follows simply from the
gauge invariance of $I(\phi_i)$~\cite{Buccella:1982nx}. In the case that $\U1_A$ is present, the monomial
$I(\phi_i)$ must be invariant with respect to all gauge symmetries except $\U1_A$. The anomalous charge
of this monomial must be negative~\cite{Cleaver:1997jb}.
Let us explain this in  more detail and propose a method to find such HIMs.

In supersymmetric theories with a single \U1 gauge theory, the so-called $D$--term potential is given by
\begin{equation}\label{eq:VD}
 V_D~\propto~\left[\sum_i q_i\,|\phi_i|^2\right]^2\;.
\end{equation}
Consider as a first example a \U1 gauge theory with two fields $\phi_\pm$ carrying the charges
$\pm1$. Since $V_D^{1/2}\propto|\phi_+|^2-|\phi_-|^2$, $V_D$ vanishes at the ground state
$|\vevof{\phi_+}|=|\vevof{\phi_-}|$. That is, one has a $D$--flat direction, parametrized by
$c=|\vevof{\phi_+}|=|\vevof{\phi_-}|$. Note that, if one writes the HIM 
$I(\phi_\pm)=\phi_+\phi_-$, which is the only possible holomorphic and invariant monomial,
eq.~\eqref{eq:BuccellaCondI} yields the same result. 

Consider now a theory with one field ($\phi_1$) with charge $2$ and two fields ($\phi_2,\phi_3$) with
charges $-1$.  Then we have many flat directions, described by the roots of the equation $2 |\vevof{\phi_1}|^2 -
|\vevof{\phi_2}|^2 - |\vevof{\phi_3}|^2 = 0$. The solutions are the three directions
\bse
\bea
&|\vevof{\phi_1}|=|\vevof{\phi_2}|=|\vevof{\phi_3}| = c\,;& \\
&|\vevof{\phi_2}|=0\,,\,|\vevof{\phi_3}|=\sqrt{2}|\vevof{\phi_1}| = c'\,;&\\
&|\vevof{\phi_3}|=0\,,\,|\vevof{\phi_2}|=\sqrt{2}|\vevof{\phi_1}| = c''&
\eea
\ese
with complex parameters $c,c'$ and $c''$.
It is convenient to associate these directions to the HIMs
\be
\phi_1\,\phi_2\,\phi_3\;,\quad \phi_1\,\phi_3^2\;,\quad \phi_1\,\phi_2^2\;,
\ee
respectively.

From this example, we notice that a monomial $I(\phi_i)=\phi_{1}^{n_1}\,\phi_{2}^{n_2}\cdots\phi_k^{n_k}$
represents a flat direction defined by the relation 
\be\index{D-flat direction}
 \frac{|\vevof{\phi_1}|}{\sqrt{n_1}}~=~\frac{|\vevof{\phi_2}|}{\sqrt{n_2}}~=~\dots
 ~=~\frac{|\vevof{\phi_k}|}{\sqrt{n_k}} \quad\text{and}\quad|\vevof{\phi_j}|~=~0~~\text{for}~~n_j~=~0\;.
\ee
It is, however, clear that there is only a finite number of linearly independent $D$--flat directions. In
the previous example, the third direction is not independent of the other two. In other words, the
requirement $V_D=0$ poses only one constraint on the three real variables ($|\vevof{\phi_i}|^2$) entering
eq.~\eqref{eq:VD}. The space of absolute values $|\vevof{\phi_i}|$ is two-dimensional. The power of using the
monomials is that checking whether certain monomials are linearly independent or not is fairly simple. One
identifies with each monomial the vector of exponents, $n=(n_1,n_2\dots )$. The flat directions are
independent if and only if the vectors are linearly independent. In the previous example one would get
the vectors $(1,1,1)$, $(1,0,2)$, and $(1,2,0)$, out of which only two are linearly independent.

It is also clear how to obtain these vectors: all of them are orthogonal to the vector of charges
$q=(q_1, q_2,\dots)$. That is, the problem of finding the above monomials (and thus the $D$--flat
directions) is reduced to the problem of finding vectors $n$ with the following properties: 
\begin{enumerate}
 \item $q\cdot n=0$\,,
 \item $n_i\in \mathbbm{N}_0$\,.
\end{enumerate}
The property that the $n_i$ be integer-valued does not pose a constraint in our models: since the charges
are rational, one can rescale any $n$ having the first property such as to have integer entries. However,
the requirement that the entries be nonnegative, which reflects that the monomials ought to be
holomorphic, is a constraint.

The discussion so far can easily be extended to $\U1^m$ theories. Here the $D$--term potential is
\begin{equation}
 V_D~\propto~\sum\limits_{j=1}^m\left[\sum_i q_i^{(j)}\,|\phi_i|^2\right]^2\;,
\end{equation}
where $q_i^{(j)}$ is the  charge of the field $\phi_i$ under the $j^\mathrm{th}$ \U1 factor. Now a $D$--flat
direction has to satisfy the above constraints for each \U1 factor separately. Again, it is advantageous
to represent $D$--flat directions by holomorphic gauge invariant monomials. Then the vector $n$ of
exponents has to be orthogonal to every charge vector $q^{(j)}=(q_1^{(j)},q_2^{(j)},\dots)$. In other
words, $n$ has to be in the kernel of the charge matrix $Q$, 
\begin{equation}
 Q\cdot n~=~0\;,\quad\text{with}\quad
 Q~=~\left(\begin{array}{ccc}
 q_1^{(1)} & q_2^{(1)} & \dots\\
 q_1^{(2)} & q_2^{(2)} & \dots\\
 \vdots & \vdots & \vdots\\
 q_1^{(n)} & q_2^{(n)} & \dots\\
 \end{array}\right)\;.
\end{equation}
Hence, the problem of finding the $D$--flat directions of a $\U1^m$ gauge theory is reduced to the task
of calculating the kernel of the charge matrix $Q$, and to forming linear combinations of elements of
this kernel in such a way that the entries are nonnegative integers. The maximal linear independent set
of such linear combinations is in one-to-one correspondence with the independent $D$--flat directions.
It is straightforward to see that the results obtained so far generalize to the nonabelian case~\cite{Buccella:1982nx}.

We can now review the issue of cancelling the FI term. For an anomalous \U1, the $D$--term
potential $V_D$ follows from the FI $D$--term, eq.~\eqref{eq:AuxFieldsFID},
\begin{equation}
 V_D^A ~\propto~ \left[\sum\limits q_i^A\,|\phi_i|^2+\xi\right]^2\;.
\end{equation}
Recalling that $\xi>0$ in our conventions, to cancel the FI term one thus has to find a holomorphic monomial,
\begin{equation}
 I~=~\phi_1^{n_1}\,s_2^{n_2}\,\dots
\end{equation}
with net negative charge under $\U1_A$, i.e.\
\begin{equation}
 \sum_i n_i\,q_i^A~<~0\;.
\end{equation}

To summarize, the $D$--flat directions are in one-to-one correspondence with holomorphic gauge invariant
monomials. In the abelian case, such monomials can be identified with elements of the kernel of the
charge matrix $Q$ with nonnegative integer entries. Cancellation of the FI term requires the existence of
a holomorphic monomial with net negative charge under $\U1_A$, which is gauge invariant with respect to
all other group factors. 

\subsection[$F$--Flatness]{$\bs{F}$--Flatness}
\label{subsec:FFlat}\index{F-flatness}

Let us now turn to the discussion of $F$--flatness. Since the superpotential $W$ is nonrenormalizable,
studying this question in detail is somewhat less general than $D$--flatness because most of the
statements one can obtain are order-dependent.

However, ensuring that a vacuum configuration is $F$--flat (ignoring momentarily $D$--flatness) is mostly
trivial because the number of equations 
\be\label{eq:FFlatRep}
\vevof{F_i}=\lvevof{\depa{W}{\phi_i}}=0\qquad\forall\,i
\ee
coincides with the number of variables and, in general, when the superpotential is a nontrivial
polynomial, some of their solutions are nontrivial.  

In particular, nontrivial solutions can always be found in orbifold compactifications. Let us illustrate
it in the case of \Z{N} orbifolds. If the superpotential $W_0$ at order $x$ in the fields $\phi_i$ is
allowed by string selection rules, then an `extended' superpotential $W\sim W_0 + W_0^{N+1} + W_0^{2N+1}
+ \ldots$ is also admissible. 

As an example, assume a \Z3 orbifold toy-model with two particles, $\phi_1$ and $\phi_2$. Suppose further that
the superpotential\footnote{Note that we omit all coefficients.} $W_0=\phi_1 \phi_2^2$ is
allowed. Clearly, the solution to eq.~\eqref{eq:FFlatRep} is given by $\vevof{\phi_2}=0$ and $\vevof{\phi_1}$
arbitrary. Suppose now that $\vevof{\phi_2}=0$ is unwanted for some reason. An allowed extension of $W_0$
would be given by $W\sim W_0+W_0^4=\phi_1 \phi_2^2+ \phi_1^4 \phi_2^8$. In this scenario, the previous
solution is still a valid option, but there is also an additional nontrivial $F$--flat vacuum
configuration, parametrized by $\vevof{\phi_2^6}=-1/4\,\vevof{\phi_1^3}$ and arbitrary
$\vevof{\phi_1^3}\neq0$. Notice that $\vevof{\phi_2}$ in this vacuum configuration is nontrivial.
This discussion makes manifest the order-dependence of $F$--flatness. In this sense, verifying
$F$--flatness at a given order in the superpotential is physically not so relevant in orbifold models.

After arriving to an $F$--flat vacuum configuration, the natural question is whether recovering
$D$--flatness imposes additional conditions and, therefore, overconstrains the choice of the vacuum
parameters. It has been shown~\cite{Buchmuller:2006ik} that in orbifold models, given a solution to the
$F$--term equations~\eqref{eq:FFlatRep}, complexified gauge transformations scale this solution to give a
family of solutions. Remarkably, particular rescalings succeed in rendering all the $D$--terms (including
the FI $D$--term) zero.

\subsection*{$\bs{\vevof{W}=0}$}

Up to now, our considerations are valid only for globally supersymmetric models at perturbative level. An
admissible vacuum configuration arising from string theory, however, must also be consistent with local
\susy. In fact, supergravity as well as nonperturbative effects can modify the properties of a particular
vacuum configuration. 

In an attempt to deal with those additional effects, the condition $\vevof{W}=0$ is commonly imposed in
the same footing as $F$--flatness (see e.g. refs.~\cite{Casas:1988se,Cleaver:1997jb}). Strictly speaking,
this condition has nothing to do with preservation of \susy. Since in supergravity the gravitino mass is
given by 
\be\label{eq:m32W}
m_{3/2}\propto |\vevof{W}|^2\,,
\ee
$\vevof{W}=0$ implies $m_{3/2}=0$ and therefore, through the relation $\Lambda=-3\,M_\mathrm{Pl}^2\,
m_{3/2}^2$, a vanishing cosmological constant $\Lambda$. Recall that in the context of supergravity, a
nonvanishing gravitino mass does not necessarily imply breakdown of supergravity. That is, imposing the
condition $\vevof{W}=0$ amounts to requiring the existence of a Minkowski vacuum (similar to the
vacuum we are living in) rather than unbroken supersymmetry. 

A caveat is in order. To compute e.g. the gravitino mass, eq.~\eqref{eq:m32W}, and thus to
discriminate a vacuum according to its nature (Minkowski, De Sitter or anti-De Sitter), it is fundamental
to consider the complete superpotential $W$, including, in particular, contributions due to
nonperturbative effects. However, the superpotential considered in previous sections does not include
those contributions because, at this level, they are hardly controllable in orbifold constructions. 
Therefore, requiring (the incomplete) $\vevof{W}$ to vanish does not affect directly the {\vev}s of the fields $\phi_i$,
contrary to the $D$-- and $F$--flatness conditions that we discussed above.

\subsection[A \susy\ Vacuum of the Orbifold-\mssm]{A \textbf{\textsc{susy}} Vacuum of the Orbifold-\textbf{\textsc{mssm}}}
\label{eq:NiceVacuumSeesaw}\index{orbifold-mssm@orbifold-\mssm}

We are now in position to verify whether the orbifold-\mssm\ introduced in section~\ref{sec:SeesawModel}
possesses a supersymmetric vacuum. By following the method described above, we find the HIM
\begin{equation}\label{eq:AnMSSMMonomial}
\mbox{\footnotesize
$I ~=~  s^0_{1}\, s^0_{2}\, \left(s^0_{3}\right)^3 \,\left(s^0_{5}\right)^3\,\left(s^0_{8}\right)\,
\left(s^0_{22}\right)\,\left(s^0_{35}\right)^2\, \left(s^0_{41}\right)^3\, \left(s^0_{43}\right)^4 \,\left(s^0_{46}\right)^3 \,
 h_{2}^4\, h_{3}\, h_{5}^5\, h_{9}^2 \, h_{13}^2\, h_{14}^2\, h_{20}\, h_{21}^3 \,h_{22}^6$}
\end{equation}
with net anomalous charge $\sum_iq^A_i=-52/3$. We further identify that some fields share the same gauge quantum numbers:
\be
s^0_5\leftrightarrow s^0_{12}\leftrightarrow s^0_{9}\leftrightarrow s^0_{16}\,,\qquad s^0_8\leftrightarrow
s^0_{15}\,,\qquad s^0_{22}\leftrightarrow s^0_{24}\,. 
\ee
Therefore, we can consider a vacuum configuration where the \sm\ singlets\footnote{In
table~\ref{tab:SpectrumSeesawModel} the two \sm\ singlets $s^0_{1,\,2}$ have been denoted by
$\chi_{1,\,2}$. The reason will become transparent when we introduce \BmL in section~\ref{sec:RParity}.} 
\begin{equation}\label{eq:stilde}
\mbox{\footnotesize $\{\widetilde{s}_i\}=\{s^0_1, s^0_2, s^0_{3}, s^0_{5},
  s^0_8, s^0_9, s^0_{12}, s^0_{15}, s^0_{16}, s^0_{22}, s^0_{24}, s^0_{35}, s^0_{41}, s^0_{43}, s^0_{46},
  h_{2}, h_3, h_{5}, h_{9}, h_{13}, h_{14}, h_{20}, h_{21}, h_{22}\}$}  
\end{equation}
develop nonzero {\vev}s while the expectation values of all other fields vanish. Fields $s^0_i$ are singlets under
all nonabelian gauge factors and hypercharge, but carry (hidden) \U1 charges; fields $h_i$ are doublets
under the hidden \SU2 and charged under the \U1's excepting hypercharge (cf. tables~\ref{tab:spectrum}
and~\ref{tab:SpectrumSeesawModel}). 

Let us assume that all particles attain (almost) the same \vev\ $\vevof{\ti s_i}\approx s$.
From eq.~\eqref{eq:AuxFieldsFID} with $\tr\mathsf{t}_A=170/3$ for the orbifold \mssm, we see that the FI
$D$--term is given by
\be
\vevof{D_A}\approx-\h{52}{3}s^2 +\h{g\,M_\mathrm{Pl}^2}{192\pi^2}\h{170}{3}\,.
\ee
Therefore, in order for the chosen vacuum to be $D$--flat, the expectation values of the \sm\ singlets
should be $s \approx \sqrt{g}\,M_\mathrm{Pl}\x10^{-2}\approx \text{(few)}\x 10^{16}$ GeV.

In the vacuum configuration~\eqref{eq:stilde}, the gauge symmetry group $\maG_{4D}$ given in
eq.~\eqref{eq:Seesaw4DGG} breaks spontaneously, as expected, to 
\begin{equation}
 G_{SM}\times G_\mathrm{hidden}\;,
\end{equation}
where $G_\mathrm{hidden}=\SO8$. Notice that the rank of $G_\mathrm{hidden}$ is four. This will turn out
to be very general in models with realistic features and will lead in the next section to the conjecture
that \susy\ breaking through gaugino condensation occurs generally at an intermediate energy scale for
realistic heterotic orbifold models.

For completeness, let us address $F$--flatness in the orbifold-\mssm.
To verify whether this vacuum is also $F$--flat, we would need the complete superpotential $W$ (i.e. up to 
arbitrary order in \sm\ singlets). Realistically, we have to stop computing the superpotential at a given
order. We consider here, for simplicity, the order-six superpotential
\bea
W &=& \mbox{\small $s^0_{32} h_{5} (s^0_{5} h_{1} + s^0_{12} h_{2} ) 
         +(s^0_{15} h_{15} + s^0_{8} h_{13})(s^0_{42}+ s^0_{43})( h_{23} + h_{25})$}\nonumber\\
 &&\mbox{\small $ +\ (s^0_{22} h_{14} + s^0_{24} h_{16})(s^0_{42}+ s^0_{43})( h_{18} + h_{20})$}\\
 &&\mbox{\small $ +\ h_{22} (s^0_{5}  h_{3} + s^0_{12} h_{4})\Big( s^0_{41} (s^0_{26} + s^0_{28}) 
         + s^0_{32} (s^0_{42} + s^0_{43}) +  s^0_{35}( s^0_{45} +  s^0_{46})\Big)$\,.}\nonumber
\eea
The resulting $F$--terms are provided in eqs.~\eqref{eq:FTermsSeesaw}. We observe that with nonvanishing
{\vev}s for the fields~\eqref{eq:stilde}, there are still some nonzero $F$--terms, implying that some of
those fields should have trivial {\vev}s. However, as we
have mentioned before, one can argue that, if we go to higher orders in $W$, nontrivial solutions to
all $F_i=0$ equations can be found. As a matter of fact, we find that if we go to order eight in the
superpotential, such nontrivial solutions exist.

\section[Supersymmetry Breakdown]{Supersymmetry Breakdown}
\label{sec:SUSYBreaking}\index{supersymmetry breaking}

In the previous section, we have found that supersymmetric vacua can be achieved in \mssm\
candidates (at least at perturbative-level). However, as \susy\ is broken in nature, realistic models
should admit spontaneous \susy\ breaking at an intermediate scale. Remarkably, it is known that in most $\maN=1$ vacua
\susy\ is broken spontaneously by nonperturbative effects. Our understanding of nonperturbative breaking
of string theory is as yet very limited, but below the string scale we can work in the effective quantum
field theory. Indeed, there is a reasonably coherent understanding of nonperturbative breaking of \susy\
in field theory, and the low--energy theories emerging, in particular, from heterotic orbifolds are
typically of the type in which this breaking happens. This topic is quite involved as there are several
symmetry-breaking mechanisms, such as gaugino condensation, instantons or composite goldstinos.

In orbifold models with realistic features, there are frequently additional (nonabelian) gauge
symmetries that remain unbroken even after all exotics have acquired large masses. This hidden
sector\footnote{Notice that a requirement to call it ``hidden sector'' is that the observable matter
 (e.g. the \mssm\ particle spectrum) be uncharged under $G_\mathrm{hidden}$.}
$G_\mathrm{hidden}$ usually contains little or no matter at all. These are precisely the ingredients 
that can trigger spontaneous \susy\  breaking via hidden sector gaugino
condensation~\cite{Nilles:1982ik,Ferrara:1982qs,Derendinger:1985kk,Dine:1985rz}.

\index{gaugino condensation}
Gaugino condensation occurs when one or more gauge groups in the hidden sector become strongly coupled at an
intermediate scale $\Lambda$. To determine the exact scale in which it happens, we need to know the
running of the coupling of the hidden sector. It is given by
\be\label{eq:RunningHidden}\index{hidden sector, RGE}
g^2_{G_\mathrm{hidden}}(\mu)\approx \h{1}{g^{-2}(M_\mathrm{GUT})-
 \beta_\mathrm{hidden}\,\ln(M_\mathrm{GUT}^2/\mu^2)}\,,\qquad \beta_\mathrm{hidden}=\h{b_0}{16\pi^2}\,,
\ee
where $g$ is the four-dimensional (string) coupling constant given by $g^{-2}=e^{-2\vevof{\phi}}=\re\vevof{S}$,
and $b_0$ is the well known beta-coefficient which depends on the gauge group $G_\mathrm{hidden}$ and
the (hidden) matter content. The Landau pole of the hidden sector is determined by
$g^2_{G_\mathrm{hidden}}\rightarrow \infty$. Hence, we obtain 
\begin{equation}\label{eq:ScaleGC}
 \Lambda\ \approx\ M_\mathrm{GUT}\,\exp\left\{-\frac{1}{2\beta}\frac{1}{g^2(M_\mathrm{GUT})}\right\}\,,
\end{equation}
where we have omitted the label $\phantom{.}_\mathrm{hidden}$ to keep the results short. Note that the
scale $\Lambda$ is below the string (and \gut) scale, but above the scale where any of the observable
gauge groups become strong. Just as with quarks in {\sc qcd}, the strong attraction causes the gauginos
to condense~\cite{Nilles:1982ik}, 
\be\index{condensation scale}
\vevof{\lambda\lambda} \approx \Lambda^3\,.
\ee
As in {\sc qcd}, this condensate breaks a chiral symmetry, but in the pure supersymmetric gauge theory
(containing only gauge boson and gauginos) it does not break \susy.

In string theory, the fields of the hidden sector couple to one special moduli, namely the dilaton
$S$. In particular, the auxiliary field of $S$ couples to the (hidden) gauginos inducing, at scales below
$\Lambda$, a nonperturbative effective interaction (compare with eq.~\eqref{eq:ScaleGC})
\be\label{eq:FSGC}
F_S\,\vevof{\lambda\bar\lambda}\approx F_S\,\Lambda^3\approx F_S \, M_\mathrm{GUT}^3\,e^{-\frac{3}{2\beta}S}\,.
\ee
It is usual to define the parameter $a\equiv 3/2\beta$ that, in a way, can be used to determine whether
\susy\ breaking occurs at a realistic energy scale (see e.g. ref.~\cite{Casas:1996zi}).

The existence of such coupling implies an effective nonperturbative superpotential
\be\label{eq:npDilatonSuperpotential}\index{dilaton superpotential}
W\approx  M_\mathrm{GUT}^3  e^{-aS}\,,\qquad a=\frac{3}{2\beta}\,.
\ee
This superpotential breaks supersymmetry. Despite the fact that a given model could be supersymmetric at all
orders of perturbation theory, nonperturbatively, the term
\be\label{eq:FSGC2}
F_S^\dagger = -\depa{W}{S}\approx a M_\mathrm{GUT}^3  e^{-aS}
\ee
is nonzero and, thus, breaks \susy. 
Furthermore, the $F$--term in eq.~\eqref{eq:FSGC2} leads to $S\rightarrow\infty$ at the minimum of
the resulting scalar potential. This is the notorious  dilaton run-away problem that appears in models
with a single gaugino condensate and a classical (universal) K\"ahler potential
\be\index{Kahler potential, universal@K\"ahler potential, universal}
K=-\ln(S+\ol S)\,.
\ee 
To solve this problem, one can either employ multiple gaugino condensates or nonperturbative
corrections to the K\"ahler potential. In orbifold models, nevertheless, the first option is generically
ruled out since there is mostly just one unbroken gauge group factor in the hidden sector or the
condensation scale of a possible second hidden gauge factor is too low (as is the case for e.g. \SU2 or
\SU3). Thus, we are left with the second option only. It consists in amending the  classical K\"ahler
potential for the dilaton by a nonperturbative functional form, such that
\begin{equation}\index{Kahler potential@K\"ahler potential}
 K~=~-\ln (S + \ol{S}) + \Delta K_\mathrm{np} \;. 
\end{equation}
The functional form of $\Delta K_\mathrm{np}$ has been studied in the
literature~\cite{Shenker:1990uf2,Banks:1994sg,Casas:1996zi,Binetruy:1996xj,Binetruy:1996nx,Binetruy:1997vr}.
For a favorable choice of the parameters, this correction allows one to stabilize the dilaton at a
realistic value, $\re S\simeq 2$, while breaking
supersymmetry~\cite{Casas:1996zi,Binetruy:1996nx,Binetruy:1997vr,Barreiro:1997rp,Buchmuller:2004xr}. 
The $T$--moduli can be stabilized at the same time by including $T$--dependence in the superpotential
required by $T$--duality~\cite{Font:1990nt,Nilles:1990jv}. In simple examples, the overall $T$--modulus
is stabilized at the self--dual point such that $F_T=0$. This leads to  dilaton dominated \susy\
breaking. There are many problems attached to moduli stabilization that, in principle, can have some
influence on our results. However, they are beyond the scope of this thesis.

If the dilaton is stabilized at a realistic value $\re\vevof{S}\approx 2$, gaugino condensation translates into
\susy\ breaking with the gravitino mass determined by the \vev\ of the (hidden) gaugino
condensate~\cite{Ferrara:1982qs} 
\be\index{gravitino mass}
m_{3/2}\approx \frac{\vevof{\lambda\lambda}}{M_\mathrm{Pl}^2}\approx \h{\Lambda^3}{M_\mathrm{Pl}^2}\,.
\ee
In particular, notice that 
\begin{equation}
 \Lambda ~\sim~10^{13} \,\text{GeV} 
\end{equation}
leads to the gravitino mass in the TeV range. It is clear that for certain (hidden) gauge groups and
matter content, $\Lambda$ can be in the right range.  

Let us point out that, even if the scale of gaugino condensation is adequate ($\sim 10^{13}$ GeV), there
are many factors that can affect it. In particular, there are string threshold corrections 
\cite{Ibanez:1986xy,Dixon:1990pc,Mayr:1993mq,Nilles:1997vk,Stieberger:1998yi,Kokorelis:2000hs}
which lead to different gauge couplings in the visible and hidden sectors. 
Whereas in the visible sector,
due to gauge coupling unification, one requires $g^{-2}(M_\mathrm{GUT})\approx2$, in the hidden sector
string threshold corrections can alter its value as
\begin{equation}\label{eq:DeltaCorrections}\index{coupling constant! threshold corrections}
 g^{-2}_\mathrm{hidden}(M_\mathrm{GUT}) ~\approx~ 2\,(1-\Delta)\;,
\end{equation}
where $\Delta$ parametrizes such corrections. In this case, the running of the hidden gauge coupling
changes and, hence, so do the corresponding condensation scale:
\be\label{eq:DeltaGCScale}
\Lambda\ \approx\ M_\mathrm{GUT}\,\exp\left\{-\frac{1}{2\beta}2(1-\Delta)\right\}\,.
\ee
In the next section, we will have the opportunity to see how these corrections affect the breaking of supersymmetry.

\subsection[\susy\ Breakdown in the Orbifold-\mssm]{\textbf{\textsc{susy}} Breakdown in the  Orbifold-\textbf{\textsc{mssm}}}
\label{subsec:SusyBreakingSeesaw}\index{orbifold-mssm@orbifold-\mssm}

\begin{figure}[t!]
\begin{minipage}{0.47\textwidth}
(a)\\
\CenterEps{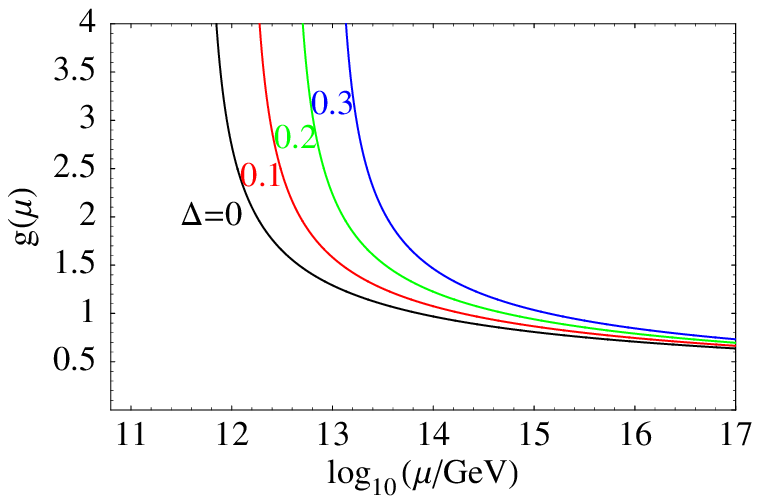}
\end{minipage}
\hskip 2mm
\begin{minipage}{0.47\textwidth}
(b)\\
\CenterEps{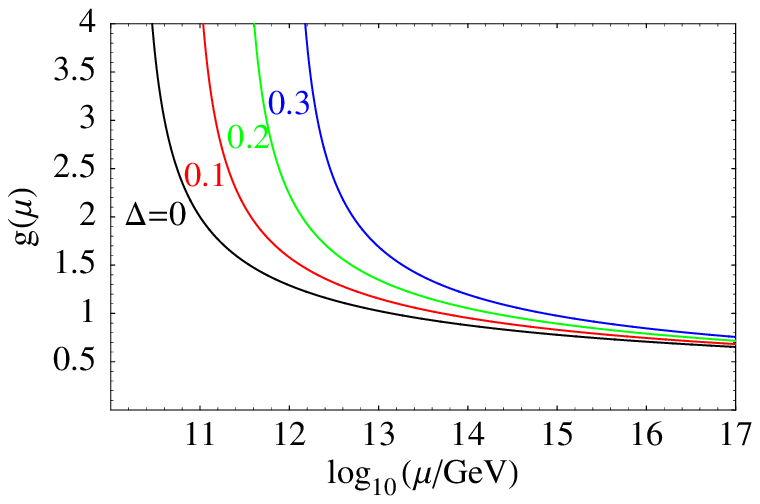}
\end{minipage}
\caption{Running of the coupling constants of (a) $G_\mathrm{hidden}=\SO8$ (orbifold-\mssm) and
  (b)  $G_\mathrm{hidden}=\SU4$~\cite{Buchmuller:2006ik} for different threshold corrections $\Delta$.  
\label{fig:gRunning}}
\end{figure}

In section~\ref{eq:NiceVacuumSeesaw}, we have seen that in an admissible vacuum configuration that
preserves \susy\ at high energies (as high as the string scale) the unbroken  hidden sector is reduced to
$G_\mathrm{hidden}=\SO8$. Furthermore, we observe that there are some \bs{8}--plets in this sector
(cf. table~\ref{tab:spectrum}). These states can be split in two sets, which form mass terms
independently (see table~\ref{Tab:spectrum}). In the vacuum configuration~\eqref{eq:stilde}, their mass
matrices are given by 
\be
\ba{ccc}
\mathcal{M}_{ww} = 
\left(
\begin{array}{ccccc}
 \widetilde{s} & \widetilde{s}^5 & 0 & \widetilde{s}^5 & \widetilde{s}^5\\
 \widetilde{s}^5 & \widetilde{s} & 0 & \widetilde{s}^5 & \widetilde{s}^5\\
 0 & 0 & 0 & \widetilde{s}^3 & \widetilde{s}^3\\
 \widetilde{s}^5 & \widetilde{s}^5 & \widetilde{s}^3 & \widetilde{s}^6 & \widetilde{s}^6\\
 \widetilde{s}^5 & \widetilde{s}^5 & \widetilde{s}^3 & \widetilde{s}^6 & \widetilde{s}^6
\end{array}
\right),
&\ &
\mathcal{M}_{f\bar f}~=~ 
\left(
\begin{array}{cc}
 0 & \widetilde{s}^3 \\
 0 & \widetilde{s}^3
\end{array}
\right)\;.
\ea\label{eq:SeesawHiddenMM}
\ee
The first of these matrices has maximal rank, but from the second matrix there is a pair $f,\bar f$ that stays
massless up to energies lower than $M_\mathrm{GUT}$.

Let us proceed to compute the gravitino mass $m_{3/2}$. We will require the beta-coefficient
that can be computed by
\be\label{eq:SO2Nb0}
b^{\SO{2N}}_0~=~6\cdot(N-1) - \#(\bs{2N}\text{--plets})\,;
\ee
in our case it is $b_0^{\SO8}=16$. Therefore, with the realistic value for the stabilization of the dilaton,
$\re \vevof{S}\approx 2$, we obtain from eq.~\eqref{eq:ScaleGC}
\be\label{eq:SeesawLambdascale}
\Lambda~\approx~ M_\mathrm{GUT} e^{-\pi^2}\approx 10^{12}\text{ GeV}\,,
\ee
what in turn yields the gravitino mass $m_{3/2}\approx 1$ GeV. This scale is phenomenologically
unacceptable because that scale has been already ruled out by experiment. However, as we mentioned
before, threshold corrections can modify this scale. In fact, by using eqs.~\eqref{eq:RunningHidden}
and~\eqref{eq:DeltaGCScale}, we find that threshold corrections enhance the scale of gaugino condensation
and, therefore, also the gravitino mass. 
In fig.~\eqref{fig:gRunning} we present the running of the gauge coupling $g(\mu)$ for two different
cases. Fig.~\eqref{fig:gRunning}(a) illustrates the influence of different values of $\Delta$ on the
behavior of the gauge coupling. We notice that for $\Delta=0$ we recover the result given in
eq.~\eqref{eq:SeesawLambdascale} whereas for $\Delta=0.3$ the condensation scale becomes almost $10^{14}$
GeV. In fig.~\eqref{fig:gRunning}(b), we compare our result with a case where the hidden sector has
an \SU4 gauge group and no massless matter in that sector. We might say that the situation in \SO8 is
somewhat better than in \SU4. However, we must recall that there are of course other factors that can
affect our estimates. For example, we have not described precisely the mechanism to stabilize the
dilaton. Further, we have used the symbol `$\approx$' in many of our equations because those values are
not precise. Therefore, in this study the most important result is that the scale of \susy\ lies around
the phenomenologically interesting interval.

\subsection[\susy\ Breakdown in the (Mini-)Landscape]{\textbf{\textsc{susy}} Breakdown in the (Mini-)Landscape}
\label{subsec:SusyBreakingML}

After having examined some of the most important aspects of gaugino condensation, we would like to
continue our study on the set of \mssm-candidates obtained in the last chapter through a search guided by
grand unification. We have found 196 orbifold models (cf. table~\ref{tab:Summary2}) with the following
properties:\\ 
\phantom{.}\hskip 5mm $\bullet$ \sm\ gauge group times additional gauge factors,\\
\phantom{.}\hskip 5mm $\bullet$ nonanomalous hypercharge of the Georgi-Glashow type, i.e. consistent with gauge
\phantom{.}coupling unification,\\
\phantom{.}\hskip 5mm $\bullet$ three \mssm\ matter generations plus vectorlike exotics,\\
\phantom{.}\hskip 5mm $\bullet$ all vectorlike exotics are decoupled from the massless spectrum.\\
In order to save some computation time, we have imposed an additional constraint on the models. Namely,
we have demanded one trilinear coupling of the type $q\,\ell\,\bar u$ which might be responsible for the
heavy mass of the top quark. Although this condition is not arbitrary, it is also not imperative since
a heavy top quark might also appear through alternative methods. However, for consistency, we will stick
to that constraint in this section.

The strategy we have followed consists in first finding a set of \mssm\ candidates and then studying
common features that could lead to some sort of low--energy predictions. In this section, we concentrate
on the question of \susy\ breaking. Particularly, we would like to figure out whether our \mssm\
candidates yield a reasonable scale of \susy\ breaking via hidden sector gaugino condensation. 

With that purpose, we will impose an additional criterion in our search
\begin{dingautolist}{"0C8}
 \item Select models where exotics decouple +  gaugino condensation
\end{dingautolist} 
At this step, we select models in which the decoupling of the \sm\ exotic states is possible
without breaking the largest gauge group in the hidden sector. We find that all or almost all of the
matter states  charged  under this group also attain large masses which allow for spontaneous \susy\ 
breaking via gaugino condensation.

\renewcommand{\arraystretch}{1.2}
\begin{table}[t!]
\centerline{
\begin{tabular}{|l||l|l||l|l|}
\hline
 Criterion & $V^{\SO{10},1}$ & $V^{\SO{10},2}$ & $V^{\E6,1}$ & $V^{\E6,2}$\\
\hline\hline
\rowgrayh
 \ding{"0C6}  heavy top                                &72 & 37 & 3 & 2\\
 \ding{"0C7}  exotics decouple at order 8              &56 & 32 & 3 & 2 \\
\rowgrayh
 \ding{"0C8}  exotics decouple + gaugino condensation  &47 & 25 & 3 & 2 \\
\hline
\end{tabular}
}
\caption{A subset of the \mssm\ candidates.
\label{tab:Summary3}}
\end{table}
\renewcommand{\arraystretch}{1}

Among the models satisfying all our criteria,~\ding{"0C1}--\ding{"0C8}, we consider the most promising \mssm\
candidates. Our results are presented in table~\ref{tab:Summary3}. We find it remarkable that out of ${\maO}(10^4)$ inequivalent 
models, ${\maO}(10^2)$ pass all of our requirements, including a hidden sector that allows \susy\
breaking. In this sense, the region of the heterotic landscape endowed with local \SO{10} and \E6
{\gut}s is particularly attractive.

A comment is in order. We impose by hand the requirement that gaugino condensation  be allowed. By
assigning {\vev}s to all \sm\ singlets, i.e. without verifying explicitly $D$-- and $F$--flatness,
the hidden sector gauge group is broken by matter {\vev}s charged under
this group. Similarly, the {\sm} gauge group is broken in more general ``vacuum'' configurations where
many other fields can also acquire {\vev}s. Clearly, most of the string landscape is not relevant to our
physical world. It is only possible to obtain useful predictions from the landscape once certain criteria
are imposed. Here we require that gaugino condensation be allowed so that \susy\ can be broken. Since the
largest hidden sector group factor would dominate {\susy} breaking, we focus on vacua in which this
factor is preserved by matter {\vev}s. Within the set of  our promising models, we can now  study
predictions for the scale of \susy\ breaking.  

Our {\mssm} candidates  have the necessary ingredients for supersymmetry breaking via gaugino
condensation in the hidden sector. In particular, they contain nonabelian gauge groups under which little
or no matter states are charged. The corresponding gauge interactions become strong at some
intermediate scale which   can lead to spontaneous supersymmetry breakdown. The specifics  depend on the
moduli stabilization mechanism, but the main features such as the scale of supersymmetry breaking hold
more generally. In particular, the gravitino mass is given by
\begin{equation}
m_{3/2} \approx \frac{\Lambda^3}{M_\mathrm{Pl}^2} \;,
\end{equation}
while the proportionality constant is model-dependent.

The gaugino condensation scale $\Lambda$ is given by the renormalization group invariant scale of
the condensing gauge group, 
\begin{equation}
 \Lambda~\sim~ M_\mathrm{GUT}\,\exp \left\{ -\frac{1}{2\beta}\,\frac{1}{g^2(M_\mathrm{GUT})}\right\} \,.
\end{equation}
With an appropriate mechanism, the dilaton can be stabilized at a realistic value $\re\vevof{S}\approx2$
while breaking \susy. As already mentioned, we see that for $\Lambda \sim 10^{13}$ GeV, the gravitino
mass  lies in the TeV range which is favored by phenomenology. {\susy} breaking is communicated to the
observable sector by gravity~\cite{Nilles:1982ik}. 

\begin{figure}[t!]
\begin{minipage}{0.47\textwidth}
(a)\\
\CenterEps[0.9]{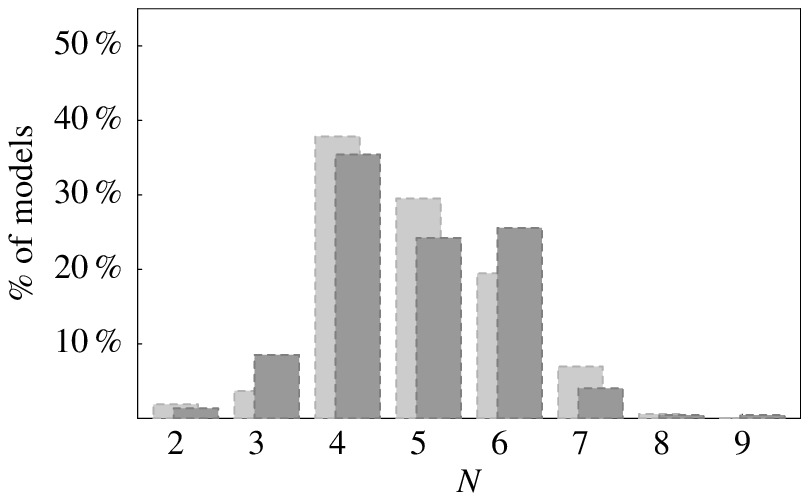}
\end{minipage}
\hskip 2mm
\begin{minipage}{0.47\textwidth}
(b)\\
\CenterEps[0.9]{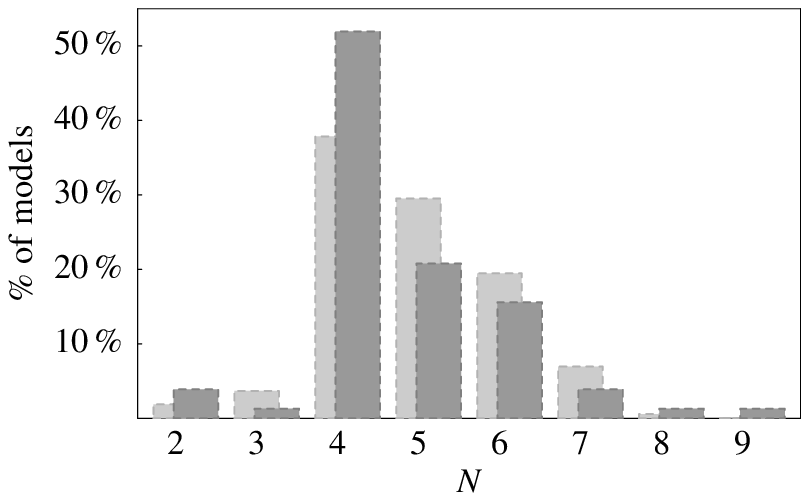}
\end{minipage}
\caption{(a) Number of models vs.\ the size of largest gauge group in the hidden sector. $N$ labels
$\SU{N}$, $\SO{2N}$, $\E{N}$ groups. The background corresponds to step~\ding{"0C1}, while the foreground
corresponds to step \ding{"0C5} (see table~\ref{tab:Summary}). (b) As before, but with models of step
\ding{"0C7} at the foreground. 
\label{fig:histogram1}}
\end{figure}

\begin{figure}[!t!]
\centerline{\includegraphics{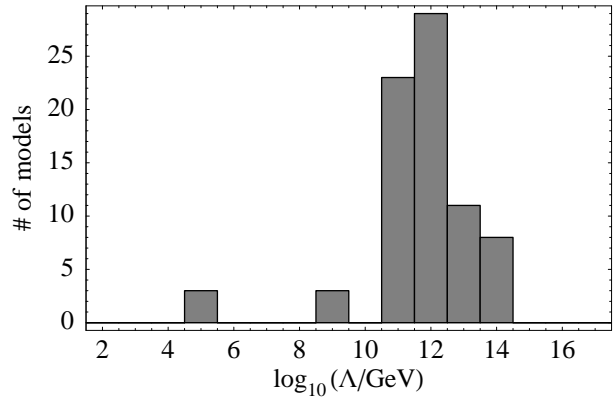}}
\caption{Number of models vs.\ scale of gaugino condensation. \label{gc}}
\end{figure}

In fig.~\ref{fig:histogram1}(a), we display the frequency of occurrence  of various gauge groups in the
hidden sector. The preferred size ($N$) of the gauge groups depends on the conditions imposed on the
spectrum. When all inequivalent models with two Wilson lines are considered, $N=4,5,6$ appear with similar
likelihood and $N=4$ is somewhat preferred. If we require the massless spectrum to be the {\mssm} +
vectorlike matter, the fractions of models with $N=4,5,6$ become even closer. However, if we further
require a heavy top quark and the decoupling of exotics at order eight, $N=4$ is clearly preferred
(figure~\ref{fig:histogram1}(b)). In this case, $\SU{4}$ and $\SO{8}$  provide the dominant
contribution. Since all or almost all matter charged under these groups is decoupled, this leads to
gaugino condensation at an  intermediate scale.\footnote{We note that before step \ding{"0C7}, gaugino
condensation does not occur in many cases due to the presence of hidden sector matter.}
Possible scales of gaugino condensation are shown in fig.~\ref{gc}.

The correlation between the observable and hidden sectors comes about for a few reasons. First, it is due
to modular invariance  which ties the gauge shifts and Wilson lines in the two sectors. Second, the gauge
shifts and  Wilson lines in the hidden sector affect properties of the massless spectrum, for instance, 
through the masslessness equations.

We see that among the promising models, just as in the orbifold-\mssm\ presented in
section~\ref{subsec:SusyBreakingSeesaw}, intermediate scale supersymmetry breaking is preferred. The 
underlying reason is that realistic spectra require complicated Wilson lines, which break the hidden
sector gauge group. The surviving gauge factors are neither too big (unlike in Calabi--Yau 
compactifications with the standard embedding) nor too small.

There are significant uncertainties in the estimation of the supersymmetry-breaking scale.
First, the identification of $\Lambda$
with the renormalization group invariant scale  is not precise. A factor of a few uncertainty in this relation
leads to two orders of magnitude uncertainty in $m_{3/2}$. Also, there could be significant string
threshold corrections  which can affect the estimate although we see from fig.~\eqref{fig:gRunning}
that threshold corrections might enhance the scale of \susy\ breaking.  Thus, the resulting
``prediction'' for the superpartner masses should be understood within 2-3 orders of magnitude.

%%%%%%%%%%%%%%%%HETEROTIC ROAD%%%%%%%%%%%%%%%%%%%%%%%%%%%%%%%%%%%%%%%
\section[$R$-Parity and Proton Decay]{$\bs{R}$-Parity and Proton Decay}
\label{sec:RParity} 

An essential property of the \mssm\ that a realistic model must exhibit is $R$-parity. This has the
advantage of greatly reducing the number of arbitrary parameters in the superpotential,  forbidding
dimension three and four baryon or lepton number violating operators, and preserving a viable dark matter
candidate, i.e. the LSP. However, obtaining a conserved $R$-parity in string constructions sets
frequently an insurmountable hurdle, which must however be overcome in order to reach the \mssm.  

\begin{figure}[t!]
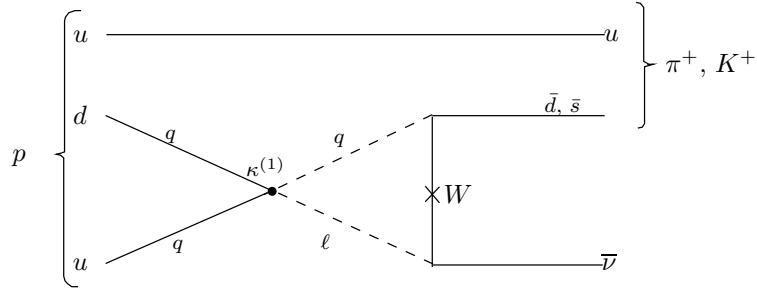

\centerline{\input RPreservingDecay.pstex_t}
\caption{Possible contribution to proton decay, involving the dimension five operator
  $q\,q\,q\,\ell$. The proton decays as $p\rightarrow K^+\bar\nu\text{ or }\pi^+\bar\nu$.} 
\label{fig:RPreservingDecay}
\end{figure}

Our strategy for accomplishing this is, in principle, quite simple. We identify first a $\U1_{\BmL}$
gauge symmetry and then give {\vev}s to some \sm\ singlets that break $\U1_{\BmL}$ to the discrete
subgroup 
\be\index{matter parity}
 \Z2^{\calM}~:~(-1)^{3(\BmL)}\,.
\ee
This unbroken discrete symmetry is the so-called family reflection symmetry or {\it matter
parity}~\cite{Dimopoulos:1981dw}, which, due to its properties, can be considered an $R$-parity.
This is a global \Z2 symmetry which is {\it even} on the
Higgs doublets and {\it odd} on all {\sm} quark and lepton fields. Further, it forbids dangerous baryon
or lepton number violating operators of dimension three and four:
\begin{equation} \label{eq:blviol}\index{proton decay!dim 4 operators}
\ol{u}\,\ol{d}\,\ol{d}\;,\quad q\,\ol{d}\, \ell\;,\quad \ell\, \ell\,\ol{e}\quad\text{and} 
\quad \ell\, h_u\;.
\end{equation}
On the other hand, it allows quark and lepton Yukawa couplings as well as the Majorana neutrino mass
operator $\bar\nu \bar\nu$. 
However, there are certain dimension five operators that are allowed by this symmetry:
\be\label{eq:Dim5Operators}\index{proton decay!dim 5 operators}
\kappa^{(1)}_{ijkl} q_i\,q_j\,q_k\,\ell_l\,,\qquad \kappa^{(2)}_{ijkl} \bar{u}_i\,\bar{u}_j\,\bar{d}_k\,\bar{e}_l\,,
\ee
which can induce proton decay (see fig.~\eqref{fig:RPreservingDecay}). $\kappa^{(1)}$ and $\kappa^{(2)}$ are
coupling constants. In the \mssm\, $\kappa^{(1)}$ is constrained by~\cite{Hinchliffe:1992ad}
\be\label{eq:Dim5ProtonDecayConstraints}\index{proton decay!constraints}
\h{\kappa^{(1)}_{1121}}{\Lambda^{cut}}\approx \h{\kappa^{(1)}_{1122}}{\Lambda^{cut}}
\lesssim 16\pi^2\h{M_\textrm{SUSY}}{M_\textrm{GUT}^2}
\ee
and $\kappa^{(2)}$ can take arbitrary values as long as they are consistent with perturbation
theory. Here, $\Lambda^{cut}$ denotes an intermediate cut-off scale and $M_\textrm{SUSY}\approx 10^{4}$
GeV is the scale at which \susy\ is supposed to break.

\index{B-L}
We build the $\U1_{\BmL}$ symmetry as a superposition of all gauge \U1's of the \mssm\ candidates
excepting the anomalous one.\footnote{In fact, this constraint might be relaxed. Yet in 
that case, one has to deal with the anomalies more carefully.} In our \mssm\ candidates, a natural choice for
the generator of $\U1_{\BmL}$ follows from the standard breaking 
\be\label{eq:SO10-GSMxBmL}
\SO{10}\rightarrow\SU5\x\U1_{X}\rightarrow G_{SM}\x\U1_{X}\,,
\ee
where the generators of hypercharge and $B-L$ are given by
\bea\index{hypercharge, standard}\index{B-L, standard}
\mathsf{t}_Y & =& \left(-\2,\,-\2,\,\3,\,\3,\,\3\right),\,\label{eq:StdY2}\\
\mathsf{t}_{\BmL} &=& \frac{2}{5}\left(2\,\mathsf{t}_Y+\mathsf{t}_X\right)~=~\left(0,\,0,\,\h{2}{3},\,\h{2}{3},\,\h{2}{3}\right)\label{eq:StdBmL}
\eea
with $\mathsf{t}_X=(1^5)$ generating the additional $\U1_X$. Under these symmetries, including $\SU3_c\x\SU2_L$,
an \mssm\ matter generation has the quantum numbers
\begin{equation}\label{eq:StdYBmLCharges}
\mbox{\footnotesize $
\ba{rcccccc}
\bs{16}~=& (\bs{3}, \bs{2})_{\6,\3}+& (\bsb{3}, \bs{1})_{-\h{2}{3},-\3}+&(\bsb{3}, \bs{1})_{\3,-\3}+&  
(\bs{1}, \bs{2})_{-\2,-1}+& (\bs{1}, \bs{1})_{1,1}+& (\bs{1}, \bs{1})_{0,1} \,,\\[2mm]
&q&\bar u & \bar d & \ell & \bar e & \bar\nu
\ea$}
\end{equation}
where the subscripts denote hypercharge and \BmL charge, respectively. Further, the Higgs doublets $\phi$
and $\bar\phi$ have  \U1 charges $(-1/2,0)$ and $(1/2,0)$, respectively.

Unfortunately, the symmetry generated by $\mathsf{t}_{\BmL}$ mixes with $\U1_A$ in our \mssm\
candidates and is, hence, ruled out. We have to search for a different ``suitable'' definition of
$\U1_{\BmL}$ which has to satisfy two conditions:\\
\phantom{.}\hskip 4mm (i)\  it must assign the standard \BmL charges, eq.~\eqref{eq:StdYBmLCharges} to all
\sm\ particles (including a pair of Higgs fields), and\\
\phantom{.}\hskip 4mm (ii) it must give most \sm\ singlets a value that satisfies $3(\BmL)=0\mod2$.

Condition (i) is clear, but the second condition requires some argumentation. Firstly, \sm\ singlets with
charges $3(\BmL)=0\mod2$ can obtain {\vev}s for decoupling exotics, as well as giving effective quark and
lepton Yukawa couplings. Secondly, if there are some singlets (let us call them $\chi_i$) satisfying
$3(\BmL)=0\mod2$ and with nontrivial $\BmL$ charge, they will break $\U1_{\BmL}$, but will leave the
$R$-parity, $\Z2^{\calM}$, unbroken. Moreover, note that if singlets with  $3(\BmL)=1\mod 2$ develop
nonvanishing {\vev}s, $R$-parity is broken and dimension four baryon/lepton number violating operators
are typically generated. We would like to remark that this approach to get $R$-parity is not
unique. Alternatives arise from considering other internal symmetries, such as R-charge\footnote{We refer
here to one of the selections rules. See section~\ref{sec:YukawaCouplings}.} conservation
(inherited from Lorentz invariance in ten dimensions)~\cite{Araki:2007ss} or
$T$--duality~\cite{Gaillard:2004aa}.   

Searching for generators of $\U1_{\BmL}$ is not difficult, but quite cumbersome. One has to
express $\mathsf{t}_{\BmL}$ as a superposition of all \U1's, excepting the anomalous \U1,
\be\label{eq:CombinationForBmL}
\mathsf{t}_{\BmL}=\sum_{i\neq A} x_i\mathsf{t}_{i}\,,
\ee
and solve (non)linear equations on the real parameters $x_i$ guaranteeing that the conditions (i) and (ii)
explained above be fulfilled. The precise description of the method is provided in appendix D of
ref.~\cite{Lebedev:2007hv}. 

\subsection[Supersymmetric \mssm\ Candidates]{Supersymmetric \textbf{\textsc{mssm}} Candidates}
\label{subsec:SusyMSSMCandidates}

Before going any further, there is a question that we have to clarify. In
section~\ref{sec:TowardsRealistic}, it was assumed that all \sm\ singlets can acquire
{\vev}s. Nevertheless, we have learnt in section~\ref{sec:SUSYVacua} a strategy to identify those
singlets which admit a nonzero {\vev} in a consistent supersymmetric vacuum. In the present case, since
\BmL must be broken spontaneously, it results impossible to postpone the task of verifying $D$--flatness
any longer. 

\renewcommand{\arraystretch}{1.3}
\begin{table}[t!]
\centerline{
\begin{tabular}{|l||rr|rr|}
\hline
           & \multicolumn{2}{c|}{$V^{\SO{10},1}$} & \multicolumn{2}{c|}{$V^{\SO{10},2}$} \\
\cline{2-5}
 Criterion & \ding{"0C6}  & \rojo{\xcancel{\negro{\ding{"0C6}}}} 
           & \ding{"0C6}  & \rojo{\xcancel{\negro{\ding{"0C6}}}} \\
\hline\hline
\rowgrayh
 \ding{"0C5} spectrum $=$ 3 generations $+$ vectorlike & 72 & 56 & 37 & 53 \\
 \ding{"0C7} exotics decouple at order 8               & 56 & 50 & 32 & 53 \\
\rowgrayh
 \ding{"0C7}' exotics decouple in \susy\ configurations & 55 & 50 & 32 & 53 \\
\hline
\end{tabular}
}
\caption{A subset of the \mssm\ candidates. The number of models with ``heavy top'' are listed
under~\ding{"0C6}. \rojo{\xcancel{\negro{\ding{"0C6}}}} denotes no ``heavy top''. Only in one model, it
was not possible to find a supersymmetric vacuum configuration where all exotics decouple.
\label{tab:Summary4}}
\end{table}
\renewcommand{\arraystretch}{1.1}

Let us focus on the models based on the shifts $V^{\SO{10},1}$ and $V^{\SO{10},2}$, that is, in
those that concentrate most promising models. According to our previous discussion, all we have to do is
to verify whether there exist holomorphic invariant monomials (HIMs) with net negative anomalous
charge. All fields entering the monomials are allowed to attain nonvanishing {\vev}s and, therefore,
contribute to the masses of the exotics.

We recall our results from chapter~\ref{ch:MLI} and complement them with our new findings in
table~\ref{tab:Summary4}. In particular, observe that we have incorporated the nontrivial
criterion~\ding{"0C7}'. We have verified that all vectorlike exotics acquire large masses. These results
are crucial for the discussion on breaking of $\U1_{\BmL}$.

Notice that we have not discussed $F$--flatness or any other additional constraint
(e.g. $\vevof{W}=0$). As we mentioned in section~\ref{sec:SUSYVacua}, the reason is that other conditions
are not relevant since they depend on technical particularities of the models, such as the order of the
perturbative potential at hand. On the other hand, we do not have any prescription to write the full
superpotential including all possible nonperturbative contributions. Additionally, we have shown that
once one has found a $D$--flat configuration, it is always possible to find a nontrivial solution to the
$F$--term equations.

\subsection[$R$-Parity Invariant \mssm\ Candidates]{$\bs{R}$-Parity Invariant \textbf{\textsc{mssm}} Candidates}

For simplicity, we focus here only on those models from table~\ref{tab:Summary4} with a cubic top Yukawa
coupling, that is, on those 55+32 \mssm\ candidates satisfying simultaneously criteria \ding{"0C6} and
\ding{"0C7}'. 

We notice that the shift $V^{\SO{10},1}$ seems still to be more promising in comparison to
$V^{\SO{10},2}$. With our method, we find that 34 of the 55 (5 of the 32) \mssm\ candidates from the
first (second) shift vector admit nonanomalous definitions of \BmL. Each {\it model}, nevertheless, correspond
to a big family of {\it configurations} with distinct acceptable $\U1_{\BmL}$ symmetries. There are two
reason for this to happen:

$\bullet$ In many cases there are vectorlike exotics with \sm\ gauge charges identical to those of
quarks and leptons. Further, without \BmL, each lepton doublet $\ell$ could very well be a Higgs boson
$\phi$.  Thus there are different ways to choose which of these states have standard \BmL\ charges. Each
choice can lead to a different (and nonanomalous) definition of \BmL.

$\bullet$ For each choice of \sm\ particles above, there may be more than one \BmL definition. The
system of equations that determine the values of the variables $x_i$ in eq.~\eqref{eq:CombinationForBmL}
might be underconstrained. Therefore, in some cases, there are (continuous) families of $\BmL$ generators
(i.e. some of the $x_i$'s are free or fulfill equations with an undetermined number of solutions). 

These ambiguities, although disturbing at first sight, might also be considered a rich source of new
physics. However, it is clear that we need to implement new criteria in order to reduce
the number of \BmL configurations. We require the presence of \sm\ singlets with charges $\BmL =0,\pm
2,\pm2/3,\pm 2/5,\ldots,\pm 6/7$ in the spectrum\footnote{Note that we have included also charges that
violate the condition $3(\BmL)=0\mod2$. This is also admissible because other discrete symmetries
resulting from the breaking of $\U1_{\BmL}$ might be as good as $\Z2^{\calM}$. Further details can be
found in appendix C of ref.~\cite{Lebedev:2007hv}}. Configurations with
singlets having \BmL charges 
consistent with $3(\BmL)=0\mod2$, but different from the values just mentioned are automatically
disregarded. 

Including only these possibilities, we find  3447 (144) consistent \BmL generators from the first (second)
\SO{10} shifts.  The question is now whether all these configurations are inequivalent or not. It can be
argued that, if one considers one single model (shift vector $+$ Wilson lines + hypercharge generator)
with several \BmL configurations, we could distinguish the inequivalent configurations by comparing the
sets of \sm\ singlets with charges fulfilling $3(\BmL)=0\mod2$. If these sets (and not the charges)
coincide in two models, one might state that they are equivalent and, thus, that one of them can be
ignored. This is indeed the approach we follow.\footnote{Nevertheless, soon we found that some configurations with
identical sets of singlets with correct \BmL charges led to different charges in the \mssm\ matter fields
and therefore different phenomenological properties. It seems necessary to try a different approach.}

In this approach, we find that there are ``only'' 85 (8) inequivalent \BmL configurations from
$V^{\SO{10},1}$ ($V^{\SO{10},2}$). Further, as we are interested in models containing solely the \sm\
gauge group times a hidden sector for gaugino condensation, one can verify whether all extra unbroken
\U1's break when fields with proper \BmL charges attain {\vev}s.
Requiring the absence of extra unbroken \U1's reduces this set to 42 (0) acceptable configurations. 
Finally, demanding that all
exotics decouple along $D$--flat directions leads to 15 (0) acceptable solutions with an exact low energy
$R$-parity. We list the corresponding shifts, Wilson lines and generators of hypercharge and \BmL\ in
table~\ref{tab:NiceBmLModels}. These input parameters can be used to determine all other properties of
the models.

This result is specific to our $(\BmL)$-based strategy and we expect, in general, more acceptable models
to exist. We point out that there is a big uncertainty in this number and, therefore, it should be taken
as a lower bound.

\subsection[$R$-Parity in the Orbifold \mssm]{$\bs{R}$-Parity in the Orbifold-\textbf{\textsc{mssm}}}

The orbifold-\mssm\ allows us to define a suitable $\BmL$ generator which leads to the standard charges
for the \sm\ matter, eq.~\eqref{eq:StdYBmLCharges}. The generator is given by 
\begin{equation}
\mathsf{t}_{\BmL}
~=~\left(-1,\,-1,\,0,\,0,\,0,\,\tfrac{2}{3},\,\tfrac{2}{3},\,\tfrac{2}{3}\right)\left(-2x+ \tfrac{1}{2},\,-\tfrac{1}{2},\,0,\,-x,\,-x,\,0,\,0,\,0\right) \;,
\end{equation}
with arbitrary $x$. Let us take $x=\2$. Then the \BmL generator we shall use is
\begin{equation}
 \mathsf{t}_{B-L}~=~
 \left( -1 , -1 , 0 , 0 , 0 , \tfrac{2}{3} , \tfrac{2}{3} , \tfrac{2}{3} \right)\,,
 \left( -\tfrac{1}{2} , -\tfrac{1}{2} , 0 , -\tfrac{1}{2} , -\tfrac{1}{2} , 0 , 0 , 0 \right)\;.
\end{equation}
An interesting feature is that the spectrum contains a pair of fields with
$\BmL$ charges $\pm2$, so that, if the $\BmL$ gauge symmetry is broken by {\vev}s of these fields, the matter
parity $(-1)^{3(\BmL)}$ is conserved and proton decay due to dimension three and four operators is
suppressed. In table~\ref{Tab:spectrum}, we have summarized the matter spectrum of the model. Since now
we include \BmL, we can now distinguish between fields with identical \sm\ quantum numbers; in
particular, we can distinguish now between Higgs and lepton doublets. 

The vacuum configuration~\eqref{eq:stilde} is still admissible in presence of $\U1_{\BmL}$. In this configuration,
\sm\ singlets with \BmL\ charges $\pm2$, labeled $\chi_1\equiv s^0_1$ and $\chi_2\equiv s^0_2$, attain
nonzero {\vev}s and thus break \BmL by two units, such that, e.g. particles with \BmL charge $1$ are
equivalent to particles with \BmL charge equal to $-1$. This example applies to the right--handed
neutrinos to be studied in the next section.

Let us now proceed to enumerate some of the properties that the orbifold-\mssm\ acquires after breaking $\BmL$.

\noindent
{\bf Absence of proton decay operators of dimension three and four.} This is, of course, not surprising
since we have imposed \BmL just to avoid these operators.

\noindent
{\bf Cancellation of the $\bs{\mu}$--term.} The Higgs mass terms are
\begin{equation}
 \bar\phi_i\,(\mathcal{M}_{\bar\phi\phi})_{ij}\,\phi_j\;,
 \quad\text{where}\quad
 \mathcal{M}_{\bar\phi\phi}~=~
 \left(
 \begin{array}{cccc}
 \widetilde{s}^4 & 0 & 0 & \widetilde{s} \\
 \widetilde{s} & \widetilde{s}^3 & \widetilde{s}^3 & \widetilde{s}^6 \\
 \widetilde{s}^5 & 0 & 0 & \widetilde{s}^3 \\
 \widetilde{s} & 0 & 0 & \widetilde{s}^3
 \end{array}
 \right)\;.
\end{equation}
The up-type Higgs $h_u$ is a linear combination of $\bar\phi_1$,
$\bar\phi_3$ and $\bar\phi_4$,
\begin{equation}
 h_u~\sim~\widetilde{s}^{2}\bar\phi_1+\bar\phi_3+\widetilde{s}^4\,\bar\phi_4\;,
\end{equation}
while the down-type Higgs is composed out of $\phi_2$ and $\phi_3$,
\begin{equation}
 h_d~\sim~\,\phi_2+\phi_3\;.
\end{equation}
Let us remark that this spoils our argument about the existence of ``heavy top''.

Notice that in this vacuum configuration the $\mu$--term, being defined as
the smallest eigenvalue of $\mathcal{M}_{\bar\phi\phi}$,
\begin{equation}
 \mu~=~\left.\frac{\partial^2 W}{\partial h_d\,\partial h_u}\right|_{h_u=h_d=0}\,,
\end{equation}
vanishes term by term up to order $\widetilde{s}^6$, at which we work.
This is a way to deal with the supersymmetric $\mu$--problem.

\noindent
{\bf Charged fermion Yukawa matrices.} The up-Higgs Yukawa couplings decompose into
\begin{equation}
 W_\mathrm{Yukawa}~\supset~\sum\limits_{k=1}^4
 (Y_u)_{ij}^{(k)}\,q_i\,\bar u_j\,\bar\phi_k\,,
\end{equation}
where the matrices $Y_u^{(i)}$ are given in appendix~\ref{ch:OrbifoldMSSMDetails}. Thus, the physical
$3\times3$ up-Higgs Yukawa matrix is 
\begin{equation}
 Y_u~\sim~\widetilde{s}^2\,Y_u^{(1)}+Y_u^{(3)}+\widetilde{s}^4\,Y_u^{(4)}~=~
 \left(
 \begin{array}{ccc}
 0 & 0 & \widetilde{s}^8 \\
 0 & 0 & \widetilde{s}^8 \\
 \widetilde{s}^5 & \widetilde{s}^5 & \widetilde{s}^2
 \end{array}
 \right)\,.
\end{equation}
Note that due to the Higgs mixing the top quark Yukawa coupling for
this vacuum configuration is given by $\widetilde{s}^2$. Therefore, the
corresponding $\widetilde{s}$ {\vev}s are required to be quite large.

The down-Higgs Yukawa couplings decompose into
\begin{equation}
 W_\mathrm{Yukawa}~\supset~\sum\limits_{k=1}^4
 (Y_d)_{ij}^{(k)}\,q_i\,\bar d_j\,\phi_k\,,
\end{equation}
where again the Yukawa matrices are provided in appendix~\ref{ch:OrbifoldMSSMDetails}.

The physical $3\times3$ down-Higgs Yukawa matrix emerges by
integrating out a pair of vectorlike $d-$ and $\bar d-$quarks,
\begin{equation}
 Y_d~=~
 \left(
 \begin{array}{ccc}
 1 & \widetilde{s}^3 & 0 \\
 1 & \widetilde{s}^3 & 0 \\
 \widetilde{s} & \widetilde{s}^4 & 0
 \end{array}
 \right)\,.
\end{equation}
We note that both the up and down quarks are massless at order six in
\sm\ singlets. However, we have checked that the up quark becomes
massive at order seven and the down quark gets a mass at order eight.

Analogously, the physical $3\times3$ matrix emerges by integrating out a pair of
vector-like $\ell-$ and $\bar\ell-$leptons,
\begin{equation}
 Y_e~=~
 \left(
 \begin{array}{ccc}
 1 & 1 & \widetilde{s} \\
 \widetilde{s} & \widetilde{s} & \widetilde{s}^2 \\
 0 &  0 &  \widetilde{s}^6 \\
 \end{array}
 \right)\,.
\end{equation}

\noindent
{\bf Dimension five baryon and lepton number violating operators.} We have looked for effective dimension
five baryon and lepton number 
violating operators in this model.   We find that to order $\tilde
s^6$  no such operators exist.   However, these operators can be
generated once the exotics  $\delta_i, \ \bar \delta_i$ are
integrated out. Fortunately, a clever choice of {\vev}s for the fields
$\{\widetilde s_i\}$ can guarantee sufficient suppression of all
induced $q\,q\,q\,\ell$ operators, consistent with current bounds on
proton decay, eq.~\eqref{eq:Dim5ProtonDecayConstraints}.

\noindent
{\bf $\boldsymbol{\mu}$--term and Minkowski space.} In many \mssm\ candidates, the $\mu$--term is
identified with the singlet-superpotential, $W(\ti s_i)$. This occurs if the product of the Higgs
fields $h_u\,h_d$ fulfills trivially all selection rules. In such models, one can show that if $W(\ti
s_i)$ is endowed with an approximate $R$-symmetry, it follows directly that 
$\mu=\vevof{W(\ti s_j)}=0$ and, in the context of supergravity, also $D_iW=0$~\cite{Kappl:2008ie}. 
That is, one obtains a supersymmetric Minkowski vacuum. Since string theories do not admit global
symmetries at all orders in the superpotential, the $R$-symmetry of the superpotential is broken at
higher orders in the superpotential. This implies that the effective $\mu$-term is highly suppressed.

Unfortunately, this is not the case in our orbifold-\mssm. The problem is that $h_u\,h_d$ does not
satisfy all selection rules, which follows from the fact that in the orbifold-\mssm\ there are four Higgs
pairs and they mix. The fact that we do obtain a Minkowski vacuum in this model is therefore not linked to
a symmetry of the theory. This has the consequence that supergravity may lead to a deep anti-De Sitter
vacuum. 

Further details of the orbifold-\mssm\ regarding the effect of our \BmL choice on the neutrino masses
will be discussed in the next section. The detailed mass matrices are displayed in appendix~\ref{ch:OrbifoldMSSMDetails}.

%%%%%%%%%%%%%%%%SEESAW FROM THE HETEROTIC STRING%%%%%%%%%%%%%%%%%%%%%%
\section{Neutrino Masses}
\label{sec:Seesaw}\index{seesaw mechanism}

The seesaw mechanism~\cite{Minkowski:1977sc,Yanagida:1979xx, GellMann:1979xx} is perhaps the most 
attractive way to explain the smallness of the neutrino masses. Its essential ingredients are  heavy
Majorana neutrinos and their Yukawa couplings to the left--handed neutrinos. The supersymmetric seesaw
mechanism is described by the superpotential
\begin{equation}\label{eq:seesawW}
 W~=~Y_\nu^{ij} \,\bar\phi\,\ell_i\, N_j + \frac{1}{2} M_{jk} N_j\, N_k \;,
\end{equation}
where $\bar\phi$ and $\ell_i$ ($i=1,2,3$) are the up--type Higgs and lepton doublets, and $N_j$
($1\le j\le n$) are some heavy standard model singlets.  At low energies, this
leads to three light neutrinos with masses of order $(Y_\nu\,\vevof{\bar\phi})^2/ M$,  where $Y_\nu$ and
$M$ represent typical values of $Y_\nu^{ij}$ and $M_{jk}$, respectively. For $Y_\nu \sim 1$ and $M\sim
10^{16}\,\mathrm{GeV}$, one has $m_\nu \sim 10^{-3}\mathrm{eV}$.  The scales of the atmospheric and solar
neutrino oscillations,~\cite{Maltoni:2004ei}
\be
\sqrt{\Delta m^2_\mathrm{atm}}\simeq0.04\,\mathrm{eV}\,,\qquad \sqrt{\Delta  m^2_\mathrm{sol}}\simeq0.008\,\mathrm{eV}\,,
\ee
are suspiciously close to this scale. This hints at {\gut} structures behind the seesaw.  

In conventional {\gut}s, $N_j$ are members of {\gut} matter multiplets, e.g.\ a $\bs{16}$--plet of
\SO{10}, and $M_{jk}$ are related to a {\vev} of a large {\gut} representation, e.g.\ a
$\ol{\bs{126}}$--plet of \SO{10}.  In this case the Majorana mass terms originate from the coupling 
$\boldsymbol{16}\,\boldsymbol{16}\,\overline{\boldsymbol{126}}$ (cf. e.g.~\cite{Mohapatra:2006gs}).

\subsection{Seesaw Mechanism with Several Neutrinos}
\label{subsec:OrbifoldSeesaw}

Even though in our scheme we have local {\gut}s, the Yukawa couplings do not necessarily preserve the
symmetry of these {\gut}s. The symmetry of the nonlocal coupling in ten dimensions is an intersection of
the local gauge groups at the vertices. 
This implies, for example, that the Majorana mass terms for the neutrino components of the
$\boldsymbol{16}$--plets can originate from the coupling 
\begin{equation}
 \nu_{\boldsymbol{16}}\times\nu_{\boldsymbol{16}}  \times(\text{\sm\ singlets})\;,
\end{equation}
where the singlets belong neither to $\accentset{(-\!\!-\!\!-)}{\boldsymbol{16}}$ nor
to $\accentset{(-\!\!-\!\!-\!\!-\!\!-)}{\boldsymbol{126}}$ 
of \SO{10}. 

Furthermore, any {\sm} singlet can play the role of the right--handed neutrino as long as it has a Yukawa
coupling to the lepton doublets and a large Majorana mass. These are abundant in orbifold models and
typical models contain  $\mathcal{O}(100)$ such singlets.  

\mssm\ candidates contain an anomalous \U1 which induces the FI $D$-term,
\begin{equation}
D_A~=~\frac{g\,M_\mathrm{Pl}^2}{192 \pi^2}~ \tr \mathsf{t}_A 
+ \sum_i q_i\, \vert \phi_i \vert^2 \;,
\end{equation}
where $\mathsf{t}_A$ is the anomalous \U1 generator, $q_i$ are the anomalous charges of fields
$\phi_i$ and $g$ is the gauge coupling. This triggers spontaneous gauge symmetry breaking while
preserving supersymmetry~\cite{Dine:1987xk}. Some of the fields charged under the anomalous \U1 (and, in
addition, under other gauge groups) develop nonzero {\vev}s  thereby  reducing gauge symmetry.  The scale
of these {\vev}s is set by the FI term which is somewhat below the string scale. This  eventually
determines the seesaw scale.  In general, any {\sm} singlets can get large {\vev}s as long as it is
consistent with supersymmetry, and one can obtain the standard model gauge  symmetry times that
of the hidden sector, 
\begin{equation}
 \maG_{4D}~\longrightarrow ~G_{SM}\times G_\mathrm{hidden} \;.
\end{equation} 
The singlet {\vev}s are not necessarily associated with  flat directions in the field space and generally
correspond to isolated solutions to supersymmetry equations~\cite{Buchmuller:2006ik}. The hidden matter
gauge group  $G_\mathrm{hidden}$ can be responsible for spontaneous supersymmetry breaking. In fact,
within the class of models with the {\mssm} spectrum, gaugino condensation in the hidden sector favors
TeV--scale soft masses for the observable fields, as we have seen in section~\ref{subsec:SusyBreakingML}.

The nonzero singlet {\vev}s lead to the mass terms for the vectorlike states,
\begin{equation} 
 W~=~ x_i\, \bar x_j\, \langle s_a\, s_b\, \dots \rangle \;,
\end{equation}
where $x_i , \bar x_j$ are the vectorlike exotics and $\vevof{s_k}$ are the {\sm} singlet {\vev}s in
string units. Such a coupling must be consistent with string selection rules. In
section~\ref{sec:RParity}, it has been shown that many \Z6-II  models satisfy this requirement and all of
the vectorlike exotics can be decoupled. This results in the {\mssm} spectrum at low energies.

Similarly, the singlet {\vev}s induce ``Majorana'' mass terms for the {\sm} singlets
as well as the neutrino Yukawa couplings of eq.~\eqref{eq:seesawW},
\begin{equation}
M_{ij} \sim \langle s_a\, s_b\, \dots \rangle ~~,~~ Y_\nu^{ij} \sim  \langle s_\alpha \, s_\beta\, \dots \rangle~,
\end{equation}
as long as it is consistent with string selection rules.

Identification of right--handed neutrinos is intimately related to the issue of  baryon/lepton number
violation.  In generic vacua, any {\sm} singlet can play the role of the right--handed
neutrino. However, such vacua also suffer from excessive $R$-parity violating interactions. The simplest
way to suppress these interactions is to identify a $\BmL$ gauge symmetry and enforce either its
approximate conservation  or conservation of its  discrete  (matter parity)  subgroup. In local
{\gut}s, the $\BmL$ generator  resembles the standard {\gut} $\BmL$, but also requires extra \U1
components beyond \SO{10}. It is nonanomalous and produces the standard $\BmL$ charges for the {\sm}
matter. If $\BmL$ is broken by {\sc vev}s of fields carrying even charges under $\BmL$, the matter parity
$(-1)^{3(\BmL)}$ is conserved. This forbids dangerous  $R$-parity violating interactions and requires the
right-handed neutrino to carry the charge $q_{\BmL}=\pm 1$. Another possibility is that  $\U1_{\BmL}$ is
broken at an intermediate  scale $M_{\BmL}$ such that all  $R$-parity violating couplings are suppressed
by $M_{\BmL}/M_\mathrm{Pl}$. In this case,  Majorana mass terms for the right--handed neutrinos are
allowed only upon $\BmL$ breaking, which lowers  the seesaw scale to intermediate
energies. In what follows, we consider 
these possibilities in specific heterotic orbifold models.

\subsection[Seesaw Mechanism on the Orbifold-\mssm]{Seesaw Mechanism on the  Orbifold-\textbf{\textsc{mssm}}}
\label{subsec:SeesawInSeesaw}

The $\BmL$ generator is identified with 
\be\label{eq:SeesawBmL}
 \mathsf{t}_{\BmL}~=~ \left(-1 ,-1 , 0 , 0 , 0 , \tfrac{2}{3} , \tfrac{2}{3} , \tfrac{2}{3} \right)\,
 \left( -\tfrac{1}{2}, -\tfrac{1}{2} , 0 , -\tfrac{1}{2} , -\tfrac{1}{2} , 0 , 0 , 0 \right)\;.
\ee
In general supersymmetric configurations, many {\sm} singlets get nonzero {\vev}s. Choosing a subset of
such singlets with 0 or $\pm 2$  $\BmL$ charges, the  unbroken gauge symmetry is  
\begin{equation}
G_\mathrm{SM} \times G_\mathrm{hidden} \;,
\end{equation}
where $ G_\mathrm{hidden}= \SO8$, while all of the exotic states get large masses and decouple.
This vacuum preserves the matter parity $(-1)^{3(\BmL)}$.

We find that there are 39 right--handed neutrinos   defined by $q_{\BmL}= \pm1$, two of which are members
of the localized  $\boldsymbol{16}$--plets. They  have Yukawa couplings to the lepton doublets and large
Majorana mass terms. We have calculated the $3\times 39$ Yukawa matrix  $Y_\nu$ and $39\times 39$
Majorana mass matrix  $M$ of eq.~\eqref{eq:seesawW} up to order six in the singlet {\vev}s.  That is, for
each matrix element, we have determined at which order  in the superpotential a nonzero coupling is
allowed by string selection rules. Each entry depends on the quantum numbers and the localization of the
Majorana neutrinos, and involves products of different singlets and moduli-dependent Yukawa couplings. We
then assume that the main hierarchy in these entries comes from products of singlet {\vev}s so that these
matrices can be treated as textures.

The effective mass matrix for the left--handed neutrinos,
\begin{equation}
 M_\mathrm{eff}~=~ - v_u^2\,Y_\nu\, M^{-1}\, Y_\nu^T 
\end{equation}
with $v_u$ being the up--type Higgs {\vev}, 
can be represented by the texture  
\begin{equation}
M_\mathrm{eff}~\sim~ - \frac{v_u^2}{M_*}  \left( 
\begin{matrix}\label{texture}
1 & s & s \\
s & s^2 & s^2 \\
s & s^2 & s^2 
\end{matrix}
\right) \;.
\end{equation}
Here $s < 1$ represents a generic singlet {\vev} in string units and $M_*$ is the effective seesaw scale.
$Y_\nu$ contains entries with powers of $s$ between 1 and 5, while the dependence of the eigenvalues of
$M$  ranges from $s$ to $s^8$ (with no massless eigenstates at generic points in moduli space). This
results in a strong $s$-dependence of the effective seesaw scale $M_*$. This scale is further suppressed
by the large multiplicity of heavy singlets $N$, $M_* \propto N^{-x}$ with $0<x<2$. The value of $x$
depends on the texture. For example, when all the singlets contribute equally, $x=2$, whereas $x=0$ if
only a fixed number of neutrinos have nonnegligible couplings. For the present model, we find
\begin{equation}
 M_* ~\sim~ 0.1\, s^5\, M_{\rm str} ~\sim~ 10^{14}\,\mathrm{GeV}\;,
\end{equation}
for the string scale $M_{\rm str} = 2\cdot 10^{17}\,\mathrm{GeV}$ and $s\sim 0.3$.
The obtained texture~\eqref{texture} is of course model dependent.

The corresponding charged lepton Yukawa matrix is of the form 
\begin{equation}
Y_e~\sim~ \left( 
\begin{matrix}
 1 & 1 & c \\
 c & c & c^2 \\
 0 & 0 & 0 
\end{matrix}
\right) \;,
\end{equation}
where ``0'' denotes absence of the coupling up to order six in the singlet {\vev}s $c$. Such zeros are
expected to be filled in at higher orders. Here we are again using a single expansion parameter although
in  practice there are many variables.

These crude estimates show that reasonable fermion masses can in principle be obtained. Inserting order
one coefficients in the textures, one finds that the eigenvalues scale as
\begin{equation}
m_{\nu_i}~\sim ~(1, s^2, s^2)\,\frac{v_u^2}{M_*}\;,\quad  m_{e_i}~\sim~(1,c,0)\,v_d \;,
\end{equation}
where $v_d$ is the down--type Higgs {\vev}. For $s\sim 0.3$ and $c\sim 0.1$ the textures reproduce
roughly the observed lepton mass hierarchy. 
The above texture favors the normal neutrino mass hierarchy and can accommodate small and large mixing angles.
Further details of the model are available in appendix~\ref{ch:OrbifoldMSSMDetails}.

%% file: localGuts.pstex_t
\begin{picture}(0,0)%
\includegraphics{localGuts.pstex}%
\end{picture}%
\setlength{\unitlength}{4144sp}%
\begingroup\makeatletter\ifx\SetFigFont\undefined%
\gdef\SetFigFont#1#2#3#4#5{%
  \reset@font\fontsize{#1}{#2pt}%
  \fontfamily{#3}\fontseries{#4}\fontshape{#5}%
  \selectfont}%
\fi\endgroup%
\begin{picture}(6620,2047)(-152,-1271)
\put(453,611){\makebox(0,0)[lb]{\smash{{\SetFigFont{12}{14.4}{\rmdefault}{\mddefault}{\updefault}\SU5}}}}
\put(-152,-453){\makebox(0,0)[lb]{\smash{{\SetFigFont{12}{14.4}{\rmdefault}{\mddefault}{\updefault}$\maG_g=\SO{10}$}}}}
\put(4086,-226){\makebox(0,0)[lb]{\smash{{\SetFigFont{14}{16.8}{\rmdefault}{\mddefault}{\updefault}$\bs{\maG_{4D}=G_{SM}}$}}}}
\put(2886,-251){\makebox(0,0)[lb]{\smash{{\SetFigFont{14}{16.8}{\rmdefault}{\mddefault}{\updefault}$\bs{\Longrightarrow}$}}}}
\put(3503,-1141){\makebox(0,0)[lb]{\smash{{\SetFigFont{12}{14.4}{\rmdefault}{\mddefault}{\updefault}\sm\ generation}}}}
\put(3503,644){\makebox(0,0)[lb]{\smash{{\SetFigFont{12}{14.4}{\rmdefault}{\mddefault}{\updefault}\sm\ generation}}}}
\put(4953,644){\makebox(0,0)[lb]{\smash{{\SetFigFont{12}{14.4}{\rmdefault}{\mddefault}{\updefault}\sm\ generation}}}}
\put(986,-785){\makebox(0,0)[lb]{\smash{{\SetFigFont{12}{14.4}{\rmdefault}{\mddefault}{\updefault}$\rojo{\bs{16}}$}}}}
\put(1683,365){\makebox(0,0)[lb]{\smash{{\SetFigFont{10}{12.0}{\rmdefault}{\mddefault}{\updefault}$(\bs{4},\bs{2},\bs{1})$}}}}
\put(4953,-1141){\makebox(0,0)[lb]{\smash{{\SetFigFont{12}{14.4}{\rmdefault}{\mddefault}{\updefault}\sm\ generation}}}}
\put(1673,155){\makebox(0,0)[lb]{\smash{{\SetFigFont{10}{12.0}{\rmdefault}{\mddefault}{\updefault}$(\bsb{4},\bs{1},\bs{2})$}}}}
\put(1636,-829){\makebox(0,0)[lb]{\smash{{\SetFigFont{10}{12.0}{\rmdefault}{\mddefault}{\updefault}$\bs{10}+\bsb{5}+\bs{1}$}}}}
\put(1936,607){\makebox(0,0)[lb]{\smash{{\SetFigFont{11}{13.2}{\rmdefault}{\mddefault}{\updefault}$\SU4\x\SU2^2$}}}}
\put(2026,-1141){\makebox(0,0)[lb]{\smash{{\SetFigFont{11}{13.2}{\rmdefault}{\mddefault}{\updefault}$\SU5_\mathrm{f\,l}\x\U1$}}}}
\put(923,370){\makebox(0,0)[lb]{\smash{{\SetFigFont{12}{14.4}{\rmdefault}{\mddefault}{\updefault}$\bs{10}+\bsb{5}$}}}}
\end{picture}%

%% file: Z6II_geometryT1.pstex_t
\begin{picture}(0,0)%
\includegraphics{Z6II_geometryT1.pstex}%
\end{picture}%
\setlength{\unitlength}{4144sp}%
\begingroup\makeatletter\ifx\SetFigFont\undefined%
\gdef\SetFigFont#1#2#3#4#5{%
  \reset@font\fontsize{#1}{#2pt}%
  \fontfamily{#3}\fontseries{#4}\fontshape{#5}%
  \selectfont}%
\fi\endgroup%
\begin{picture}(6030,1489)(324,-718)
\put(1472,-659){\makebox(0,0)[lb]{\smash{{\SetFigFont{10}{12.0}{\rmdefault}{\mddefault}{\updefault}$e_1$}}}}
\put(4302,-669){\makebox(0,0)[lb]{\smash{{\SetFigFont{10}{12.0}{\rmdefault}{\mddefault}{\updefault}$e_3$}}}}
\put(5985,-669){\makebox(0,0)[lb]{\smash{{\SetFigFont{10}{12.0}{\rmdefault}{\mddefault}{\updefault}$e_5$}}}}
\put(4623,604){\makebox(0,0)[lb]{\smash{{\SetFigFont{10}{12.0}{\rmdefault}{\mddefault}{\updefault}$e_6$}}}}
\put(2592,604){\makebox(0,0)[lb]{\smash{{\SetFigFont{10}{12.0}{\rmdefault}{\mddefault}{\updefault}$e_4$}}}}
\put(2025,602){\makebox(0,0)[lb]{\smash{{\SetFigFont{10}{12.0}{\rmdefault}{\mddefault}{\updefault}$e_2$}}}}
\put(324,307){\makebox(0,0)[lb]{\smash{{\SetFigFont{10}{12.0}{\rmdefault}{\mddefault}{\updefault}$T_1$}}}}
\end{picture}%

%% file: Z6II_gutsT1.pstex_t
\begin{picture}(0,0)%
\includegraphics{Z6II_gutsT1.pstex}%
\end{picture}%
\setlength{\unitlength}{4144sp}%
\begingroup\makeatletter\ifx\SetFigFont\undefined%
\gdef\SetFigFont#1#2#3#4#5{%
  \reset@font\fontsize{#1}{#2pt}%
  \fontfamily{#3}\fontseries{#4}\fontshape{#5}%
  \selectfont}%
\fi\endgroup%
\begin{picture}(5835,1556)(324,-795)
\put(5911,-746){\makebox(0,0)[lb]{\smash{{\SetFigFont{10}{12.0}{\rmdefault}{\mddefault}{\updefault}$\blue{A_5}$}}}}
\put(4231,-741){\makebox(0,0)[lb]{\smash{{\SetFigFont{10}{12.0}{\rmdefault}{\mddefault}{\updefault}$\verde{A_3}$}}}}
\put(2436,419){\makebox(0,0)[lb]{\smash{{\SetFigFont{10}{12.0}{\rmdefault}{\mddefault}{\updefault}$\verde{A_3}$}}}}
\put(4236,129){\makebox(0,0)[lb]{\smash{{\SetFigFont{10}{12.0}{\rmdefault}{\mddefault}{\updefault}$2\x\bs{16}$}}}}
\put(324,307){\makebox(0,0)[lb]{\smash{{\SetFigFont{10}{12.0}{\rmdefault}{\mddefault}{\updefault}$T_1$}}}}
\end{picture}%

%% file: R-parity_violating_decay.pstex_t
\begin{picture}(0,0)%
\includegraphics{R-parity_violating_decay.pstex}%
\end{picture}%
\setlength{\unitlength}{3522sp}%
\begingroup\makeatletter\ifx\SetFigFont\undefined%
\gdef\SetFigFont#1#2#3#4#5{%
  \reset@font\fontsize{#1}{#2pt}%
  \fontfamily{#3}\fontseries{#4}\fontshape{#5}%
  \selectfont}%
\fi\endgroup%
\begin{picture}(4360,1924)(-166,-1396)
\put(1182,-169){\makebox(0,0)[lb]{\smash{{\SetFigFont{8}{9.6}{\rmdefault}{\mddefault}{\updefault}$\bar u$}}}}
\put(2817,-159){\makebox(0,0)[lb]{\smash{{\SetFigFont{8}{9.6}{\rmdefault}{\mddefault}{\updefault}$\bar d$}}}}
\put(2842,-1134){\makebox(0,0)[lb]{\smash{{\SetFigFont{8}{9.6}{\rmdefault}{\mddefault}{\updefault}$\ell$}}}}
\put(2112,-754){\makebox(0,0)[lb]{\smash{{\SetFigFont{8}{9.6}{\rmdefault}{\mddefault}{\updefault}$\bar d$}}}}
\put(242,251){\makebox(0,0)[lb]{\smash{{\SetFigFont{9}{10.8}{\rmdefault}{\mddefault}{\updefault}$d$}}}}
\put(242,-284){\makebox(0,0)[lb]{\smash{{\SetFigFont{9}{10.8}{\rmdefault}{\mddefault}{\updefault}$u$}}}}
\put(237,-1274){\makebox(0,0)[lb]{\smash{{\SetFigFont{9}{10.8}{\rmdefault}{\mddefault}{\updefault}$u$}}}}
\put(3777,-289){\makebox(0,0)[lb]{\smash{{\SetFigFont{9}{10.8}{\rmdefault}{\mddefault}{\updefault}$\bar d$}}}}
\put(3767,-1274){\makebox(0,0)[lb]{\smash{{\SetFigFont{9}{10.8}{\rmdefault}{\mddefault}{\updefault}$e^+$}}}}
\put(3782,256){\makebox(0,0)[lb]{\smash{{\SetFigFont{9}{10.8}{\rmdefault}{\mddefault}{\updefault}$d$}}}}
\put(4194, 54){\makebox(0,0)[lb]{\smash{{\SetFigFont{9}{10.8}{\rmdefault}{\mddefault}{\updefault}$\pi^0$}}}}
\put(1167,-1142){\makebox(0,0)[lb]{\smash{{\SetFigFont{9}{10.8}{\rmdefault}{\mddefault}{\updefault}$q$}}}}
\put(-166,-545){\makebox(0,0)[lb]{\smash{{\SetFigFont{9}{10.8}{\rmdefault}{\mddefault}{\updefault}$p$}}}}
\end{picture}%

%% file: RPreservingDecay.pstex_t
\begin{picture}(0,0)%
\includegraphics{RPreservingDecay.pstex}%
\end{picture}%
\setlength{\unitlength}{4144sp}%
\begingroup\makeatletter\ifx\SetFigFont\undefined%
\gdef\SetFigFont#1#2#3#4#5{%
  \reset@font\fontsize{#1}{#2pt}%
  \fontfamily{#3}\fontseries{#4}\fontshape{#5}%
  \selectfont}%
\fi\endgroup%
\begin{picture}(3908,1726)(222,-1130)
\put(583,-1019){\makebox(0,0)[lb]{\smash{{\SetFigFont{10}{12.0}{\rmdefault}{\mddefault}{\updefault}$u$}}}}
\put(3748,-1019){\makebox(0,0)[lb]{\smash{{\SetFigFont{10}{12.0}{\rmdefault}{\mddefault}{\updefault}$\ol\nu$}}}}
\put(222,-366){\makebox(0,0)[lb]{\smash{{\SetFigFont{10}{12.0}{\rmdefault}{\mddefault}{\updefault}$p$}}}}
\put(3760,352){\makebox(0,0)[lb]{\smash{{\SetFigFont{10}{12.0}{\rmdefault}{\mddefault}{\updefault}$u$}}}}
\put(4130,170){\makebox(0,0)[lb]{\smash{{\SetFigFont{10}{12.0}{\rmdefault}{\mddefault}{\updefault}$\pi^+,\,K^+$}}}}
\put(587,-142){\makebox(0,0)[lb]{\smash{{\SetFigFont{10}{12.0}{\rmdefault}{\mddefault}{\updefault}$d$}}}}
\put(587,346){\makebox(0,0)[lb]{\smash{{\SetFigFont{10}{12.0}{\rmdefault}{\mddefault}{\updefault}$u$}}}}
\put(3405,-72){\makebox(0,0)[lb]{\smash{{\SetFigFont{8}{9.6}{\rmdefault}{\mddefault}{\updefault}$\bar d,\,\bar s$}}}}
\put(2803,-620){\makebox(0,0)[lb]{\smash{{\SetFigFont{10}{12.0}{\rmdefault}{\mddefault}{\updefault}$W$}}}}
\put(1135,-237){\makebox(0,0)[lb]{\smash{{\SetFigFont{8}{9.6}{\rmdefault}{\mddefault}{\updefault}$q$}}}}
\put(1179,-891){\makebox(0,0)[lb]{\smash{{\SetFigFont{8}{9.6}{\rmdefault}{\mddefault}{\updefault}$q$}}}}
\put(2145,-255){\makebox(0,0)[lb]{\smash{{\SetFigFont{8}{9.6}{\rmdefault}{\mddefault}{\updefault}$q$}}}}
\put(2066,-894){\makebox(0,0)[lb]{\smash{{\SetFigFont{8}{9.6}{\rmdefault}{\mddefault}{\updefault}$\ell$}}}}
\put(1621,-457){\makebox(0,0)[lb]{\smash{{\SetFigFont{8}{9.6}{\rmdefault}{\mddefault}{\updefault}$\kappa^{(1)}$}}}}
\end{picture}%

%% file: Statistics.tex
%%%%%%%%%%%%%%%%%%%%%%%%%%%%%%%%%%%%%%%%%%%%%%%%%%%%%%%%%%%%
%CHAPTER 6
%%%%%%%%%%%%%%%%%%%%%%%%%%%%%%%%%%%%%%%%%%%%%%%%%%%%%%%%%%%%
\chapter{Beyond the Mini-Landscape}
\label{ch:MLII}

\begin{center}
\begin{minipage}[t]{14cm}
We broaden our search of promising models. We drop the requirement of  having a local \SO{10} or \E6
\gut\ and look for \Z6-II orbifold models with three Wilson lines that yield the exact spectrum of the
\mssm.  We find that the vast majority of these models present local {\gut}s at some fixed points.
Besides constructing new models, this helps us understand whether (and how) the ``intelligent'' search
strategy  based on local {\gut}s is more efficient than a ``blind scan''.  In addition, we
investigate whether \mssm\ candidates also arise from other \Z{N} orbifolds without 
discrete torsion. We find no model in \Z3 and \Z4 orbifolds that fulfill all our 
phenomenological criteria. In \Z6-I and \Z7 there is only a small set of realistic 
models. Finally, we comment on \SO{32} orbifolds and their phenomenological potential.
\end{minipage}
\end{center}

\section[Three Wilson Lines in the `Fertile Patch']{Three Wilson Lines in the `Fertile Patch'}
\label{sec:3WLGuts}
\index{Z6II models with@\Z6-II models with 3 WL!Mini-Landscape shifts}

In the Mini-Landscape study of chapter~\ref{ch:MLI}, we have analyzed \Z6-II orbifold
models with up to two Wilson lines and local \SO{10} and \E6 structures. There are only four shift
vectors that admit these local {\gut}s. Having fixed these shifts, we have scanned over possible Wilson
lines to get the \sm\ gauge group and other desirable features. 

In this section, we extend our previous analysis by allowing for three Wilson lines,
which is the maximal possible number of Wilson lines in the \Z6-II orbifold.
An immediate consequence of this is that all three matter generations  obtained
in this case would be distinct. In concrete, one generation would be located at the fixed point 
corresponding to the origin in the $T_1$ sector whereas the other two generations should be formed by pieces
distributed irregularly in the bulk and at other fixed points. Furthermore, we relax the requirement of
the hypercharge embedding into a local \SO{10} or \E6 {\gut}, while still having the correct \gut\ 
hypercharge normalization. 

Our results are presented in table~\ref{tab:Summary_3WL}.  Note the difference
in step \ding{"0C2} compared to that in  the two Wilson line case: now we do not
require  the hypercharge embedding  in \SU5$~\subset~$\SO{10}  at this step,
whereas  at step \ding{"0C4} we require   $\U1_{Y}\subset \SU5 $ with \SU5 not
necessarily being inside \SO{10} (or \E6). This allows us to retain more models while
keeping the standard \gut\ hypercharge normalization.

Compared to the two-Wilson-line case, the total number of inequivalent models has grown
from $3\times 10^4$ to $10^6$. In the end, however, we retain only 81 models. Thus the efficiency is much
lower than that in the two-Wilson-line case. It is interesting that most of the models at step \ding{"0C7} come form
the \E6 local \gut\ with the gauge shift $V^{\E6,1}$. The fact that \E6 models contribute much
more in the three-Wilson-line  than two-Wilson-line case is understood by symmetry breaking: it is
easier to get to the \sm\ gauge group from \E6 using three Wilson lines.

\begin{table}[t!]
\begin{center}
\begin{tabular}{|l||r|r||r|r|}
\hline
 Criterion & $V^{\SO{10},1}$ & $V^{\SO{10},2}$ & $V^{\E6,1}$ & $V^{\E6,2}$\\
\hline\hline
\rowgrayh
\ding{"0C1} ineq. models with 3 WL
  & $942,469$ & $246,779$ & $8,815$ & $37,407$ \\[0.2cm]
\ding{"0C2} SU(3)$\times$SU(2) gauge group
  & $373,412$ & $89,910$  & $2,321$ & $13,857$\\[0.2cm]
\rowgrayh
\ding{"0C3} 3 net $(\boldsymbol{3},\boldsymbol{2})$
  & $5,853$   & $2,535$   & $352$ & $745$\\[0.2cm]
\ding{"0C4} nonanomalous $\U1_{Y}\subset \SU5 $
  & $2,620$   & $1,294$   & $314$ & $420$\\[0.2cm]
\rowgrayh
\ding{"0C5} spectrum $=$ 3 generations $+$ vectorlike
  & $45$ & $19$ & $123$ & $0$\\[0.2cm]
\ding{"0C6}  heavy top
  & $44$ & $1$ & $123$ & $0$\\[0.2cm]
\rowgrayh
\ding{"0C7}  exotics decouple at order 8
  & $20$ & $1$ & $60$ & $0$ \\
\hline
\end{tabular}
\end{center}
\caption{Statistics of $\Z6-$II orbifold models based on the shifts
 $V^{\SO{10},1},V^{\SO{10},2},V^{\E6,1},V^{\E6,2}$ with three
 nontrivial Wilson lines.}
 \label{tab:Summary_3WL}
\end{table}

\section{General Models with three Wilson Lines}
\label{sec:statistics3WL}
\index{Z6II models with@\Z6-II models with 3 WL!all models}

By following the statistical method proposed in sec.~\ref{sec:StatisticalMethod}, we perform a
search of models with three Wilson lines based on all 61 admissible shift vectors of \Z6-II orbifolds,
not only on the Mini-Landscape shifts. In the procedure described in sec.~\ref{sec:StatisticalMethod}, we
have seen that the total number of models of a certain class can be estimated by
\begin{equation}
\label{eq:estim_N}
  N~\simeq~\left(n_0\,t \right)^2\;,
\end{equation}
in terms of the predominant size of random-generated sets of models $n_0$ and a fudge parameter
$t$ that can be determined from previous results. 

\begin{figure}[h!]
\psfrag{n}{\footnotesize$n$}
\psfrag{num}{\footnotesize\# of samples}
\begin{minipage}{0.48\textwidth}
(a)\; $V^{SO(10),1}$\\[3mm]
\CenterEps[0.55]{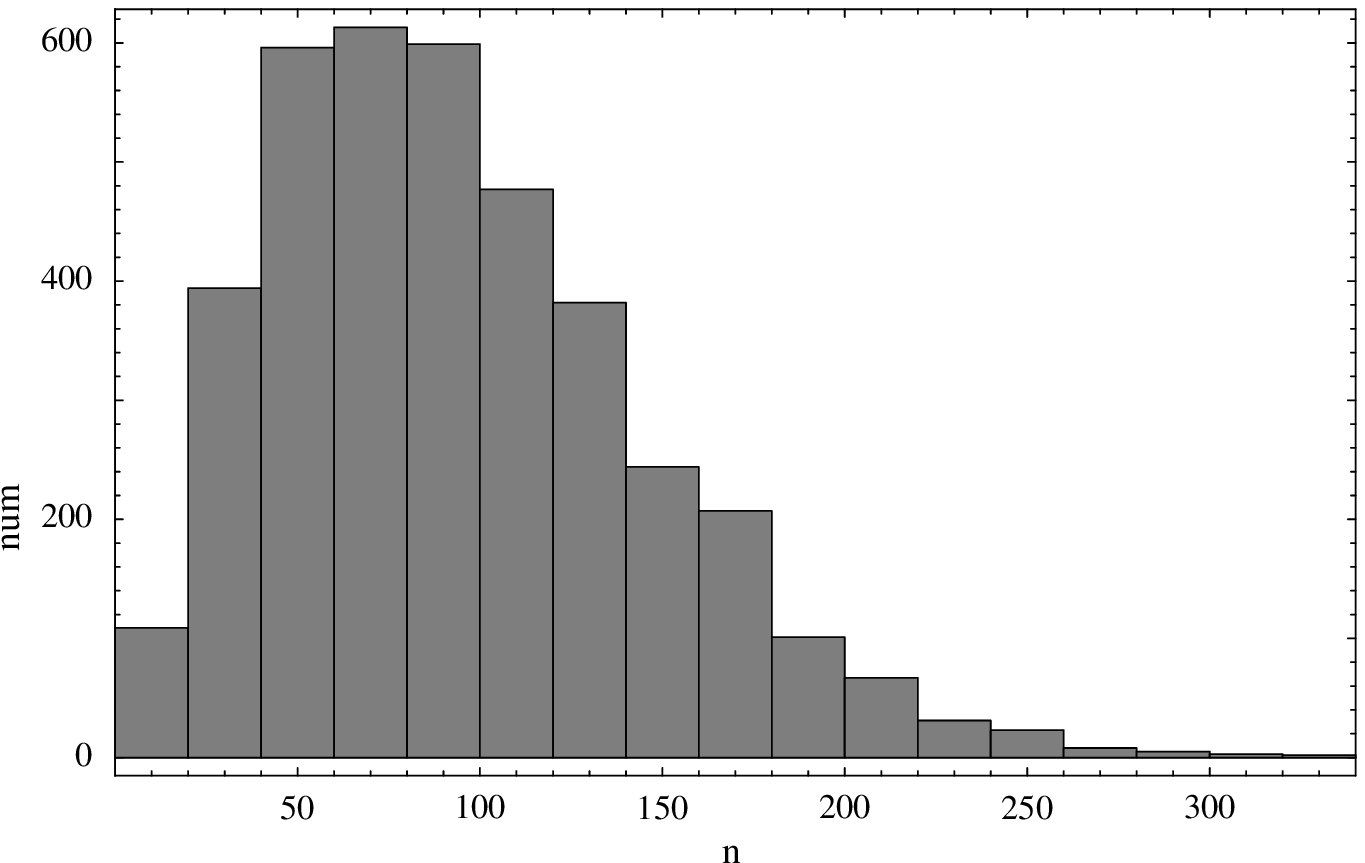}
\end{minipage}
\hskip 2mm
\begin{minipage}{0.48\textwidth}
(b)\;$V^{SO(10),2}$\\[3mm]
\CenterEps[0.55]{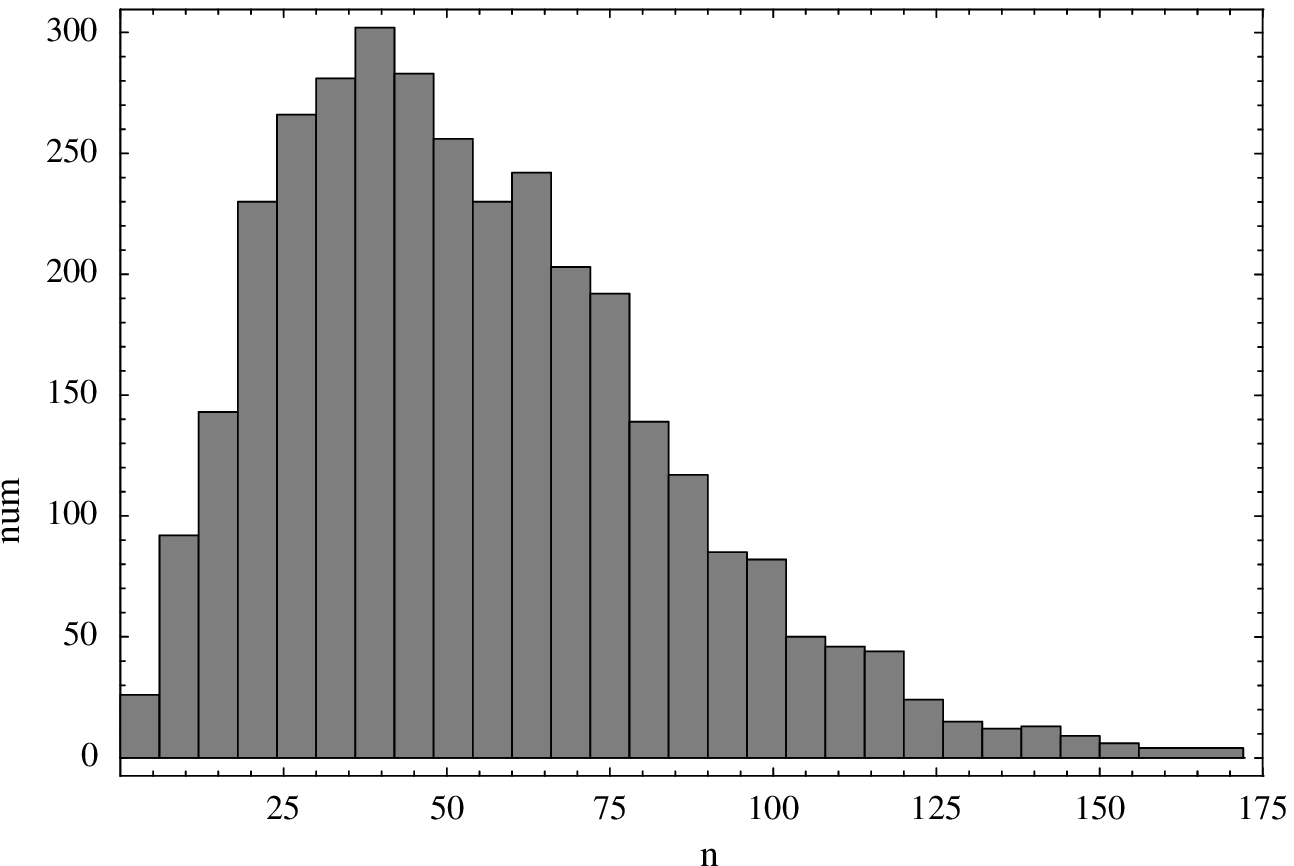}
\end{minipage}
\caption{Distribution of set sizes of random generated models with 2 Wilson lines. We find (a) $n_0\sim
  73.5$ and (b) $n_0\sim 42.3$. Comparing eq.~\eqref{eq:estim_N} and the total number of models in each
  case (see table~\ref{tab:Summary}), the fudge parameter $t$ is set to (a) $t\sim 2.02$ and (b) $t\sim 2.09$.}  
\label{fig:2WL}
\end{figure}

The parameter $t$ can be set 
by comparing the number of models $N$ with two Wilson lines of our previous scan (see
table~\ref{tab:Summary}) and the value $n_0$ obtained through the statistical method. For example,
consider the models with shift vector $V^{SO(10),1}$ and two Wilson lines. We have found that there are
$N\sim22,000$ inequivalent models. On the other hand, after generating about 4,000 sets of models, the
resulting sample size distribution looks as in fig.~\ref{fig:2WL}(a). The predominant set size is
$n_0\sim 73.5$, implying $t\sim 2.02$. This value can be improved if we also consider the other three
shift vectors, for which we already know the amount of inequivalent models. Table~\ref{tab:2WLSettingt}
displays the obtained values for $t$ and $n_0$. The variations of $t$ are taken into account if one
considers the weighted average of $t$
\begin{equation}
\label{eq:global_t}
t = \frac{\sum_{(V)}N^{(V)} t^{(V)}}{\sum_{V's}N^{(V)}} \sim 2.16\;,
\end{equation}
where $(V)$ denotes any of the Mini-Landscape vectors. We shall take this value as the best estimate of
$t$. It will be used to estimate the total number of inequivalent models in other cases. The fifth column
of table~\ref{tab:2WLSettingt} correspond to the estimated number of inequivalent models when one replaces
$t\sim2.16$ in eq.~\eqref{eq:estim_N}. We see that there is an overall uncertainty of about 1\%. Other
uncertainties arise from our method to distinguish the inequivalent models. We have found
that comparing only the nonabelian quantum numbers of the matter states and the number of singlets is not
enough. Two models considered equivalent from these criteria might still differ in the localization and
\U1 charges of the matter states, leading e.g. to different Yukawa couplings. Empirically, the resulting
uncertainty is found to be within a factor of 2.

\begin{table}[t!]
\begin{center}
\begin{tabular}{|c|r|r|r|r|}
\hline
Shift & \# ineq. models & \multicolumn{1}{c|}{$n_0$} & \multicolumn{1}{c|}{$t$} & \multicolumn{1}{c|}{$N$ from eq.~\eqref{eq:estim_N}}\\
      & (cf. table~\ref{tab:Summary}) &       &     & \multicolumn{1}{c|}{with $t\sim 2.16$}\\
\hline\hline
\rowgrayh
$V^{SO(10),1}$ & $22,000$ &  $73.5$ & $2.02$ & $25,183$\\
$V^{SO(10),2}$ &  $7,800$ &  $42.3$ & $2.09$ &  $8,341$\\
\rowgrayh
$V^{E_6,1}$    &    $680$ &  $11.4$ & $2.29$ &    $606$\\
$V^{E_6,2}$    &  $1,700$ &  $17.7$ & $2.33$ &  $1,460$\\
\hline
\end{tabular}
\end{center}
\caption{Determining the value of the parameter $t$. For each of the Mini-Landscape shifts, we compare
 the number of inequivalent models with two Wilson lines found previously and the statistically
 predominant sample size $n_0$.
\label{tab:2WLSettingt}} 
\end{table}

\begin{figure}[h!]
\psfrag{n}{\footnotesize$n$}
\psfrag{num}{\footnotesize\# of samples}
\CenterEps[0.7]{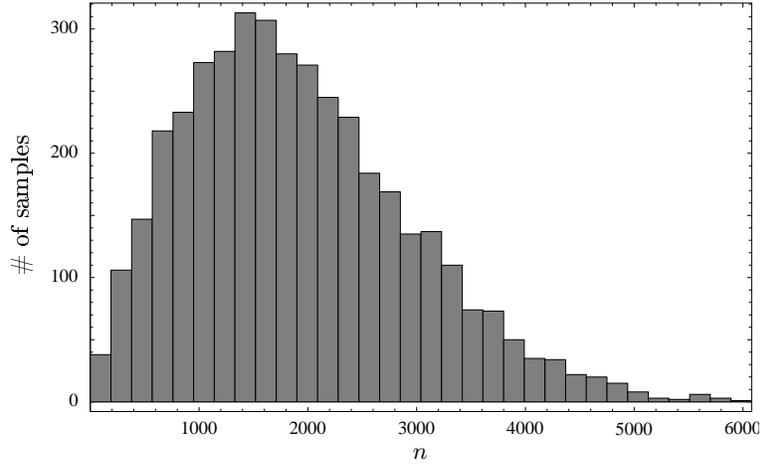}
\caption{Distribution of set sizes of random generated models with 3 Wilson lines 
  The maximum occurs at $n_{0}\sim 1510.3$. Eq.~\eqref{eq:estim_N} implies
  that the estimated total number of inequivalent models is about $1.06\times 10^7$.}
\label{fig:3WL_All}
\end{figure}

\vskip 4mm
One of the key results of this study is the total number of models for the \Z6-II orbifold. We can apply
the statistical method to models with arbitrary shift and up to three Wilson lines.
As illustrated in fig.~\ref{fig:3WL_All}, we find $n_0\sim1510.3$. Hence, the total number of
inequivalent models is around $1.06\times 10^7$.
 This means particularly that the 223 promising models found in the Mini-Landscape constitute about
$0.002\%$ of all inequivalent \Z6-II orbifold models from the \E8\x\E8 heterotic string.

The explicit construction of all \Z6-II models is rather time--consuming. However, the statistical method
used here allows us to arrive to a very large set of models.
Out of the $10^7$ models with up to three Wilson lines, we have constructed explicitly all possible
models with two Wilson lines (a total of $\sim499,000$ models) and a sample of $5\times 10^6$  models
with three Wilson lines. 
This resulted in 267 \mssm\ candidates (i.e. models satisfying all criteria \ding{"0C0}-\ding{"0C7}),
$193$ with two Wilson lines and $74$ with three Wilson lines.\footnote{Here we only obtain 74 {\mssm}s
  with three Wilson lines which is fewer than the number in table~\ref{tab:Summary_3WL}. This is because
  our sample of $5\times 10^6$ models does not contain all models with local 
\SO{10} and \E6. } Most of them originate from  \E6, \SO{10} and \SU5 local
{\gut}s as shown in table~\ref{tab:LocalGUTStructurePromisingModels}. Note that
models with \SU5 local structure do  not have a complete localized family,
rather only part of it. The additional states come from other sectors of the
model.  The conclusion is that any model with  the exact \mssm\ spectrum, gauge
coupling unification and a heavy top quark is likely to have come from some
local \gut.

\begin{table}[h!]
\centerline{
\begin{tabular}{|c|c|r|r|}
\hline
 Local {\gut} & ``family'' &  2 WL & 3  WL \\
\hline\hline
\rowgrayh
\E6          & $\bs{27}$ & $14$  & $53$ \\[2mm]
\SO{10}      & $\bs{16}$ & $87$  &  $7$ \\[2mm]
\rowgrayh
\SU6         & $\bs{15}$+$\bs{\bar 6}$ &  $2$  &  $4$ \\[2mm]
\SU5         & $\bs{10}$ & $51$  & $10$ \\[2mm]
\rowgrayh
rest &                 & $39$  &  $0$ \\[4mm]
\hline
total &                 & $193$ & $74$ \\
\hline
\end{tabular}
}
\caption{Local {\gut} structure  of the  \mssm\ candidates. These gauge groups
appear at some fixed point(s) in the $T_1$ twisted sector. The \SU5 local \gut\
does not  produce a complete family, so additional ``non-{\gut}'' states are
required.}
\label{tab:LocalGUTStructurePromisingModels}
\end{table}

\subsection{Hidden Sector Statistics of \textbf{\textsc{mssm}} Candidates}

\begin{figure}[!t!]
\begin{minipage}{0.48\textwidth}
(a)\\
\CenterEps[0.75]{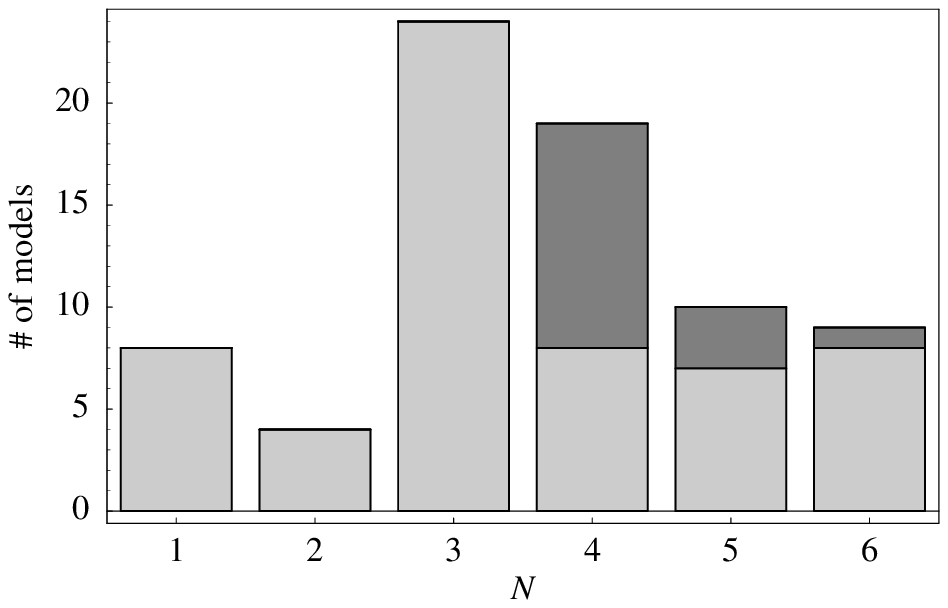}
\end{minipage}
\hskip 2mm
\begin{minipage}{0.48\textwidth}
(b)\\
\CenterEps[0.75]{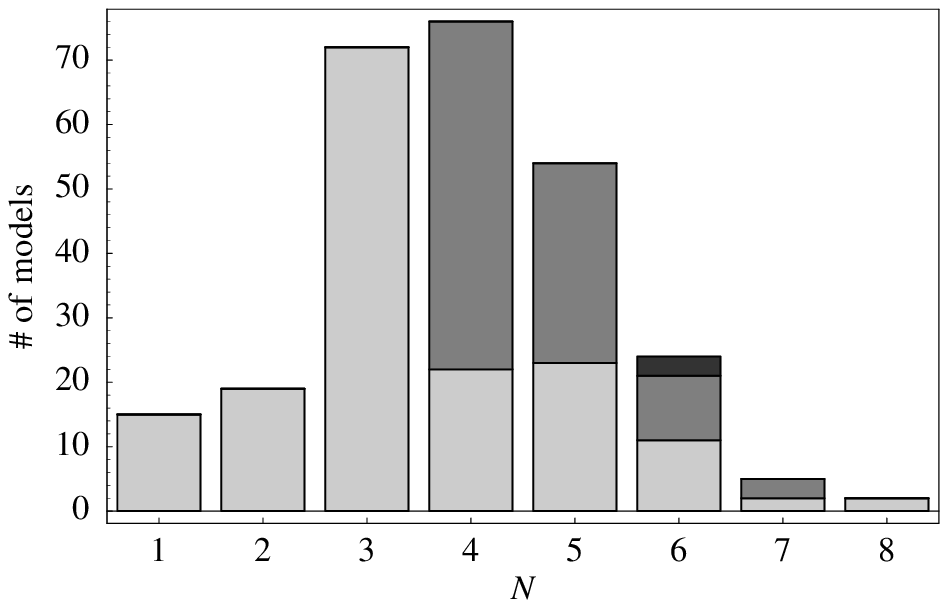}
\end{minipage}
\caption{ Hidden sector `rank' distribution for (a) models with 3 nontrivial
Wilson lines only, and (b) the combined set of models with 2 and 3 
Wilson lines. \SU{N}/\SO{2N}/\E{N} are given by light/dark/darker bins.
\label{fig:ComparingHiddenSector3WLVs2-3WL}}
\end{figure}

From the Mini-Landscape we obtained an interesting correlation between the scale of {\sc susy} breaking
and other realistic properties of promising orbifold models. We would like to examine whether this property
holds in the current study. As we have seen, the size of the hidden sector determines the scale of
gaugino condensation $\Lambda$  and consequently the scale of soft \susy\ breaking masses $m_{3/2}$
\begin{equation}
 m_{3/2}~\sim~\frac{\Lambda^3}{M_\mathrm{Pl}^2}\;,
\end{equation}
where $M_\mathrm{Pl}$ denotes the Planck scale. If the largest hidden sector
gauge factor  is too big, e.g. \E6  or \E8, the gaugino condensation scale is
too high and supersymmetry is irrelevant to low energy physics. If it is too
low, the model is ruled out by experiment. For orbifold models with three Wilson lines, we present  the
statistics of the hidden sector gauge groups in fig.~\ref{fig:ComparingHiddenSector3WLVs2-3WL}(a). There $N$ labels the
``size'' of the gauge groups  $\SU{N}$ and $\SO{2N}$. Although the peak of this
distribution is at $N=3$ corresponding to \SU3, a significant fraction of the
models have $N=4,5$ which  leads  (in the absence of hidden matter) to  gaugino
condensation at an intermediate scale. If \susy\ breaking is due to gaugino
condensation, the corresponding soft masses are in the TeV range as favored by
phenomenology.

\begin{figure}[!h!]
\CenterEps[0.75]{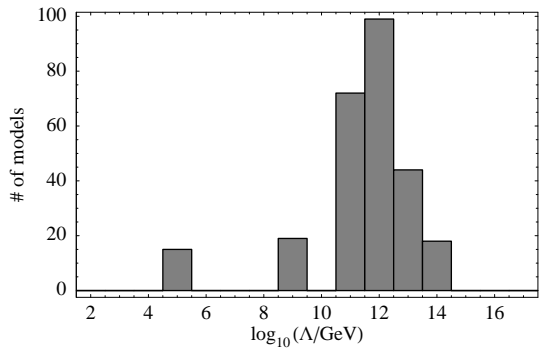}
\caption{Number of \mssm\ candidates with 2 and 3 Wilson lines vs. the scale of gaugino condensation. 
\label{fig:HiddenSector2WLAnd3WLCombined}}
\end{figure}

Combining all models with two and three Wilson lines, we get the distribution peaking at $N=4$
displayed in fig.~\ref{fig:ComparingHiddenSector3WLVs2-3WL}(b).  The corresponding gaugino
condensation scales are plotted in fig.~\ref{fig:HiddenSector2WLAnd3WLCombined}.
Notice that fig.~\ref{fig:HiddenSector2WLAnd3WLCombined} is almost identical to fig.~\ref{gc}, obtained
in the case with two Wilson lines.
 It is remarkable that,
just as in the case with two Wilson lines, requiring the exact \mssm\ spectrum in the observable sector
constrains the hidden sector such that gaugino condensation at an intermediate scale is
automatically preferred. This provides a top-down motivation for TeV scales in
particle physics.

\section[Other \Z{N} Orbifolds]{Other \bs{\Z{N}} Orbifolds}
\label{sec:other_orbifolds}
\index{MSSM candidates in ZN@\mssm\ candidates in \Z{N} orbifolds}

\Z6-II orbifolds yield a rich set of \mssm\ candidates. An interesting question is whether this property
is particular to this kind of orbifolds. The fact that \Z3 is a subset of \Z6-II and, therefore, there is
a complex plane with three equivalent fixed points could be one advantage of \Z6 against \Z4
orbifolds. However, we have seen that \mssm\ candidates have rather a structure with only two equivalent
matter generations. Hence, one might argue that \Z4 could also produce good \mssm\ candidates. 

In this section, we search for realistic models in the context of \Z3, \Z4, \Z6-I and \Z7 orbifold
compactifications. We are interested in torsionless models with the properties \ding{"0C0}--\ding{"0C5},
that is, models where the chiral matter matter reproduces the exact spectrum of the \mssm\ and the exotic
particles are vectorlike. We do not verify the explicit decoupling of the exotics.

\begin{table}[h!]
\begin{center}
\begin{tabular}{|l||r|r|r|r|}
\hline
 criterion 
  & \multicolumn{1}{c|}{\Z3} & \multicolumn{1}{c|}{\Z4} & \multicolumn{1}{c|}{\Z6-I} & \multicolumn{1}{c|}{\Z7} \\
\hline
\hline
\rowgrayh
maximal \# of Wilson lines
  & \multicolumn{1}{c|}{\cellgrayh$3$} & \multicolumn{1}{c|}{\cellgrayh$4$} 
  & \multicolumn{1}{c|}{\cellgrayh$1$} & \multicolumn{1}{c|}{\cellgrayh$1$} \\[1mm]
\ding{"0C1} ineq. models
  & $\sim3,500$ & $\sim261,000$  & $\sim4,600$ & $\sim630,000$\\[0.2cm]
\rowgrayh
\ding{"0C2} SU(3)$\times$SU(2) gauge group
  & $964$   & $46,816$   & $1,193$ & $344,255$\\[0.2cm]
\ding{"0C3} 3 net $(\boldsymbol{3},\boldsymbol{2})$
  & $228$   & $683$   & $217$ & $16,536$\\[0.2cm]
\rowgrayh
\ding{"0C4} nonanomalous $\U1_{Y}\subset \SU5 $
  & $99$ & $508$ & $123$ & $4,046$\\[0.2cm]
\ding{"0C5} spectrum $=$ 3 generations $+$ vectorlike
  & $0$ & $0$ & $30$ & $1$\\[0.2cm]
\hline
\end{tabular}
\end{center}
\caption{Statistics of \Z{N} orbifolds with factorizable lattices and
maximal number of Wilson lines.
\label{tab:ZNorbifoldStatistics}}
\end{table}

Our results are presented in table~\ref{tab:ZNorbifoldStatistics}. In \Z3 and \Z4, there are several
models with three generations of quarks and leptons and nonanomalous hypercharge. However, their matter
content includes some exotics that are not vectorlike. 

In contrast, there are \Z6-I and \Z7 orbifold models that fulfill all our requirements. It is remarkable
that in \Z6-I about $0.7\%$ of the total of models have realistic properties. We observe that these
models have one family located at a fixed point (with or without local \gut). The other two families
arise from different twisted and untwisted sectors. In fact, it has been suggested that this picture 
may be phenomenologically favored~\cite{Nilles:2004ej}. In this sense, \Z6-I viable models are similar to
\Z6-II models with three Wilson lines. We find only one \Z7 orbifold model that passes our criteria.

This study tells us that, despite the fact that \Z6-II orbifolds with local {\gut}s are indeed very appealing, 
other regions of the orbifold landscape could also render interesting vacua. A complete analysis of the
models found here and of other successful geometries, such as \Z{12}~\cite{Kim:2006hw}, is beyond the
scope of this thesis and will be done elsewhere.

\section[A Word on \SO{32} Orbifolds]{A Word on \bs{\SO{32}} Orbifolds}
\index{MSSM candidates in SO@\mssm\ candidates in \SO{32} orbifolds}

There are good reasons to think that \SO{32} orbifolds are less promising than  compactifications of
the E$_8\times$E$_8$ heterotic string. First, spinor representations appear only at the orbifold fixed points
(see appendix~\ref{ch:SpinorsInSO32} for a complete discussion). Noticing that matter generations can be
embedded in spinorial representations of \SO{10} {\gut}s indicates that it is less likely to get models with
\mssm\ matter. Secondly, there appear larger gauge groups, such as \SU{15} which are more difficult to
get broken to $\mathcal{G}_{SM}$. Finally, the Wilson lines in \SO{32} orbifolds are more constrained by modular
invariance and the embedding conditions. This is due to the fact that the length of spinorial
Wilson lines in \SO{32} is larger than in E$_8\times$E$_8$.

However, these claims do not mean that the SM cannot appear from \SO{32} heterotic orbifolds. Therefore, it is
unexplainable that there has been almost no effort in trying to get realistic \SO{32} compactifications.
To amend this, we have performed a search of \mssm\ candidates in $\mathbbm{Z}_3$, $\mathbbm{Z}_4$ and 
$\mathbbm{Z}_6$-II orbifolds. In $\mathbbm{Z}_3$
and $\mathbbm{Z}_4$ orbifold models, the situation is quite similar to that of E$_8\times$E$_8$ orbifolds. At step
(5) we find some models (comparatively, less models than in E$_8\times$E$_8$), but none of them is
retained after demanding vectorlikeness of the exotics, at step (6).

In $\mathbbm{Z}_6$-II orbifolds the situation is different. We have studied models with arbitrary shift vector and up to
two Wilson lines. There are about $380,000$ inequivalent models. Among them, only $2,276$ models have the
\sm\ gauge group, including nonanomalous hypercharge and three generations of quarks and
leptons. The exotic particles can be decoupled from the low-energy spectrum in $87$ of these models.

It is noteworthy to compare our findings in \SO{32} and E$_8\times$E$_8$. First of all, in E$_8\times$E$_8$ we have found a total of
$352$ models with up to two Wilson lines at step (6), that is, about a factor $4$ more models
than in \SO{32}. This is related to the difficulty of getting spinors in \SO{32} orbifolds. On the other
hand, looking at the low energy spectrum, the features of the models in both scenarios are very similar.
In addition, we find that a good fraction (about $70\%$) of the \SO{32} models exhibit local {\gut}s too. 
From these results, it seems reasonable to argue that
there is no valid explanation to disregard semirealistic constructions appearing from the \SO{32} heterotic string.
In the following section we present some properties of one promising \SO{32} orbifold.
A complete discussion of these models and their phenomenological potential will be done elsewhere. 

\subsection{Phenomenology of SO(32) Orbifolds}

In order to illustrate the qualities of potentially realistic \SO{32} orbifolds, let us consider a
$\mathbbm{Z}_6-$II orbifold model parametrized by the following gauge embedding:
\begin{eqnarray}
V &=& \tfrac16 \left(3^2,\,-1^3,\,-2^2,\,-3^9\right)\,,\nonumber\\
A_3 &=& \tfrac13 \left(0,\,3,\,1^2,\,-1,\,0,\,-2,\,0^2,\,-3,\,1^2,\,-1,\,0,\,1^2\right)\,,\\
A_5 &=& \tfrac12 \left(1^2,\,-1,\,0,\,1,\,-1,\,1,\,-1^2,\,0^3,\,1^4\right)\,.\nonumber
\end{eqnarray}
The unbroken gauge group in four dimensions is $\mathcal{G}_{4D}=\mathcal{G}_{SM}\times$\,SO(10)\,$\times$\,U(1)$^7$, where the hypercharge 
generator takes the \gut--like form
\begin{equation}
t_Y ~=~ \left( 0^{10},\,\tfrac12^2,\,\tfrac13,\,0,\,-\tfrac13^2\right)\,. 
\end{equation}
The matter spectrum of this model is provided in table~\ref{tab:SO32Z6model}. It comprises three \mssm\ matter
families plus additional vectorlike exotics, which decouple from the low energy theory at order 8 in
singlets $\widetilde{s}=\{s^0_i,\chi_i\}$, once these states acquire VEVs. In the generic vacuum, the relevant mass matrices are given by\\
{\scriptsize 
\begin{subequations}
\noindent
\begin{minipage}{0.57\textwidth}
\begin{eqnarray}
\mathcal{M}_{\bar\ell\ell}
&\!\!\!\!\! =\! & \!\!\!\!\!
\left(
\begin{array}{cccccccc} 
\widetilde{s}^3 & \widetilde{s}^2 & \widetilde{s}^2 & \widetilde{s}^2 &  \widetilde{s}^2 & \widetilde{s}^2 & \widetilde{s}^2 & \widetilde{s}^3\\
\widetilde{s}^3 & \widetilde{s}^2 & \widetilde{s}^2 & \widetilde{s}^2 &  \widetilde{s}^2 & \widetilde{s}^2 & \widetilde{s}^2 & \widetilde{s}^3\\
\widetilde{s} & \widetilde{s}^2 & \widetilde{s}^2 & \widetilde{s}^3 &  \widetilde{s}^3 & \widetilde{s}^3 & \widetilde{s}^3 & \widetilde{s}^4\\
\widetilde{s} & \widetilde{s}^2 & \widetilde{s}^2 & \widetilde{s}^3 &  \widetilde{s}^3 & \widetilde{s}^3 & \widetilde{s}^3 & \widetilde{s}^4\\
\widetilde{s}^3 & \widetilde{s}^3 & \widetilde{s}^3 & \widetilde{s}^3 &  \widetilde{s}^3 & \widetilde{s}^3 & \widetilde{s}^3 & \widetilde{s}^3
\end{array}
\right)\;,\label{eq:MbarllSO32}\\
\mathcal{M}_{\delta\bar\delta}
&\!\!\!\!\! =\! & \!\!\!\!\!
\left(
\begin{array}{cccccccc}
 \widetilde{s}^4 & \widetilde{s} & \widetilde{s}^5 &  \widetilde{s}^3 & \widetilde{s}^5 & \widetilde{s}^3 &  \widetilde{s} & \widetilde{s}\\
 \widetilde{s}^4 & \widetilde{s}^3 & \widetilde{s}^5 &  \widetilde{s} & \widetilde{s}^5 & \widetilde{s}^3 &  \widetilde{s} & \widetilde{s}\\
 \widetilde{s}^2 & \widetilde{s}^3 & \widetilde{s}^3 &  \widetilde{s}^3 & \widetilde{s}^3 & \widetilde{s}^3 &  \widetilde{s}^4 & \widetilde{s}^4\\
 \widetilde{s}^2 & \widetilde{s}^3 & \widetilde{s}^5 &  \widetilde{s}^3 & \widetilde{s}^5 & \widetilde{s} &  \widetilde{s}^3 & \widetilde{s}^3\\
 \widetilde{s}^3 & \widetilde{s}^4 & \widetilde{s}^3 &  \widetilde{s}^4 & \widetilde{s}^3 & \widetilde{s}^4 &  \widetilde{s}^5 & \widetilde{s}^5\\
 \widetilde{s}^3 & \widetilde{s}^4 & \widetilde{s}^3 &  \widetilde{s}^4 & \widetilde{s}^3 & \widetilde{s}^4 &  \widetilde{s}^5 & \widetilde{s}^5\\
 \widetilde{s}^6 & \widetilde{s}^5 & \widetilde{s} &  \widetilde{s}^5 & \widetilde{s} & \widetilde{s}^4 &  \widetilde{s}^3 & \widetilde{s}^3\\
 \widetilde{s}^6 & \widetilde{s}^5 & \widetilde{s} &  \widetilde{s}^5 & \widetilde{s} & \widetilde{s}^4 &  \widetilde{s}^3 & \widetilde{s}^3
\end{array}
\right),\label{eq:MdeltabardeltaSO32}
\end{eqnarray}
\end{minipage}
\begin{minipage}{0.43\textwidth}
\begin{eqnarray}
\mathcal{M}_{e\bar{e}} \!\!
&\!\! =\!\! & \!\!\!\!
\left(
\begin{array}{cccc}
 \widetilde{s}^3 & \widetilde{s}^3 & \widetilde{s}^4 & \widetilde{s}^4
\end{array}
\right)\;,\label{eq:MebeSO32}\\
\mathcal{M}_{\bar q q}\!\!
&\!\! =\!\! & \!\!\!\!
\left(
\begin{array}{cccc}
 \widetilde{s}^3 & \widetilde{s}^3 & \widetilde{s}^3 & \widetilde{s}^3
\end{array}
\right)\;,\label{eq:MbqqSO32}\\
\mathcal{M}_{u\bar{u}}\!\!
&\!\! =\!\! & \!\!\!\!
\left(
\begin{array}{cccc}
 \widetilde{s}^3 & \widetilde{s}^3 & \widetilde{s}^4 & \widetilde{s}^4
\end{array}
\right)\;,\label{eq:MubuSO32}\\
\mathcal{M}_{\phi\bar\phi}\!\!
&\!\! =\!\! & \!\!\!\!
\left(
\begin{array}{ccccc}
 \widetilde{s}^3 & \widetilde{s}^3 & \widetilde{s}^3 & \widetilde{s} & \widetilde{s}\\
 \widetilde{s}^3 & \widetilde{s}^5 & \widetilde{s}^3 & \widetilde{s}^5 & \widetilde{s}^5\\
 \widetilde{s}^3 & \widetilde{s}^3 & \widetilde{s}^3 & \widetilde{s} & \widetilde{s}\\
 \widetilde{s}^5 & \widetilde{s}^4 & \widetilde{s}^5 & \widetilde{s}^3 & \widetilde{s}^3\\
 \widetilde{s}^5 & \widetilde{s}^4 & \widetilde{s}^5 & \widetilde{s}^3 & \widetilde{s}^3
\end{array}
\right)\;,
\end{eqnarray}
\end{minipage}

\noindent
\begin{minipage}{0.57\textwidth}
\begin{eqnarray}
\mathcal{M}_{d\bar d}
&\!\!\!\!\! =\! & \!\!\!\!\!
\left(
\begin{array}{cccccc}
 \widetilde{s}   & \widetilde{s}^3 & \widetilde{s} & \widetilde{s}   & \widetilde{s}   & \widetilde{s}\\
 \widetilde{s}^5 & \widetilde{s} & \widetilde{s}^3 & \widetilde{s}^3 & \widetilde{s}^3 & \widetilde{s}^3\\
 \widetilde{s}^5 & \widetilde{s} & \widetilde{s}^3 & \widetilde{s}^3 & \widetilde{s}^3 & \widetilde{s}^3
\end{array}
\right)\;,\label{eq:MdbardSO32}
\end{eqnarray}
\end{minipage}
\begin{minipage}{0.43\textwidth}
\begin{eqnarray}
\mathcal{M}_{w\bar{w}}\!\!
&\!\! =\!\! & \!\!\!\!
\left(
\begin{array}{cccc}
 \widetilde{s}^2 & \widetilde{s} & \widetilde{s}^2 & \widetilde{s}^4\\
 \widetilde{s}^5 & \widetilde{s}^5 & \widetilde{s}^5 & \widetilde{s}^5\\
 \widetilde{s}^2 & \widetilde{s}^4 & \widetilde{s}^2 & \widetilde{s}\\
 \widetilde{s}^5 & \widetilde{s}^5 & \widetilde{s}^5 & \widetilde{s}^5\\
\end{array}
\right)\;.\label{eq:MwbwSO32}
\end{eqnarray}
\end{minipage}
\end{subequations}
}
Due to their size, we omit here $\mathcal{M}_{s^+s^-}$ and $\mathcal{M}_{mm}$.

Furthermore, the additional gauge bosons of U(1)$^7$
also get massive, leaving only $\mathcal{G}_{\rm hidden}=$ SO(10) unbroken that serves to achieve a
proper scale of \susy\ breaking through gaugino condensation.
Among the nonvanishing couplings, we find one unsuppressed trilinear coupling $q \bar\phi \bar{u}$, 
which indicates that the top quark acquires a mass hierarchically larger than the mass of the other quarks.

\begin{table}[t!]
\begin{center}
\begin{tabular}{|c|l|c|c|c|l|c|}
\hline
\# & {\small Representation} & Label & & \# & {\small Representation} & Label\\
\rowgrayh
\hline
 4 &
$\left(\boldsymbol{3},\boldsymbol{2};\boldsymbol{1}\right)_{(1/6,1/3)}$
 & $q_i$ 
 & &
 1 &
$\left(\overline{\boldsymbol{3}},\boldsymbol{2};\boldsymbol{1}\right)_{(-1/6,-1/3)}$
 & $\bar q_i$
\\
 4 &
$\left(\overline{\boldsymbol{3}},\boldsymbol{1};\boldsymbol{1}\right)_{(-2/3,-1/3)}$
 & $\bar u_i$
& & 
 1 &
$\left(\boldsymbol{3},\boldsymbol{1};\boldsymbol{1}\right)_{(2/3,1/3)}$
 & $u_i$
 \\
\rowgrayh
 4 &
$\left(\boldsymbol{1},\boldsymbol{1};\boldsymbol{1}\right)_{(1,1)}$
 & $\bar e_i$ 
 & &
 1 &
$\left(\boldsymbol{1},\boldsymbol{1};\boldsymbol{1}\right)_{(-1,-1)}$
 & $e_i$ 
 \\
 6 &
$\left(\overline{\boldsymbol{3}},\boldsymbol{1};\boldsymbol{1}\right)_{(1/3,-1/3)}$
 & $\bar d_i$ 
 & &
 3 &
$\left(\boldsymbol{3},\boldsymbol{1};\boldsymbol{1}\right)_{(-1/3,1/3)}$
 & $d_i$ 
 \\
\rowgrayh
 8 &
$\left(\boldsymbol{1},\boldsymbol{2};\boldsymbol{1}\right)_{(-1/2,-1)}$
 & $\ell_i$ 
 & &
 5 &
$\left(\boldsymbol{1},\boldsymbol{2};\boldsymbol{1}\right)_{(1/2,1)}$
 & $\bar \ell_i$ 
 \\
 5 &
$\left(\boldsymbol{1},\boldsymbol{2};\boldsymbol{1}\right)_{(-1/2,0)}$
 & $\phi_i$ 
 & &
 5 &
$\left(\boldsymbol{1},\boldsymbol{2};\boldsymbol{1}\right)_{(1/2,0)}$
 & $\bar \phi_i$ 
 \\
\rowgrayh
 8 &
$\left(\overline{\boldsymbol{3}},\boldsymbol{1};\boldsymbol{1}\right)_{(1/3,2/3)}$
 & $\bar\delta_i$ 
 & &
 8 &
$\left(\boldsymbol{3},\boldsymbol{1};\boldsymbol{1}\right)_{(-1/3,-2/3)}$
 & $\delta_i$ 
 \\
 20 &
$\left(\boldsymbol{1},\boldsymbol{1};\boldsymbol{1}\right)_{(1/2,*)}$
 & $s^+_i$ 
 & &
 20 &
$\left(\boldsymbol{1},\boldsymbol{1};\boldsymbol{1}\right)_{(-1/2,*)}$
 & $s^-_i$ 
 \\
\rowgrayh
 17 &
$\left(\boldsymbol{1},\boldsymbol{1};\boldsymbol{1}\right)_{(0,1)}$
 & $\bar n_i$ 
 & &
 14 &
$\left(\boldsymbol{1},\boldsymbol{1};\boldsymbol{1}\right)_{(0,-1)}$
 & $n_i$ 
 \\
 4 &
$\left(\boldsymbol{3},\boldsymbol{1};\boldsymbol{1}\right)_{(-1/6,*)}$
 & $w_i$ 
 & &
 4 &
$\left(\overline{\boldsymbol{3}},\boldsymbol{1};\boldsymbol{1}\right)_{(1/6,*)}$
 & $\bar{w}_i$ 
 \\
\rowgrayh
 1 &
$\left(\boldsymbol{1},\boldsymbol{1};\boldsymbol{10}\right)_{(0,0)}$
 & $h_i$ 
 & &
 16 &
$\left(\boldsymbol{1},\boldsymbol{2};\boldsymbol{1}\right)_{(0,*)}$
 & $m_i$ 
 \\
 6 &
$\left(\boldsymbol{1},\boldsymbol{1};\boldsymbol{1}\right)_{(0,\pm2)}$
 & $\chi_i$ 
 & &
 68 &
$\left(\boldsymbol{1},\boldsymbol{1};\boldsymbol{1}\right)_{(0,0)}$
 & $s^0_i$ 
 \\
\hline
\end{tabular}
\caption{Spectrum. The quantum numbers under SU(3)$_C\times$SU(2)$_L\times$\,SO(10) are shown in boldface;
  hypercharge and $B\!-\!L$ charge appear as subscripts. Note that the states $s_i^\pm$, $w_i$, $\bar w_i$
  and $m_i$ have different $B-L$ charges for different $i$, which we do not explicitly list. }
\label{tab:SO32Z6model}
\end{center}
\end{table}

It is possible to identify the nonanomalous $B\!-\!L$ generator
\begin{equation}
t_{B\!-\!L} ~=~ \left( 0^9,2,\,-1^2,\,\tfrac13,\,0,\,-\tfrac13^2\right)
\end{equation}
under which some particles $\chi_i$ have $B\!-\!L$ charges $\pm2$. 
The existence of U(1)$_{B\!-\!L}$ has three direct consequences: first,
it allows us to distinguish between lepton doublets $\ell_i$ and Higgs fields $\phi_i$, and between
$\mathcal{G}_{\rm SM}\times$\,U(1)$_{B\!-\!L}$ singlets $\{s^0_i,\,h_i\}$ and right--handed neutrinos $\bar{n}_i$; second, dimension four proton 
decay operators are forbidden; and third, couplings such as $(s^0)^n \chi_i \bar{n}_j \bar{n}_k$
grant Majorana masses for the right--handed neutrinos.
Note that, as before, if the fields $\chi_i$ attain VEVs, U(1)$_{B\!-\!L}$ is spontaneously broken
by two units to a $\mathbbm{Z}_2$ matter parity. Consequently, not only $\bar{n}_i$ but also $n_i$ shall
correspond to right--handed neutrinos, giving a total of 31 right--handed neutrinos. We have verified that
the corresponding mass matrix has full rank, implying that the masses of theses states is just below the
string scale. As discussed in detail in
section~\ref{sec:Seesaw}, the larger the number $N$ of heavy singlets the smaller the effective mass induced for
the seesaw mechanism, $M_* \propto N^{-x}$ with $0<x<2$. The largest left--handed neutrino mass
is then given by
\begin{equation}
  M_\mathrm{eff}~\sim~v_u^2 / M_*\,,
\end{equation}
where $v_u$ corresponds to the VEV of the physical up--Higgs.

Another interesting property of this model is provided by the hidden sector, $\mathcal{G}_{\rm hidden}=$
SO(10). As discussed in section~\ref{sec:SUSYBreaking}, assuming dilaton stabilization and gaugino
condensation, the gravitino mass is fixed by the beta function of the hidden local interactions. In the
current model, from eq.~\eqref{eq:SO2Nb0} follows that $b^{\mathrm{SO}(10)}_0 = 23$ and, consequently,
\begin{equation}
  \label{eq:m32SO32}
m_{3/2}\approx \frac{\Lambda^3}{M_\mathrm{Pl}^2}\approx\frac{2.1\times 10^{13}\ {\rm
    GeV}}{M_\mathrm{Pl}^2}\approx 9\ {\rm TeV}\,,
\end{equation}
which fits well the usual expectations.

\subsection{SO(32) Orbifolds, String Dualities and Prospects}

We have seen in the previous section, that many appealing properties observed before in E$_8\times$E$_8$
models reappear in \SO{32} constructions.
Let us first contrast here the features of the \SO{32} model with those of the Orbifold-\mssm\ introduced in
section~\ref{sec:SeesawModel}. By mere comparison of the number of massless states, one sees that they
are different. From the point of view of the spectrum, a phenomenological disadvantage of the
\SO{32} model is the number of quark doublets $q_i$. In the Orbifold-\mssm\ there are only three of these
states, two of them localized at the fixed points. This allows us to associate a geometrical picture to
the origin of the hierarchy between the heavy family (in particular, the heavy top) and the two lighter
generations. This picture is enhanced by the fact that a contribution to the physical Higgs $\bar{\phi}$
appears in the untwisted sector, yielding naturally an unsuppressed mass term for the top quark. In the
\SO{32} model, this intuition is lost and the existence of a large mass for the top quark seems rather accidental.

In fact, one can compare all models with phenomenologically attractive properties in both heterotic
cases. We find that, at massless level, the spectra are all inequivalent. This can be easily explained by
the fact that some (\sm) matter generations in the E$_8\times$E$_8$ stem directly from the gauge fields
whereas in \SO{32} this cannot occur as spinors are not embedded into the adjoint representation
$\boldsymbol{496}$. 

However, it is well known that the heterotic E$_8\times$E$_8$ and \SO{32} theories are related by T-duality.
Therefore, one would expect that a large-volume E$_8\times$E$_8$ orbifold compactification can yield an
identical theory in \SO{32} when compactifying in an orbifold with small volume. Yet at the stage of the
current analysis, due to the features mentioned before, it seems clear that the promising models we find
are truly different. It might be interesting to verify whether the $\mathbbm{Z}_6$--II promising models
in one heterotic theory can be found in a different $\mathbbm{Z}_N$ or $\mathbbm{Z}_N\times\mathbbm{Z}_M$
compactification. We note that it is also possible that an orbifold model in one theory could be only
identified with a model from the dual theory just after blowing up the orbifold
singularities~\cite{Nibbelink:2009sp}. Therefore, it is natural to wonder whether the dual model can be
found among orbifold compactifications at all, or one should rather investigate other constructions such
as fermionic models or Calabi-Yau compactifications.

Another interesting relation is provided by S-duality. It connects the weakly coupled \SO{32} heterotic string 
with the strongly coupled type I theory and vice versa~\cite{Polchinski:1995df}. Some effort in the
construction of semirealistic four-dimensional compactifications has been
done~\cite{Angelantonj:1996uy,Kakushadze:1997wx} via a $\mathbbm{Z}_3$ orbifold. However, as far as we
know, none of these constructions groups all the properties of the heterotic orbifold models discussed in
this work. Therefore, establishing the connection between heterotic \SO{32} orbifolds and type I
compactifications requires first a bigger effort in the context of type I.

%% file: Appendices.tex
%%%%%%%%%%%%%%%%%%%%%%%%%%%%%%%%%%%%%%%%%%%%%%%%%%%%%%%%%%%%%%%%%%%%
%%% Appendix A
%%%%%%%%%%%%%%%%%%%%%%%%%%%%%%%%%%%%%%%%%%%%%%%%%%%%%%%%%%%%%%%%%%%%
\chapter{Form of Shift Vectors and Wilson lines}
\label{ch:ansatz}

Even though there is a straightforward prescription to obtain all \Z{N} shift vectors (see
section~\ref{sec:DynkinMethod}), it is necessary to implement a method which can be easily extended to
\ZZ{N}{M} orbifolds and orbifolds with Wilson lines. An option is an ansatz that describes any shift or
Wilson line. To start with, let us specify an ansatz for \Z{N} shifts of \SO{32} orbifolds. We will see in
section~\ref{sec:GeneralAnsatz} that the resulting ansatz can be then trivially generalized.

\section[\Z{N} Shift Vectors of \SO{32} Orbifolds]{$\bs{\Z{N}}$ Shift Vectors of $\bs{\SO{32}}$ Orbifolds}
\label{sec:AnsatzSO32}

We will call {\em vectorial} shifts to those shift vectors of order $N$ ($N V\in\Lambda$) whose entries
have a maximal denominator of $N$. {\em Spinorial} shifts will be those other which have a denominator of
$2N$ and an odd numerator. As an example, consider the \Z4 shift vectors of
table~\ref{tab:Shifts_Z4SO32}. The first 12 of them are vectorial, whereas the last four are
spinorial. It is convenient to describe separately these two cases and further distinguish between \Z{N}
shifts with $N$ even and $N$ odd.

To obtain the general form of a shift, we take for granted that two shift vectors are equivalent if they are
related by lattice vectors or by Weyl reflections, i.e. by any permutation of the entries and pairwise
sign flips.\footnote{\Z{N} orbifold models without Wilson lines do not change under lattice translations,
unless one introduces lattice-valued Wilson lines. In such case, the new model can be a {\it brother
model}. Further details are provided in section~\ref{subsec:BrotherModels}.} 

\subsubsection{a) Vectorial Shifts and $\bs{N}$ Even}
Invariance under lattice translations implies that the entries $V_i$ of a shift of order $N$ are
constrained by $-(N-1)/N \leq V_i \leq 1$. Therefore, we can start with the ansatz
\begin{equation}
V = \h{1}{N}\big(-(N-1)^{n_{-(N-1)}},\,-(N-2)^{n_{-(N-2)}},\ldots,\,0^{n_0}, 1^{n_1}, \ldots ,N^{n_{N}}\big)
\label{eq:SO32Ansatz1}
\end{equation}
for a vectorial shift. As usual, the exponent $n_i$ of an entry $i$ counts how many times that entry is
repeated; for instance, the entry $-(N-1)$ appears $n_{-1(N-1)}$ times.
Therefore, the exponents $n_i$ are integers satisfying 
\be
\sum_i n_i = 16\;.
\label{eq:nCondition}
\ee
Since pairwise sign-flips are allowed, we see that all negative entries can be made nonnegative as long as
at least one $n_{-i}$ is even ($i=1,\ldots,N$). However, if there is an odd number of negative entries
and $n_0=0$, one negative entry will remain. Combining both results, we get
\begin{equation}
V = \h{1}{N}\big((-j)^\al,\,0^{n_0}, 1^{n_1},\ldots,j^{n_j-\al},\ldots,(N-1)^{n_{N-1}}\big)\,,\;\;\qquad 1\leq j\leq N-1
\label{eq:SO32Ansatz2}
\end{equation}
with $\al=0,1$. This ansatz contains still some redundancies. Lattice translations of the shifts amounts
to adding $\pm 1$ to an even number of entries of $V$. One can thus apply $V_i\rightarrow V_i-1$ for
an even number of entries fulfilling $V_i>\h{N}{2}$ and then flip their signs. By this operation, a \Z{N}
vectorial shift can be finally expressed by
\begin{equation}
V = \frac{1}{N}\left( (\pm j)^{\alpha}, -(N-j)^{\beta}, 0^{n_0}, 1^{n_1}, \ldots ,
  (N-j)^{n_{(N-j)}-\alpha-\beta},\ldots, \left(\frac{N}{2}\right)^{n_{\frac{N}{2}}} \right)\,,
\label{eq:SO32Ansatz3}
\end{equation}
where  $\al,\beta=0,1$ such that $\al+\beta=0,1$ and $j$ can take the values $\{\h{N}{2}+1,\ldots,N-1\}$.

As a side remark, using the shift $V$ of eq.~\eqref{eq:SO32Ansatz3} and the \SO{32} simple roots $\al_i$
provided in table~\ref{tab:RootsSO32}, one can compute $\al_i\cdot V$. Since those roots with $\al_i\cdot
V\neq 0$ are projected out while the others form the Dynkin diagram of the unbroken gauge group
$\maG_{4D}$, one finds that the symmetry breakdown induced by a generic vectorial shift of \SO{32}
orbifolds is given by
\begin{equation}
\SO{32}\longrightarrow\SO{2n_0}\x\U{n_1}\x\ldots\x\U{n_{\left(\frac{N}{2}-1\right)}}\x\SO{2n_{N/2}}\,,
\end{equation}
recalling that $\U{n}=\SU{n}\x\U1$.

\subsubsection{b) Spinorial Shifts and $\bs{N}$ Even}
An analogous analysis to that presented in the previous case leads to the conclusion that a generic
spinorial shift can be written as
\begin{equation}
V = \frac{1}{2N}\left((\pm j)^{\alpha}, -(2N-j)^{\beta}, 1^{n_1}, 3^{n_3}, \ldots ,
  (2N-j)^{n_{(2N-j)}-\alpha-\beta},\ldots, (N-1)^{n_{N-1}} \right)\,,
\label{eq:SO32Ansatz4}
\end{equation}
with $j\in\{N+1,N+3,\ldots,2N-1\}$. It is again not hard to confirm that a shift vector of this kind
leads to the breaking
\begin{equation}
\SO{32} \longrightarrow \U{n_1}\x\U{n_3}\x\ldots\times\U{n_{N-3}}\x\U{n_{N-1}}\,.
\end{equation}

\subsubsection{c)  $\bs{N}$ Odd}
The general form of the shift vectors changes slightly. First, as explained in ref.~\cite{Choi:2004wn},
in this case it is enough to determine either the vectorial or the spinorial shifts, since one spinorial
shift can always be transformed into a vectorial one by the action of Weyl reflections and lattice
vectors. Choosing the vectorial form, we obtain the following shift vector of odd order:
{\small
\begin{equation}
V = \frac{1}{N}\left( (\pm j)^\al, -(N-j)^\beta, 0^{n_0}, 1^{n_1}, \ldots ,(N-j)^{n_{(N-j)}-\al-\beta},
\ldots, \left(\frac{N-1}{2}\right)^{n_{\left(\frac{N-1}{2}\right)}}\right)\,,
\label{eq:SO32Ansatz5}
\end{equation}}
where $j\in\{\frac{N+1}{2},\frac{N+3}{2},\ldots,N\}$. The resulting four dimensional gauge group is
\begin{equation}
\SO{32}\longrightarrow\SO{2n_0}\x \U{n_1}\x\ldots\x\U{n_{\left(\frac{N-3}{2}\right)}}\x\U{n_{\left(\frac{N-1}{2}\right)}}\,.
\end{equation}

The shifts obtained by the ans\"atze given in eqs.~\eqref{eq:SO32Ansatz3}--\eqref{eq:SO32Ansatz5} are
further restrained by $N V\in\Lambda$ and the requirement that follows from invariance of the orbifold
partition function: 
\be
N\left( V^2 - v^2 \right) = 0 \text{ mod } 2
\ee
for a given twist vector $v$ of order $N$. Only those shifts satisfying these conditions are consistent
with orbifold compactifications.
The ans\"atze~\eqref{eq:SO32Ansatz3}--\eqref{eq:SO32Ansatz5} constitute, as we will see in the next
section, the basic constructing blocks of general \ZZ{N}{M} shifts and Wilson lines.

\section{A General Ansatz}
\label{sec:GeneralAnsatz}

Now we are in position to propose an ansatz which covers all orbifolds, including Wilson lines. 
Consider a model with a given shift vector $V$. Independently of whether $V$ belongs to an \E8\x\E8 or an
\SO{32} orbifold, Weyl reflexions and lattice translations allow to make its entries as small as possible
and to gather them in blocks $V_\text{block i}$, just as in the
ans\"atze~\eqref{eq:SO32Ansatz3}--\eqref{eq:SO32Ansatz5}:
\begin{equation}
V = \frac{1}{N}\left(V_\text{block 1},\,V_\text{block 2},\,V_\text{block 3},\ldots\right)\,,
\label{eq:BlockV}
\end{equation}
where each subvector $V_\text{block i}$ contains $m_i$ identical integer or half-integer entries, such
that $\sum_i m_i = 16$. Our task is then to find a universal second shift $V_2$ of order $M$, in the case
of \ZZ{N}{M} orbifolds, or/and the general form of a Wilson line $A_\al$ of order $N_\al$, trying to
avoid redundancies at maximum. From the block-structure of the shift vector, we can expect $V_2$ and
$A_\al$ also to be splitted in blocks of length $m_i$ as
\bse
\bea
\label{eq:BlockV2}
V_2 &=& \left(V_{2,\text{block 1}},\,V_{2,\text{block 2}},\,V_{2,\text{block 3}},\ldots\right)\,,\\
\label{eq:BlockA}
A_\al &=& \left(A_{\al,\text{block 1}},\,A_{\al,\text{block 2}},\,A_{\al,\text{block 3}},\ldots\right)\,.
\eea
\label{eq:BlockV2A}
\ese
This would mean that each block $V_{2,\text{block i}}$ and $A_{\al,\text{block i}}$ is independent and
can, therefore, be described separately. In other words, we have just to find an ansatz for a subvector
$V_{2,\text{block i}}$ or $A_{\al,\text{block i}}$ of the corresponding order ($M$ or $N_\al$) and then
to apply it consecutively over all blocks. 

In the previous section, we have seen that the
ans\"atze~\eqref{eq:SO32Ansatz3}--\eqref{eq:SO32Ansatz5} describe the minimal form of a shift of order 
$N$. They can be used in order to build the blocks $V_{2,\text{block i}}$ and $A_{\al,\text{block i}}$,
by replacing $N$ by $M$ or $N_\al$ and demanding that
\be
\sum_j n_j = m_i
\label{eq:nCondition2}
\ee
for each block. We have additionally to guarantee that the vectors obtained in this way lie on the
lattice $\Lambda$, i.e.\ one has to impose $MV_2,\,N_\al A_\al\in\Lambda$. This condition implies that,
for even $N$, all blocks must be of the same type (either vectorial or spinorial).
Clearly, admissible orbifold models are only those with shift(s) and Wilson lines fulfilling the modular
invariance conditions provided in eq.~\eqref{eq:NewModInv}.

%%%%%%%%%%%%%%%%%%%%%%%%%%%%%%%%%%%%%%%%%%%%%%%%%%%%%%%%%%%%%%%%%%%%
%%% Appendix B
%%%%%%%%%%%%%%%%%%%%%%%%%%%%%%%%%%%%%%%%%%%%%%%%%%%%%%%%%%%%%%%%%%%%
\chapter{Spinors in SO(32) Orbifolds}
\label{ch:SpinorsInSO32}

Early works on orbifold compactifications considered the \SO{32} heterotic string to be
phenomenologically disfavored. One of the reasons being that the theory by itself does not include
spinors in its massless spectrum. This situation was shown to be preserved in the simplest orbifold
compactifications, leading to the somewhat na\"ive conclusion that, if spinors can be found from the
\SO{32} heterotic string at all, they must appear in a quite unnatural way. 

Only recently, the interest in four-dimensional heterotic \SO{32} orbifold constructions has been
revived~\cite{Giedt:2003an,Nibbelink:2007rd}. In the case of \Z{N} orbifolds, a complete classification
of gauge embeddings in the absence of background fields has been
achieved~\cite{Choi:2004wn,Nilles:2006np}. Interestingly, that classification has shown that spinors of
\SO{2n} gauge groups appear rather frequently in the twisted sectors of \SO{32} orbifolds. 
Spinors of \SO{10}, in particular, are found locally at fixed points of the first twisted sector of many
orbifold models. 
Moreover, that classification reveals that the amount of available \SO{32} orbifold models is comparable, at
the same level, to that of its more famous brother: the $\E8\times\E8$ string. Thus, it is not adventurous
rather to conclude that model building based on the heterotic \SO{32} string theory might be as
interesting as that based on the $\E8\times\E8$ theory.

On the other hand, the appearance of spinors on models derived from the \SO{32} heterotic string might
also be important for a possible understanding of the \SO{32} heterotic type I duality in four spacetime
dimensions. We know that spinors do not appear in the perturbative type I theory. Thus, the mentioned
duality requires the implementation of nonperturbative effects.

\section[\SO{10} Spinors and Shift Vectors]{$\bs{\SO{10}}$ Spinors and Shift Vectors}
Let us investigate here the possibility of having the 16-dimensional spinor representation of SO(10) in
\Z{N} orbifolds of the \SO{32} heterotic string theory.
In the standard basis of the roots of \SO{10}, the highest weight of the \bs{16}--plet is given by the
five-dimensional vector~\footnote{As usual, powers of an entry represent how often that entry is repeated
 in the vector.} 
\be
\bs{16}_\text{HW}=\left(\2^5\right)\,. 
\label{eq:16HW}
\ee
To be included in the spectrum of an orbifold compactification, this vector must be part of a
16-dimensional solution $p_\text{sh}$ to the mass formula for massless states,
eq.~\eqref{eq:MasslessnessUntwisted} or eq.~\eqref{eq:MasslessnessTwisted}. Therefore, $p_\text{sh}$ must
be written as
\begin{equation}
p_\text{sh}~=~p+V_g~=~\left(\2^5,\, a_1,\,\ldots,\, a_{11} \right)\,,
\label{eq:gral_p}
\end{equation}
where $p\in\Lambda_{\text{Spin(32)}/\Z2}$, $V_g$ is the local shift of a constructing element
$g$ and $a_i\in\maR$ are selected so that $p_\text{sh}$ fulfills $N
p_\text{sh}\in\Lambda_{\text{Spin(32)}/\Z2}$ and the mass formula.
There can be more than one combination of different $a_i$'s for which the resulting $p_\text{sh}$
satisfies all the conditions. 

A first consequence of eq.~\eqref{eq:gral_p} is that one cannot get \bs{16}--plets of
\SO{10} in the untwisted sector of any orbifold compactification. The reason is that the only
$p_\text{sh}$ available for untwisted massless states correspond to the roots of \SO{32}, which are
expressed by the 480 vectors $(\underline{\pm 1,\,\pm 1,\;0^{14}})$. 

As a second consequence, one finds that it is not possible to get \bs{16}--plets of \SO{10} in the
\Z3 orbifold. This can be understood as follows. Since in \Z3 orbifolds $3 p_\text{sh}$ must lie on the
lattice $\Lambda_{\text{Spin(32)}/\Z2}$ and the first five entries of $3 p_\text{sh}$ are
half-integer, all $a_i$'s must also be half-integer, $3a_i\in \mathbb{Z}+1/2$. Assuming in the best of
the cases that $3a_i=1/2$, it follows that $p_\text{sh}^2\ge 14/9$. However, masslessness of the states
in \Z3 orbifolds constrains $p_\text{sh}$ to have at most squared length $4/3$. 

It is possible to find the form of the local shift vectors producing \SO{10} spinors. From
eq.~\eqref{eq:gral_p}, for all \Z{N} orbifolds with $N>3$ the local shift(s) giving rise to the \bs{16}
representation of \SO{10} can be written as 
\begin{equation}
V_g~=~p_\text{sh}-p\,
\label{eq:shift16}
\end{equation}
with $p_\text{sh}$ given as in eq.~\eqref{eq:gral_p}.

Hence, we can determine the shift vector $V$ of an orbifold model containing \SO{10} spinors
in the first twisted sector, relevant for phenomenology, as we have discussed in
section~\ref{sec:OrbiGUT}. In absence of Wilson lines, the local shift vector $V_g$ at any fixed
point of the first twisted sector coincides with $V$, therefore 
\begin{equation}
V~=~p_\text{sh}-p \stackrel{p = 0}{\longrightarrow}p_\text{sh},
\label{eq:ShiftWith16InT1}
\end{equation}
where we have chosen $p=0$ because two shifts are equivalent if they differ by an arbitrary lattice
vector. This shift vector is automatically modular invariant. For each allowed $p_\text{sh}$ from
eq.~\eqref{eq:gral_p}, one can verify that the shift vector $V$ obtained through
eq.~\eqref{eq:ShiftWith16InT1} leads to a four-dimensional \SO{10} gauge group by applying the patterns 
provided in appendix~\ref{sec:AnsatzSO32}.

As an example, we consider the \Z4 orbifold without Wilson lines. The only possible shift
vector consistent with $4V\in\Lambda_{\text{Spin(32)}/\Z2}$, the masslessness condition
eq.~\eqref{eq:MasslessnessTwisted} and eq.~\eqref{eq:ShiftWith16InT1} is
\begin{equation}
V ~=~p_\text{sh}~=~\left(\h{1}{2}^5,\,\h{1}{4}^2,\,0^{9} \right)\,,
\label{eq:Z4shiftwith16}
\end{equation}
which is one of the admissible shift vectors for the \Z4 orbifold, listed in table~\ref{tab:Shifts_Z4SO32}.

%%%%%%%%%%%%%%%%%%%%%%%%%%%%%%%%%%%%%%%%%%%%%%%%%%%%%%%%%%%%%%%%%%%%
%%% Appendix C
%%%%%%%%%%%%%%%%%%%%%%%%%%%%%%%%%%%%%%%%%%%%%%%%%%%%%%%%%%%%%%%%%%%%
\chapter{$\bs{\Z6}$-II on Nonfactorizable Lattices}
\label{ch:Z6IINonFactorizable}\index{nonfactorizable lattice}

\Z6-II orbifold compactifications on the lattice $\G2\times\SU3\times\SO4$ have proven to lead to
realistic models. The question here is how the properties of such models are influenced by the use of one
of the other allowed lattices of \Z6-II. In the best of the cases, the possible changes shall be of
relevance when dealing with unsolved phenomenological issues such as proton decay.

\section{Lattices and Spectrum}

The allowed lattices of \Z6-II orbifolds are presented in table~\ref{tab:Z6IIAllLattices}. Note that, with
exception of the lattice (A), $\G2\times\SU3\times\SO4$, all lattices are nonfactorizable. These lattices
are independent in the sense that they cannot be deformed continuously into one another. Therefore, one
knows that orbifolds on those lattices will have certainly distinct properties.

The choice of the lattice of the compact space will affect some other properties. Firstly, one finds that
the number of fixed points of the orbifold varies according to 
\be
\ba{clccc}
&\text{\bf Lattice} & \bs{T_1}  & \bs{T_2} & \bs{T_3}\\
(A)&\G2\times\SU3\times\SO4  & 12 & 6 & 8 \\
(B)&\SU3\times\SO8           & 12 & 3 & 8 \\
(C)&\SU3\times\SO7\times\SU2 & 12 & 3 & 8 \\
(D)&\SU6\times\SU2           & 12 & 3 & 4
\ea\nonumber
\ee
where, as usual, $T_i$ denotes the $i^\mathrm{th}$ twisted sector. Secondly, the number and order of the allowed
Wilson lines changes. In the factorizable lattice (A), we have seen that there are two order 2 Wilson lines
corresponding to the \SO4 torus, and one independent order 3 Wilson line in the \SU3 torus. This
contrasts e.g. to the lattice $\SU6\x\SU2$ in which only two Wilson lines are allowed: an order 2 Wilson
line in the \SU2 torus and an order 6 Wilson line in the \SU6 torus.\footnote{The order 6 Wilson line has
been usually disregarded in the literature.} The number and order of the Wilson lines in all lattices are
shown in table~\ref{tab:Z6IIAllLattices}. 

These two features have crucial effects on the matter spectrum of orbifold models. Evidently, the number
of states of orbifolds on nonfactorizable lattices will be reduced 
compared to the number of states on the usual lattice (A). What is somehow more interesting is that the
missing states in nonfactorizable orbifolds form vectorlike pairs. 

Consider, for instance, the spectrum of the orbifold-\mssm\  listed in
table~\ref{tab:SeesawDiffLattices}. We see that the number of exotics is reduced by
compactifying on nonfactorizable lattices. All nonfactorizable lattices (B,C,D) yield
in the observable sector only three \mssm\ generations plus two pairs of Higgses. That means that
unwanted quark triplets and two Higgs pairs have disappeared. Notice that we lose simultaneously some
needed particles, such as some \sm\ singlets $s^0,\,h$, used to decouple exotics, and some
right-handed neutrinos $n,\,\bar n,\,\eta,\,\bar\eta$. This is not a problem as long as the remaining
exotics can still be decoupled.

\begin{table}[t!]
\small{
\begin{center}
\begin{tabular}{|c|@{}c@{}|c|c|}
\hline
 & 6D Lattice $\Gamma$ & Twist $v$  & Conditions on Wilson lines \\
\hline\hline
\rowgrayh
(A)&$\G2\times\SU3\times\SO4$  & $\h16(0,\,1,\,2,\,-3)$ &  $3A_3\approx 2A_5\approx 2A_6\approx 0;\;A_1\approx A_2\approx 0,\, A_3\approx A_4$ \\
(B)&$\SU3\times\SO8$           & $\h16(0,\,2,\,1,\,-3)$ &  $3A_1\approx 2A_5\approx 0;\;A_1\approx A_2,\,A_3\approx A_4\approx 0,\,A_5\approx A_6$ \\
\rowgrayh
(C)&$\SU3\times\SO7\times\SU2$ & $\h16(0,\,2,\,1,\,-3)$ &  $3A_1\approx 2A_5\approx 2A_6\approx 0;\;A_1\approx A_2,\, A_3\approx A_4\approx 0$ \\ 
(D)&$\SU6\times\SU2$           & $\h16(0,\,2,\,1,\,-3)$ &  $6A_1\approx 2A_6\approx 0;\;A_1\approx A_2\approx A_3\approx A_4\approx A_5$\\
\hline
\end{tabular}
\end{center}}
\vskip -2mm
\caption{Allowed lattices for \Z6-II orbifolds and constraints on the Wilson lines. With exception of
  (A), all lattices are nonfactorizable.}
\label{tab:Z6IIAllLattices}
\end{table}

This situation is similar to what is observed between brother models of $\Z2\times\Z2$ orbifolds. In that
case, nonvanishing discrete torsion on orbifolds with a given torus lattice projects out some of the states
living at particular fixed points, reducing thereby the effective number of fixed points. Hence, the massless
spectra of models with discrete torsion on a given lattice turn out to be identical to the spectra of models without
discrete torsion on a different lattice~\cite{Ploger:2007iq}. A natural question is then whether discrete torsion also
relates the spectra of \Z6-II orbifolds compactified on different lattices. Unfortunately, a close
inspection of the parameters of discrete torsion in \Z6-II orbifolds convinces us that discrete
torsion (as introduced in ref.~\cite{Ploger:2007iq}) does not have any effect on the spectrum of the
models. An interesting question is whether discrete torsion does not exhaust all the available degrees
of freedom that alter orbifold models.
\subsection{A Comment on the String Selection Rules}

Compactifying on nonfactorizable lattices affects more than the spectrum. The computation of Yukawa couplings in
factorizable orbifold models is based on certain selection rules stated in a form that appears to be
related with  the factorizability of the lattice. Particularly, conservation of $R$-charge in \Z6-II factorizable
orbifolds is expressed as
\be\label{eq:Z6IIRChargeConservationCopy}
\sum_i R^1_i = -1\mod 6\,,\;\;\;\sum_i R^2_i = -1\mod 3\,,\;\;\;\sum_i R^3_i = -1\mod 2\,,
\ee
where the sum runs over the states of a given coupling. $R^a_i$ corresponds to the R-charge of
the $i^{\mathrm{th}}$ particle of the coupling on the $a^\mathrm{th}$ complex plane of the compact
space. 

In the literature~\cite{Font:1988tp,Choi:2006qh}, it is usually stated that R-charge conservation,
eq.~\eqref{eq:Z6IIRChargeConservationCopy}, follows from the symmetries of the three complex
planes of a factorizable orbifold. In the case of factorizable \Z6-II orbifolds, it is clear that the
three independent (discrete) rotations generated by the Cartan generators $H^a$ of the underlying \SO6
($\simeq \SU4$) Lorentz symmetry leave the compactification lattice (A) invariant. Therefore, the original
\SO6 Poincar\'e symmetry of the compact space is broken to $\Z6\times\Z3\times\Z2$ in the factorizable
orbifold. 

We might assume that this also holds for compactifications on other lattices and then we could apply
the same R-charge conservation given in eq.~\eqref{eq:Z6IIRChargeConservationCopy}. However, in this
case, the relation between R-charge conservation and complex-plane rotation invariance is violated in
nonfactorizable lattices. For example, consider the lattice \SU6\x\SU2.
One can show that the only element of \SO6 containing only Cartan generators that
leaves the lattice invariant is
\be
\bs{R^t}\equiv e^{2\pi\I\6(2 H_1 + H_2 - 3 H_3)}
\ee
and powers of it. 
We can then conjecture that the corresponding ``R-charge conservation rule'', in analogy to the
factorizable case, is
\be\label{eq:RChargeZ6IISU6xSU2}
\sum_i R^t_i \equiv \sum_i (2R^1 + R^2 - 3 R^3)_i = 0\mod 6\,.
\ee
Nevertheless, one notices immediately one problem: charges $R^t_i$ of bosonic and fermionic superpartners
coincide, i.e. the \Z6 symmetry generated by $2 H_1 + H_2 - 3 H_3$ is not an R-symmetry. This issue will
be investigated elsewhere.

\section[\Z6-II on \SU6\x\SU2]{(D) $\bs{\Z6}$-II on $\bs{\SU6\x\SU2}$}
\begin{figure}[t!]
\centerline{\input Z6II_SU6_geometry.pstex_t}
\caption{The root lattice of $\SU6\times\SU2$ for \Z6-II orbifolds in complex coordinates.}
\label{fig:Z6IISU6Geometry}
\end{figure}

The basis vectors of the root lattice of \SU6\x\SU2 are expressed in the six-dimensional orthogonal
coordinates by 
\be\ba{rl}
e_1 =& (1,\,0,\,1/\sqrt{3},\,0,\,\sqrt{2/3},\,0)\,,\\
e_2 =& (-1/2,\,\sqrt{3}/2,\,1/2\sqrt{3},\,1/2,\,-\sqrt{2/3},\,0)\,,\\
e_3 =& (-1/2,\,-\sqrt{3}/2,\,-1/2\sqrt{3},\,1/2,\,\sqrt{2/3},\,0)\,,\\
e_4 =& (1,\,0,\,-1/\sqrt{3},\,0,\,-\sqrt{2/3},\,0)\,,\\
e_5 =& (-1/2,\,\sqrt{3}/2,\,-1/2\sqrt{3},\,-1/2,\,\sqrt{2/3},\,0)\,,\\
e_6 =& (0,\,0,\,0,\,0,\,0,\,\sqrt{2})\,.
\ea\ee
Notice that the twist vector in this basis is given by $v=(0,\,1/3,\,1/6,\,-1/2)$. The lattice is illustrated
in fig.~\ref{fig:Z6IISU6Geometry}.

\begin{table}[!t!]
{\footnotesize
\begin{minipage}{0.48\textwidth}
\begin{tabular}{|lc||@{}r|@{}r|@{}r|@{}r|}
\hline
                       &         & \multicolumn{4}{c|}{Lattice $\Gamma$}\\
\cline{3-6}
{\footnotesize \ \ \ \ \sm\ Repr.} & Label             & (A)& (B)& (C)& (D)\\
\hline\hline
\rowgrayh
( \bs{1},  \bs{2})$_{(-1/2,0)}$   & $\phi$             &  4 &  2 &  2 &  2\\
\rowgrayh
( \bs{1},  \bs{2})$_{(1/2,0)}$    &  $\bar\phi$        &  4 &  2 &  2 &  2\\
( \bsb{3}, \bs{1})$_{(1/3,-1/3)}$ &  $\bar d$          &  4 &  3 &  3 &  3\\
( \bs{3},  \bs{1})$_{(-1/3,1/3)}$ &  $d$               &  1 &  0 &  0 &  0\\
\rowgrayh
( \bs{1},  \bs{2})$_{(-1/2,-1)}$  &  $\ell$            &  4 &  3 &  3 &  3\\
\rowgrayh
( \bs{1},  \bs{2})$_{(1/2,1)}$    &  $\bar\ell$        &  1 &  0 &  0 &  0\\
( \bs{3},  \bs{2})$_{(1/6,1/3)}$  &  $q$               &  3 &  3 &  3 &  3\\
\rowgrayh
( \bs{1},  \bs{1})$_{(1,1)}$      &  $\bar e$           &  3 &  3 &  3 &  3\\
( \bsb{3}, \bs{1})$_{(-2/3,-1/3)}$&  $\bar u$           &  3 &  3 &  3 &  3\\
\hline\hline
\rowgrayh
( \bs{1},  \bs{1})$_{(0,1)}$    & $\bar n,\bar\eta$    & 21 & 14 & 14 & 12\\
\rowgrayh
( \bs{1},  \bs{1})$_{(0,-1)}$   & $n,\eta$             & 18 & 11 & 11 &  9\\
( \bs{1},  \bs{1})$_{(0,0)}$    & $h,s^0,w$            & 98 & 60 & 60 & 52\\
\rowgrayh
( \bs{1},  \bs{1})$_{(0,-2)}$   & $\chi_1$             &  1 &  1 &  1 &  1\\
\rowgrayh
( \bs{1},  \bs{1})$_{(0,2)}$    & $\chi_2$             &  1 &  1 &  1 &  1\\
( \bs{1},  \bs{1})$_{(0,1/2)}$  & $\bar f$             & 16 &  8 &  8 &  8\\
( \bs{1},  \bs{1})$_{(0,-1/2)}$ & $f$                  & 16 &  8 &  8 &  8\\
\hline
\end{tabular}
\end{minipage}
\hskip 2mm
\begin{minipage}{0.48\textwidth}
\begin{tabular}{|lc||@{}r|@{}r|@{}r|@{}r|}
\hline
                       &         & \multicolumn{4}{c|}{Lattice $\Gamma$}\\
\cline{3-6}
{\footnotesize \ \ \ \ \sm\ Repr.} & Label           & (A)& (B)& (C)& (D)\\
\hline\hline
\rowgrayh
( \bs{1},  \bs{1})$_{(1/2,1)}$  & $s^+,x^+$            & 14 & 14 & 14 & 10\\
\rowgrayh
( \bs{1},  \bs{1})$_{(-1/2,-1)}$& $s^-,x^-$            & 14 & 14 & 14 & 10\\
( \bs{1},  \bs{1})$_{(1/2,0)}$  & $s^+$                &  8 &  8 &  8 &  6\\
( \bs{1},  \bs{1})$_{(-1/2,0)}$ & $s^-$                &  8 &  8 &  8 &  6\\
\rowgrayh
( \bs{1},  \bs{1})$_{(1/2,2)}$  & $s^+$                &  2 &  2 &  2 &  2\\
\rowgrayh
( \bs{1},  \bs{1})$_{(-1/2,-2)}$& $s^-$                &  2 &  2 &  2 &  2\\
\hline\hline
( \bs{3},  \bs{1})$_{(-1/3,-2/3)}$&  $\delta$          &  3 &  1 &  1 &  1\\
( \bsb{3}, \bs{1})$_{(1/3,2/3)}$ &  $\bar\delta$       &  3 &  1 &  1 &  1\\
\rowgrayh
( \bsb{3}, \bs{1})$_{(-1/6,2/3)}$ &  $\bar v$          &  2 &  2 &  2 &  0\\
\rowgrayh
( \bs{3},  \bs{1})$_{(1/6,-2/3)}$ &  $v$               &  2 &  2 &  2 &  0\\
( \bsb{3}, \bs{1})$_{(-1/6,-1/3)}$&  $\bar v$          &  2 &  2 &  2 &  2\\
( \bs{3},  \bs{1})$_{(1/6,1/3)}$  &  $v$               &  2 &  2 &  2 &  2\\
\rowgrayh
( \bs{1},  \bs{2})$_{(0,-1)}$     &  $m$               &  2 &  2 &  2 &  2\\
\rowgrayh
( \bs{1},  \bs{2})$_{(0,1)}$      &  $m$               &  2 &  2 &  2 &  2\\
( \bs{1},  \bs{2})$_{(0,0)}$      &  $y$               &  4 &  4 &  4 &  4\\
\hline
\end{tabular}
\vskip 4mm \phantom{.}
\end{minipage}
}
\caption{Multiplicity of the \sm\ fields of the orbifold-\mssm\ in the different admissible
compactification lattices of \Z6-II. The hidden-sector quantum numbers have been omitted. The labels
coincide with those of table~\ref{tab:SpectrumSeesawModel}.}
\label{tab:SeesawDiffLattices}
\end{table}

%%%%%%%%%%%%%%%%%%%%%%%%%%%%%%%%%%%%%%%%%%%%%%%%%%%%%%%%%%%%%%%%%%%%
%%% Appendix D
%%%%%%%%%%%%%%%%%%%%%%%%%%%%%%%%%%%%%%%%%%%%%%%%%%%%%%%%%%%%%%%%%%%%
%\begin{landscape}
\chapter{Orbifold Tables}
\label{ch:Tables}
\renewcommand{\arraystretch}{1.4}
\index{Wilson lines!order of}\vskip -3mm
\begin{table}[h!]
\begin{center}
\begin{tabular}{|c|c|c|}
\hline
 $P$ & 6D Lattice $\Gamma$ & Conditions on the Wilson lines \\
\hline
\hline
\rowgrayh
 $\Z{3}$     & $\SU3^3$                    & $3A_1\approx 3A_3\approx 3A_5\approx 0;\;A_1\approx A_2,\,A_3\approx A_4,\,A_5\approx A_6$ \\
 $\Z{4}$     & $\SU4^2$                    & $4A_1\approx 4A_4\approx 0;\;A_1\approx A_2\approx A_3,\, A_4\approx A_5\approx A_6$ \\
             & $\SU4\times\SO5\times\SU2$  & $4A_1\approx 2A_5\approx 2A_6\approx 0;\;A_1\approx A_3\approx A_4,\, A_4\approx 0$ \\
             & $\SO5^2\times\SU2^2$        & $2A_2\approx 2A_4\approx 2A_5\approx 2A_6\approx 0;\;A_1\approx A_3\approx 0$ \\
\rowgrayh
 $\Z{6}$-I   & $\G2^2\times\SU3$           & $3A_5\approx 0;\;A_1\approx A_2\approx A_3\approx A_4\approx 0,\,A_5\approx A_6$ \\
 $\Z{6}$-II  & $\G2\times\SU3\times\SO4$   & $3A_3\approx 2A_5\approx 2A_6\approx 0;\;A_1\approx A_2\approx 0,\, A_3\approx A_4$ \\
             & $\SU6\times\SU2$            & $6A_1\approx 2A_6\approx 0;\;A_1\approx A_2\approx A_3\approx A_4\approx A_5$ \\
\hline
\end{tabular}
\end{center}
\caption{Wilson lines are constrained in orbifold compactifications due to the compactification lattice
  $\Gamma$. The symbol `$\approx$' means equivalence up to translations in the gauge lattice
  $\Lambda$, thus $N_\al A_\al\approx 0$ means that $N_\al A_\al\in\Lambda$. Some frequent typos in the current literature have been corrected here.}  
\label{tab:ZNWLConstraints}
\end{table}
%\end{landscape}
\begin{table}[ht!]
\small{
\begin{center}
\begin{tabular}{|c|c|c|}
\hline
\rowgrayh
 $P$ & 6D Lattice $\Gamma$  & Conditions on the Wilson lines \\
\hline
\hline
$\Z2 \times\Z2$    &1: $\SU2^6$                    & $2A_1\approx 2A_2\approx 2A_3\approx 2A_4\approx 2A_5\approx 2A_6\approx 0$ \\
                   &2: $\SU3\times\SU2^4$          & $2A_2\approx2A_3\approx2A_4\approx2A_5\approx2A_6\approx0;\;A_1\approx0$\\
                   &3: $\SU3^2\times\SU2^2$-I      & $2A_3\approx2A_4\approx2A_5\approx2A_6\approx0;\;A_1\approx A_2\approx0$\\
                   &4: $\SU4\times\SU2^3$          & $2A_1\approx 2A_2\approx 2A_3\approx 2A_4\approx 2A_6\approx0;\; A_1\approx A_5$ \\
                   &5: $\SU3^2\times\SU2^2$-II     & $2A_2\approx2A_3\approx2A_5\approx2A_6\approx0;\;A_1\approx A_4\approx0$\\
                   &6: $\SU4\times\SU3\times\SU2$  & $2A_1\approx2A_3\approx2A_4\approx2A_6\approx0;\;A_1\approx A_5;\;A_2\approx0$\\
                   &7: $\SU4^2$                    & $2A_1\approx2A_2\approx2A_3\approx2A_4\approx0;\;A_1\approx A_5;\;A_2\approx A_6$\\
\hline
\end{tabular}
\caption{Sample of constraints on the Wilson lines of \ZZ{N}{M} orbifolds. The notation is the same as in table~\ref{tab:ZNWLConstraints}. There exist further lattices that are not listed here.}
\label{tab:ZNxZMWLConstraints}
\end{center}}
\end{table}

\begin{table}[!h!]
\small{
\begin{center}
\begin{tabular}{|c|c||c|c|c|}
\hline
\multicolumn{2}{|c||}{\Z{N} orbifolds} & \multicolumn{3}{c|}{$\Z{N}\times\Z{M}$ orbifolds}\\
Point group $P$ & Twist vector $v$           & Point group $P$   &  Twist vector $v_1$       & Twist vector $v_2$ \\
\hline\hline
\rowgrayh
$\Z{3}  $    & $\frac{1}{3}(0,\,1,\,1,\,-2)$ & $\Z{2}\times\Z2$  & $\2(0,\,1,\,0,\,-1)$  & $\2(0,\,0,\,1,\,-1)$\\
$\Z{4}  $    & $\frac{1}{4}(0,\,1,\,1,\,-2)$ & $\Z2\times\Z4$    & $\2(0,\,1,\,0,\,-1)$  & $\h14(0,\,0,\,1,\,-1)$\\
\rowgrayh
$\Z6$-I      & $\frac{1}{6}(0,\,1,\,1,\,-2)$ & $\Z3\times\Z3$    & $\h13(0,\,1,\,0,\,-1)$& $\h13(0,\,0,\,1,\,-1)$\\
$\Z6$-II     & $\frac{1}{6}(0,\,1,\,2,\,-3)$ & $\Z2\times\Z6-$I  & $\2(0,\,1,\,0,\,-1)$  & $\h16(0,\,0,\,1,\,-1)$\\
\rowgrayh
$\Z7$        & $\frac{1}{7}(0,\,1,\,2,\,-3)$ & $\Z2\times\Z6-$II & $\2(0,\,1,\,0,\,-1)$  & $\h16(0,\,1,\,1,\,-2)$\\
$\Z8$-I      & $\frac{1}{8}(0,\,1,\,2,\,-3)$ & $\Z4\times\Z4$    & $\h14(0,\,1,\,0,\,-1)$& $\h14(0,\,0,\,1,\,-1)$\\
\rowgrayh
$\Z8$-II     & $\frac{1}{8}(0,\,1,\,3,\,-4)$ & $\Z3\times\Z6$    & $\h13(0,\,1,\,0,\,-1)$& $\h16(0,\,0,\,1,\,-1)$\\
$\Z{12}$-I   & $\h{1}{12}(0,\,1,\,4,\,-5)$   & $\Z6\times\Z6$    & $\h16(0,\,1,\,0,\,-1)$& $\h16(0,\,0,\,1,\,-1)$\\
\rowgrayh
$\Z{12}$-II  & $\h{1}{12}(0,\,1,\,5,\,-6)$   &  & & \\
\hline
\end{tabular}
\caption{Twist vectors for \Z{N} and \ZZ{N}{M} orbifolds.}
\label{tab:AllTwists}
\end{center}}
\end{table}

\begin{table}[t!]
\centering
\footnotesize{
\begin{tabular}{|c|l|c|l|l|}
\hline
$V^{(\#)}$ & Shift vector $V$ & 4D gauge group $\maG_{4D}$ & $U$ sector & $T_2$ sector \\
\hline\hline
\rowgrayh
 0 & $(\,0^8\,)(\,0^8\,)$
   & $\E8\times\E8$ &   & $27\times 9(\bs{1},\,\bs{1})$   \\  
 1 & $\left(\h23,\,\h13^2,\,0^6\right)\Big(0^8\Big)$
   & $\E6\times\SU3\times\E8$ & $3(\bs{27},\,\bsb{3},\,\bs{1})$ & $27\times 1(\bs{27},\,\bs{1},\,\bs{1})$\\
  &&                       & & $27\times 3(\bs{1},\,\bs{3},\,\bs{1})$   \\  
\rowgrayh
 2 & $\left(\h23,\,\h13^2,\,0^6\right)\left(\h23,\,\h13^2,\,0^6\right)$
   & $\E6\times\SU3\times\E6\times\SU3$ & $3(\bs{27},\,\bsb{3},\,\bs{1},\,\bs{1})$
                                        & $27\times 1(\bs{1},\,\bs{3},\,\bs{1},\,\bs{3})$  \\  
\rowgrayh
   &&                                   & $3(\bs{1},\,\bs{1},\,\bs{27},\,\bsb{3})$  & \\
 3 & $\Big(\h23,\,0^7\Big)\left(\h13^2,\,0^6\right)$
   & $\SO{14}\times\E7$ & $3(\bs{64},\,\bs{1})$ & \\
  &&                    & $3(\bs{14},\,\bs{1})$ & $27\times 1(\bs{14},\,\bs{1})$\\
  &&                    & $3(\bs{1},\,\bs{56})$ & $27\times 4(\bs{1},\,\bs{1})$\\
  &&                    & $3(\bs{1},\,\bs{1})$  & \\
\rowgrayh
 4 & $\left(\h13^4,\,\h23,\,0^3\right)\Big(\h23,\,0^7\Big)$
   & $\SU9\times\SO{14}$ & $3(\bs{84},\,\bs{1})$  &   \\  
\rowgrayh
   &&                    & $3(\bs{1},\,\bs{64})$  
                         & $27\times 1(\bsb{9},\,\bs{1})$\\
\rowgrayh
   &&                    & $3(\bs{1},\,\bs{14})$  & \\
\hline
\end{tabular}}
\caption{Admissible shift vectors and corresponding matter content for the \Z3 orbifold of the
  $\E8\times\E8$ heterotic string.}
\label{tab:Shifts_Z3E8}
\end{table}

\begin{table}[!h!]
\begin{center}
\scriptsize{
\begin{tabular}{|r|cl|c|}
\hline
$V^{(\#)}$ & \phantom{.}& Shift vector $V$ & 4D gauge group $\maG_{4D}$\\
\hline\hline
\rowgrayh
1 && $\4\left( 1^2,\, 2,\, 0^{13}\right)$ &  $\SO{26}\times\SU2\times\U1_A\times\U1$ \\
2 && $\4\left( 1^2,\, 2^3,\, 0^{11}\right)$ &  $\SO{22}\times\SU4\times\SU2\times\U1$ \\
\rowgrayh
3 && $\4\left( 1^6,\phantom{, 0^{1}} 0^{10}\right)$ & $\SO{20}\times\SU6\times\U1_A$ \\
4 && $\4\left( 1^5,\, 3,\, 0^{10}\right)$ &  $\SO{20}\times\SU6\times\U1_A$ \\
\rowgrayh
5 && $\4\left( 1^2,\, 2^5,\, 0^9\right)$ &  $\SO{18}\times\SO{10}\times\SU2\times\U1_A$ \\
6 && $\4\left( 1^6,\, 2^2,\, 0^8\right)$ &  $\SO{16}\times\SU2^2\times\SU6\times\U1_A$ \\
\rowgrayh
7 && $\4\left( 1^2,\, 2^7,\, 0^7\right)$ &  $\SO{14}\times\SO{14}\times\SU2\times\U1$ \\
8 && $\4\left( 1^6,\, 2^4,\, 0^6\right)$ &  $\SO{12}\times\SO8\times\SU6\times\U1_A$ \\
\rowgrayh
9 &&  $\4\left( 1^{10},\, 2\phantom{^2},\, 0^5\right)$ &  $\SO{10}\times\SU{10}\times\U1_A\times\U1$ \\
10&& $\4\left( 1^{10},\, 2^3,\, 0^3\right)$ &  $\SU{4}^2\times\SU{10}\times\U1$ \\
\rowgrayh
11&& $\4\left( 1^{14},\phantom{, 0^{1}} 0^2\right)$ &  $\SU{2}^2\times\SU{14}\times\U1$ \\
12&& $\4\left( 1^{13},\, 3,\, 0^2\right)$  &  $\SU{2}^2\times\SU{14}\times\U1_A$ \\
\rowgrayh
13&& $\8\left(1^{14},\,\text{-}1,\,3\phantom{^3}\right)$ &  $\SU{15}\times\U1_A\times\U1$ \\
14&& $\8\left(1^{10},\,\text{-}1,\,3^5\right)$ &  $\SU{11}\times\SU5\times\U1_A\times\U1$ \\
\rowgrayh
15&& $\8\left(1^6,\,\text{-}1,\,3^9\right)$ &  $\SU{7}\times\SU9\times\U1_A\times\U1$ \\
16&& $\8\left(1^2,\,\text{-}1,\,3^{13}\right)$ &  $\SU{3}\times\SU{13}\times\U1_A\times\U1$ \\
\hline
\end{tabular}}
\caption{All admissible shift vectors for the \Z4 orbifold of the \SO{32} heterotic
  string. Shift vectors leading to the same four-dimensional gauge group correspond to models with
  different matter content. Further details are provided in our website~\cite{WebTables:2006xx}.}
\label{tab:Shifts_Z4SO32}
\end{center}
\end{table}

\begin{table}[!h!]
\begin{center}
\begin{tabular}{|l|r|l|l|}
\hline
Orbifold & Torsion $\varepsilon$ & Shift vector $V_1$ & Shift vector $V_2$\\
\hline\hline
\rowgrayh
$\Z2\times\Z2$ &
$1$ &
$\displaystyle \left(\tfrac{1}{2},0,-\tfrac{1}{2},0,0,0,0,0\right)$ & 
$\displaystyle \left(0,\tfrac{1}{2},-\tfrac{1}{2},0 ,0, 0, 0, 0\right)$\\
\rowgrayh
& $-1$ &
$\displaystyle \left(\tfrac{1}{2}, -1,-\tfrac{1}{2},1,0,0,0,0\right)$ & 
$\displaystyle \left(1,\tfrac{1}{2},-\tfrac{1}{2},0,1,0,0,0\right)$\\
$\Z2\times\Z4$ &
$1$ &
$\displaystyle \left(0,\tfrac{1}{2},-\tfrac{1}{2},0,0,0,0,0\right)$ &
$\displaystyle \left(\tfrac{1}{4},0,-\tfrac{1}{4},0,0,0,0,0\right)$ \\
& $-1$ &
$\displaystyle \left(2,\tfrac{1}{2},-\tfrac{1}{2},0,0,0,0,0\right)$ &
$\displaystyle \left(\tfrac{1}{4},-1,-\tfrac{1}{4},1,0,0,0,0\right)$ \\ 
\rowgrayh
$\Z2\times\Z6$-I &
$1$ &
$\displaystyle \left(0,\tfrac{1}{2},-\tfrac{1}{2},0 ,0, 0, 0, 0\right)$&
$\displaystyle \left(\tfrac{1}{6},0,-\tfrac{1}{6},0,0,0,0,0\right)$ \\
\rowgrayh
& $-1$ &
$\displaystyle \left(3,\tfrac{1}{2},-\tfrac{1}{2},0,1,0,0,0\right)$ &
$\displaystyle \left(\tfrac{1}{6},-1,-\tfrac{1}{6},1,0,0,0,0\right)$ \\
$\Z2\times\Z6$-II &
$1$ &
$\displaystyle \left(\tfrac{1}{2},0,-\tfrac{1}{2},0 ,0, 0, 0, 0\right)$&
$\displaystyle \left(\tfrac{1}{6},\tfrac{1}{6},-\tfrac{1}{3},0,0,0,0,0\right)$ \\
\hline
\end{tabular}
\caption{\ZZ{N}{M} models with discrete torsion and standard embedding are equivalent to
 models without discrete torsion and non-standard embedding. We write the torsion phase factor as
 $\varepsilon=\mathrm{e}^{-2\pi\I\, V_2\cdot\Delta V_1}$. The components of the shifts within the second
 \E8 all vanish. This result also applies to orbifold models in SO(32), where the second half of the
 shift vector also vanishes.}
\label{tab:DiscreteTorsionBrothers}
\end{center}
\end{table}

\begin{table}[!h!]
\centering
\begin{minipage}{.45\linewidth}
\footnotesize{
\begin{tabular}{|rl|}
\hline
\multicolumn{2}{|l|}{Simple roots of \SO{32}}\\
\hline\hline
$\al_1$ = & $\left(1,-1,0,0,0,0,0,0,0,0,0,0,0,0,0,0\right)$\\
\rowgrayh
$\al_2$ = & $\left(0,1,-1,0,0,0,0,0,0,0,0,0,0,0,0,0\right)$\\
$\al_3$ = & $\left(0,0,1,-1,0,0,0,0,0,0,0,0,0,0,0,0\right)$\\
\rowgrayh
$\al_4$ = & $\left(0,0,0,1,-1,0,0,0,0,0,0,0,0,0,0,0\right)$\\
$\al_5$ = & $\left(0,0,0,0,1,-1,0,0,0,0,0,0,0,0,0,0\right)$\\
\rowgrayh
$\al_6$ = & $\left(0,0,0,0,0,1,-1,0,0,0,0,0,0,0,0,0\right)$\\
$\al_7$ = & $\left(0,0,0,0,0,0,1,-1,0,0,0,0,0,0,0,0\right)$\\
\rowgrayh
$\al_8$ = & $\left(0,0,0,0,0,0,0,1,-1,0,0,0,0,0,0,0\right)$\\
$\al_9$ = & $\left(0,0,0,0,0,0,0,0,1,-1,0,0,0,0,0,0\right)$\\
\rowgrayh
$\al_{10}$ = & $\left(0,0,0,0,0,0,0,0,0,1,-1,0,0,0,0,0\right)$\\
$\al_{11}$ = & $\left(0,0,0,0,0,0,0,0,0,0,1,-1,0,0,0,0\right)$\\
\rowgrayh
$\al_{12}$ = & $\left(0,0,0,0,0,0,0,0,0,0,0,1,-1,0,0,0\right)$\\
$\al_{13}$ = & $\left(0,0,0,0,0,0,0,0,0,0,0,0,1,-1,0,0\right)$\\
\rowgrayh
$\al_{14}$ = & $\left(0,0,0,0,0,0,0,0,0,0,0,0,0,1,-1,0\right)$\\
$\al_{15}$ = & $\left(0,0,0,0,0,0,0,0,0,0,0,0,0,0,1,-1\right)$\\
\rowgrayh
$\al_{16}$ = & $\left(0,0,0,0,0,0,0,0,0,0,0,0,0,0,1,\ \;1\right)$\\
\hline
\end{tabular}}
\end{minipage}
\hspace{1mm}
\begin{minipage}{.45\linewidth}
\footnotesize{
\begin{tabular}{|rl|}
\hline
\multicolumn{2}{|l|}{Fundamental weights of \SO{32}}\\
\hline\hline
$\al_1^*$ = & $\left(1, 0, 0, 0, 0, 0, 0, 0, 0, 0, 0, 0, 0, 0, 0, 0\right)$\\
\rowgrayh
$\al_2^*$ = & $\left(1, 1, 0, 0, 0, 0, 0, 0, 0, 0, 0, 0, 0, 0, 0, 0\right)$\\
$\al_3^*$ = & $\left(1, 1, 1, 0, 0, 0, 0, 0, 0, 0, 0, 0, 0, 0, 0, 0\right)$\\ 
\rowgrayh
$\al_4^*$ = & $\left(1, 1, 1, 1, 0, 0, 0, 0, 0, 0, 0, 0, 0, 0, 0, 0\right)$\\
$\al_5^*$ = & $\left(1, 1, 1, 1, 1, 0, 0, 0, 0, 0, 0, 0, 0, 0, 0, 0\right)$\\ 
\rowgrayh
$\al_6^*$ = & $\left(1, 1, 1, 1, 1, 1, 0, 0, 0, 0, 0, 0, 0, 0, 0, 0\right)$\\ 
$\al_7^*$ = & $\left(1, 1, 1, 1, 1, 1, 1, 0, 0, 0, 0, 0, 0, 0, 0, 0\right)$\\ 
\rowgrayh
$\al_8^*$ = & $\left(1, 1, 1, 1, 1, 1, 1, 1, 0, 0, 0, 0, 0, 0, 0, 0\right)$\\ 
$\al_9^*$ = & $\left(1, 1, 1, 1, 1, 1, 1, 1, 1, 0, 0, 0, 0, 0, 0, 0\right)$\\
\rowgrayh
$\al_{10}^*$ = & $\left(1, 1, 1, 1, 1, 1, 1, 1, 1, 1, 0, 0, 0, 0, 0, 0\right)$\\ 
$\al_{11}^*$ = & $\left(1, 1, 1, 1, 1, 1, 1, 1, 1, 1, 1, 0, 0, 0, 0, 0\right)$\\ 
\rowgrayh
$\al_{12}^*$ = & $\left(1, 1, 1, 1, 1, 1, 1, 1, 1, 1, 1, 1, 0, 0, 0, 0\right)$\\ 
$\al_{13}^*$ = & $\left(1, 1, 1, 1, 1, 1, 1, 1, 1, 1, 1, 1, 1, 0, 0, 0\right)$\\ 
\rowgrayh
$\al_{14}^*$ = & $\left(1, 1, 1, 1, 1, 1, 1, 1, 1, 1, 1, 1, 1, 1, 0, 0\right)$\\ 
$\al_{15}^*$ = & $\left(\2,\!\2,\!\2,\!\2,\!\2,\!\2,\!\2,\!\2,\!\2,\!\2,\!\2,\!\2,\!\2,\!\2,\!\2,\!\text{-}\2\right)$\\
\rowgrayh
$\al_{16}^*$ = & $\left(\2,\!\2,\!\2,\!\2,\!\2,\!\2,\!\2,\!\2,\!\2,\!\2,\!\2,\!\2,\!\2,\!\2,\!\2,\!\ \2\right)$\\
\hline
\end{tabular}}
\end{minipage}
\caption{Simple roots $\al_i$ and fundamental weights $\al^*_i$ of \SO{32}}
\label{tab:RootsSO32}
\end{table}

\begin{table}[!h!]
\begin{center}
\begin{minipage}{.45\linewidth}
\begin{tabular}{|rl|}
\hline
\multicolumn{2}{|l|}{Simple roots of \E8}\\
\hline\hline
$\al_1$ = & $\left(0,1,-1,0,0,0,0,0\right)$\\
\rowgrayh
$\al_2$ = & $\left(0,0,1,-1,0,0,0,0\right)$\\
$\al_3$ = & $\left(0,0,0,1,-1,0,0,0\right)$\\
\rowgrayh
$\al_4$ = & $\left(0,0,0,0,1,-1,0,0\right)$\\
$\al_5$ = & $\left(0,0,0,0,0,1,-1,0\right)$\\
\rowgrayh
$\al_6$ = & $\left(0,0,0,0,0,0,1,-1\right)$\\
$\al_7$ = & $\left(\2,-\2,-\2,-\2,-\2,-\2,-\2,\2\right)$\\
\rowgrayh
$\al_8$ = & $\left(0,0,0,0,0,0,1,\ 1\right)$\\
\hline
\end{tabular}
\end{minipage}
\hspace{1mm}
\begin{minipage}{.35\linewidth}
\begin{tabular}{|rl|}
\hline
\multicolumn{2}{|l|}{Fundamental weights of \E8}\\
\hline\hline
$\al_1^*$ = & $\left(1, 1, 0, 0, 0, 0, 0, 0\right)$\\
\rowgrayh
$\al_2^*$ = & $\left(2, 1, 1, 0, 0, 0, 0, 0\right)$\\
$\al_3^*$ = & $\left(3, 1, 1, 1, 0, 0, 0, 0\right)$\\ 
\rowgrayh
$\al_4^*$ = & $\left(4, 1, 1, 1, 1, 0, 0, 0\right)$\\
$\al_5^*$ = & $\left(5, 1, 1, 1, 1, 1, 0, 0\right)$\\ 
\rowgrayh
$\al_6^*$ = & $\left(\h{7}{2}, \2, \2, \2, \2, \2, \2, -\2\right)$\\ 
$\al_7^*$ = & $\left(2, 0, 0, 0, 0, 0, 0, 0\right)$\\ 
\rowgrayh
$\al_8^*$ = & $\left(\h{5}{2}, \2, \2, \2, \2, \2, \2, \2\right)$\\ 
\hline
\end{tabular}
\end{minipage}
\end{center}
\caption{Simple roots $\al_i$ and fundamental weights $\al^*_i$ of \E8}
\label{tab:RootsE8}
\end{table}

%%%%%%%%%%%%%%%%%%%%%%%%%%%%%%%%%%%%%%%%%%%%%%%%%%
%      APPENDIX  E
%%%%%%%%%%%%%%%%%%%%%%%%%%%%%%%%%%%%%%%%%%%%%%%%%%
\chapter{Supersymmetric $\bs{\BmL}$ Configurations}
\label{ch:15Configurations}\index{B-L configurations}
\vskip -1.8cm
\renewcommand{\arraystretch}{1.07}
{\footnotesize
\setlength{\LTcapwidth}{0.9\textwidth}
\begin{longtable}{|@{}r@{}|rrrrrrrr@{}|@{}rrrrrrrr@{}|}
\hline
\endfirsthead
\hline
\endhead
\hline
\multicolumn{17}{|r|}{{\it continued ...}$\phantom{_{6}^6}$}\\
\hline
\endfoot
\caption{Shift vectors, Wilson lines, hypercharge generator and \BmL generator of 15 \BmL configurations
of the \mssm\ candidates based on the shift vectors with local \SO{10} structure.
All properties of the corresponding models can be computed through this data.} 
\label{tab:NiceBmLModels}
\endlastfoot
\hline 
$V$& $-\h{1}{3}$ &  $\h{1}{2}$ &  $\h{1}{2}$ &  $0$ &  $0$ &  $0$ &  $0$ &  $0$ &  $\h{1}{2}$ & $-\h{1}{6}$ & $-\h{1}{2}$ & $-\h{1}{2}$ & $-\h{1}{2}$ & $-\h{1}{2}$ & $-\h{1}{2}$ &  $\h{1}{2}$ \\
$A_3$&  $\h{1}{2}$ & $-\h{1}{6}$ &  $\h{1}{2}$ & $-\h{1}{6}$ & $-\h{1}{6}$ & $-\h{1}{6}$ & $-\h{1}{6}$ & $-\h{1}{6}$ & $\h{16}{3}$ &  $-5$ &$-\h{17}{3}$ &  $-5$ &$-\h{17}{3}$ &$-\h{17}{3}$ &$-\h{13}{3}$ & $\h{14}{3}$ \\
$A_5$& $-\h{1}{2}$ & $-\h{1}{2}$ &  $0$ &  $\h{1}{2}$ &  $\h{1}{2}$ &  $0$ &  $0$ &  $0$ & $\h{13}{4}$ &$-\h{17}{4}$ &$-\h{13}{4}$ &$-\h{13}{4}$ &$-\h{13}{4}$ &$-\h{13}{4}$ & $-\h{9}{4}$ &  $\h{9}{4}$ \\
$\mathsf{t}_Y$&  $0$ &  $0$ &  $0$ &  $\h{1}{2}$ &  $\h{1}{2}$ & $-\h{1}{3}$ & $-\h{1}{3}$ & $-\h{1}{3}$ &  $0$ &  $0$ &  $0$ &  $0$ &  $0$ &  $0$ &  $0$ &  $0$ \\
$\mathsf{t}_{\BmL}$&  $1$ &  $0$ &  $1$ &  $0$ &  $0$ & $-\h{2}{3}$ & $-\h{2}{3}$ & $-\h{2}{3}$ &$-\h{11}{2}$ & $-\h{1}{2}$ &  $\h{5}{2}$ & $-\h{1}{2}$ &  $\h{5}{2}$ &  $\h{5}{2}$ & $\h{11}{2}$ & $-\h{5}{2}$ \\
\hline\hline
$V$& $-\h{1}{3}$ &  $\h{1}{2}$ &  $\h{1}{2}$ &  $0$ &  $0$ &  $0$ &  $0$ &  $0$ &  $\h{1}{2}$ & $-\h{1}{6}$ & $-\h{1}{2}$ & $-\h{1}{2}$ & $-\h{1}{2}$ & $-\h{1}{2}$ & $-\h{1}{2}$ &  $\h{1}{2}$ \\
$A_3$&  $\h{1}{2}$ & $-\h{1}{6}$ &  $\h{1}{2}$ & $-\h{1}{6}$ & $-\h{1}{6}$ & $-\h{1}{6}$ & $-\h{1}{6}$ & $-\h{1}{6}$ & $-\h{2}{3}$ &  $0$ &  $\h{2}{3}$ &  $0$ &  $\h{2}{3}$ &  $\h{2}{3}$ & $\h{32}{3}$ &$-\h{32}{3}$ \\
$A_5$& $-\h{1}{2}$ & $-\h{1}{2}$ &  $0$ &  $\h{1}{2}$ &  $\h{1}{2}$ &  $0$ &  $0$ &  $0$ & $\h{13}{4}$ &$-\h{17}{4}$ &$-\h{13}{4}$ &$-\h{13}{4}$ &$-\h{13}{4}$ &$-\h{13}{4}$ & $-\h{9}{4}$ &  $\h{9}{4}$ \\
\hline
\end{longtable}
}

%%%%%%%%%%%%%%%%%%%%%%%%%%%%%%%%%%%%%%%%%%%%%%%%%%%%%%%%%%%%%%%%%%%%
%%% Appendix F
%%%%%%%%%%%%%%%%%%%%%%%%%%%%%%%%%%%%%%%%%%%%%%%%%%%%%%%%%%%%%%%%%%%%
\chapter{An Orbifold-\textbf{\textsc{mssm}}: Details}
\label{ch:OrbifoldMSSMDetails}

In this appendix we display in full detail the most important properties of the so-called {\it
orbifold-\mssm} introduced in chapter~\ref{ch:OrbifoldPhenomenology}. This model has been subject of
analysis also in our previous works, refs.~\cite{Lebedev:2006kn,Buchmuller:2007zd,Lebedev:2007hv}.

\section{Model Definitions and Spectrum}

The model is defined by its gauge embedding, i.e.\ shift and Wilson lines~\cite{Lebedev:2006kn,Buchmuller:2007zd}
\begin{subequations}
\begin{eqnarray}
V^{\SO{10},1} & = &\left(\tfrac{1}{3},\,-\tfrac{1}{2},\,-\tfrac{1}{2},\,0,\,0,\,0,\,0,\,0\right)\left(\tfrac{1}{2},\,-\tfrac{1}{6},\,-\tfrac{1}{2},\,-\tfrac{1}{2},\,-\tfrac{1}{2},\,-\tfrac{1}{2},\,-\tfrac{1}{2},\,\tfrac{1}{2}\right)\,,\\[1mm]
A_{3} & = &\left(-\tfrac{1}{2},\,-\tfrac{1}{2},\,\tfrac{1}{6},\tfrac{1}{6},\,\tfrac{1}{6},\,\tfrac{1}{6},\tfrac{1}{6},\,\tfrac{1}{6}\right)\left(\tfrac{10}{3},\,0,\,-6,\,-\tfrac{7}{3},\,-\tfrac{4}{3},\,-5,\,-3,\,3\right)\,,\\[1mm]
A_{5} & =
&\left(\tfrac{1}{4},\,-\tfrac{1}{4},\,-\tfrac{1}{4},-\tfrac{1}{4},\,-\tfrac{1}{4},\,\tfrac{1}{4},\tfrac{1}{4},\,\tfrac{1}{4}\right)\left(1,\,-1,\,-\tfrac{5}{2},\,-\tfrac{3}{2},\,-\tfrac{1}{2},\,-\tfrac{5}{2},\,-\tfrac{3}{2},\,\tfrac{3}{2}\right).
\end{eqnarray}
\end{subequations}
\noindent
The unbroken gauge group after compactification is
\begin{equation}
\maG_{4D} ~=~[\SU3\times\SU2]\times[\SO8\times\SU2]\times\U1^8\;.
\end{equation}
The \U1 generators are chosen to be
\begin{subequations}\label{eq:U1generators}
\begin{eqnarray}
\mathsf{t}_1 & = & \mathsf{t}_Y~=~
\left(0,0,0,-\hh12,-\hh12,\hh{1}{3},\hh{1}{3},\hh{1}{3}\right) \, (0,0,0,0,0,0,0,0) \;,\\
\mathsf{t}_2 & = & (1,0,0,0,0,0,0,0) \, (0,0,0,0,0,0,0,0) \;,\\
\mathsf{t}_3 & = & (0,1,0,0,0,0,0,0) \, (0,0,0,0,0,0,0,0) \;,\\
\mathsf{t}_4 & = & (0,0,1,0,0,0,0,0) \, (0,0,0,0,0,0,0,0) \;,\\
\mathsf{t}_5 & = & (0,0,0,1,1,1,1,1) \, (0,0,0,0,0,0,0,0) \;,\\
\mathsf{t}_6 & = & (0,0,0,0,0,0,0,0) \, (1,0,0,0,0,0,0,0) \;,\\
\mathsf{t}_7 & = & (0,0,0,0,0,0,0,0) \, (0,1,0,0,0,0,0,0) \;,\\
\mathsf{t}_8 & = & (0,0,0,0,0,0,0,0) \, (0,0,0,1,1,0,0,0)\;.
\end{eqnarray}
\end{subequations}
The anomalous \U1 is generated by 
\begin{equation}
 \mathsf{t}_A~=~\sum c_i\,\mathsf{t}_i\;,\quad
 \text{where}\quad
 c_i~=~\left(0,\frac{7}{3},-1,-\frac{5}{3},\frac{1}{3},\frac{2}{3},-\frac{2}{3},-\frac{2}{3}\right)
 \;.
\end{equation}
The sum of anomalous charges is
\begin{equation}
 \tr\mathsf{t}_A~=~\frac{170}{3}~>~0\;.
\end{equation}

The model allows us to define a suitable $B-L$ generator,
\begin{equation}
 \mathsf{t}_{B-L}~=~
 \left( -1 , -1 , 0 , 0 , 0 , \frac{2}{3} , \frac{2}{3} , \frac{2}{3} \right)\,,
 \left( -\frac{1}{2} , -\frac{1}{2} , 0 , -\frac{1}{2} , -\frac{1}{2} , 0 , 0 , 0 \right)\;.
\end{equation}
with two important properties (cf.\ table~\ref{Tab:spectrum}):
\begin{itemize}
\item the spectrum includes the families of quarks and leptons plus vectorlike exotics with respect to
  $G_{SM}\times\U1_{\BmL}\,$, and
\item there are \sm\ singlets with \BmL\ charge $\pm2$, labeled as $\chi_i$.
\end{itemize}

\begin{table}[h]
{\small
\begin{center}
\begin{minipage}{0.4\textwidth}
\renewcommand{\arraystretch}{1.4}
\begin{tabular}{|r|l|c|}
\hline
\# & Representation & Label \\
\hline\hline
\rowgrayh
 3 & $\left(\boldsymbol{3},\boldsymbol{2};\boldsymbol{1},\boldsymbol{1}\right)_{(1/6,1/3)}$ & $q_i$\\
 3 & $\left(\boldsymbol{1},\boldsymbol{1};\boldsymbol{1},\boldsymbol{1}\right)_{(1,1)}$  & $\bar e_i$\\
\rowgrayh
 4 & $\left(\overline{\boldsymbol{3}},\boldsymbol{1};\boldsymbol{1},\boldsymbol{1}\right)_{(1/3,-1/3)}$ & $\bar d_i$\\
 4 & $\left(\boldsymbol{1},\boldsymbol{2};\boldsymbol{1},\boldsymbol{1}\right)_{(-1/2,-1)}$  & $\ell_i$\\
\rowgrayh
 4 & $\left(\boldsymbol{1},\boldsymbol{2};\boldsymbol{1},\boldsymbol{1}\right)_{(-1/2,0)}$ & $\phi_i$\\
 3 & $\left(\overline{\boldsymbol{3}},\boldsymbol{1};\boldsymbol{1},\boldsymbol{1}\right)_{(1/3,2/3)}$  & $\bar\delta_i$\\
\rowgrayh
20 & $\left(\boldsymbol{1},\boldsymbol{1};\boldsymbol{1},\boldsymbol{1}\right)_{(1/2,*)}$ & $s^+_i$\\
15 & $\left(\boldsymbol{1},\boldsymbol{1};\boldsymbol{1},\boldsymbol{1}\right)_{(0,1)}$ & $\bar n_i$\\
\rowgrayh
 3 & $\left(\boldsymbol{1},\boldsymbol{1};\boldsymbol{1},\boldsymbol{2}\right)_{(0,1)}$ & $\bar \eta_i$\\
20 & $\left(\boldsymbol{1},\boldsymbol{1};\boldsymbol{1},\boldsymbol{2}\right)_{(0,0)}$ & $h_i$\\
\rowgrayh
 2 & $\left(\boldsymbol{1},\boldsymbol{1};\boldsymbol{1},\boldsymbol{2}\right)_{(1/2,1)}$ & $x^+_i$\\
 2 & $\left(\boldsymbol{1},\boldsymbol{1};\boldsymbol{1},\boldsymbol{1}\right)_{(0,\pm2)}$ & $\chi_i$\\
\rowgrayh
 4 & $\left(\overline{\boldsymbol{3}},\boldsymbol{1};\boldsymbol{1},\boldsymbol{1}\right)_{(-1/6,*)}$ & $\bar v_i$\\
 2 & $\left(\boldsymbol{1},\boldsymbol{1};\boldsymbol{8},\boldsymbol{1}\right)_{(0,-1/2)}$ & $f_i$\\
\rowgrayh
 5 & $\left(\boldsymbol{1},\boldsymbol{1};\boldsymbol{8},\boldsymbol{1}\right)_{(0,0)}$ & $w_i$\\
\hline
\end{tabular}
\end{minipage}
\begin{minipage}{0.4\textwidth}
\renewcommand{\arraystretch}{1.5}
\begin{tabular}{|r|l|c|}
\hline
\# & Representation & Label \\
\hline\hline
\rowgrayh
 3 & $\left(\overline{\boldsymbol{3}},\boldsymbol{1};\boldsymbol{1},\boldsymbol{1}\right)_{(-2/3,-1/3)}$ & $\bar u_i$\\
 4 & $\left(\boldsymbol{1},\boldsymbol{2};\boldsymbol{1},\boldsymbol{1}\right)_{(0,*)}$ & $m_i$ \\
\rowgrayh
 1 & $\left(\boldsymbol{3},\boldsymbol{1};\boldsymbol{1},\boldsymbol{1}\right)_{(-1/3,1/3)}$ & $d_i$ \\
 1 & $\left(\boldsymbol{1},\boldsymbol{2};\boldsymbol{1},\boldsymbol{1}\right)_{(1/2,1)}$  & $\bar \ell_i$ \\
\rowgrayh
 4 & $\left(\boldsymbol{1},\boldsymbol{2};\boldsymbol{1},\boldsymbol{1}\right)_{(1/2,0)}$ & $\bar \phi_i$ \\
 3 & $\left(\boldsymbol{3},\boldsymbol{1};\boldsymbol{1},\boldsymbol{1}\right)_{(-1/3,-2/3)}$ & $\delta_i$ \\
\rowgrayh
 20 & $\left(\boldsymbol{1},\boldsymbol{1};\boldsymbol{1},\boldsymbol{1}\right)_{(-1/2,*)}$ & $s^-_i$ \\
 12 & $\left(\boldsymbol{1},\boldsymbol{1};\boldsymbol{1},\boldsymbol{1}\right)_{(0,-1)}$ & $n_i$ \\
\rowgrayh
 3 & $\left(\boldsymbol{1},\boldsymbol{1};\boldsymbol{1},\boldsymbol{2}\right)_{(0,-1)}$ & $\eta_i$ \\
 2 & $\left(\boldsymbol{1},\boldsymbol{2};\boldsymbol{1},\boldsymbol{2}\right)_{(0,0)}$ & $y_i$ \\
\rowgrayh
 2 & $\left(\boldsymbol{1},\boldsymbol{1};\boldsymbol{1},\boldsymbol{2}\right)_{(-1/2,-1)}$ & $x^-_i$ \\
 18 & $\left(\boldsymbol{1},\boldsymbol{1};\boldsymbol{1},\boldsymbol{1}\right)_{(0,0)}$ & $s^0_i$ \\
\rowgrayh
 4 & $\left(\boldsymbol{3},\boldsymbol{1};\boldsymbol{1},\boldsymbol{1}\right)_{(1/6,*)}$ & $v_i$ \\
 2 & $\left(\boldsymbol{1},\boldsymbol{1};\boldsymbol{8},\boldsymbol{1}\right)_{(0,1/2)}$ & $\bar f_i$ \\
\hline
\end{tabular}
\end{minipage}
\caption{Spectrum summary. The quantum numbers under
$\SU3\times\SU2\times[\SO8\times\SU2]$ are shown in boldface; hypercharge and \BmL\ charge
appear as subscripts. Note that the states $s_i^\pm$, $m_i$ and $v_i$ have
different B-L charges for different $i$, which we do not explicitly list. Further details are given in
section~\ref{sec:DetailsSpectrum}.}
\label{Tab:spectrum}
\end{center}}
\end{table}

\section[\mssm\ Configuration with $R$-Parity]{\textbf{\textsc{mssm}} Configuration with $\boldsymbol{R}$-Parity}

Consider a vacuum configuration where the fields
\begin{equation}
\ba{rl}
 \{\widetilde{s}_i\} ~=~ &\{
 \chi_{1}, \chi_{2}, s^0_{3}, s^0_{5}, s^0_8, s^0_9, s^0_{12}, s^0_{15}, s^0_{16}, s^0_{22}, s^0_{24}, s^0_{35}, 
 s^0_{41}, s^0_{43}, s^0_{46}, h_{2},\\
 &\phantom{....} h_3, h_{5}, h_{9}, h_{13}, h_{14}, h_{20}, h_{21}, h_{22} \}
\ea
\label{eq:stildeCopy}
\end{equation}
develop a nonzero \vev\ while the expectation values of all other fields vanish.
The emerging effective theory has the following properties:
\begin{enumerate}
\item the unbroken gauge symmetries are
\begin{equation}
 G_{SM}\times G_\mathrm{hidden}\;\qquad \text{with }\quad\ G_\mathrm{hidden}=\SO8\;.
\end{equation}
\item since \BmL\ is broken by two units, there is an effective matter parity $\mathbbm{Z}_2^{\cal M}$.
\item the Higgs mass terms are
\begin{equation}
 \bar\phi_i\,(\mathcal{M}_{\bar\phi\phi})_{ij}\,\phi_j
 \quad\text{where}\quad
\mbox{\footnotesize $
 \mathcal{M}_{\bar\phi\phi}~=~
 \left(
 \begin{array}{cccc}
 \widetilde{s}^4 & 0 & 0 & \widetilde{s} \\
 \widetilde{s} & \widetilde{s}^3 & \widetilde{s}^3 & \widetilde{s}^6 \\
 \widetilde{s}^5 & 0 & 0 & \widetilde{s}^3 \\
 \widetilde{s} & 0 & 0 & \widetilde{s}^3
 \end{array}
 \right)\;.
$}
\end{equation}
The up-type Higgs $h_u$ is a linear combination of $\bar\phi_1$, $\bar\phi_3$
and $\bar\phi_4$,
\begin{equation} 
 h_u~\sim~\widetilde{s}^{2}\bar\phi_1+\bar\phi_3+\widetilde{s}^4\,\bar\phi_4\;,
\end{equation}
while the down-type Higgs is composed out of $\phi_2$ and $\phi_3$,
\begin{equation} 
 h_d~\sim~\,\phi_2+\phi_3\;.
\end{equation}
The vacuum configuration is such that the $\mu$--term, being defined as the
smallest eigenvalue of $\mathcal{M}_{\bar\phi\phi}$,
\begin{equation}
 \mu~=~\left.\frac{\partial^2 W}{\partial h_d\,\partial h_u}\right|_{h_u=h_d=0}
\end{equation}
vanishes up to order $\widetilde{s}^6$, at which we work.
\item we check that switching on $\{\widetilde{s}_i\}$--fields allows us to cancel the FI term
without inducing $D$-terms (cf.\ section~\ref{sec:ModelDflat}).
\item all exotics decouple (cf.\ section~\ref{sec:exoticsMasses}).
\item neutrino masses are suppressed via the seesaw mechanism (cf.\ section~\ref{sec:NeutrinoMasses}).
\end{enumerate}

Furthermore, the up--Higgs Yukawa couplings decompose into 
\begin{equation}
 W_\mathrm{Yukawa}~\supset~\sum\limits_{k=1}^4
 (Y_u)_{ij}^{(k)}\,q_i\,\bar u_j\,\bar\phi_k\,,
\end{equation}
where
\begin{eqnarray}
\mbox{\footnotesize $
 Y_u^{(1)}~=~
 \left(
 \begin{array}{ccc}
 0 & 0 & \widetilde{s}^6 \\
 0 & 0 & \widetilde{s}^6 \\
 \widetilde{s}^3 & \widetilde{s}^3 & 1
 \end{array}
 \right),\; 
$}
&
\mbox{\footnotesize $
 Y_u^{(2)}~=~
 \left(
 \begin{array}{ccc}
 0 & 0 & 0 \\
 0 & 0 & 0 \\
 0 & 0 &  \widetilde{s}^6
 \end{array}
 \right),\; 
$}
& \\
\mbox{\footnotesize $
 Y_u^{(3)}~=~
 \left(
 \begin{array}{ccc}
 0 & 0 & 0 \\
 0 & 0 & 0 \\
 0 & 0 & \widetilde{s}^6
 \end{array}
 \right),\; 
$}
&
\mbox{\footnotesize $
 Y_u^{(4)}~=~
 \left(
 \begin{array}{ccc}
 0 & 0 & 0 \\
 0 & 0 & 0 \\
 0 & 0 & \widetilde{s}^6
 \end{array}
 \right)\;
$}
 & .  \nonumber
\end{eqnarray}
Thus, the physical $3\times3$ up--Higgs Yukawa matrix is
\begin{equation}
\mbox{\footnotesize $
 Y_u~\sim~\widetilde{s}^2\,Y_u^{(1)}+Y_u^{(3)}+\widetilde{s}^4\,Y_u^{(4)}~=~
 \left(
 \begin{array}{ccc}
 0 & 0 & \widetilde{s}^8 \\
 0 & 0 & \widetilde{s}^8 \\
 \widetilde{s}^5 & \widetilde{s}^5 & \widetilde{s}^2
 \end{array}
 \right)\,.
$}
\end{equation}
It does not have maximal rank, which means that the up--quark is massless at order six in \sm singlets. 
However, at order seven $Y_d$ takes the form
\begin{equation}
\mbox{\footnotesize $
 Y_u~\sim~\widetilde{s}^2\,Y_u^{(1)}+Y_u^{(3)}+\widetilde{s}^4\,Y_u^{(4)}~=~
 \left(
 \begin{array}{ccc}
 \widetilde{s}^7 & \widetilde{s}^7 & \widetilde{s}^8 \\
 \widetilde{s}^7 & \widetilde{s}^7 & \widetilde{s}^8 \\
 \widetilde{s}^5 & \widetilde{s}^5 & \widetilde{s}^2
 \end{array}
 \right)\,,
$}
\end{equation}
providing then masses for all up--type quarks. The down--Higgs Yukawa couplings decompose into
\begin{equation}
 W_\mathrm{Yukawa}~\supset~\sum\limits_{k=1}^4
 (Y_d)_{ij}^{(k)}\,q_i\,\bar d_j\,\phi_k\,,
\end{equation}
where
\begin{subequations}
\begin{eqnarray}
\mbox{\footnotesize $
 Y_d^{(1)}~=~
 \left(
 \begin{array}{cccc}
 \widetilde{s}^4 & \widetilde{s}^4 & \widetilde{s}^5 & \widetilde{s}^5 \\
 \widetilde{s}^4 & \widetilde{s}^4 & \widetilde{s}^5 & \widetilde{s}^5 \\
 \widetilde{s}^5 & \widetilde{s}^5 & \widetilde{s}^6 & \widetilde{s}^6
 \end{array}
 \right),\; 
 $}&
\mbox{\footnotesize $
 Y_d^{(2)}~=~
 \left(
 \begin{array}{cccc}
 1 & \widetilde{s}^4 & 0 & 0 \\
 \widetilde{s}^4 & 1 & 0 & 0 \\
 \widetilde{s} & \widetilde{s} & 0 & 0
 \end{array}
 \right),\; 
$}
& \\
\mbox{\footnotesize $
 Y_d^{(3)}~=~
 \left(
 \begin{array}{cccc}
 1 & \widetilde{s}^4 & 0 & 0 \\
 \widetilde{s}^4 & 1 & 0 & 0 \\
 \widetilde{s} & \widetilde{s} & 0 & 0
 \end{array}
 \right),\; 
$}
&
 Y_d^{(4)}~=~0 \;& .  \nonumber
\end{eqnarray}
\end{subequations}
The physical $3\times3$ down--Higgs Yukawa matrix emerges by integrating out a pair of vectorlike $d-$
and $\bar d$--quarks,
\begin{equation}
\mbox{\footnotesize $
 Y_d~=~
 \left(
 \begin{array}{ccc}
 1 & \widetilde{s}^3 & 0 \\
 1 & \widetilde{s}^3 & 0 \\
 \widetilde{s} & \widetilde{s}^4 & 0
 \end{array}
 \right)\,,
$}
\end{equation}
As before, one quark is massless at order six in \sm\ singlets. Yet, at order eight $Y_d$ provides masses
for all down--quarks:
\begin{equation}
\mbox{\footnotesize $
 Y_d~=~
 \left(
 \begin{array}{ccc}
 1 & \widetilde{s}^3 & 0 \\
 1 & \widetilde{s}^3 & \widetilde{s}^8 \\
 \widetilde{s} & \widetilde{s}^4 & \widetilde{s}^8
 \end{array}
 \right)\,.
$}
\end{equation}

The charged lepton Yukawa couplings decompose into
\begin{equation}
 W_\mathrm{Yukawa}~\supset~\sum\limits_{k=1}^4
 (Y_e)_{ij}^{(k)}\,\ell_i\,\bar e_j\,\phi_k\,,
\end{equation}
where
\begin{subequations}
\begin{eqnarray}
\mbox{\footnotesize $
 Y_e^{(1)}~=~
 \left(
 \begin{array}{ccc}
 \widetilde{s}^4 & \widetilde{s}^4 & \widetilde{s}^5 \\
 \widetilde{s}^4 & \widetilde{s}^4 & \widetilde{s}^5 \\
 0 & 0 & 0 \\
 0 & 0 & 0
 \end{array}
 \right),\; 
$}
&
\mbox{\footnotesize $
 Y_e^{(2)}~=~
 \left(
 \begin{array}{ccc}
 1 & \widetilde{s}^4 & \widetilde{s}\\
 \widetilde{s}^4 & 1 & \widetilde{s}\\
 0 & 0 & \widetilde{s}^6  \\
 0 & 0 & \widetilde{s}^6
 \end{array}
 \right),\; 
$}
& \\
\mbox{\footnotesize $
 Y_e^{(3)}~=~
 \left(
 \begin{array}{ccc}
 1 & \widetilde{s}^4 & \widetilde{s}\\
 \widetilde{s}^4 & 1 & \widetilde{s}\\
 0 & 0 & \widetilde{s}^6  \\
 0 & 0 & \widetilde{s}^6
 \end{array}
 \right),\; 
$}
&
\mbox{\footnotesize $
 Y_e^{(4)}~=~
 \left(
 \begin{array}{ccc}
 0 &  0 &  \widetilde{s}^5 \\
 0 &  0 &  \widetilde{s}^5 \\
 0 &  0 &  \widetilde{s}^6 \\
 0 &  0 &  \widetilde{s}^6
 \end{array}
 \right)\; 
$}
& . \nonumber
\end{eqnarray}
\end{subequations}
The physical $3\times3$ matrix emerges by integrating out a pair of
vectorlike $\ell-$ and $\bar\ell-$leptons,
\begin{equation}
\mbox{\footnotesize $
 Y_e~=~
 \left(
 \begin{array}{ccc}
 1 & 1 & \widetilde{s} \\
 \widetilde{s} & \widetilde{s} & \widetilde{s}^2 \\
 0 &  0 &  \widetilde{s}^6 \\
 \end{array}
 \right)\,.
$}
\end{equation}

\section[$D$--Flatness]{$\boldsymbol{D}$--Flatness}
\label{sec:ModelDflat}

One can write down gauge invariant monomials which carry net negative anomalous charge.
An example for such a monomial involving all $\widetilde{s}_i$  is
\begin{equation}\label{eq:SeesawConfig}
\mbox{\footnotesize $
\ba{rl}
I(\ti{s}_i)~=~ & \chi_{1}\, \chi_{2}\, \left(s^0_{3}\right)^3 \,\left(\begin{array}{c}s^0_{5}\\ s^0_9\\ s^0_{12}\\
    s^0_{16}\end{array}\right)^3\,
\left(\begin{array}{c}s^0_{8}\\ s^0_{15}\end{array}\right)\, \left(\begin{array}{c}s^0_{22}\\ s^0_{24}\end{array}\right)\,
\left(s^0_{35}\right)^2\, \left(s^0_{41}\right)^3\, \left(s^0_{43}\right)^4 \,\left(s^0_{46}\right)^3 \,\\
 &\phantom{....}\qquad \qquad\qquad\quad\times\ \, h_{2}^4\, h_{3}\, h_{5}^5\, h_{9}^2 \, h_{13}^2\, h_{14}^2\, h_{20}\, h_{21}^3 \,h_{22}^6\,\;.
\ea
$}
\end{equation}
Here, gauge equivalent expressions are arranged vertically, e.g.\ $s^0_{22}$ and $s^0_{24}$ carry the same charges (cf.\
table~\ref{tab:SpectrumSeesawModel}). The monomial carries anomalous charge $\sum_iq^A_i=-52/3$.

\section[$F$--Flatness]{$\boldsymbol{F}$--Flatness}
\label{sec:ModelFflat}
Provided the superpotential at order six
{\footnotesize
\bea
W &=& s^0_{32} h_{5} (s^0_{5} h_{1} + s^0_{12} h_{2} ) 
         +(s^0_{15} h_{15} + s^0_{8} h_{13})(s^0_{42}+ s^0_{43})( h_{23} + h_{25})\nonumber\\
 && +\ (s^0_{22} h_{14} + s^0_{24} h_{16})(s^0_{42}+ s^0_{43})( h_{18} + h_{20})\\
 && +\ h_{22} (s^0_{5}  h_{3} + s^0_{12} h_{4})\Big( s^0_{41} (s^0_{26} + s^0_{28}) 
         + s^0_{32} (s^0_{42} + s^0_{43}) +  s^0_{35}( s^0_{45} +  s^0_{46})\Big)\,,\nonumber
\eea}
the $F$--terms of this model are
{\footnotesize
\bea
F_{s^0_{5}}& = &h_{1} h_{5} s^0_{32}+ h_{22} h_{3} ((s^0_{26}+s^0_{28}) s^0_{41}+s^0_{32} (s^0_{42}+s^0_{43})+s^0_{35} (s^0_{45}+s^0_{46}))\,,\label{eq:FTermsSeesaw}\\
F_{s^0_{8}}& = &h_{13} (h_{23}+h_{25}) (s^0_{42}+s^0_{43})\,,\nonumber\\
F_{s^0_{12}}& = &h_{2} h_{5} s^0_{32}\,,\nonumber\\
F_{s^0_{15}}& = &h_{15} (h_{23}+h_{25}) (s^0_{42}+s^0_{43})\,,\nonumber\\
F_{s^0_{22}}& = &h_{14} (h_{18}+h_{20}) (s^0_{42}+s^0_{43})\,,\nonumber\\
F_{s^0_{24}}& = &h_{16} (h_{18}+h_{20}) (s^0_{42}+s^0_{43})\,,\nonumber\\
F_{s^0_{26}}& = &h_{22} h_{3} s^0_{41} s^0_{5}\,,\nonumber\\
F_{s^0_{28}}& = &h_{22} h_{3} s^0_{41} s^0_{5}\,,\nonumber\\
F_{s^0_{32}}& = &(h_{22} h_{3} (s^0_{42}+s^0_{43}) s^0_{5}+ h_{5} (h_{2} s^0_{12}+h_{1} s^0_{5}))\,,\nonumber\\
F_{s^0_{35}}& = &h_{22} h_{3} (s^0_{45}+s^0_{46}) s^0_{5}\,,\nonumber\\
F_{s^0_{41}}& = &h_{22} h_{3} (s^0_{26}+s^0_{28}) s^0_{5}\,,\nonumber\\
F_{s^0_{42}}& = &(h_{18}+h_{20}) (h_{14} s^0_{22}+h_{16} s^0_{24})+ h_{22} h_{3} s^0_{32} s^0_{5}+(h_{23}+h_{25}) (h_{15} s^0_{15}+h_{13} s^0_{8})\,,\nonumber\\
F_{s^0_{43}}& = &(h_{18}+h_{20}) (h_{14} s^0_{22}+h_{16} s^0_{24})+ h_{22} h_{3} s^0_{32} s^0_{5}+(h_{23}+h_{25}) (h_{15} s^0_{15}+h_{13} s^0_{8})\,,\nonumber\\
F_{s^0_{45}}& = &h_{22} h_{3} s^0_{35} s^0_{5}\,,\nonumber\\
F_{s^0_{46}}& = &h_{22} h_{3} s^0_{35} s^0_{5}\,,\nonumber\\
F_{h_{1}}& = &h_{5} s^0_{32} s^0_{5}\,,\nonumber\\
F_{h_{2}}& = &h_{5} s^0_{12} s^0_{32}\,,\nonumber\\
F_{h_{3}}& = &h_{22} ((s^0_{26}+s^0_{28}) s^0_{41}+  s^0_{32} (s^0_{42}+s^0_{43})+s^0_{35} (s^0_{45}+s^0_{46})) s^0_{5}\,,\nonumber\\
F_{h_{5}}& = &s^0_{32} (h_{2} s^0_{12}+h_{1} s^0_{5})\,,\nonumber\\
F_{h_{13}}& = &(h_{23}+h_{25}) (s^0_{42}+s^0_{43}) s^0_{8}\,,\nonumber\\
F_{h_{14}}& = &(h_{18}+h_{20}) s^0_{22} (s^0_{42}+s^0_{43})\,,\nonumber\\
F_{h_{15}}& = &(h_{23}+h_{25}) s^0_{15} (s^0_{42}+s^0_{43})\,,\nonumber\\
F_{h_{16}}& = &(h_{18}+h_{20}) s^0_{24} (s^0_{42}+s^0_{43})\,,\nonumber
\eea
\bea
F_{h_{18}}& = &(h_{14} s^0_{22}+h_{16} s^0_{24}) (s^0_{42}+s^0_{43})\,,\nonumber\\
F_{h_{20}}& = &(h_{14} s^0_{22}+h_{16} s^0_{24}) (s^0_{42}+s^0_{43})\,,\nonumber\\
F_{h_{22}}& = &h_{3} ((s^0_{26}+s^0_{28}) s^0_{41}+s^0_{32} (s^0_{42}+s^0_{43})+s^0_{35} (s^0_{45}+s^0_{46})) s^0_{5}\,,\nonumber\\
F_{h_{23}}& = &(s^0_{42}+s^0_{43}) (h_{15} s^0_{15}+h_{13} s^0_{8})\,,\nonumber\\
F_{h_{25}}& = &(s^0_{42}+s^0_{43}) (h_{15} s^0_{15}+h_{13} s^0_{8})\,.\nonumber
\eea}

At order six in the \sm\ singlets, the vaccum configuration~\eqref{eq:stildeCopy} leaves some nonzero
$F$--terms which can only be cancelled if some singlets have trivial {\vev}s. However, as discussed in
section~\ref{subsec:FFlat}, at order eight in the superpotential, one does find nontrivial solutions. We
do not list them here due to their length.

\section{Mass Matrices}
\label{sec:exoticsMasses}
Provided the exotics' mass terms  $x_i\,\mathcal{M}_{x\bar x}^{ij}\,\bar x_j$, all exotic particles of
the type $x_i$ get large masses if the mass matrix $\mathcal{M}_{x\bar x}$ has full rank.
In the following, we list the structure of the mass matrices of all exotic particles.

{\scriptsize 
\begin{subequations}
\noindent
\begin{minipage}{0.47\textwidth}
\begin{eqnarray}
\mathcal{M}_{\bar\ell\ell}
& = &
\left(
\begin{array}{cccc}
 \widetilde{s}^2 & \widetilde{s}^2 & \widetilde{s}^3 & \widetilde{s}^3
\end{array}
\right)\;,\label{eq:Mbarll}\\
\mathcal{M}_{d\bar d}
& = &
\left(
\begin{array}{cccc}
 \widetilde{s}^6 & \widetilde{s}^6 & \widetilde{s}^3 & \widetilde{s}^3
\end{array}
\right)\;,\label{eq:Mdbard}
\eea
\end{minipage}
\begin{minipage}{0.5\textwidth}
\begin{eqnarray}
\mathcal{M}_{x^+x^-} & = &
\left(
\begin{array}{cc}
 \widetilde{s}^5 & \widetilde{s}^5 \\
 \widetilde{s}^5 & \widetilde{s}^5
\end{array}
\right)\;,\label{eq:Mxpxm}
\eea
\end{minipage}

\vskip -3mm
\noindent
\begin{minipage}{0.47\textwidth}
\begin{eqnarray}
\mathcal{M}_{mm} 
& = & 
\left(
\begin{array}{cccc}
 0 & 0 & \widetilde{s}^6 & \widetilde{s}^6 \\
 0 & 0 & \widetilde{s}^6 & \widetilde{s}^6 \\
 \widetilde{s}^6 & \widetilde{s}^6 & 0 & \widetilde{s}^6 \\
 \widetilde{s}^6 & \widetilde{s}^6 & \widetilde{s}^6 & 0
\end{array}
\right)\;,\label{eq:Mmm}\\
\mathcal{M}_{v\bar v} & = &
\left(
\begin{array}{cccc}
 \widetilde{s} & \widetilde{s}^5 & 0 & 0\\
 \widetilde{s}^5 & \widetilde{s} & 0 & 0\\
 0 & 0 & \widetilde{s}^5 & \widetilde{s}^5\\
 0 & 0 & \widetilde{s}^5 & \widetilde{s}^5
\end{array}
\right)\;,\label{eq:Mvbarv}\\
\mathcal{M}_{yy} 
& = & 
\left(
\begin{array}{cc}
 \widetilde{s}^1 & \widetilde{s}^5 \\
 \widetilde{s}^5 & \widetilde{s}^1
\end{array}
\right)\;,\label{eq:Myy}
\eea
\end{minipage}
\begin{minipage}{0.5\textwidth}
\begin{eqnarray}
\mathcal{M}_{\delta\bar\delta} 
& = & 
\left(
\begin{array}{ccc}
 \widetilde{s}^3 & \widetilde{s}^3 & \widetilde{s}^3 \\
 \widetilde{s}^3 & \widetilde{s}^3 & \widetilde{s}^3 \\
 0 & \widetilde{s}^3 & \widetilde{s}^3
\end{array}
\right)\;,\label{eq:Mdeltabardelta}\\
\mathcal{M}_{f\bar f} & = &
\left(
\begin{array}{cc}
 0 & \widetilde{s}^3 \\
 0 & \widetilde{s}^3
\end{array}
\right)\;,\label{eq:Mfbarf}\\
\mathcal{M}_{ww} & = &
\left(
\begin{array}{ccccc}
 \widetilde{s} & \widetilde{s}^5 & 0 & \widetilde{s}^5 & \widetilde{s}^5\\
 \widetilde{s}^5 & \widetilde{s} & 0 & \widetilde{s}^5 & \widetilde{s}^5\\
 0 & 0 & 0 & \widetilde{s}^3 & \widetilde{s}^3\\
 \widetilde{s}^5 & \widetilde{s}^5 & \widetilde{s}^3 & \widetilde{s}^6 & \widetilde{s}^6\\
 \widetilde{s}^5 & \widetilde{s}^5 & \widetilde{s}^3 & \widetilde{s}^6 & \widetilde{s}^6
\end{array}
\right),\label{eq:Mww}
\eea
\end{minipage}

\vskip -2mm
\begin{equation}
\label{eq:Mspsm}
\mathcal{M}_{s^+s^-}\!=\!
\mbox{\scriptsize $
\!\!\left(
\begin{array}{cccccccccccccccccccc}
 0 & 0 & 0 & 0 & 0 & 0 & 0 & 0 & 0 & 0 & 0 & 0 & 0 & \widetilde{s}^6 & \widetilde{s}^6 & \widetilde{s}^6 & 0 & \widetilde{s}^6 & \widetilde{s}^6 & \widetilde{s}^6 \\
 \widetilde{s}^6 & 0 & \widetilde{s}^6 & \widetilde{s}^6 & 0 & \widetilde{s}^6 & 0 & 0 & 0 & 0 & 0 & \widetilde{s}^6 & \widetilde{s}^6 & 0 & 0 & 0 & \widetilde{s}^6 & 0 & 0 & 0 \\
 \widetilde{s}^6 & 0 & 0 & \widetilde{s}^6 & 0 & 0 & 0 & 0 & 0 & 0 & \widetilde{s}^6 & 0 & \widetilde{s}^6 & 0 & 0 & 0 & \widetilde{s}^6 & 0 & 0 & 0 \\
 0 & 0 & 0 & 0 & 0 & 0 & 0 & 0 & 0 & 0 & 0 & 0 & 0 & \widetilde{s}^6 & \widetilde{s}^6 & \widetilde{s}^6 & 0 & \widetilde{s}^6 & \widetilde{s}^6 & \widetilde{s}^6 \\
 \widetilde{s}^6 & 0 & \widetilde{s}^6 & \widetilde{s}^6 & 0 & \widetilde{s}^6 & 0 & 0 & \widetilde{s}^6 & 0 & 0 & 0 & \widetilde{s}^6 & 0 & 0 & 0 & \widetilde{s}^6 & 0 & 0 & 0 \\
 \widetilde{s}^6 & 0 & 0 & \widetilde{s}^6 & 0 & 0 & 0 & \widetilde{s}^6 & 0 & 0 & 0 & 0 & \widetilde{s}^6 & 0 & 0 & 0 & \widetilde{s}^6 & 0 & 0 & 0 \\
 0 & \widetilde{s} & 0 & 0 & \widetilde{s}^5 & 0 & \widetilde{s} & 0 & 0 & \widetilde{s}^5 & 0 & 0 & 0 & \widetilde{s}^6 & \widetilde{s}^6 & \widetilde{s}^5 & 0 & \widetilde{s}^6 & \widetilde{s}^6 & \widetilde{s}^5 \\
 \widetilde{s}^5 & 0 & \widetilde{s}^6 & \widetilde{s}^5 & 0 & \widetilde{s}^6 & 0 & 0 & 0 & 0 & 0 & 0 & 0 & 0 & 0 & 0 & 0 & 0 & 0 & 0 \\
 \widetilde{s}^6 & 0 & \widetilde{s}^6 & \widetilde{s}^6 & 0 & \widetilde{s}^6 & 0 & \widetilde{s}^6 & 0 & 0 & 0 & 0 & \widetilde{s}^6 & 0 & 0 & 0 & \widetilde{s}^6 & 0 & 0 & 0 \\
 0 & \widetilde{s}^5 & 0 & 0 & \widetilde{s} & 0 & \widetilde{s}^5 & 0 & 0 & \widetilde{s} & 0 & 0 & 0 & \widetilde{s}^6 & \widetilde{s}^6 & \widetilde{s}^5 & 0 & \widetilde{s}^6 & \widetilde{s}^6 & \widetilde{s}^5 \\
 \widetilde{s}^5 & 0 & \widetilde{s}^6 & \widetilde{s}^5 & 0 & \widetilde{s}^6 & 0 & 0 & 0 & 0 & 0 & 0 & 0 & 0 & 0 & 0 & 0 & 0 & 0 & 0 \\
 \widetilde{s}^6 & 0 & \widetilde{s}^6 & \widetilde{s}^6 & 0 & \widetilde{s}^6 & 0 & 0 & 0 & 0 & \widetilde{s}^6 & 0 & \widetilde{s}^6 & 0 & 0 & 0 & \widetilde{s}^6 & 0 & 0 & 0 \\
 0 & 0 & \widetilde{s}^6 & 0 & 0 & \widetilde{s}^6 & 0 & \widetilde{s}^6 & 0 & 0 & \widetilde{s}^6 & 0 & \widetilde{s}^5 & 0 & 0 & 0 & \widetilde{s}^5 & 0 & 0 & 0 \\
 0 & 0 & 0 & 0 & 0 & 0 & \widetilde{s}^5 & 0 & 0 & \widetilde{s}^5 & 0 & 0 & 0 & \widetilde{s}^5 & \widetilde{s}^5 & \widetilde{s}^5 & 0 & \widetilde{s}^5 & \widetilde{s}^5 & \widetilde{s}^5 \\
 0 & 0 & 0 & 0 & 0 & 0 & \widetilde{s}^5 & 0 & 0 & \widetilde{s}^5 & 0 & 0 & 0 & \widetilde{s}^5 & \widetilde{s}^5 & \widetilde{s}^5 & 0 & \widetilde{s}^5 & \widetilde{s}^5 & \widetilde{s}^5 \\
 0 & 0 & 0 & 0 & 0 & 0 & \widetilde{s}^6 & 0 & 0 & \widetilde{s}^6 & 0 & 0 & 0 & \widetilde{s}^5 & \widetilde{s}^5 & \widetilde{s}^5 & 0 & \widetilde{s}^5 & \widetilde{s}^5 & \widetilde{s}^5 \\
 0 & 0 & \widetilde{s}^6 & 0 & 0 & \widetilde{s}^6 & 0 & \widetilde{s}^6 & 0 & 0 & \widetilde{s}^6 & 0 & \widetilde{s}^5 & 0 & 0 & 0 & \widetilde{s}^5 & 0 & 0 & 0 \\
 0 & 0 & 0 & 0 & 0 & 0 & \widetilde{s}^5 & 0 & 0 & \widetilde{s}^5 & 0 & 0 & 0 & \widetilde{s}^5 & \widetilde{s}^5 & \widetilde{s}^5 & 0 & \widetilde{s}^5 & \widetilde{s}^5 & \widetilde{s}^5 \\
 0 & 0 & 0 & 0 & 0 & 0 & \widetilde{s}^5 & 0 & 0 & \widetilde{s}^5 & 0 & 0 & 0 & \widetilde{s}^5 & \widetilde{s}^5 & \widetilde{s}^5 & 0 & \widetilde{s}^5 & \widetilde{s}^5 & \widetilde{s}^5 \\
 0 & 0 & 0 & 0 & 0 & 0 & \widetilde{s}^6 & 0 & 0 & \widetilde{s}^6 & 0 & 0 & 0 & \widetilde{s}^5 & \widetilde{s}^5 & \widetilde{s}^5 & 0 & \widetilde{s}^5 & \widetilde{s}^5 & \widetilde{s}^5
\end{array}
\right)\;.$}
\end{equation}
\end{subequations}
}

\section{Neutrino Masses}
\label{sec:NeutrinoMasses}

We consider vacua where \SU2 is broken. This means that the $\eta_i$ and
$\bar\eta_i$ give rise to further \sm\ singlets with $q_{\BmL}=\pm1$,
{\footnotesize 
\begin{equation}
 \bar\eta_1
 ~=~
 \left(\begin{array}{c}
 \bar n_{16}\\
 \bar n_{17}\end{array}\right)\;,\dots
 \bar\eta_3
 ~=~
 \left(\begin{array}{c}
 \bar n_{20}\\
 \bar n_{21}\end{array}\right)
 \quad\text{and}\quad
 \eta_1
 ~=~
 \left(\begin{array}{c}
  n_{13}\\
  n_{14}\end{array}\right)\;,\dots
 \eta_3
 ~=~
 \left(\begin{array}{c}
  n_{17}\\
  n_{18}\end{array}\right)
 \;.
\end{equation}
}
The neutrino mass matrix has full rank. It is given by
{\small
\begin{equation}
 \mathcal{M}_{\bar\nu\bar\nu}~=~\left(\begin{array}{cc}
 \mathcal{M}_{\bar n\bar n} & \mathcal{M}_{n\bar n}^T\\
 \mathcal{M}_{n\bar n} & \mathcal{M}_{nn}
 \end{array}\right)\;,
\end{equation}
}
where
\begin{subequations}
\begin{equation}
\mathcal{M}_{nn} \! =\!
\mbox{\scriptsize $
\left(
\begin{array}{cccccccccccccccccc}
 0 & 0 & 0 & \widetilde{s}^2 & 0 & \widetilde{s}^6 & 0 & 0 & 0 & 0 & 0 & 0 & 0 & 0 & 0 & 0 & 0 & 0 \
\\
 0 & 0 & 0 & \widetilde{s}^6 & 0 & \widetilde{s}^2 & 0 & 0 & 0 & 0 & 0 & 0 & 0 & 0 & 0 & 0 & 0 & 0 \
\\
 0 & 0 & 0 & 0 & 0 & 0 & 0 & 0 & 0 & 0 & 0 & 0 & 0 & 0 & 0 & 0 & 0 & 0 \\
 \widetilde{s}^2 & \widetilde{s}^6 & 0 & \widetilde{s}^6 & 0 & \widetilde{s}^6 & 0 & 0 & 0 & 0 & 0 & 0 & 0 & 0 & 0 & 0 & 0 & \
0 \\
 0 & 0 & 0 & 0 & 0 & 0 & 0 & 0 & 0 & 0 & 0 & 0 & 0 & 0 & 0 & 0 & 0 & 0 \\
 \widetilde{s}^6 & \widetilde{s}^2 & 0 & \widetilde{s}^6 & 0 & \widetilde{s}^6 & 0 & 0 & 0 & 0 & 0 & 0 & 0 & 0 & 0 & 0 & 0 & \
0 \\
 0 & 0 & 0 & 0 & 0 & 0 & 0 & 0 & 0 & 0 & 0 & 0 & 0 & 0 & 0 & 0 & 0 & 0 \\
 0 & 0 & 0 & 0 & 0 & 0 & 0 & 0 & 0 & 0 & 0 & 0 & 0 & 0 & 0 & 0 & 0 & 0 \\
 0 & 0 & 0 & 0 & 0 & 0 & 0 & 0 & 0 & 0 & 0 & 0 & 0 & 0 & 0 & 0 & 0 & 0 \\
 0 & 0 & 0 & 0 & 0 & 0 & 0 & 0 & 0 & 0 & 0 & 0 & 0 & 0 & 0 & 0 & 0 & 0 \\
 0 & 0 & 0 & 0 & 0 & 0 & 0 & 0 & 0 & 0 & 0 & 0 & 0 & 0 & 0 & 0 & 0 & 0 \\
 0 & 0 & 0 & 0 & 0 & 0 & 0 & 0 & 0 & 0 & 0 & 0 & 0 & 0 & 0 & 0 & 0 & 0 \\
 0 & 0 & 0 & 0 & 0 & 0 & 0 & 0 & 0 & 0 & 0 & 0 & 0 & 0 & 0 & 0 & 0 & 0 \\
 0 & 0 & 0 & 0 & 0 & 0 & 0 & 0 & 0 & 0 & 0 & 0 & 0 & 0 & 0 & 0 & 0 & 0 \\
 0 & 0 & 0 & 0 & 0 & 0 & 0 & 0 & 0 & 0 & 0 & 0 & 0 & 0 & 0 & 0 & 0 & 0 \\
 0 & 0 & 0 & 0 & 0 & 0 & 0 & 0 & 0 & 0 & 0 & 0 & 0 & 0 & 0 & 0 & 0 & 0 \\
 0 & 0 & 0 & 0 & 0 & 0 & 0 & 0 & 0 & 0 & 0 & 0 & 0 & 0 & 0 & 0 & 0 & 0 \\
 0 & 0 & 0 & 0 & 0 & 0 & 0 & 0 & 0 & 0 & 0 & 0 & 0 & 0 & 0 & 0 & 0 & 0
\end{array}
\right)\;,$}
\ee
\be
\mathcal{M}_{n\bar n}\! =\!
\mbox{\scriptsize $
\left(
\begin{array}{ccccccccccccccccccccc}
 0 & \widetilde{s}^3 & 0 & 0 & \widetilde{s}^6 & 0 & \widetilde{s}^6 & 0 & \widetilde{s}^6 & \widetilde{s}^6 & \widetilde{s}^6 & \widetilde{s}^6 & \widetilde{s}^6 & \widetilde{s}^6 & \widetilde{s}^2 & \
\widetilde{s}^6 & \widetilde{s}^6 & 0 & 0 & 0 & 0 \\
 0 & \widetilde{s}^6 & 0 & 0 & \widetilde{s}^3 & 0 & 0 & \widetilde{s}^6 & \widetilde{s}^6 & \widetilde{s}^6 & \widetilde{s}^6 & \widetilde{s}^6 & \widetilde{s}^6 & \widetilde{s}^6 & \widetilde{s}^6 & \
\widetilde{s}^6 & \widetilde{s}^6 & 0 & 0 & 0 & 0 \\
 \widetilde{s}^6 & 0 & 0 & \widetilde{s}^6 & 0 & 0 & \widetilde{s}^6 & \widetilde{s}^6 & 0 & 0 & 0 & 0 & \widetilde{s}^6 & \widetilde{s}^6 & 0 & 0 \
& 0 & 0 & 0 & 0 & 0 \\
 \widetilde{s} & \widetilde{s}^2 & \widetilde{s}^3 & \widetilde{s}^5 & \widetilde{s}^4 & \widetilde{s}^4 & \widetilde{s}^6 & \widetilde{s}^6 & 0 & 0 & \widetilde{s}^6 & \widetilde{s}^6 & \widetilde{s}^6 & \widetilde{s}^6 \
& \widetilde{s}^6 & 0 & 0 & \widetilde{s}^6 & \widetilde{s}^6 & \widetilde{s}^6 & \widetilde{s}^6 \\
 \widetilde{s}^6 & 0 & 0 & \widetilde{s}^6 & 0 & 0 & \widetilde{s}^6 & \widetilde{s}^6 & 0 & 0 & 0 & 0 & \widetilde{s}^6 & \widetilde{s}^6 & 0 & \
0 & 0 & 0 & 0 & 0 & 0 \\
 \widetilde{s}^5 & \widetilde{s}^4 & \widetilde{s}^4 & \widetilde{s} & \widetilde{s}^2 & \widetilde{s}^3 & \widetilde{s}^6 & \widetilde{s}^6 & 0 & 0 & \widetilde{s}^6 & \widetilde{s}^6 & \widetilde{s}^6 & \widetilde{s}^6 \
& \widetilde{s}^6 & 0 & 0 & \widetilde{s}^6 & \widetilde{s}^6 & \widetilde{s}^6 & \widetilde{s}^6 \\
 0 & 0 & \widetilde{s}^6 & 0 & 0 & 0 & \widetilde{s}^6 & \widetilde{s}^6 & \widetilde{s}^6 & \widetilde{s}^6 & \widetilde{s}^6 & \widetilde{s}^6 & 0 & 0 & \widetilde{s}^3 & \widetilde{s}^6 & \
\widetilde{s}^6 & 0 & 0 & 0 & 0 \\
 0 & 0 & \widetilde{s}^6 & 0 & 0 & 0 & \widetilde{s}^6 & \widetilde{s}^6 & \widetilde{s}^6 & \widetilde{s}^6 & \widetilde{s}^6 & \widetilde{s}^6 & 0 & 0 & \widetilde{s}^3 & \widetilde{s}^6 & \
\widetilde{s}^6 & 0 & 0 & 0 & 0 \\
 \widetilde{s}^6 & 0 & 0 & \widetilde{s}^6 & 0 & 0 & \widetilde{s}^6 & \widetilde{s}^6 & 0 & 0 & 0 & 0 & \widetilde{s}^5 & \widetilde{s}^5 & 0 & 0 & \
0 & 0 & 0 & 0 & 0 \\
 \widetilde{s}^6 & 0 & 0 & \widetilde{s}^6 & 0 & 0 & \widetilde{s}^6 & \widetilde{s}^6 & 0 & 0 & 0 & 0 & \widetilde{s}^5 & \widetilde{s}^5 & 0 & \
0 & 0 & 0 & 0 & 0 & 0 \\
 0 & \widetilde{s}^3 & 0 & 0 & \widetilde{s}^3 & 0 & \widetilde{s}^2 & \widetilde{s}^6 & \widetilde{s}^3 & \widetilde{s}^3 & 0 & 0 & \widetilde{s}^2 & \
\widetilde{s}^6 & 0 & 0 & 0 & 0 & 0 & 0 & 0 \\
 0 & \widetilde{s}^3 & 0 & 0 & \widetilde{s}^3 & 0 & 0 & 0 & 0 & 0 & \widetilde{s}^3 & \widetilde{s}^3 & 0 & 0 & 0 & \
0 & 0 & 0 & 0 & 0 & 0 \\
 0 & 0 & 0 & 0 & 0 & 0 & \widetilde{s}^6 & \widetilde{s}^6 & \widetilde{s}^6 & \widetilde{s}^6 & \widetilde{s}^6 & \widetilde{s}^6 & \widetilde{s}^6 & \widetilde{s}^6 & 0 & \widetilde{s}^6 & \
\widetilde{s}^6 & \widetilde{s}^3 & \widetilde{s}^3 & \widetilde{s}^3 & \widetilde{s}^3 \\
 0 & 0 & 0 & 0 & 0 & 0 & \widetilde{s}^6 & \widetilde{s}^6 & \widetilde{s}^6 & \widetilde{s}^6 & \widetilde{s}^6 & \widetilde{s}^6 & \widetilde{s}^6 & \widetilde{s}^6 & 0 & \widetilde{s}^6 & \
\widetilde{s}^6 & \widetilde{s}^3 & \widetilde{s}^3 & \widetilde{s}^3 & \widetilde{s}^3 \\
 0 & 0 & 0 & 0 & 0 & 0 & 0 & 0 & 0 & 0 & \widetilde{s}^6 & \widetilde{s}^6 & 0 & 0 & 0 & \widetilde{s}^3 & \
\widetilde{s}^3 & 0 & 0 & 0 & 0 \\
 0 & 0 & 0 & 0 & 0 & 0 & 0 & 0 & 0 & 0 & \widetilde{s}^6 & \widetilde{s}^6 & 0 & 0 & 0 & \widetilde{s}^3 & \
\widetilde{s}^3 & 0 & 0 & 0 & 0 \\
 0 & 0 & 0 & 0 & 0 & 0 & 0 & 0 & 0 & 0 & \widetilde{s}^6 & \widetilde{s}^6 & 0 & 0 & 0 & \widetilde{s}^3 & \
\widetilde{s}^3 & 0 & 0 & 0 & 0 \\
 0 & 0 & 0 & 0 & 0 & 0 & 0 & 0 & 0 & 0 & \widetilde{s}^6 & \widetilde{s}^6 & 0 & 0 & 0 & \widetilde{s}^3 & \
\widetilde{s}^3 & 0 & 0 & 0 & 0 
\end{array}
\right) $}
\ee
\be
\mathcal{M}_{\bar n\bar n}\! =\!
\mbox{\scriptsize $
\left(
\begin{array}{ccccccccccccccccccccc}
 0 & 0 & 0 & 0 & 0 & 0 & 0 & 0 & 0 & 0 & 0 & 0 & 0 & 0 & 0 & 0 & 0 & 0 & 0 & \
0 & 0 \\
 0 & 0 & \widetilde{s}^4 & 0 & 0 & 0 & 0 & 0 & 0 & 0 & 0 & 0 & 0 & 0 & 0 & \widetilde{s}^3 & \widetilde{s}^3 & 0 & 0 \
& 0 & 0 \\
 0 & \widetilde{s}^4 & 0 & 0 & 0 & 0 & 0 & 0 & 0 & 0 & 0 & 0 & 0 & 0 & \widetilde{s}^3 & 0 & 0 & 0 & \
0 & 0 & 0 \\
 0 & 0 & 0 & 0 & 0 & 0 & 0 & 0 & 0 & 0 & 0 & 0 & 0 & 0 & 0 & 0 & 0 & 0 & 0 & \
0 & 0 \\
 0 & 0 & 0 & 0 & 0 & \widetilde{s}^4 & 0 & 0 & 0 & 0 & 0 & 0 & 0 & 0 & 0 & \widetilde{s}^3 & \widetilde{s}^3 & 0 \
& 0 & 0 & 0 \\
 0 & 0 & 0 & 0 & \widetilde{s}^4 & 0 & 0 & 0 & 0 & 0 & 0 & 0 & 0 & 0 & 0 & 0 & 0 & 0 & 0 \
& 0 & 0 \\
 0 & 0 & 0 & 0 & 0 & 0 & 0 & 0 & 0 & 0 & 0 & 0 & 0 & 0 & 0 & 0 & 0 & 0 & 0 & \
0 & 0 \\
 0 & 0 & 0 & 0 & 0 & 0 & 0 & 0 & 0 & 0 & 0 & 0 & 0 & 0 & 0 & 0 & 0 & 0 & 0 & \
0 & 0 \\
 0 & 0 & 0 & 0 & 0 & 0 & 0 & 0 & 0 & 0 & 0 & 0 & 0 & 0 & 0 & 0 & 0 & 0 & 0 & \
0 & 0 \\
 0 & 0 & 0 & 0 & 0 & 0 & 0 & 0 & 0 & 0 & 0 & 0 & 0 & 0 & 0 & 0 & 0 & 0 & 0 & \
0 & 0 \\
 0 & 0 & 0 & 0 & 0 & 0 & 0 & 0 & 0 & 0 & 0 & 0 & 0 & 0 & 0 & 0 & 0 & 0 & \
0 & 0 & 0 \\
 0 & 0 & 0 & 0 & 0 & 0 & 0 & 0 & 0 & 0 & 0 & 0 & 0 & 0 & 0 & 0 & 0 & 0 & \
0 & 0 & 0 \\
 0 & 0 & 0 & 0 & 0 & 0 & 0 & 0 & 0 & 0 & 0 & 0 & 0 & 0 & 0 & 0 & 0 & 0 & 0 & \
0 & 0 \\
 0 & 0 & 0 & 0 & 0 & 0 & 0 & 0 & 0 & 0 & 0 & 0 & 0 & 0 & 0 & 0 & 0 & 0 & 0 & \
0 & 0 \\
 0 & 0 & \widetilde{s}^3 & 0 & 0 & 0 & 0 & 0 & 0 & 0 & 0 & 0 & 0 & 0 & 0 & 0 & 0 & 0 & 0 \
& 0 & 0 \\
 0 & \widetilde{s}^3 & 0 & 0 & \widetilde{s}^3 & 0 & 0 & 0 & 0 & 0 & 0 & 0 & 0 & 0 & 0 & 0 & 0 & 0 \
& 0 & 0 & 0 \\
 0 & \widetilde{s}^3 & 0 & 0 & \widetilde{s}^3 & 0 & 0 & 0 & 0 & 0 & 0 & 0 & 0 & 0 & 0 & 0 & 0 & 0 \
& 0 & 0 & 0 \\
 0 & 0 & 0 & 0 & 0 & 0 & 0 & 0 & 0 & 0 & 0 & 0 & 0 & 0 & 0 & 0 & 0 & 0 & 0 & \
0 & 0 \\
 0 & 0 & 0 & 0 & 0 & 0 & 0 & 0 & 0 & 0 & 0 & 0 & 0 & 0 & 0 & 0 & 0 & 0 & 0 & \
0 & 0 \\
 0 & 0 & 0 & 0 & 0 & 0 & 0 & 0 & 0 & 0 & 0 & 0 & 0 & 0 & 0 & 0 & 0 & 0 & 0 & \
0 & 0 \\
 0 & 0 & 0 & 0 & 0 & 0 & 0 & 0 & 0 & 0 & 0 & 0 & 0 & 0 & 0 & 0 & 0 & 0 & 0 & \
0 & 0
\end{array}
\right)\;.$}
\ee
\end{subequations}

The neutrino Yukawa couplings decompose into 
\begin{equation}
 W_\mathrm{Yukawa}~\supset~\sum\limits_{k=1}^4
 (Y_n)_{ij}^{(k)}\,\ell_i\,n_j\,\bar\phi_k
 +(Y_{\bar n})_{ij}^{(k)}\,\ell_i\,\bar n_j\,\bar\phi_k
\end{equation}
where 

{\footnotesize 
\begin{subequations}
\begin{eqnarray}
 Y_n^{(1)} & = &
\left(
\begin{array}{ccccccccccccccccccccc}
 0 & 0 & 0 & 0 & 0 & \widetilde{s}^3 & 0 & 0 & 0 & 0 & 0 & 0 & 0 & 0 & 0 & 0 & 0 & 0  \\
 0 & 0 & 0 & \widetilde{s}^3 & 0 & 0 & 0 & 0 & 0 & 0 & 0 & 0 & 0 & 0 & 0 & 0 & 0 & 0  \\
 0 & 0 & 0 & 0 & 0 & 0 & 0 & 0 & 0 & 0 & 0 & 0 & 0 & 0 & 0 & 0 & 0 & 0  \\
 0 & 0 & 0 & 0 & 0 & 0 & 0 & 0 & 0 & 0 & 0 & 0 & 0 & 0 & 0 & 0 & 0 & 0  \\
\end{array}
\right)\;,\\
 Y_n^{(2)} & = &
\left(
\begin{array}{ccccccccccccccccccccc}
 0 & 0 & 0 & \widetilde{s}^6 & 0 & \widetilde{s}^6 & 0 & 0 & 0 & 0 & 0 & 0 & 0 & 0 & 0 & 0 & 0 & 0  \\
 0 & 0 & 0 & \widetilde{s}^6 & 0 & \widetilde{s}^6 & 0 & 0 & 0 & 0 & 0 & 0 & 0 & 0 & 0 & 0 & 0 & 0  \\
 0 & 0 & 0 & 0 & 0 & 0 & 0 & 0 & 0 & 0 & 0 & 0 & 0 & 0 & 0 & 0 & 0 & 0  \\
 0 & 0 & 0 & 0 & 0 & 0 & 0 & 0 & 0 & 0 & 0 & 0 & 0 & 0 & 0 & 0 & 0 & 0  \\
\end{array}
\right)\;,\\
 Y_n^{(k>2)} & = & 0\;,
\eea
\bea
 Y_{\bar n}^{(1)}\!\! &\!\! =\!\! &\!\!
 \left(
\begin{array}{ccccccccccccccccccccc}
 0 & \widetilde{s}^5 & 0 & 0 & \widetilde{s}^4 & 0 & 0 & 0 & \widetilde{s}^6 & \widetilde{s}^6 & \widetilde{s}^5 & \widetilde{s}^5 & 0 & 0 & 0 & 0 & 0 & 0 & 0 & 0 & 0 \\
 0 & \widetilde{s}^4 & 0 & 0 & \widetilde{s}^5 & 0 & 0 & 0 & \widetilde{s}^6 & \widetilde{s}^6 & \widetilde{s}^5 & \widetilde{s}^5 & 0 & 0 & 0 & 0 & 0 & 0 & 0 & 0 & 0 \\
 0 & \widetilde{s}^5 & 0 & 0 & \widetilde{s}^5 & 0 & 0 & 0 & 0 & 0 & 1 & \widetilde{s}^4 & 0 & 0 & 0 & 0 & 0 & 0 & 0 & 0 & 0 \\
 0 & \widetilde{s}^5 & 0 & 0 & \widetilde{s}^5 & 0 & 0 & 0 & 0 & 0 & \widetilde{s}^4 & 1 & 0 & 0 & 0 & 0 & 0 & 0 & 0 & 0 & 0
\end{array}
\right)\\
 Y_{\bar n}^{(2)}\!\! \!\!\! &\!\! =\!\!\! & \!\!
\left(
\begin{array}{ccccccccccccccccccccc}
 0 & 0 & 0 & 0 & 0 & 0 & 0 & 0 & 0 & 0 & \widetilde{s}^5 & \widetilde{s}^5 & 0 & 0 & 0 & \widetilde{s}^5 & \widetilde{s}^5 & 0 & 0 & 0 & 0 \\
 0 & 0 & 0 & 0 & 0 & 0 & 0 & 0 & 0 & 0 & \widetilde{s}^5 & \widetilde{s}^5 & 0 & 0 & 0 & \widetilde{s}^4 & \widetilde{s}^4 & 0 & 0 & 0 & 0 \\
 0 & \widetilde{s}^2 & \widetilde{s}^6 & 0 & \widetilde{s}^2 & \widetilde{s}^6 & 0 & 0 & 0 & 0 & \widetilde{s}^6 & \widetilde{s}^6 & 0 & 0 & 0 & \widetilde{s}^5 & \widetilde{s}^5 & 0 & 0 & 0 & 0 \\
 0 & \widetilde{s}^2 & \widetilde{s}^6 & 0 & \widetilde{s}^2 & \widetilde{s}^6 & 0 & 0 & 0 & 0 & \widetilde{s}^6 & \widetilde{s}^6 & 0 & 0 & 0 & \widetilde{s}^5 & \widetilde{s}^5 & 0 & 0 & 0 & 0
\end{array}
\right)\\
 Y_{\bar n}^{(3)}\!\! \!\!\! &\!\! =\!\!\! & \!\!
\left(
\begin{array}{ccccccccccccccccccccc}
 0 & \widetilde{s}^6 & 0 & 0 & \widetilde{s}^6 & 0 & 0 & 0 & 0 & 0 & \widetilde{s} & \widetilde{s} & 0 & 0 & 0 & \widetilde{s}^5 & \widetilde{s}^5 & 0 & 0 & 0 & 0 \\
 0 & \widetilde{s}^6 & 0 & 0 & \widetilde{s}^6 & 0 & 0 & 0 & 0 & 0 & \widetilde{s} & \widetilde{s} & 0 & 0 & 0 & \widetilde{s}^5 & \widetilde{s}^5 & 0 & 0 & 0 & 0 \\
 0 & \widetilde{s}^2 & \widetilde{s}^6 & 0 & \widetilde{s}^2 & \widetilde{s}^6 & 0 & 0 & 0 & 0 & \widetilde{s}^6 & \widetilde{s}^6 & 0 & 0 & 0 & 0 & 0 & 0 & 0 & 0 & 0 \\
 0 & \widetilde{s}^2 & \widetilde{s}^6 & 0 & \widetilde{s}^2 & \widetilde{s}^6 & 0 & 0 & 0 & 0 & \widetilde{s}^6 & \widetilde{s}^6 & 0 & 0 & 0 & 0 & 0 & 0 & 0 & 0 & 0
\end{array}
\right)\\
 Y_{\bar n}^{(4)}\!\! \!\!\! &\!\! =\!\!\! & \!\!
\left(
\begin{array}{ccccccccccccccccccccc}
 0 & \widetilde{s}^6 & 0 & 0 & \widetilde{s}^6 & 0 & 0 & 0 & 0 & 0 & \widetilde{s} & \widetilde{s} & 0 & 0 & 0 & \widetilde{s}^5 & \widetilde{s}^5 & 0 & 0 & 0 & 0 \\
 0 & \widetilde{s}^6 & 0 & 0 & \widetilde{s}^6 & 0 & 0 & 0 & 0 & 0 & \widetilde{s} & \widetilde{s} & 0 & 0 & 0 & \widetilde{s}^5 & \widetilde{s}^5 & 0 & 0 & 0 & 0 \\
 0 & \widetilde{s}^2 & \widetilde{s}^6 & 0 & \widetilde{s}^2 & \widetilde{s}^6 & 0 & 0 & 0 & 0 & \widetilde{s}^6 & \widetilde{s}^6 & 0 & 0 & 0 & 0 & 0 & 0 & 0 & 0 & 0 \\
 0 & \widetilde{s}^2 & \widetilde{s}^6 & 0 & \widetilde{s}^2 & \widetilde{s}^6 & 0 & 0 & 0 & 0 & \widetilde{s}^6 & \widetilde{s}^6 & 0 & 0 & 0 & 0 & 0 & 0 & 0 & 0 & 0
\end{array}
\right).
\end{eqnarray}
\end{subequations}
}

\noindent
With these couplings, one can calculate the effective $\ell$ bilinear
\begin{equation}
 \calM_{\ell\ell}^\mathrm{eff}~=~\2 Y_\nu\,\mathcal{M}^{-1}_{\bar\nu\bar\nu}\,Y_\nu^{T}\;,
\end{equation}
where $Y_\nu~=~(Y_n,Y_{\bar n})$.  By integrating out the heavy $\ell$, one arrives at the $3\x 3$
effective neutrino mass operator $\kappa$, which is related to the light neutrino mass matrix via
\be
m_{\nu}=v_u^2\,\kappa
\ee
with $v_u$ being the up--type Higgs \vev. By using the method explained in
section~\ref{subsec:OrbifoldSeesaw}, we find that $m_{\nu}$ is (very roughly) given by
\begin{equation}
M_\mathrm{eff}~\sim~ - \frac{v_u^2}{M_*}  \left(
\begin{matrix}
1 & s & s \\
s & s^2 & s^2 \\
s & s^2 & s^2
\end{matrix}
\right) \;,
\end{equation}
and the effective seesaw scale is given by
\begin{equation}
 M_* ~\sim~ 0.1\, s^5\, M_{\rm str} \;,
\end{equation}
where $M_{\rm str} = 2\cdot 10^{17}\,\mathrm{GeV}$ is the string scale taken as the overall scale of
$\mathcal{M}_{\bar\nu\bar\nu}$.

\section{Detailed Spectrum}
\label{sec:DetailsSpectrum}
In this section, we display the properties of each state in the spectrum. The spectrum listed
here differs only aesthetically with respect to the results presented in
ref.~\cite{Lebedev:2007hv}. The reason being that the twist vector we have used in that work and here
differ by a minus sign.

{\scriptsize
\renewcommand{\arraystretch}{1.1}
\setlength{\LTcapwidth}{0.9\textwidth}
\begin{longtable}{|@{}l@{}|@{}c@{}|@{}rrrrrrrr|@{}c@{}|@{}rrc@{}|@{}c@{}|c|}
\caption{The spectrum of the orbifold-\mssm\ in terms of \emph{left-chiral} states. The U(1) charges refer to the
  basis of generators, eq.~\eqref{eq:U1generators}. $R_i$ denote bosonic $R$-charges. $\gamma(\vartheta)$
  is the so-called $\gamma$-phase defined in sec.~\ref{sec:YukawaCouplings}. $T_{k(n_1,n_2,n_3,n_4,n_5,n_6)}$ denotes the $k$-th twisted sector
  corresponding to the constructing element $g=(\vartheta^k,\,n_\alpha e_\alpha)$.} 
\label{tab:SpectrumSeesawModel}
\\
\hline
\ Sector & {\scriptsize Representation} & 
     $q_{Y}$ & $q_{1}$ & $q_{2}$ & $q_{3}$ & $q_{4}$ & $q_{5}$ & $q_{6}$ & $q_{7}$ & $q_{B-L}$ & $R_1$ & $R_2$ & $R_3$ & $\gamma(\vartheta)$ & \\
\hline\hline
\endfirsthead
\hline
\ Sector & {\scriptsize Representation} & 
     $q_{Y}$ & $q_{1}$ & $q_{2}$ & $q_{3}$ & $q_{4}$ & $q_{5}$ & $q_{6}$ & $q_{7}$ & $q_{B-L}$ & $R_1$ & $R_2$ & $R_3$ & $\gamma(\vartheta)$ &  \\
\hline
\hline
\endhead
\hline
\multicolumn{16}{|r|}{{\it continued ...}$\phantom{_{6}^6}$}\\
\hline
\endfoot
\endlastfoot
\ $U_1$  & $(\boldsymbol{1}, \boldsymbol{1}, \boldsymbol{1}, \boldsymbol{1})$ & $0$ & $-\frac{1}{2 }$ & $-\frac{1}{2 }$ & $\frac{1}{2 }$ & $\frac{5}{2 }$ & $0$ & $0$ & $0$ &  $2$ &  $-1$ & $0$ & $0$ & $0$ & $\chi_{2}$\\
 & $(\boldsymbol{1}, \boldsymbol{1}, \boldsymbol{1}, \boldsymbol{1})$ & $0$ & $1$ & $1$ & $0$ & $0$ & $0$ & $0$ & $0$ &  $-2$ &  $-1$ & $0$ & $0$ & $0$ & $\chi_{1}$\\
 & $(\boldsymbol{3}, \boldsymbol{2}, \boldsymbol{1}, \boldsymbol{1})$ & $\frac{1}{6 }$ & $-\frac{1}{2 }$ & $\frac{1}{2 }$ & $-\frac{1}{2 }$ & $\frac{1}{2 }$ & $0$ & $0$ & $0$ &  $ \frac{1}{3}$ &  $-1$ & $0$ & $0$ & $0$ & $q_{3}$\\
 & $(\boldsymbol{1}, \boldsymbol{1}, \boldsymbol{8}, \boldsymbol{1})$ & $0$ & $0$ & $0$ & $0$ & $0$ & $\frac{1}{2 }$ & $-\frac{1}{2 }$ & $-1$ &  $ \frac{1}{2}$ &  $-1$ & $0$ & $0$ & $0$ & $\bar{f}_{1}$\\
\hline
\ $U_2$  & $(\boldsymbol{1}, \boldsymbol{1}, \boldsymbol{1}, \boldsymbol{1})$ & $1$ & $\frac{1}{2 }$ & $-\frac{1}{2 }$ & $-\frac{1}{2 }$ & $\frac{1}{2 }$ & $0$ & $0$ & $0$ &  $1$ &  $0$ & $-1$ & $0$ & $0$ & $\bar{e}_{3}$\\
 & $(\boldsymbol{\overline{3}}, \boldsymbol{1}, \boldsymbol{1}, \boldsymbol{1})$ & $-\frac{2}{3 }$ & $\frac{1}{2 }$ & $-\frac{1}{2 }$ & $-\frac{1}{2 }$ & $\frac{1}{2 }$ & $0$ & $0$ & $0$ &  $- \frac{1}{3}$ &  $0$ & $-1$ & $0$ & $0$ & $\bar{u}_{3}$\\
 & $(\boldsymbol{1}, \boldsymbol{1}, \boldsymbol{1}, \boldsymbol{1})$ & $0$ & $\frac{1}{2 }$ & $\frac{1}{2 }$ & $\frac{1}{2 }$ & $\frac{5}{2 }$ & $0$ & $0$ & $0$ &  $0$ &  $0$ & $-1$ & $0$ & $0$ & $s^0_{3}$\\
\hline
\ $U_3$  & $(\boldsymbol{1}, \boldsymbol{2}, \boldsymbol{1}, \boldsymbol{1})$ & $-\frac{1}{2 }$ & $0$ & $0$ & $-1$ & $1$ & $0$ & $0$ & $0$ &  $0$ &  $0$ & $0$ & $-1$ & $0$ & $\phi_{1}$\\
\hline
\end{longtable}
}

%% file: Z6II_SU6_geometry.pstex_t
\begin{picture}(0,0)%
\includegraphics{Z6II_SU6_geometry.pstex}%
\end{picture}%
\setlength{\unitlength}{4144sp}%
\begingroup\makeatletter\ifx\SetFigFont\undefined%
\gdef\SetFigFont#1#2#3#4#5{%
  \reset@font\fontsize{#1}{#2pt}%
  \fontfamily{#3}\fontseries{#4}\fontshape{#5}%
  \selectfont}%
\fi\endgroup%
\begin{picture}(6425,2276)(-45,-1979)
\put(-45,177){\makebox(0,0)[lb]{\smash{{\SetFigFont{12}{14.4}{\familydefault}{\mddefault}{\updefault}$\mathbbm{C}^3$}}}}
\put(5548,-144){\makebox(0,0)[lb]{\smash{{\SetFigFont{11}{13.2}{\rmdefault}{\mddefault}{\updefault}$e_6$}}}}
\put(5796,-1332){\makebox(0,0)[lb]{\smash{{\SetFigFont{11}{13.2}{\rmdefault}{\mddefault}{\updefault}$e_1,\,e_3,\,e_5$}}}}
\put(4673,-1332){\makebox(0,0)[lb]{\smash{{\SetFigFont{11}{13.2}{\rmdefault}{\mddefault}{\updefault}$e_2,\,e_4$}}}}
\put(3254,-687){\makebox(0,0)[lb]{\smash{{\SetFigFont{11}{13.2}{\rmdefault}{\mddefault}{\updefault}$e_2$}}}}
\put(2675,-692){\makebox(0,0)[lb]{\smash{{\SetFigFont{11}{13.2}{\rmdefault}{\mddefault}{\updefault}$e_3$}}}}
\put(3402,-1304){\makebox(0,0)[lb]{\smash{{\SetFigFont{11}{13.2}{\rmdefault}{\mddefault}{\updefault}$e_1$}}}}
\put(2527,-1301){\makebox(0,0)[lb]{\smash{{\SetFigFont{11}{13.2}{\rmdefault}{\mddefault}{\updefault}$e_4$}}}}
\put(2753,-1690){\makebox(0,0)[lb]{\smash{{\SetFigFont{11}{13.2}{\rmdefault}{\mddefault}{\updefault}$e_5$}}}}
\put(1527,-1318){\makebox(0,0)[lb]{\smash{{\SetFigFont{11}{13.2}{\rmdefault}{\mddefault}{\updefault}$e_1,\,e_4$}}}}
\put(189,-372){\makebox(0,0)[lb]{\smash{{\SetFigFont{11}{13.2}{\rmdefault}{\mddefault}{\updefault}$e_2,\,e_5$}}}}
\put(427,-1939){\makebox(0,0)[lb]{\smash{{\SetFigFont{11}{13.2}{\rmdefault}{\mddefault}{\updefault}$e_3$}}}}
\end{picture}%